\documentclass[11pt,a4paper]{article}

\pdfoutput=1

\usepackage{jheppub}

\newcommand{\beq}{\begin{equation}}
\newcommand{\eeq}{\end{equation}}
\newcommand{\bea}{\begin{eqnarray}}
\newcommand{\eea}{\end{eqnarray}}
\newcommand{\ba}{\begin{array}}
\newcommand{\ea}{\end{array}}
\newcommand{\bc}{\begin{center}}
\newcommand{\ec}{\end{center}}
\newcommand{\bit}{\begin{itemize}}
\newcommand{\eit}{\end{itemize}}
\newcommand{\eq}[1]{Eq.~(\ref{#1})}
\newcommand{\Ref}[1]{Ref.~\cite{#1}}

\makeatletter
\def\@fpheader{\relax}
\makeatother

\title{Fermionic extensions of the Standard Model in light of the Higgs couplings}

\author{Nicolas Bizot,}
\author{Michele Frigerio}

\affiliation{Laboratoire Charles Coulomb (L2C), UMR 5221 CNRS-Universit\'e de Montpellier, \\
Place Eug\'ene Bataillon, F-34095 Montpellier, France}

\emailAdd{nicolas.bizot@umontpellier.fr}
\emailAdd{michele.frigerio@umontpellier.fr}

\abstract{As the Higgs boson properties settle, the constraints on the Standard Model extensions tighten. 
We consider all possible new fermions that can couple to the Higgs, 
inspecting sets of up to  four chiral multiplets. 
We confront them with direct collider searches, electroweak precision tests,
and current knowledge of the Higgs couplings. 
The focus is on scenarios that may depart from the decoupling
limit of very large masses and vanishing mixing, as they offer the best prospects for detection.
We identify exotic chiral families that may receive a mass from the Higgs only, still in agreement with 
the $h\gamma\gamma$ signal strength. 
A mixing $\theta$ between the Standard Model and non-chiral fermions induces 
order $\theta^2$ deviations in the Higgs couplings.
The mixing can be as large as $\theta\sim  0.5$
in case of custodial protection of the $Z$ couplings or accidental cancellation in the oblique parameters.
We also notice some intriguing effects for much smaller values of $\theta$, especially in the lepton sector.
Our survey includes a number of unconventional pairs of vector-like and Majorana fermions coupled through the Higgs, 
that may induce order one corrections to the Higgs radiative couplings.
We single out the regions of parameters where $h\gamma\gamma$ and $hgg$ are unaffected, while the $h\gamma Z$ signal strength 
is significantly modified, turning a few times larger than in the Standard Model in two cases.
The second run of the LHC will effectively test most of these scenarios.}

\arxivnumber{1508.01645}

\begin{document}
\maketitle
\flushbottom

\section{Introduction}

When the Large Hadron Collider (LHC) began its data taking, possible extensions of the Standard Model (SM) at the TeV scale were
already severely constrained: electroweak (EW) 
precision measurements accurately confirmed the structure of the gauge sector \cite{Flacher:2008zq,Schael:2013ita,Agashe:2014kda}, 
a number of flavour violating observables showed no significant deviation from the SM predictions  \cite{Raidal:2008jk,Isidori:2010kg}, 
all direct searches of non-standard particles at LEP and Tevatron gave null results 
\cite{Schael:2013ita,Toback:2014tea,Nakamura:2010zzi}. 
After the first run of the LHC, the lower bounds on the masses of new particles increased substantially \cite{twiki-ATLAS,twiki-CMS}. 
The crucial discovery of the Higgs boson \cite{Aad:2012tfa,Chatrchyan:2012ufa} and the measurement of some of its properties 
 \cite{Khachatryan:2014jba,Aad:2015zhl,Aad:2015gba}
supported the minimal realisation of electroweak symmetry breaking (EWSB), as predicted by the SM. 
Thus, the room for SM extensions further compressed.

In this phase, it is essential to reassess the possibilities still open for non-standard physics close to the TeV scale. 
In this paper, we focus on new spin-$1/2$ degrees of freedom. In particular, we will assume that the scalar and gauge sector is the SM one,
with one standard Higgs doublet.
In general, additional dynamics in the EWSB sector may well be present, including corrections to the Higgs boson couplings to the electroweak gauge bosons,
as well as additional scalar or vector states: here we do not consider these possibilities, 
assuming that such dynamics takes place at sufficiently higher scale,
or it is sufficiently weakly coupled to the SM. In other words, we will study effective field theories containing the SM degrees of freedom
plus additional fermions only,
being agnostic on the ultraviolet completion at higher energy. 

When wondering what fundamental fermions exist in Nature, one may notice that
the fermion field content of the SM appears whimsical in some respects.   
Each fermion family is a set of five chiral multiplets, $l_L$, $e_R$, $q_L$, $u_R$ and $d_R$, whose gauge quantum numbers are not explained within the SM, with the 
remarkable property to be anomaly-free. The number of families, three, is unexplained too. 
All SM fermions are massless before EWSB and, when the Higgs develops a vacuum expectation value (vev), they acquire a mass proportional to their
Yukawa coupling to the Higgs doublet. The structure of the Yukawa coupling matrices is not determined by the SM symmetries either. 
Of course, some of these issues may find a convincing interpretation in very high energy extensions of the SM, such as grand unification or flavour theories,
but in this paper we will take a phenomenological point of view and centre on the TeV scale only. 

Additional chiral fermions that are massless before EWSB 
are definitely worth to look for, as their mass is bound to the TeV scale;
the classical example of a chiral fourth family, with the same field content as the SM ones, was ruled out long ago \cite{ALEPH:2005ab}, 
but we will show that more exotic possibilities exist. 
On the other hand, chiral fermions that transform in a real representation or form vector-like pairs, with respect to the SM gauge group, can acquire 
a mass before EWSB. While in general such mass can be much larger than the EW scale, in a number of well-motivated extensions of the SM 
there are new, fermionic degrees of freedom close to the TeV scale. Here are some familiar examples:
\bit
\item Non-zero neutrino masses may be generated through the mixing with heavier fermions, typically sterile neutrinos. The latter may have a mass 
close to the EW scale. However very different scenarios are possible, spanning from the eV scale to the grand unification scale.
\item The dark matter energy density may be carried by new, weakly interacting fermions. If thermally produced, their mass should be close to the EW scale.
\item If the quantum stability of the EW scale is guaranteed by supersymmetry broken at the TeV scale, the SM gauge bosons shall be accompanied by gauginos, and the scalar bosons by higgsinos.
\item If the weak scale is stabilised by dimensional transmutation, via a new strongly-coupled sector that condenses at the TeV scale, a number of composite spin-$1/2$ resonances
may be present in the low energy spectrum. For example, in the scenario of partial compositeness \cite{Kaplan:1991dc,Contino:2004vy},
the SM chiral fermions, or at least the heaviest ones, are accompanied by
vector-like composite partners with the same gauge quantum numbers.
\eit

Given the diversity of phenomenological and theoretical motivations, and the wide-ranging discovery reach of the LHC, in section \ref{criteria} we undergo a classification of the 
fermionic extensions of the SM, as general as possible. Theoretical consistency requires the absence of gauge anomalies. In addition, all new fermions should acquire a large mass to comply 
with null direct searches. As we want to explore the new constraints that materialised after the Higgs discovery, we limit ourselves to those fermions that interact with the Higgs doublet via Yukawa couplings.
We provide the full list of SM extensions with these properties, formed by up to four new chiral multiplets, 
and comment on larger sets of new fermions.
We note in passing that there may be alternative phenomenological motivations to study fermions that do not interact with the Higgs, for example to
avoid the flavour problem altogether, or to look for generic dark matter candidates;
some complementary classifications along this line can be found e.g. in Refs.~\cite{Cirelli:2005uq,DelNobile:2009st,DiLuzio:2015oha}.

The rest of the paper is dedicated to the phenomenology of the fermionic SM extensions, in particular to identify the regions of parameters that survive to 
three broad classes of constraints: (i) EW precision tests; (ii) collider direct searches; (iii) Higgs boson couplings.
Our purpose is to provide a comprehensive, comparative description of all possible sets of fermions, that are presently allowed and may have
an observable effect at the second run of the LHC. Such analysis has several limitations
that one should keep in mind:
\bit
\item We compute only the leading order corrections to the Higgs and gauge boson couplings due to the extra fermions, and we roughly 
extract the collider bounds on their masses and couplings
from the available experimental papers. A precision analysis would require a dedicated study for each given set of new fermions
(and it is already available for several specific cases).
\item 
We assume that the new fermions do not mix with the first and second SM families, in order to avoid
the strong constraints coming from flavour observables (tree-level flavour changing neutral currents are absent). 
Indeed, the mixing with the third family is sufficient to characterise the corrections to the EW and Higgs observables, that are our main subject of interest.
In addition, the Higgs couplings to the light families are presently unconstrained. 
Note that the mixing with the third family can still induce flavour violating processes at one loop, especially in $B$-meson decays or oscillations, when the top or bottom quarks mix with new fermions.
However, the corrections are suppressed with respect to the SM by the masses of the new fermions and their mixing with the SM ones. 
The constraints are typically mild, but in some cases may be complementary to those discussed in this paper, see e.g. Refs.~\cite{Picek:2008dd,Cacciapaglia:2011fx,Han:2014qia}.
\item The new fermions are not supposed to form an ultraviolet complete theory, consistent up to a scale much larger than the TeV. 
Therefore, we do not impose constraints coming from the coupling evolution at high energies, such as vacuum stability, absence of Landau poles,
or gauge unification. We have no pretension to determine the full theory. 
\item We do not restrict the possible SM extensions using cosmological considerations, that rely in most cases on specific assumptions on the early Universe evolution.
For example, we do not impose bounds on the relic abundance of the new fermions, based on the assumption of an initial thermal abundance. 
\eit
As a matter of fact, these points can be addressed only in a model-dependent manner. In specific, well-motivated scenarios, 
it would be worth to perform
more precise computations and include  additional constraints from the other sectors of the theory.

In section \ref{chiral} we discuss purely  chiral sets of fermions, that is, fermions that are massless before EWSB.
Fermions with an EW-invariant mass, that is, either a Majorana or a vector-like mass term, are discussed in
section \ref{phenoL}, if they are colourless, and in section \ref{phenoQ}, if they are coloured.
With a little abuse of terminology, we will call `leptons' all the colourless fermions, even when they do not mix with the SM leptons,
and `quarks' all the coloured ones, even when they do not transform in the fundamental representation of the colour group $SU(3)_c$. 
Finally, in section \ref{conclu} we recapitulate the most interesting results of our analysis.

For each sets of new fermions, we were confronted with the need to compute EW precision observables and Higgs couplings.
Thus, we took the opportunity to collect all the relevant formulas in the appendices, that generalise well-known results to the case of a generic fermionic extension of the SM.
In appendix \ref{appendixA}, we present the fermion-gauge boson couplings, the corrections to the $S$ and $T$ parameters, as well as to the $Zf\bar{f}$ vertex.
In appendix \ref{appendix B}, we discuss the fermion-Higgs boson couplings, both the tree-level and the loop-induced ones, and 
we briefly summarise the present experimental constraints on the
Higgs couplings.

\numberwithin{equation}{section}

\section{Minimal fermionic extensions of the SM}\label{criteria}

Let us consider the extension of the SM by additional fermions, classified according to their transformation under the SM gauge 
group $SU(3)_c\times SU(2)_w\times U(1)_Y$, whose irreducible representations can be denoted by $(R_c,R_w,Y)$. 
The SM extension is defined by the most general renormalizable Lagrangian involving the SM fields and a given set of extra chiral fermion multiplets.

We wish each set of new fermions  (i) to be phenomenologically viable, (ii) to be theoretically self-consistent, 
{\it and} (iii) to modify the Higgs couplings. This leads to the following series of requirements:
\bit

\item[(i)] {\it No massless fermions after EWSB, except for the three SM neutrinos and gauge singlets.} Indeed, massless fermions are
phenomenologically forbidden, unless they have no colour ($R_c=1$), no electric charge ($Q=T_{3}+Y=0$), and no 
coupling to the $Z$-boson ($T_{3} - \tan^2\theta_w Y =0$).
The latter two conditions imply $Y=T_{3}=0$. These conditions would allow for a massless
neutral component in the chiral multiplet $(1,R_w,0)$ with $R_w$ odd,
however the gauge symmetries permit a Majorana mass term for such a multiplet.

\item[(ii)] {\it No SM gauge anomalies.} The fermionic extensions of the SM under consideration are intended as effective theories valid up to the multi-TeV scale,
therefore they should cancel all SM anomalies self-consistently. (Extra fermions much heavier than the EW scale play no role in the anomaly cancellation, 
since they form vector-like pairs with respect to the SM gauge group.)
Since the SM field content is anomaly-free by itself, the anomaly-cancellation conditions must be imposed on the set of new fermions only.

As we require the absence of massless coloured states, the new fermions form a (reducible) real representation of $SU(3)_c$, therefore
the $SU(3)_c$-cubic anomaly is automatically vanishing. Denoting the new fermion representations by $(R_{ci},R_{wi},Y_i)$, for $i=1,\dots,n$,
the remaining anomaly-cancellation conditions read
\beq
\ba{ccl}
SU(3)_c-SU(3)_c-U(1)_Y &: & \sum_{i=1}^n N_{wi} C(R_{ci}) Y_i = 0 ~,\\
SU(2)_w-SU(2)_w-U(1)_Y &: & \sum_{i=1}^n N_{ci} C(R_{wi}) Y_i = 0 ~,\\
U(1)_Y-U(1)_Y-U(1)_Y &: & \sum_{i=1}^n N_{ci} N_{wi}  Y_i^3 = 0 ~,\\
grav-grav-U(1)_Y &: & \sum_{i=1}^n N_{ci} N_{wi} Y_i = 0 ~,\\
\ea
\label{anom}
\eeq
where
$N_{w}\equiv \dim(R_w)$, $N_{c}\equiv \dim(R_c)$ -- this notation is redundant for $SU(2)$ but not for $SU(3)$ --  
and the index $C(R)$ of a given representation is defined by ${\rm Tr}(T_R^aT_R^b)=C(R)\delta_{ab}$, with the index of the fundamental conventionally normalised to 
$C(N)=1/2$ for $SU(N)$.
In the case of $SU(2)$ one has $C(R_w)= N_w(N_w^2-1)/12$. In the case of $SU(3)$
each representation $R_c$ is characterised by two integer Dynkin labels $(a_1,a_2)$ with $a_i\ge 0$, and one has $N_c=(1+a_1)(1+a_2)(1+a_1/2+a_2/2)$ and
$C(R_c) = N_c(a_1^2+3a_1+a_1a_2+3a_2+a_2^2)/24$.

Additionally, the $SU(2)_w$ gauge group has a global anomaly, that cancels only when the sum
$\sum_{i=1}^n N_{ci} C(R_{wi})$ is an integer number ~\cite{Witten:1982fp}.
Note that $C(R_w)$ is half-integer for $N_w = 2+4n$, $n=0,1,2,\dots$, and integer in all other cases.
As for the previous anomalies, this condition must be satisfied by the fermions below the multi-TeV scale. (Heavier fermions, decoupled from the EW scale, 
necessarily give an integer contribution to the sum: only an even number of multiplets with $N_w$ even can acquire a vector-like mass.)

\item[(iii)] {\it Non-zero corrections to the Higgs boson couplings.}  
This corresponds to consider only new fermions with a Yukawa coupling to the SM Higgs doublet. 
More precisely, any subset of new fermions that satisfies the
requirements (i) and (ii) by itself -- any subset with no massless states nor anomalies -- 
should have a non-zero Yukawa coupling to the Higgs.
Otherwise, such subset would interact with the SM only through gauge interactions; strictly speaking, 
it can still affect the Higgs boson couplings at the two-loop level, but here we neglect such small effects.

\eit
We stress that the three requirements above are independent, in the sense that none is automatically implied by the others. 
In particular: vector-like fermions are automatically massive and anomaly-free, but they may not couple to the Higgs doublet; 
chiral fermions that have non-zero masses, such as an extra family of quarks, can be anomalous;
an anomaly-free set of fermions, such as zero-hypercharge fermions, may contain some massless components.

In the following we classify the sets of $n$ chiral fermions that satisfy the requirements (i), (ii) and (iii), 
for $n=1,2,3,4,$ and we briefly comment on larger sets.
For convenience, we will mark with the symbol $\square$ each viable set that is identified.

\subsection{One multiplet}\label{one}

If we add to the SM only one new chiral fermion $\psi\sim(R_c,R_w,Y)$, the only possibility to avoid massless states with non-zero SM gauge charges is the presence of a Majorana mass term
$m_\psi\psi\psi$,
that requires $R_c = \overline{R_c}$, $Y=0$ and $N_w$ odd. Such multiplet is anomaly-free.
The additional requirement to couple to the Higgs doublet, $H\sim(1,2,1/2)$, restricts the possibilities
to $R_c=1$ and $R_w=1$ or $3$, that is, 
\beq
\square\qquad N \sim (1,1,0) \qquad{\rm or}\qquad \Sigma \sim (1,3,0)~.
\eeq
In both cases  a Yukawa coupling  is allowed among the new fermion, $H$ and the SM lepton doublet $l \sim (1,2,-1/2)$.

Since $N$ ($\Sigma$) forms by itself a self-consistent extension of the SM  that modifies the Higgs couplings, $n$ replicas of $N$ (and/or $\Sigma$)
also define a set of new fermions satisfying all our criteria. 
We will analyse their phenomenology in section \ref{seesaw}.
Of course, there may also be consistent sets of $n$ new fermions that are partly formed by replicas of $N$ or $\Sigma$, and partly by different multiplets,
as we will see in the following sections.

\subsection{Two multiplets}\label{two}

Let us classify the possible pairs of chiral fermions $\psi_{1}$ and $\psi_{2}$ that can be added consistently to the SM and that modify
the Higgs couplings. The fermion $\psi_1$ can satisfy all requirements without $\psi_2$ only if it transforms as $(1,1,0)$ or $(1,3,0)$, as shown in section \ref{one}.
In this case the three obvious possibilities are 
\beq
\square\qquad N_1+N_2~,\qquad \Sigma_1+\Sigma_2~,\qquad N+\Sigma~.
\eeq

For all other representations, a coupling between $\psi_1$ and $\psi_2$ is necessary:
either there is  a vector-like mass term $m_{12}\psi_1\psi_2$, or a Yukawa coupling $\psi_1\psi_2 H(\tilde H)$.
The latter possibility leads to an inconsistent mass spectrum: one has $N_{w1}=N_{w2}+1$ (or vice versa) 
and the unbalanced component of $\psi_1$  has $T_{3}\ne0$, therefore it cannot be massless. 
Then, either $\psi_1$ admits a Majorana mass term or it couples to a SM fermion too. One can check that, in both cases, either another unwanted massless state is left,
or a SM neutrino acquires a large mass too. 
An alternative way to exclude the case of the Yukawa coupling $\psi_1\psi_2 H(\tilde H)$ is
to solve the anomaly system (\ref{anom}) for $n=2$: one obtains $N_{c1}=N_{c2}$, $C(R_{c1})=C(R_{c2})$, $R_{w1}=R_{w2}$ and $Y_1=-Y_2$.
Thus, we conclude that the two chiral fermions should form a vector-like  pair, 
\beq
\square\qquad \psi_1\sim (R_c,R_w,Y) ~,\qquad \psi_2\sim (\overline{R_c},R_w,-Y) ~.
\eeq 
In order to modify the Higgs couplings, at least one among $\psi_1$ and $\psi_2$ should have a Yukawa coupling with a SM fermion and the Higgs doublet.
We take conventionally all chiral fermions to be left-handed. A SM family is formed by $l \sim (1,2,-1/2)$, $e^c\sim (1,1,1)$, $q\sim(3,2,1/6)$, $u^c\sim(\bar 3, 1, -2/3)$, and $d^c\sim(\bar 3, 1, 1/3)$.
In order to have a Yukawa coupling with these representations, the new fermions should transform under $SU(3)_c$ either as singlets or triplets. The former mix with SM leptons and can be called vector-like leptons (VLLs), 
the latter mix with SM quarks, hence the name vector-like quarks (VLQs). Under $SU(2)_w$ they can transform as singlets, doublets or triplets. 
All possible vector-like fermions with a Yukawa coupling to the SM fermions are listed in Table \ref{twotable}.

To analyse the phenomenology of vector-like fermion multiplets, it is useful to name the components with different electric charge $Q$.
The possible components of the VLLs have charges $Q(N)=0$, $Q(E)=-1$ and $Q(F)=-2$. 
Then, the self-conjugate leptons $N$ and $\Sigma$ and the four VLLs  can be written as
\begin{equation}
N,\quad 
\Sigma= \begin{pmatrix} E^c \\ N \\ E \end{pmatrix} ;\quad 
E ,\quad 
L= \begin{pmatrix} N \\ E \end{pmatrix},\quad 
\Lambda = \begin{pmatrix} E \\ F \end{pmatrix},\quad 
\Delta=\begin{pmatrix} N \\ E \\ F \end{pmatrix} ~.
\label{newleptons}\end{equation}
After EWSB, $N$, $E$ and $E^c$ can mix with the SM leptons $\nu$, $e$ and $e^c$, respectively, while $F$ does not mix with the SM.
We will discuss the phenomenology of these VLLs in section \ref{VLL}.

The possible components of the VLQs have charges $Q(X)=5/3$, $Q(T)=2/3$, $Q(B)=-1/3$ and $Q(Y)=-4/3$.
They are embedded in seven possible VLQ multiplets, 
\begin{equation}
T ,\quad 
B ,\quad 
X_T=\begin{pmatrix} X \\ T \end{pmatrix} ,\quad 
Q=\begin{pmatrix} T\\ B \end{pmatrix},\quad 
Y_B=\begin{pmatrix} B\\ Y \end{pmatrix} ,\quad 
X_Q= \begin{pmatrix} X \\ T\\ B \end{pmatrix} ,\quad 
Y_Q= \begin{pmatrix} T\\ B\\ Y \end{pmatrix} ~.
\label{newleptons1}
\end{equation}
After EWSB, $T$, $T^c$, $B$ and $B^c$ can mix with the SM quarks $t$, $t^c$, $b$ and $b^c$, respectively. On the contrary, the components $X$, $Y$ and their conjugate do not mix with the SM. 
We will discuss the phenomenology of these VLQs in section \ref{VLQ}.

\begin{table}[tb]
\renewcommand{\arraystretch}{1.3}
\bc\begin{tabular}{|c|c|c|}
\hline
$\psi_{L1}$&$\psi_{L2}$ & coupling to\\
\hline\hline
$E \sim (1,1,-1)$ & $E^c\sim(1,1,+1)$ & $l,e^c$\\
\hline
$L\sim(1,2,-\frac12)$ & $L^c \sim (1,2,+\frac12)$ & $l,e^c$\\
\hline
$\Lambda\sim (1,2,-\frac32)$ & $\Lambda^c\sim (1,2,+\frac32)$& $e^c$\\
\hline
$\Delta\sim(1,3,-1)$ & $\Delta^c\sim (1,3,+1)$& $l$\\
\hline\hline
$T \sim(3,1,+\frac23)$ & $T^c \sim (\overline 3,1,-\frac23)$ & $q,u^c$\\
\hline
$B \sim (3,1,-\frac13)$ & $B^c\sim (\overline 3,1,+\frac13)$ & $q,d^c$\\
\hline
$X_T \sim(3,2,+\frac76)$ & $X_T^c\sim(\overline 3,2,-\frac76)$ & $u^c$\\
\hline
$Q \sim (3,2,+\frac16)$ & $Q^c\sim (\overline 3,2,-\frac16)$ & $q,d^c,u^c$\\
\hline
$Y_B \sim(3,2,-\frac56)$ & $Y_B^c\sim(\overline 3,2,+\frac56)$ & $d^c$\\
\hline
$X_Q \sim(3,3,+\frac23)$ & $X_Q^c\sim(\overline 3,3,-\frac23)$ & $q$\\
\hline
$Y_Q \sim(3,3,-\frac13)$ & $Y_Q^c\sim(\overline 3,3,+\frac13)$ & $q$\\
\hline
\end{tabular}\ec
\caption{Vector-like pairs of left-handed chiral fermions, that provide a consistent extension of the SM and modify the Higgs boson couplings.} 
\label{twotable}
\end{table}

\subsection{Three multiplets}\label{three}

Let us classify the possible sets of three chiral fermions $\psi_{1,2,3}$ that can be added to the SM consistently with the requirements of section \ref{criteria}.
Of course, there is the trivial possibility to combine smaller sets that are already consistent on their own:
\bit
\item[$\square$] Three copies of $N\sim(1,1,0)$ and/or of  $\Sigma\sim(1,3,0)$. 
\item[$\square$] A vector-like fermion from Table \ref{twotable} plus one copy of $N$ or $\Sigma$. As the latter couples to the SM lepton doublet $l$, 
there may be a non-trivial interplay with the VLLs  
$E$, $L$ and $\Delta$, that also couple to $l$. 
We will discuss this case in section \ref{MVL}.
\eit
To explore all the other possibilities, note that
there are two patterns for the colour representations of $\psi_{1,2,3}$, that guarantee the absence of massless coloured states:
\bit
\item $R_{c1}=\overline{R_{c1}}$ plus a vector-like pair $R_{c2}=\overline{R_{c3}}\ne R_{c1}$. 
The only choice allowing for couplings to the Higgs is $R_{c1}=1$ and $R_{c2}=3$. 
In this case the two subsets $\psi_1$ and  $\psi_{2,3}$ do not interact with each other, therefore each should be a consistent extension of the SM by itself,
reducing to the trivial possibilities already listed above.
\item $R_{c1}=R_{c2}=\overline{R_{c3}}$. These can be either real or complex representations. One may have considered three real representations of $SU(3)_c$
not all equal to each other; in this case, however, it is not possible to allow Yukawa couplings to the Higgs and to give a mass to all coloured components, at the same time.
\eit
In the only non-trivial case, $R_{c1}=R_{c2}=\overline{R_{c3}}$ or permutations, the anomaly conditions in \eq{anom} reduce to
\beq\left\{
\ba{ccl}
N^3_{w1}Y_1+N^3_{w2}Y_2+N^3_{w3} Y_3 = 0 ~,\\
N_{w1}Y_1+N_{w2}Y_2+N_{w3} Y_3 = 0 ~,\\
N_{w1}  Y_1^3 +N_{w2}  Y_2^3 + N_{w3}  Y_3^3= 0 ~.\\
\ea\right.
\label{anom3}
\eeq
The possible solutions of this system (with $N_{wi}$ positive integers) are
\bit
\item[(a)] $Y_i=0$ and $N_{wi}$ arbitrary, for $i=1,2,3$.
\item[(b)] $Y_1=0$ with $N_{w1}$ arbitrary, $Y_2=-Y_3\ne 0$ with $N_{w2}=N_{w3}$, or permutations.
\item[(c)] For any pair of integer numbers $n,m\ge 0$,
\beq\left\{\ba{lll}
N_{w1}=1+n~, & N_{w2}=2+n+m~, & N_{w3} = 3+2n+m~, \\ 
Y_1\ne 0 {\rm~arbitrary}, & 
Y_2=-Y_1 \dfrac{N_{w1}(N_{w3}^2-N_{w1}^2)}{N_{w2}(N_{w3}^2-N_{w2}^2)}~, &
Y_3=Y_1 \dfrac{N_{w1}}{N_{w3}}\dfrac{N_{w2}^2-N_{w1}^2}{N_{w3}^2-N_{w2}^2}~, 
\ea\right.\eeq
or permutations.
\eit
Let us consider these solutions in turn, to analyse whether they can satisfy the other requirements of section \ref{criteria}.
\bit
\item[(a)] In order to couple to the Higgs boson, one needs $R_{ci}=1$ and $R_{wi} = 1$ or $3$ for $i=1,2,3$, that is three copies of $N$ or $\Sigma$: 
a trivial possibility already considered.

\item[(b)] Suppose first that $\psi_2$ and $\psi_3$ form a vector-like pair, that is, $R_{c1}=R_{c2}=\overline{R_{c3}}$. But, to avoid unpaired complex representations
of $SU(3)_c$, one needs $R_{c1}$ to be real. 
Barring the trivial cases where $\psi_1$ and $\psi_{2,3}$ form self-consistent extensions of the SM separately, 
the necessary condition to modify the Higgs couplings is to allow for a Yukawa coupling between $\psi_1$ and $\psi_{2,3}$. This leads to
\beq
\square\qquad \psi_1\sim (R_c,R_w,0)~,~~  \psi_2 \sim (R_c,R_w\pm1,\frac12)~,~~ \psi_3\sim (R_c,R_w\pm1,-\frac12)~,\quad R_c=\overline{R_c} ~. 
\label{vechi}\eeq
The choice of $R_w$ is arbitrary up to the $SU(2)_w$ global anomaly: 
for $N_w=2+4n$, one needs $N_{c}$ to be even.
The colourless case $R_c=1$ will be discussed in section \ref{MVL}.
The coloured case $R_c=8,27,\dots$ will be discussed in section \ref{VLMQ}.

Next, suppose that $\psi_2$ and $\psi_3$ do not form a vector-like pair, that is, a complex representation $R_{c1}=\overline{R_{c2}}=\overline{R_{c3}}$.
In order to have the same number of colour-conjugate representations one needs $N_{w1}=2N_{w2}$. A Yukawa coupling is also needed among $\psi_1$ and
$\psi_{2,3}$ to provide masses, so the only possibility is  
\beq
\square\qquad \psi_1\sim (R_c,2,0)~,~~  \psi_2\sim(\overline{R_c},1,\frac12)~,~~ \psi_3\sim (\overline{R_c},1,-\frac12)~,\quad R_c \ne \overline{R_c}~. 
\label{chichi}\eeq
The $SU(2)_w$ global anomaly further requires that $N_c$ must be even.
Note that this is the minimal, consistent set of chiral fermions 
that has no vector-like mass terms, rather
it acquires a mass from the Yukawa couplings only. 
The phenomenology of purely chiral fermions is discussed in section \ref{chiral}.

\item[(c)] For $(n,m)\ne(0,0)$, one has $N_{w3}\ge 4$. Then, $\psi_3$ does not couple to the SM nor to $\psi_1$, and even the possible Yukawa coupling to $\psi_2$ cannot provide
a mass to all the $N_{w3}$ components of $\psi_3$. Therefore, let us take $(n,m)=(0,0)$, that implies
\beq
\psi_1\sim (R_c,1,Y)~,~~  \psi_2\sim(R_c,2,-\frac45 Y)~,~~ \psi_3\sim (\overline{R_c},3,\frac15 Y)~.
\eeq
By requiring an equal number of components with opposite electric charge, one finds $Y=\pm1/2$ or $Y=\pm1/6$. In both cases one can check that some
components of the new fermions remain massless, therefore no consistent SM extension of this type exists.
\eit

\subsection{Four multiplets}\label{four}

In the previous sections we derived the list of all the consistent sets of $n$ new chiral fermions, with $n\le 3$, discussing in detail how
to implement the requirements of section \ref{criteria}. Here we provide the complete list for $n=4$, without displaying the lengthy and involved
analysis needed to prove this result.

First of all, there are a number of possibilities to combine smaller subsets of new fermions that are already consistent by themselves. It is worth to list them
for bookkeeping and to point out those combinations with special phenomenological relevance:
\bit
\item[$\square$] Four copies of $N$ and/or $\Sigma$.
\item[$\square$] Two copies of $N$ and/or $\Sigma$ plus a vector-like fermion from Table \ref{twotable}. 
In particular, the set $(N,\Sigma,L,L^c)$ corresponds to the neutralinos and charginos of the minimal supersymmetric SM: bino, wino and the two higgsinos.
This case is discussed in section \ref{MVL}.

\item[$\square$] Two vector-like fermions $\Psi_1$ and $\Psi_2$ from Table \ref{twotable}. A non-trivial interplay occurs when $\Psi_1$ and $\Psi_2$ couple both to a given SM fermion, 
as indicated in the last column of Table \ref{twotable}, and/or when there is a Yukawa coupling between $\Psi_1$ and $\Psi_2$: this happens for 
$E$ or $\Delta$ with $L$ or $\Lambda$; $T$ or $X_Q$ with $X_T$ or $Q$; $B$ or $Y_Q$ with $Q$ or $Y_B$.
In models of partial compositeness, a SM fermion acquires its mass by mixing with two vector-like composite fermions,
with the same quantum numbers as the SM left- and right-handed components: $Q$ and $T$ for the top quark, $Q$ and $B$ for the bottom quark,
$L$ and $E$ for the tau lepton.
This case is discussed in section \ref{2VLL} for leptons and \ref{2VLQ} for quarks.

\item[$\square$] One copy of $N$ or $\Sigma$ and a set of three fermions from \eq{vechi} or \eq{chichi}. A non-trivial interplay occurs in the case $(1,3,0)+(1,5,0)+(1,4,1/2)+(1,4,-1/2)$,
discussed in section \ref{MVL}.
\eit

Let us come to the consistent sets of four fermions that are not the union of two smaller self-consistent sets.
We found that non-trivial solutions are possible only when the four colour representations $R_{ci}$ are all equal or conjugate to each other. After all
requirements of section \ref{criteria} are taken into account, only two possible patterns emerge: 
\bit

\item For arbitrary $R_c$ and $R_w$, a viable set of four multiplets is provided by  
\beq
\square\qquad (R_c,R_w-1,0) + (R_c,R_w+1,0) + (\overline{R_c},R_w,1/2) + (\overline{R_c},R_w,-1/2)~,
\label{chi4}
\eeq 
with one exception: if $N_w$ is odd, then either $C(R_w-1)$ or $C(R_w+1)$ is half-integer, therefore one needs $N_c$ even
to cancel the global $SU(2)_w$ anomaly.
 
The case $R_c=1$ can be described as two Majorana leptons plus a vector-like lepton (see section \ref{MVL}). 
The case $R_c=\overline{R_c}\ne 1$ is the analogue for coloured fermions  (section \ref{VLMQ}).
Finally, the case $R_c \ne \overline{R_c}$ is purely chiral, with no masses before EWSB (section \ref{chiral}).

\item For arbitrary $R_c$, $R_w$ and $Y$, a viable set of four multiplets is 
\beq
\square\qquad (R_c,R_w,Y)+(\overline{R_c},R_w,-Y)+(R_c,R_w+1,Y-1/2)+(\overline{R_c},R_w+1,-Y+1/2)~.
\eeq
For the first time we encounter a pattern where the hypercharges of the new fermions are not determined uniquely.

The case $R_c=1$ corresponds to two VLLs (section \ref{2VLL}), except when $Y=0$ with $N_w$ odd, or $Y=1/2$ with $N_w$ even: then, one has two Majorana leptons
plus one VLL (section \ref{MVL}). 
The case $R_c \ne 1$ corresponds to two VLQs (section \ref{2VLQ}), except when $R_c=\overline{R_c}$ and 
$Y=0$ with $N_w$ odd, or $Y=1/2$ with $N_w$ even: then, one has two Majorana quarks
plus one VLQ  (section \ref{VLMQ}). 

\eit
We found that all other sets of four multiplets relevant for Higgs couplings are not viable: either some component remains massless, or a gauge anomaly is present.

\subsection{Larger sets of new fermions}\label{large}

We do not attempt a general classification for $n\ge 5$ new chiral fermions. On the one hand, the general principles and the different phenomenological possibilities are already well illustrated by more minimal sets of fermions.
On the other hand, a detailed analysis is worth only in the context of a specific, well-motivated theory beyond the SM.
Here we shall mention some prominent examples that have been extensively studied, to situate them in the context of our classification.
\bit
\item 
We have shown that there are two sets of purely chiral fermions, displayed in  \eq{chichi} and in \eq{chi4}, formed by three and four multiplets, respectively.
The more traditional chiral extension of the SM is a fourth family, formed by the five multiplets $q'_L$, $t'_R$, $b'_R$, $l'_L$ and $\tau'_R$.
It was already excluded at the time of LEP, because the $Z$ invisible width forbids a fourth massless neutrino, but it could be rescued adding a sixth multiplet, 
a sterile neutrino $\nu'_R$. It is by now excluded by the measurement of the Higgs boson couplings \cite{Eberhardt:2012gv}, 
as we will review at the end of section \ref{chiral}.
\item
The minimal supersymmetric extensions of the SM predicts fermionic partners for the gauge bosons and for the two Higgs doublets. This amounts to five chiral multiplets:
a bino $\sim (1,1,0)$, a wino $\sim(1,3,0)$, two higgsinos $\sim (1,2,\pm 1/2)$ and a gluino $\sim(8,1,0)$.
The latter does not enter in our classification, since it does not couple to the Higgs doublet. Concerning the other four multiplets, supersymmetry restricts
the possible couplings among them and to the SM, therefore it corresponds to a special case in the parameter space of the SM extension by the set $(N,\Sigma,L,L^c)$.
We will briefly discuss the related phenomenology in section \ref{MVL}. Of course, our purely fermionic extension of the SM  
corresponds to the limit where the scalar supersymmetric partners are significantly heavier than neutralinos and charginos.
\item
Another scenario addressing the hierarchy problem is compositeness. An effective way to couple the SM fermions to
a composite Higgs doublet amounts to a partial
fermion compositeness: each SM chiral fermion mixes with a composite vector-like fermion with the same quantum numbers. Thus, to induce a Yukawa coupling among two SM fermions and the composite Higgs
one needs two vector-like fermions. Therefore, 
a SM extension by four chiral multiplets is suitable to study this mechanism for one SM Yukawa coupling at a time.
Indeed, partial compositeness corresponds to a special subspace of parameters, because the symmetries of composite models restrict the couplings
of the new fermions and of the SM ones. 
We will briefly discuss the phenomenology of $\tau$-compositeness in section \ref{MVL} and the case of $b$ and $t$-compositeness in section \ref{2VLQ}.
Of course, realistic models of partial compositeness require more than two vector-like fermions, e.g. to induce the Yukawa couplings of all the heavy, third family fermions.
The interplay between the Higgs and composite vector-like fermions is studied in detail e.g. in Refs.~\cite{Azatov:2012rj,Montull:2013mla,Flacke:2013fya,Chen:2014xwa,Matsedonskyi:2014mna}.
\eit

\section{Phenomenology of new chiral fermions}\label{chiral}

In this section we study purely chiral sets of new fermions, that is to say, one cannot write any fermion mass term before EWSB,
thus their masses are generated by the Higgs vev only. 
These sets do not contain vector-like pairs of chiral multiplets, that would admit a Dirac mass, nor multiplets in the representations $(R_c,R_w,0)$ with $R_c=\overline{R_c}$ 
and $R_w$ odd, that would admit a Majorana mass.
A purely chiral set, consistent with the requirements of section \ref{criteria}, constitutes a new fermion `family', very much analogue to the three SM families.

We identified two classes of purely chiral sets, formed by three and four multiplets respectively, displayed in \eq{chichi} and \eq{chi4}.
We will discuss the phenomenology of these two classes in some detail. In the last part of the section, we will investigate whether larger chiral sets of fermions
may be compatible with present Higgs data. 
\\

\noindent $\bullet$ \textbf{\textit{Three chiral multiplets.}}
The only consistent `family' formed by three chiral multiplets is 
\beq
\psi_{1L}\sim (R_c,2,0)~,\quad  \psi_{2R} \sim\left(R_c,1,\frac12\right)~,\quad \psi_{3R} \sim \left(R_c,1,-\frac12\right)~,\quad 
R_c \ne \overline{R_c}~,N_c{\rm~even}~, 
\eeq
the smallest viable representation being $R_c=6$.
The Yukawa interactions are
\begin{equation}
-{\cal L}_Y= \lambda_{12} \overline{\psi_{1L}} \tilde{H}\psi_{2R} + \lambda_{13} \overline{\psi_{1L}} H\psi_{3R} +h.c. ~.
\label{Y3chi}\end{equation}
Here and in the rest of the paper we do not display the obvious kinetic terms, that must be added for each new fermion.
After EWSB one is left with two mass eigenstates $F_{12}$ and $F_{13}$ in the same colour representation $R_c$,
with charge $Q=\pm1/2$ and mass
$m_{12}=\lambda_{12}v/\sqrt{2}$ and  $m_{13}=\lambda_{13} v/\sqrt{2}$, respectively.

The lightest new fermion is stable and forms hadrons with exotic charges, that are constrained to be heavier than several hundreds of GeVs.
Indeed, the searches for R-hadrons \cite{Aad:2012pra,Chatrchyan:2013oca} assume the existence of a stable stop or sbottom (scalar with $R_c=3$),
or of a stable gluino (fermion with $R_c=8$).
In the latter case one finds $m_{R_c=8}\gtrsim 1320$ GeV \cite{ATLAS:2014fka}, and we expect a similar (stronger)
bound for $R_c=6$ (larger colour representations), as these fermions are pair-produced through their coupling to gluons.
To roughly estimate the limit on a stable sextet, we rescale the gluino bound by computing the ratio between the sextet and octet production cross section at tree level. 
Taking into account the different colour contractions and the interference of the s, t and u-channel, we obtain a lower bound 
$m_{R_c=6} \gtrsim 1400$ GeV.
Further discussions and references on the constraints on  R-hadrons can be found e.g. in Refs.~\cite{Fairbairn:2006gg,DiLuzio:2015oha}.
This limit from direct searches already leads to some tension with the perturbativity upper bound, $m_{12,13} \ll (4\pi) v/\sqrt{2} \simeq 2.2$ TeV.

The contribution of $F_{12,13}$ to the oblique parameters reads
\begin{equation}
S\simeq \frac{N_c}{6\pi}~, \quad\quad 
T\simeq \frac{N_c}{16 \pi s_w^2 c_w^2 m_Z^2}\left( m_{12}^2+m_{13}^2-2 \frac{m_{12}^2 m_{13}^2}{m_{12}^2-m_{13}^2} \ln \frac{m_{12}^2}{m_{13}^2}\right)~.
\end{equation}
Note that for $(m_{12}-m_{13})\rightarrow 0$ one finds $T\propto(m_{12}-m_{13})^2$, because in the degenerate limit the custodial symmetry is restored:
for $\lambda_{12}=\lambda_{23}$ \eq{Y3chi} has a global $SU(2)_R$ symmetry with $\psi_{2R}$ and $\psi_{3R}$ transforming as a doublet.
The value of $S$ can lie within the $3\sigma$ ellipse, but only if $N_c = 6$
and $T\simeq 0.3$ at the same time (see figure \ref{ST}).
This can be achieved for $(m_{12}+m_{13})/2 \simeq 1500$ GeV and $(m_{12}-m_{13})\simeq 50$ GeV.
Thus, this `family' of new fermions is marginally compatible with direct searches and EW precision tests.

After the Higgs discovery, however, one can definitely exclude such a set of new fermions.
As they are heavier than the Higgs boson and do not mix with the SM fermions,
the Higgs decay width at tree-level are unchanged. However, a huge deviation occurs in the loop-induced Higgs coupling to gluons,
\begin{equation}
R_{gg} \equiv \frac{\sigma(gg\rightarrow h)}{\sigma_{SM}(gg\rightarrow h)}
\simeq \left[ 1+ 4 C(R_c) \right]^2 ~. 
\label{Rgg}\end{equation}
where we have taken the limit $m_t^2,m_{12}^2,m_{13}^2 \gg m_h^2/4$ in the loop form factor (see appendix \ref{appendix B.2}).
Even the smallest possible colour representation has $C(6)=5/2$, leading to a huge $R_{gg}=121$, totally incompatible with LHC data.
\\

\noindent $\bullet$ \textit{\textbf{Four chiral multiplets.}}
Let us move to the only consistent `family' formed by four chiral multiplets, 
\beq
\psi_{1L} \sim (R_c,R_w-1,0) ,~~  \psi_{2L} \sim(R_c,R_w+1,0) ,~~ \psi_{3R}\sim \left(R_c,R_w,\frac12\right) ,~~ \psi_{4R} \sim \left(R_c,R_w,-\frac12\right),
\label{purely-chiral}
\eeq
with $R_c \ne \overline{R_c}$ to prevent vector-like or Majorana mass terms, and with $N_c$ even if $N_w$ is odd, to prevent a global $SU(2)_w$ anomaly.
The allowed Yukawa interactions are 
\begin{equation}
-{\cal L}_Y= \lambda_{13} \overline{\psi_{1L}} \tilde{H} \psi_{3R} + \lambda_{14} \overline{\psi_{1L}} H \psi_{4R} 
+ \lambda_{23} \overline{\psi_{2L}} \tilde{H}\psi_{3R} + \lambda_{24} \overline{\psi_{2L}} H \psi_{4R}  + h.c. ~.
\label{fourchiralfermions}
\end{equation}
where the explicit $SU(2)_w$ contractions are defined in \eq{CGy}.
After EWSB the components of $\psi_{1,2L}$ combine with those of $\psi_{3,4R}$ to forms $2N_w$ mass eigenstates: 
one with charge $Q=N_w/2$; 
two mixed states with $Q=N_w/2-1,\dots,-N_w/2 + 1$; one with $Q=-N_w/2$.
As they have (half-)integer charges, they do not mix with the SM quarks and the lightest state is stable and hadronises, with collider bounds
above 1 TeV, analogue to those discussed above.
The discussion of oblique parameters is also similar to the previous case: one can be marginally consistent with data, by choosing the parameters 
to realise an approximate custodial protection.

The way to definitely exclude this set of chiral fermions is, once again, their contribution to the Higgs boson coupling to gluons.
Note that each of the $2N_w$  mass eigenstates belongs to the same colour representation $R_c$ and must be (much) heavier than the top quark, therefore one finds
\begin{equation}
R_{gg}\simeq \left[ 1+ 2(2 N_w) C(R_c) \right]^2 ~.
\end{equation}
Even in the minimal case with $N_w=2$ and $C(3)=1/2$, one finds a very large $R_{gg}\simeq 25$, incompatible with the LHC Higgs data.
\\

\noindent $\bullet$ \textit{\textbf{Larger sets of chiral multiplets.}}
Let us ask the question whether we can exclude any set of purely chiral fermions, even when it is formed by more than four multiplets.
Indeed, the Higgs coupling to gluons implies that any new  chiral fermion should be colourless, because even the minimal set of chiral coloured fermions, formed by a weak doublet and two weak singlets with $R_c=3$, 
leads to a large $R_{gg}^{min} \simeq 9$. This is not compatible with the range currently allowed by global fits, $0.5 \lesssim R_{gg} \lesssim 1.8$  at 99 \% C.L. \cite{ATLAS-CONF-2015-044} (see appendix \ref{experimental} for details).
In particular, in this way one can exclude \cite{Eberhardt:2012gv} a fourth SM family, formed by 
the six multiplets $q'_L$, $t'_R$, $b'_R$, $l'_L$, $\tau'_R$ and $\nu'_R$. Recall that 
the sterile neutrino is required to avoid an additional massless neutrino, that is forbidden by the $Z$ invisible width; 
then this set of fermions is not purely chiral, but one may invoke a lepton number symmetry
to forbid the sterile neutrino Majorana mass. 
Let us remark that coloured chiral fermions may be allowed in the case of extended Higgs sectors, not considered in the present paper.
For example, adding an Higgs triplet, it is possible to rescue  the fourth family \cite{Banerjee:2013hxa}.
Another example is provided by coloured chiral fermions receiving their mass from a second Higgs doublet \cite{Alves:2013dga}.

Coming to colourless chiral fermions, some of the mass eigenstates are necessarily charged and thus contribute to the Higgs width to photons as
\begin{equation}
R_{\gamma\gamma}= \frac{\left| {\cal{A}}_{SM}^{\gamma \gamma} + A^{\gamma\gamma}_{new}\right|^2}{\left| {\cal{A}}_{SM}^{\gamma \gamma}\right|^2}~,\qquad 
A^{\gamma\gamma}_{new} \simeq \frac{4}{3} \sum_k Q_k^2 ~,
\label{Rppchi}\end{equation}
where $R_{\gamma\gamma}$ is defined in \eq{R}, the SM amplitude is ${\cal{A}}_{SM}^{\gamma \gamma} \simeq -6.5$ and the sum runs over the new fermion mass eigenstates. 
Note that, to derive \eq{Rppchi} from Eqs.~(\ref{hpp}) and (\ref{App}), we took (i) $2m_i \gg m_h$, that is accurate enough, 
even though the lower bounds on heavy charged lepton masses are weaker than those on coloured particles;
(ii) Higgs-fermion couplings $y_i = m_i / v$ and $\tilde{y}_i =0$, 
that is the case for purely chiral fermions, because their mass matrices are proportional to the Higgs vev $v$, see \eq{y-ytilde}. 
Note also that the result is independent from potential mixing between the new fermions and the SM leptons.
The presently allowed range is $0.5 \lesssim R_{\gamma\gamma} \lesssim 1.9 $ at 99 \% C.L. \cite{ATLAS-CONF-2015-044}.
For $n$ chiral multiplets $(R_{wi},Y_i)$, one finds
\beq
{\cal A}^{\gamma\gamma}_{new} \simeq \frac 23 \sum_{i=1}^n  \sum_{k=1}^{N_{wi}}  Q_k^2 = \frac 23 \sum_{i=1}^n \frac{N_{wi}(N_{wi}^2+12Y_i^2-1)}{12}~.
\label{pp-chiral}\eeq
We added an overall factor $1/2$ to take into account that each massive fermion is formed by two chiral components. Equivalently, one may take the sum
only over fermions of a given chirality.
For example, a minimal set is formed by a weak doublet $2_L$ of hypercharge $Y$ paired with two weak singlets
$(1+1)_R$, giving a contribution ${\cal A}^{\gamma\gamma}_{new}=2(1+4Y^2)/3\ge 2/3$. 
The next-to-minimal set $(2+2)_L$ paired with $(1+1+1+1)_R$ implies $A^{\gamma\gamma}_{new} > 4/3$, that is still allowed by 
the present constraint on $R_{\gamma\gamma}$, while for example
$(3+2)_L$ gives already $A^{\gamma\gamma}_{new} > 10/3$, that is almost excluded.
Since the SM amplitude has opposite sign w.r.t. the one of new fermions, one can also envisage the contrived possibility of a large 
$A^{\gamma\gamma}_{new} \sim -2{\cal{A}}_{SM}^{\gamma \gamma} \simeq 13$.

This shows that there are purely chiral sets of $n$ fermions that satisfy the $\gamma\gamma$ constraint. However,
we have also shown before that no set exists for $n\le 4$, that satisfies the consistency requirements of section \ref{criteria}.
It is non-trivial to check whether purely chiral sets with $n>4$ could be consistent.
Consider for example the case of two weak doublets plus four weak singlets.
In order for all components to receive a mass from the Yukawa couplings to the Higgs doublet,
the hypercharges must be chosen as
\beq
\left(2,Y_1\right)~,~~\left(2,Y_2\right)~,~~\left(1,-Y_1+\frac 12\right)~,~~\left(1,-Y_1-\frac 12\right)~,~~\left(1,-Y_2+\frac 12\right)~,~~\left(1,-Y_2-\frac 12\right)~,
\label{chiral-set-6multiplets}
\eeq
in the convention where all multiplets have the same chirality.   The absence of anomalies requires \eq{anom} to hold, and this leads to $Y_1=-Y_2\equiv Y$.
As a consequence, three vector-like mass terms are allowed and such set of fermions does not qualify as purely chiral.

At this point one should recall that, in this paper, we took the point of view that all mass terms and interactions allowed by the gauge symmetries are present.
In alternative, one can easily introduce some global symmetry to forbid possible vector-like mass terms, thus imposing by hand that a given set of fermions is chiral.
For example in \eq{chiral-set-6multiplets} take a $U(1)$ symmetry with charge $+1$ for doublets and $-1$ for singlets.
If one takes this point of view,   the set of fermions in \eq{chiral-set-6multiplets} qualifies as the minimal still viable set of purely chiral fermions.
Indeed, besides being consistent with all the requirements of section \ref{criteria},  it can be compatible with direct searches, EW precision tests,  and constraints from the Higgs couplings.

Concerning direct collider searches, as the new leptons have charges $Q=Y\pm1/2$, they do not mix with the SM leptons (except for $|Y|= 1/2$ or $3/2$) and the lightest state is stable.
There are severe bounds on the number
density of such charged relics \cite{DiLuzio:2015oha}, but they depend on cosmological assumptions: e.g. for a reheating temperature below their mass, they were never produced in the early Universe.
Limits on heavy stable leptons  are of the order of a few hundreds of GeVs, and are displayed in Table~\ref{6table} for some representative values of $Q$.
For a review on heavy stable particles see Ref.~\cite{Fairbairn:2006gg}.

The contributions to the $S$ and $T$ parameters of a fermion `family' formed by one doublet and two singlets are given in Eqs.~(\ref{S-TB}) and (\ref{T-TB}).
In the present case we have two such `families' with $N_c=1$ and opposite hypercharges $\pm Y$, and
one can easily lie within the $3\sigma$ ellipse of Fig.~\ref{figST}.
For example in the custodial limit where the two mass eigenstates of each `family' are degenerate, one finds $T\simeq 0$ and $S\simeq 0.1$.
Concerning the Higgs couplings, one finds ${\cal A}^{\gamma\gamma}_{new}=4(1+4Y^2)/3$, that lies in the allowed range of $R_{\gamma\gamma}$
for $|Y|\lesssim 0.3$ and $1.4 \lesssim |Y| \lesssim 1.6$.

Finally, it is interesting to compare \eq{pp-chiral} with the analogue amplitude for the Higgs boson coupling to $\gamma Z$.
For $n$ chiral multiplets $(R_{wi},Y_i)$, one finds
\beq
{\cal A}^{\gamma Z}_{new} \simeq \frac 23 \sum_{i=1}^n  \sum_{k=1}^{N_{wi}}  Q_k \frac{T_{3k}-s_w^2 Q_k}{c_w^2} = \frac 23 \sum_{i=1}^n \frac{N_{wi}(N_{wi}^2-12Y_i^2\tan^2\theta_w-1)}{12}~,
\label{pZ-chiral}\eeq
where we used \eq{CP-even} particularised to the case of chiral fermions, in the same way we did above for the $\gamma\gamma$ case.
Note that, when summing over the mass eigenstates of equal charge $Q$, the mixing matrices disappear from the $Z$ couplings in  \eq{Zmass}, therefore one reduces to a sum
over the interaction eigenstates. For example, the set in \eq{chiral-set-6multiplets} gives ${\cal A}^{\gamma Z}_{new} \simeq 2 \left[1-\left(1+8 Y^2 \right)\tan^2\theta_w \right]/3$.

\begin{table}[tb]
\renewcommand{\arraystretch}{1.3}
\bc\begin{tabular}{|c|c|c|c|c|c|c|c|c|c|c|}
\hline
$\left|Q \right|$ & ~1/3 & ~2/3 & ~1 & ~2 & ~3 & ~4 & ~5 & ~6 & ~7 & ~8 \\ 
\hline 
bound in GeVs 
& ~ 200 & ~ 480 & ~ 574 & ~ 685 & ~ 752 & ~ 793 & ~ 796 & ~ 781 & ~ 757 & ~ 715 \\
\hline
\end{tabular}\ec
\caption{The 95 \% C.L. lower bounds on the mass of  heavy stable leptons, from the CMS collaboration \cite{Chatrchyan:2013oca}. 
The production is assumed to occur through the Drell-Yan process only. Limits are obtained for $SU(2)_w$ singlets, but they remain similar in general \cite{DiLuzio:2015oha}. 
The ATLAS collaboration obtains comparable but less stringent limits in the range $2 \le \left| Q \right| \le 6$ \cite{Aad:2015oga}. For  larger values of $Q$ see Ref.~\cite{Aad:2011mb}.}
\label{6table}
\end{table}

\section{Phenomenology of non-chiral leptons}\label{phenoL}

In this section we discuss new colourless fermions, which admit either a Majorana or a vector-like mass term before EWSB.

\subsection{Majorana leptons}\label{seesaw}

Let us consider leptons that admit a Majorana mass term. The latter requires a vanishing hypercharge,
$Y=0$. The two possibilities relevant for the Higgs couplings are sterile neutrinos $N\sim(1,1,0)$, and weak triplets $\Sigma\sim(1,3,0)$.
\\

\noindent $\bullet$ \textbf{\textit{Sterile neutrinos.}}
In the case of one sterile neutrino, 
the SM Lagrangian is extended to ${\cal L} = {\cal L}_{SM} + {\cal L}_{N}$, where
\beq
{\cal L}_{N} = - \overline{l_{L\alpha}} \lambda_{N\alpha} \tilde{H} N_R-\frac 12 \overline{N_R^c} M_N N_R + h.c. ~.
\eeq
Only one linear combination $\nu_L$ of the three active neutrinos $\nu_{L\alpha}$ couples to $N_R$, and since we are not concerned with flavour issues 
we will drop the index $\alpha$ in the following. After EWSB, $\nu_L$ and $N_R^c$ mix and combine
into two Majorana fermions $\nu_l$ and $\nu_h$ with definite masses, $m_{\nu_l}\le m_{\nu_h}$; active neutrino mass searches imply $m_{\nu_l} \lesssim 1$ eV.
If one takes the limit $m_{\nu_l}\ll m_h$, the type I seesaw mechanism is realised: $m_{\nu_l}\simeq 
\lambda_N^2v^2/(2M_N)$ and $m_{\nu_h}\simeq M_N$.
The Higgs boson couplings to $\nu_l\nu_l$ and  $\nu_h\nu_h$ are proportional to $m_{\nu_l}/v$, and the coupling to $\nu_l\nu_h$ is proportional to $\sqrt{m_{\nu_l} m_{\nu_h}}/v$, leading 
 to a negligibly small decay width, $\Gamma(h\rightarrow \nu_l\nu_h) < m_h^2  / (8\pi v^2) m_{\nu_l}\lesssim 10^{-8}$ MeV.
 Also, one can check that the decay widths of the $Z$-boson receive negligible corrections, always proportional to $m_{\nu_l}/v$. 
 For $m_{\nu_h}$ above the EW scale, one can compute the $\nu_h$ contribution to the $S$ and $T$ parameters, that turns out to be suppressed by the tiny ratio $m_{\nu_l}/m_{\nu_h}$.

In the case of two or more sterile neutrinos $N_i$, in most of the parameter space the arguments above apply to each $N_i$ separately: either the Majorana mass $M_i$ is as small as the eV scale, or 
the active-sterile mixing $\theta_{i}\equiv \lambda_{N_i} (v/\sqrt{2}) / M_{N_i}$ is suppressed, $|\theta_i^2 M_{N_i}| \sim m_{\nu_l} \lesssim 1$ eV. In either case the corrections to the Higgs and $Z/W$-boson couplings are tiny. 
The only exception occurs when much larger $\theta_{i}$ are tuned among each other, in order for the $N_i$ contributions to the light neutrino mass to cancel.
Consider for simplicity two sterile neutrinos $N_{1,2}$. 
At leading order in the mixing angles $\theta_i$ one has
\beq
m_{\nu_l}\simeq 
|\theta_1^2 M_{N_1} +\theta_2^2 M_{N_2}| \lesssim 1{\rm~eV}~.
\eeq
The two summands have a physical relative phase, therefore they can be orders of magnitude larger than $m_{\nu_l}$, if there is a strong cancellation between the two:
the active-sterile mixing can be large, no matter how large the sterile masses $m_{\nu_{h1,2}}$ are. 
Even though this scenario requires a severe tuning of parameters to lead to observable effects, it may be justified by some symmetry. 
For example, in the so-called inverse seesaw model \cite{Wyler:1982dd,Mohapatra:1986bd,GonzalezGarcia:1988rw} (see also Ref.~\cite{Law:2013gma}), 
the lepton number symmetry $U(1)_{L}$ is broken by a small mass parameter, and the cancellation occurs naturally in the limit where this parameter goes to zero.
Therefore, it is worth to analyse the phenomenological consequences of a large active-sterile mixing: 
both Higgs couplings and EW gauge boson couplings can be significantly modified.

Consider first the neutrino mass eigenstates $\nu_l$, $\nu_{h1}$, $\nu_{h2}$ in the regime $m_{\nu_l}\ll m_{\nu_{h1,2}} \ll m_h,m_Z$. 
One finds that the decay widths of the Higgs boson can be significantly modified,
in particular
$\Gamma(h\rightarrow \nu_l \nu_{hi})\simeq m_h  m_{\nu_{hi}}^2  |\theta_i|^2 / (8\pi v^2)$ and $\Gamma(h\rightarrow \nu_{hi} \nu_{hi})\simeq m_h  m_{\nu_{hi}}^2 |\theta_i|^4/ (4\pi v^2)$.
These rates can be easily as large as the total SM Higgs width, $\Gamma^{SM}_h \simeq 4$ MeV, therefore the LHC experiments already constrain $\theta_{i}$ and $m_{\nu_{hi}}$.
Note that both invisible and visible decay channels are affected, since $\nu_{h1,2}$ decay not only into light neutrinos, but also into SM particles e.g. via virtual $W$-bosons.
Detailed analyses of the parameter space and of various constraints can be found e.g. in 
Refs.~\cite{deGouvea:2007uz,Chen:2010wn,BhupalDev:2012zg,Cely:2012bz,Antusch:2015mia}. 
Note that the $Z$-boson invisible width $\Gamma^{inv}_Z$, that is measured at the few per mil level, is not significantly affected  
for $m_{\nu_{h1,2}}\lesssim 1$ MeV, with $\nu_{h1,2}$ decaying mostly invisibly into three $\nu_l$: even in the presence of large mixing, only the active components
of $\nu_{l,h1,h2}$ couple to the $Z$-boson, and one recovers the SM value of  $\Gamma^{inv}_Z$ once the sum over all neutrino pairs is taken.
On the contrary, for larger $m_{\nu_{h1,2}}$ the heavy neutrinos mediate visible $Z$-decays, 
therefore $\Gamma^{inv}_Z$ is depleted
and a significant upper bound applies on $|\theta_i|$.

Consider next the complementary regime $m_h,m_Z \lesssim m_{\nu_{h1,2}}$.
In this case the Higgs and $Z$ decay width are not modified, but a significant active neutrino fraction in $\nu_{h1,2}$ can still have observable consequences.
Direct searches of EW scale sterile neutrinos through their mixing with active neutrinos have been performed e.g. by ATLAS ~\cite{ATLAS-CONF-2012-139,Aad:2015xaa} and 
CMS  ~\cite{Chatrchyan:2012fla,Khachatryan:2015gha}. 
Here we would like to point out that the EW precision parameters $S$ and $T$ can also receive important corrections, that constrain the masses and mixing of the sterile neutrinos.
To understand this quite surprising fact, that is generally overlooked, it is convenient to write the $3\times 3$ neutrino mass matrix in the basis $(\nu_L,N_{R1}^c,N_{R2}^c)$ as
\beq
{\cal M}_\nu = U^* diag(m_{\nu_l},m_{\nu_{h1}},m_{\nu_{h2}}) U^\dagger ~,
\label{mnu}\eeq
with $U$ unitary and subject to the constraint $0=({\cal M}_\nu)_{11} \simeq U_{12}^{*2} m_{\nu_{h1}} + U_{13}^{*2} m_{\nu_{h2}}$, where we neglected the tiny $m_{\nu_l}$.
Then, the active neutrino fractions $U_{1i}$ contained in the mass eigenstates can be parametrized in full generality as follows:
$U_{13} = \theta$ taken to be real,  
$U_{12} = i \theta / \sqrt{r_h}$ with $r_h\equiv m_{\nu_{h1}}/m_{\nu_{h2}}$ and $|U_{11}|^2 = 1 -\theta^2 (1+r_h)/r_h$.
The $3\sigma$ lower bound  on $\Gamma_Z^{inv}$ implies $\theta^2 (1+r_h)/r_h \lesssim 
0.015$. We computed $T$ and $S$ using the formulas in appendix \ref{appendixA}, as a function of $\theta$, $r_h$ and $r_Z\equiv m_Z/m_{\nu_{h2}}$, 
neglecting the mass of the SM leptons and including a symmetry factor $1/2$ for loops of Majorana fermions. 
Here we report  the result in some physically interesting limits: 
\beq\ba{ll}
(a)~m_{\nu_{h2}}=m_{\nu_{h1}} \gg m_Z :
& T \simeq \dfrac{\theta^4}{4\pi s_w^2c_w^2}\dfrac{m_{\nu_{h2}}^2}{m_Z^2}~,
\quad S \simeq  \dfrac{2\theta^2}{9\pi} \left(4+ 3\log\dfrac{m_{\nu_{h2}}^2}{m_Z^2}\right);\\ \\
(b)~ m_{\nu_{h2}}\gg m_{\nu_{h1}} = m_Z : 
& T \simeq \dfrac{\theta^4}{8\pi s_w^2c_w^2}\dfrac{m_{\nu_{h2}}^2}{m_Z^2}\left(3-\log\dfrac{m_{\nu_{h2}}^2}{m_Z^2}\right) ,
\quad S \simeq  \dfrac{\theta^2}{4\pi} \dfrac{m_{\nu_{h2}}}{m_Z} ~.\\
\ea\eeq
In case $(a)$, taking the maximal allowed value $\theta^2_{max}\simeq 0.007$, the correction to $T$ grows quadratically with the sterile neutrino mass: requiring to remain in the $3\sigma$ ellipse in the  $S-T$ plane (see Fig.~\ref{figST}),
one finds the upper bound $m_{\nu_{h2}}\lesssim 8.5$ TeV. 
This sensitivity to very large scales is due to the significant fraction of the active neutrino in the heavy states; note that this non-decoupling effect requires a strong tuning among
the two sterile neutrino parameters.
In case $(b)$, the active fraction in the heaviest sterile neutrino is rather $\theta^2_{max}\simeq 0.015(m_Z/m_{\nu_{h2}})$, therefore 
$T$ grows only logarithmically with $m_{\nu_{h2}}$, while $S$ 
remains constant: one remains in the ellipse for $m_{\nu_{h2}}$ as large as the Planck scale.
\\

\noindent $\bullet$ \textbf{\textit{Weak triplets with zero hypercharge.}}
In the case of a weak triplet $\Sigma_R \sim (1,3,0)$, the SM Lagrangian is extended by
\beq
{\cal L}_{\Sigma_R} = - \sqrt{\frac 23} \overline{l_{L\alpha}} \lambda_{\Sigma\alpha} \Sigma_R \tilde{H}  - \frac 12 {\rm Tr}\left( \overline{\Sigma_R^c} M_\Sigma \Sigma_R\right) + h.c. ~,
\label{sigma}\eeq
where we adopted the matrix notation
\beq
\Sigma_R \equiv \sqrt 2 \Sigma_R^a \tau^a = \frac{1}{\sqrt 2}\left(\ba{cc} 
\Sigma_R^3 & \Sigma_R^1-i\Sigma_R^2 \\
\Sigma_R^1+i\Sigma_R^2 & -\Sigma_R^3 
\ea\right) \equiv \left(\ba{cc} 
\frac{1}{\sqrt 2}\Sigma_R^0 & -\Sigma_R^+ \\
\Sigma_R^- & -\frac{1}{\sqrt 2}\Sigma_R^0 
\ea\right) ~
\eeq
and we normalised the triplet Yukawa coupling according to appendix \ref{treeH}.
After EWSB, the neutral component $\Sigma_R^0$ and a linear combination of active neutrinos combine into two Majorana fermions $\nu$ and $\Sigma_0$, in complete analogy to the
case of $\nu_l$ and $\nu_h$ discussed above.
As usual, we will consider only the mixing with the third lepton family, taking $\lambda_{\Sigma e,\Sigma\mu}=0$ and $\lambda_{\Sigma\tau}\equiv \lambda_\Sigma$. 
Indeed, flavour changing neutral current processes 
such as $\mu\rightarrow e\gamma$ or $\mu\rightarrow 3e$ are strongly constrained ~\cite{Abada:2008ea,Abada:2007ux}.
In the limit $\lambda_{\Sigma} v \ll M_\Sigma$ one realises the so-called type III seesaw mechanism: 
$m_{\nu}\simeq \lambda_{\Sigma}^2v^2/(6M_\Sigma)$ and $m_{\Sigma_0}\simeq M_\Sigma$.
The charged components $(\Sigma_R^+)^c$ and $\Sigma_R^-$ mix with the SM charged leptons $\tau_{L}$ and $\tau_R$ respectively, to form the mass eigenstates $\tau$ and $\Sigma^-$. 
Then, three real parameters -- the Yukawa coupling $\lambda_\Sigma$, the mass $M_\Sigma$ and the SM tau Yukawa coupling $\lambda_\tau$ --
determine the mass of four physical states, $m_\nu$ and $M_{\Sigma^0}$ in the neutral sector, 
$m_\tau$ and $M_{\Sigma^-}$ in the charged sector.
The mixing angles for neutrinos and left-handed charged leptons are suppressed by $\sqrt{{m_\nu}/{M_{\Sigma^0}}}$; the mixing of right-handed charged leptons receives an additional suppression by ${m_{\tau}}/{M_{\Sigma^-}}$.

The couplings of $\Sigma_0$ to the $Z$ and Higgs bosons are exactly the same as the couplings of  $\nu_h$ discussed above.
In particular, for vanishing $m_\nu$ all $Z$ and Higgs couplings reduce to their SM values, therefore the corrections are negligibly small.
At tree level, $M_{\Sigma^-}- M_{\Sigma^0}$ also vanishes with $m_\nu$, however it is well-known that at one loop weak interactions induce a mass split,
$M_{\Sigma^-}- M_{\Sigma^0}\simeq 170$ MeV (see e.g. \Ref{Cirelli:2005uq}). 
Neglecting the tiny mixing angles, the only heavy lepton couplings are 
 $Z \Sigma^+ \Sigma^-$ and $W^+ \Sigma^- \Sigma^0$; other mixing-suppressed couplings are relevant for $\Sigma$-decays  ~\cite{Franceschini:2008pz}. 
 The contribution of $\Sigma$ to the EW precision parameters $S$ and $T$ is vanishingly small,
as EWSB is felt only through the mixing angles and through the loop-induced mass splitting among the $\Sigma$-components, and both are very small.

Coming to direct searches, LEP looked for new charged leptons pair produced and decaying to $W \nu$, setting a lower bound $M_{\Sigma^-} \gtrsim 100$ GeV \cite{Achard:2001qw}.
At the LHC heavy leptons are mostly pair-produced via  
$Z^* /\gamma^* \rightarrow \Sigma^+ \Sigma^-$  and $W^{\pm*}\rightarrow \Sigma^\pm \Sigma^0$. 
The fraction of $\Sigma$ that decays into each lepton flavour, 
$b_\alpha={\theta_{\alpha }}/({\theta_{e }+ \theta_{\mu }+ \theta_{\tau }})$,  characterises the final state. 
CMS \cite{CMS:2012ra} considered either $b_e=b_\mu=b_\tau=1/3$, $b_e=1$ or $b_\mu=1$, obtaining constraints in the range $M_{\Sigma} \geq 180-210$ GeV.  
The most stringent constraint comes from ATLAS \cite{Aad:2015cxa}, with $M_\Sigma \geq$ 325 GeV for $b_e=1$ and $M_\Sigma \geq$ 400 GeV for $b_\mu=1$.
We expect a weaker bound in the case $b_\tau=1$, that we assumed above.
Other decay channels relevant for $\Sigma$ searches at the LHC are discussed in Ref.~\cite{Franceschini:2008pz}, including displaced vertexes, as $\Sigma$ becomes long-living in the limit of very small mixing.

In the case of two or more lepton triplets $\Sigma_i$, the phenomenology is similar, except when the mixing between the SM leptons and the new leptons is not suppressed.
As in the case of sterile neutrinos, this is possible only by severely tuning the Yukawa couplings of the various $\Sigma_i$ to keep $m_\nu$ small. 
In the case of two triplets,
the neutrino mass matrix is diagonalised as in \eq{mnu}, while the charge lepton mass matrix can be written as $M_e = U_L diag(m_\tau,M_{\Sigma_1},M_{\Sigma_2}) U_R^\dagger$.
Neglecting $m_\nu$ and $m_\tau$, the left-hand mixing matrix $U_L$ coincides with the neutrino mixing matrix up to a $\sqrt{2}$ factor:
$(U_L)_{13} \simeq \sqrt{2} \theta$, $(U_L)_{12}\simeq i\sqrt{2} \theta/\sqrt{r_h}$ and
$\left| (U_L)_{11} \right|^2\simeq 1 - 2 \theta^2 (1+r_h)/r_h$, with $r_h = M_{\Sigma_1}/M_{\Sigma_2}$, while the mixing angles in $U_R$ are further suppressed by $m_\tau/M_{\Sigma_i}$.
When $\theta$ is large, the corrections to $S$ and $T$ may become significant as already discussed for sterile neutrinos.
In addition, the new charged leptons, that are necessarily above the EW scale, could contribute significantly to $h\rightarrow \gamma\gamma,\gamma Z$.
Before computing these corrections, one should notice that
a strong upper bound on the mixing comes from the $Z$ coupling to $\tau^+\tau^-$.
The LEP measurement of $\Gamma(Z\rightarrow \tau^+ \tau^-)$  \cite{Agashe:2014kda} implies $2 \theta^2 (1+r_h)/r_h \lesssim 0.004$ at $3\sigma$.

Let us describe in some detail the main corrections to $R_{\gamma\gamma}$ and $R_{\gamma Z}$, defined by \eq{R}. 
Similar analytic approximations could be used for the models analysed in the next sections as well.
For the diphoton channel, using the results of Appendix \ref{Hpp} one finds 
\begin{equation}
\ba{rcl}
R_{\gamma\gamma}
& \simeq &
\dfrac{\left| {\cal{A}}^{\gamma\gamma}_{SM} +(\left|(U_L)_{11}\right|^2-1)A_{1/2}(\tau_\tau)+\left|(U_L)_{12}\right|^2 A_{1/2}(\tau_{\Sigma_{1}}) 
+\left|(U_L)_{13}\right|^2 A_{1/2}(\tau_{\Sigma_{2}})\right|^2}
{\left| {\cal{A}}^{\gamma\gamma}_{SM}\right|^2} \\
&\simeq&
\dfrac{\left| {\cal{A}}^{\gamma\gamma}_{SM} + 2\theta^2\dfrac{1+r_h}{r_h} A_{1/2}(\tau_{\Sigma_{k}})
\right|^2}{\left| {\cal{A}}^{\gamma\gamma}_{SM}\right|^2} 
\simeq 1 - 4\theta^2\dfrac{1+r_h}{r_h} \dfrac{A_{1/2}(\tau_{\Sigma_{k}})}
{\left| {\cal{A}}^{\gamma\gamma}_{SM}\right|} \gtrsim 0.998~,
\ea
\label{ggM}
\end{equation}
where we took the maximal allowed values for the mixing angle and the form factor, $A_{1/2}(\tau_{\Sigma_k})\simeq 1.5$ for $M_{\Sigma_k}\simeq 100$ GeV. 
Note that $R_{\gamma\gamma}\simeq \mu_{\gamma\gamma}$, because the Higgs production rate and the total Higgs width are not significantly modified with respect to the SM.
Thus the diphoton signal strength can be slightly reduced (the fermionic part of the amplitude slightly increases and  interferes destructively with the $W$-loops), but only at a
few permil level.
For the $\gamma Z$ channel the new physics contribution can be written as 
${\cal{A}}^{\gamma Z}_{new} = {\cal{A}}^{\gamma Z}_{\Sigma,diag} + {\cal{A}}^{\gamma Z}_{\Sigma,off-diag}$. 
Using the results of Appendix \ref{HpZ}, the loops involving a single mass eigenstate give 
\begin{equation}\ba{rcl}
{\cal{A}}^{\gamma Z}_{\Sigma,diag} & \simeq & 
\Bigg[\left( \left|(U_L)_{11}\right|^2-1 \right) 
\dfrac{1-4s_w^2}{4c_w^2} A_{1/2}(\tau_\tau,\lambda_\tau) \\ 
& + &  \left|(U_L)_{12}\right|^2 
A_{1/2}(\tau_{\Sigma_1},\lambda_{\Sigma_1}) 
+ \left|(U_L)_{13}\right|^2 
A_{1/2}(\tau_{\Sigma_2},\lambda_{\Sigma_2})
\Bigg]~,
\ea\label{diagApz}\end{equation}
where we took the $Z$ couplings to the interaction eigenstates, thus neglecting corrections of higher order in the small mixing.
The loops involving two mass eigenstates give
\beq\ba{c}
{\cal{A}}^{\gamma Z}_{\Sigma,off-diag} \simeq 
-\sum_{k=2,3}\dfrac{|(U_L)_{11}|^2|(U_L)_{1k}|^2}{4c_w^2}   \times \\
\times \left[ \dfrac{M_{\Sigma_k}+m_\tau}{\sqrt{m_\tau M_{\Sigma_k}}}  
A_{1/2}(\tau_\tau,\lambda_\tau,\tau_{\Sigma_k},\lambda_{\Sigma_k})  -i \dfrac{M_{\Sigma_k}-m_\tau}{\sqrt{m_\tau M_{\Sigma_k}}}  
B_{1/2}(\tau_\tau,\lambda_\tau,\tau_{\Sigma_k},\lambda_{\Sigma_k})  \right]~.
\label{offApz}
\ea\eeq
Retaining only terms of order $\theta^2$ and neglecting $m_\tau/M_{\Sigma_k}$, the rate relative to the SM can be written as
 \begin{equation}
R_ {\gamma Z}\simeq 1- 4 \theta^2 \frac{1+r_h}{r_h} \frac{    
  A_{1/2}(\tau_{\Sigma_k},\lambda_{\Sigma_k})- \dfrac{1}{4c_w^2}   \sqrt{\dfrac{M_{\Sigma_k}}{m_\tau}}
A_{1/2}(\tau_\tau,\lambda_\tau,\tau_{\Sigma_k},\lambda_{\Sigma_k})}{ \left| {\cal{A}}^{\gamma Z}_{SM} \right|}~.
\label{RpZ-seesaw}
\end{equation}
We neglected the $B_{1/2}$ term, as it interferes only with the very small imaginary part of the SM amplitude.
The diagonal and off-diagonal form factors have comparable size, 
$A_{1/2}(\tau_{\Sigma_k},\lambda_{\Sigma_k})\simeq 1.3$ and $\sqrt{M_{\Sigma_k}/m_\tau}
A_{1/2}(\tau_\tau,\lambda_\tau,\tau_{\Sigma_k},\lambda_{\Sigma_k}) \simeq 1$,
where we took the large $M_{\Sigma_k}$ limit.
	Replacing the maximal allowed value for the mixing, we find $R_{\gamma Z} \gtrsim 0.998$, with a suppression at the few permil level, of the same order as for $R_{\gamma\gamma}$.
\\

Finally, let us note that, with two or more sterile neutrinos $N_i$ (or triplets $\Sigma_i$) the CP symmetry can be broken. 
In general, the Higgs couplings to the fermion mass eigenstates are not real,
and the off-diagonal couplings of the $Z$-boson can be complex as well. This does not modify any of the above results,
because the CP violating effects vanish in the limit $m_{\nu_l}/m_{\nu_{hk}}\rightarrow 0$ (and $m_\tau/m_{\Sigma_k}\rightarrow 0$ in the triplet case),
therefore they are subleading.

\subsection{One vector-like lepton}\label{VLL}

Let us consider the addition to the SM of one vector-like lepton (VLL). 
The four different possibilities are a weak singlet $E$, a weak doublet $L$ or $\Lambda$, a weak triplet $\Delta$,
whose charges are displayed in Table \ref{twotable}. 
As usual, we restrict ourself to mixing with the third SM family, i.e. with $\tau$ and $\nu_\tau$. 
The SM Lagrangian is extended by 
\beq
-{\cal L}_{\psi} = \lambda_\psi  \overline{l_L} H \psi_R + M_\psi \overline{\psi_L} \psi_R +h.c. ~,\quad \psi = E,\Delta~,
\eeq
\beq
-{\cal L}_{\psi} = \lambda_\psi  \overline{\psi_L} H (\tilde{H}) \tau_R + M_\psi \overline{\psi_L} \psi_R +h.c. ~,\quad \psi = L (\Lambda)~,
\eeq
where the $SU(2)_w$ contractions are understood (see appendix \ref{treeH} for details). 
In the case of $E$ ($L$) one could write an additional mass term $m_E \overline{E_L} \tau_R$ ($m_L \overline{l_L} L_R$), but such term can be removed
by choosing conveniently the basis for  the two fields $E_R$ and $\tau_R$ ($l_L$ and $L_L$), that have identical charges.
Thus, in each case there are only two real parameters: the vector-like mass $M_\psi$ and the Yukawa coupling $\lambda_\psi$; the mixing
among the new leptons and the SM ones vanishes for vanishing $\lambda_\psi$. There is no CP violation.

The components of each multiplet $\psi$ are listed in \eq{newleptons}. 
The doubly-charged  component $F$  does not mix as there is no SM counterpart with $Q=2$, therefore $m_F^2= M_\psi^2$. 
The $Q=1$ component $E$ mixes with the SM $\tau$ to form the two physical mass eigenstates $\tau'$ and $\tau$.
The mass matrix is given by 
 \begin{equation}
 {\cal{M}}_{e}=\begin{pmatrix}
 \lambda_{\tau}\frac{v}{\sqrt{2}} &  \kappa_\psi \lambda_\psi  \frac{v}{\sqrt{2}} \\ 0 & M_\psi
 \end{pmatrix}~,~\psi = E,\Delta~,
 \quad
  {\cal{M}}_{e}=\begin{pmatrix}
 \lambda_{\tau} \frac{v}{\sqrt{2}} & 0 \\   \kappa_\psi \lambda_\psi  \frac{v}{\sqrt{2}}   & M_\psi
 \end{pmatrix}~,~ \psi=L,\Lambda~. 
 \label{VLL-mass-matrices}
 \end{equation}
The $SU(2)_w$ Clebsch-Gordan coefficient $\kappa_\psi$ is equal to one, except in the triplet case, $\kappa_\Delta = \sqrt{1/3}$.
The rotation to the mass basis can be parametrized as 
\begin{equation}
 {\cal{M}}_{e}= V_L  \begin{pmatrix} m_\tau & 0 \\ 0 & m_{\tau'}\end{pmatrix} V_R^T~, 
 \quad V_{L,R}=\begin{pmatrix} c_{L,R} & s_{L,R} \\ -s_{L,R} & c_{L,R}
 \end{pmatrix} 
 ~.
 \label{VLL-rotation}
\end{equation}
The triangular mass matrix structure of \eq{VLL-mass-matrices} implies some strict relations among the mixing angles and the mass eigenvalues. For the case of $L$ and $\Lambda$, one finds 
\begin{equation}
\tan\theta_L = \frac{m_\tau}{m_{\tau'}} \tan\theta_R \ll \tan\theta_R~,\quad  s_L=\frac{m_\tau}{M_\psi} s_R ~, \quad c_L=\frac{m_{\tau'}}{M_\psi} c_R ~.
\label{suppressed-mixing}
\end{equation}
For the case of $E$ and $\Delta$, the same relations hold with $L\leftrightarrow R$.
Note that direct searches of charged leptons at LEP  \cite{Achard:2001qw}   require $m_{\tau'} \gtrsim 100$ GeV , therefore one angle is at least two orders of magnitude smaller than the other. In the following we will refer only to the dominant mixing angle $\theta_\psi$
for each $\psi$, dropping the subscript $L,R$ on (co)sines. 
Note that $m_{\tau'}^2 \simeq M_\psi^2/c_\psi^2 \ge M_\psi^2$. 
The neutral component $N$ does not mix with the SM neutrino $\nu_\tau$ in the case of $L$. In the case of $\Delta$ there is mixing in the neutral sector, described by 
\begin{equation}
{\cal M}_{\nu}=
\begin{pmatrix}
\sqrt{\frac 13}\lambda_\Delta v \\ M_\Delta
 \end{pmatrix}
 = U_L \begin{pmatrix} 0 \\ m_{\nu'} \end{pmatrix}~, \quad U_L =\begin{pmatrix} \widetilde{c} & \widetilde{s} \\ -\widetilde{s} & \widetilde{c}
 \end{pmatrix}~,
\end{equation}
with $\widetilde{s}^2 \simeq 2s_\Delta^2/(1+s_\Delta^2)\ge s_\Delta^2$.
One neutrino remains massless, while the second acquires a mass $m_{\nu'}^2\simeq (1+s_\Delta^2)m_{\tau'}^2\ge m_{\tau'}^2$.
In summary, the tree-level spectrum of heavy leptons satisfies
\begin{equation}
\ba{ll}
M_E \le m_{\tau'} ~~(\psi=E)~,  & m_F = M_\Delta \le m_{\tau'} \le m_{\nu'}~~(\psi=\Delta)~, \\
m_{\nu'} = M_L \le m_{\tau'}~~(\psi=L)~, &
m_{F} = M_\Lambda \le m_{\tau'}~~(\psi=\Lambda)~. \\
\ea
\end{equation}
with the mass splitting controlled by the mixing, $\Delta m^2(\psi) \sim s_\psi^2 M_\psi^2$.

Let us briefly discuss the collider bounds on $M_\psi$. In first approximation one can neglect the mass splitting. 
It is possible to recast some LHC multi-lepton searches to put bounds on VLLs. 
The limits on $M_\psi$ strongly depend on the SM generation that couples to the heavy leptons. 
For couplings only to the third one and for the doublet $L$, Ref.~\cite{Falkowski:2013jya} reports $M_L\gtrsim 280$ GeV,  
while the LEP limit remains more constraining in the case of the singlet $E$, $M_E\gtrsim 100$ GeV. 
For the exotic doublet $\Lambda$ with a doubly-charged component, Ref.~\cite{Altmannshofer:2013zba} reports $M_\Lambda \gtrsim 320$ GeV. 
To the best of our knowledge, no similar analysis is available for the triplet $\Delta$. 
We expect a bound comparable or slightly stronger than to the one for $\Lambda$.
These bounds only apply for promptly decaying particles. We will only consider this possibility, because heavy leptons  become long-lived ($c \tau \gtrsim 1 m$) 
for a tiny mixing $s_\psi \simeq 10^{-8}-10^{-9}$, and the mixing suppresses  all the deviations from the SM that we are interested in.  
More details on the collider phenomenology of $\Lambda$ and $\Delta$ can be found in Refs.~\cite{Ma:2014zda} and ~\cite{Delgado:2011iz,Ma:2013tda}, respectively.

It is mandatory to require that the Yukawa coupling $\lambda_\psi$ lies in the perturbative regime, $|\lambda_\psi|\ll 4\pi$. 
This consistency requirement translates into an upper bound
on the product of the heavy lepton mass and the mixing angle, 
\begin{equation}
|\lambda_\psi| 
\simeq \left|\frac{\sqrt{2}}{\kappa_\psi} \frac{m_{\tau'}}{v} s_\psi\right| \ll 4\pi~.
\label{pertVLL}\end{equation}
The perturbativity constraint on the SM Yukawa coupling $\lambda_\tau$ is satisfied a fortiori. 
The non-zero couplings of the physical Higgs boson to the charged mass eigenstates, using the convention of \eq{hff1}, are given by 
\begin{equation}
y_{\tau\tau}=c_{\psi}^2 \frac{m_\tau}{v}~,\quad 
y_{\tau'\tau'}=s_{\psi}^2 \frac{m_{\tau'}}{v}~,\quad
y_{\tau\tau'}=c_{\psi} s_{\psi} \frac{m_\tau+m_{\tau'}}{2 v}~,\quad 
\widetilde{y}_{\tau\tau'} = \pm c_{\psi} s_{\psi} \frac{m_{\tau'}-m_{\tau}}{2 v i}~. 
\label{tauH}
\end{equation}
where the plus (minus) sign holds in the case of $E$ and $\Delta$ ($L$ and $\Lambda$). In the case of $\Delta$, there are also non-zero couplings to neutral leptons,
\begin{equation}
y_{\nu'\nu'}=\widetilde{s}^2 \frac{m_{\nu'}}{v}~, \quad y_{\nu\nu'}=\widetilde{c} \widetilde{s} \frac{m_{\nu'}}{2 v}~,\quad 
\widetilde{y}_{\nu\nu'}=\widetilde{c} \widetilde{s} \frac{m_{\nu'}}{2 v i} ~. 
\label{neuH}\end{equation}

Important constraints come from the $Z$-decays into SM leptons. 
The couplings of the fermion mass eigenstates  to the $Z$ are defined in \eq{Zmass}.
The couplings to the SM leptons $\tau_L$, $\nu_{\tau L}$ and $\tau_R$ are modified if they mix with new leptons with a different weak isospin $T_{3}$.   
Neglecting $(m_\tau/m_Z)$-corrections, at tree level one finds
\begin{equation}
R(Z\rightarrow \tau^+ \tau^-) 
\simeq\frac{(g^V_{\tau\tau})^2+(g^A_{\tau\tau})^2}{(g^{V,SM}_{\tau\tau})^2+(g^{A,SM}_{\tau\tau})^2} ~,
\end{equation}
where $g^{V,A}_{\tau\tau}$
receive a correction of order $s_\psi^2$ with respect to the SM. 
The experimentally allowed range given in \eq{Rlep}
implies an upper bound on the mixing angle, for any VLL: we find $s_{E,\Delta} \lesssim 6.0 \cdot 10^{-2}$ and $s_{L,\Lambda}\lesssim 6.7\cdot 10^{-2}$. 
In addition, in the case of $\Delta$ there is a correction to the $Z$-coupling to neutrinos, and thus to the $Z$-invisible width, 
\begin{equation}
R(Z\rightarrow inv) 
\simeq \frac 23 + \frac 13 \frac{(g^V_{\nu\nu})^2+(g^A_{\nu\nu})^2}{(g^{V,SM}_{\nu\nu})^2+(g^{A,SM}_{\nu\nu})^2}~.
\end{equation}
The couplings $g^{V,A}_{\nu\nu}$ receive a correction of order $\tilde{s}^2$, that leads to a comparable limit $s_\Delta \lesssim 8.2 \cdot 10^{-2}$.
Extracting the couplings of $W^\pm_\mu$, $W^3_\mu$ and $B_\mu$ from \eq{Vmass}, one can calculate the $S$ and $T$ parameters with the general formulas in appendix \ref{ST}.
We find, at leading order in the mixing angle, 
\begin{equation}
T\simeq \frac{1}{16 \pi c_w^2 s_w^2} s_{\psi}^4 \left(a^T_\psi \frac{m_{\tau'}^2}{ m_Z^2}\right) ~,\quad\quad 
S\simeq \frac{1}{6 \pi} s_{\psi}^2  \left(a^S_\psi + b^S_\psi \log \frac{m_{\tau'}^2}{m_Z^2}\right) 
\label{TS1VLL}
\end{equation}
where $a^T_\psi$, $a^S_\psi$ and $b^S_\psi$ are numerical coefficients of order one. 
Taking into account the upper bound $s_\psi\lesssim 0.06$ from $Z\rightarrow \tau^+\tau^-$, as well as the perturbativity bound from \eq{pertVLL}, $(m_{\tau'}/m_Z)s_\psi \lesssim 10$,
we checked that $S$ and $T$ always lie in the allowed ellipse of Fig.~\ref{figST}.

Coming to the Higgs boson signals, we first recall that all the dominant Higgs production channels at the LHC are not affected by the new leptons, as they leave the Higgs couplings to
gluons and quarks unchanged. The total Higgs width also receives negligible corrections, as new leptons affect only the partial widths $\Gamma(h\rightarrow \alpha)$, 
for $\alpha = \tau^+\tau^-, \gamma\gamma,\gamma Z$. 
Therefore the Higgs signal is given by the ratio of partial widths in the model w.r.t. the SM, $\mu_\alpha \simeq R_\alpha$.
The tree-level Higgs decays are directly controlled by the couplings in \eq{tauH} and \eq{neuH}, in particular
\begin{equation}
R_{\tau\tau} =(1-s_{\psi}^2)^2 \gtrsim 0.99~, 
\label{1VLL-htautau}
\end{equation}
where we used the bound from $Z\rightarrow \tau^+\tau^-$.
There is also the marginal possibility that the new leptons are lighter than the Higgs boson, thus opening the channels $h\rightarrow \tau \tau'$ and $h\rightarrow \nu\nu'$. However, direct searches seem to allow the singlet $E$ only to be sufficiently light. 
Using equation (\ref{Gammaff}) and neglecting $m_\tau/m_{\tau'}$, we find
\begin{equation}
\Gamma(h\rightarrow \tau^+ \tau'^-)
\simeq\frac{c_E^2 s_E^2m^2_{\tau'} }{16 \pi v^2}  m_h  \left( 1-\frac{m_{\tau'}^2}{m_h^2} \right)^2 \lesssim 0.2 {\rm~MeV}~.
\end{equation}
Note that both the couplings $y_{\tau \tau'}$ and $\tilde{y}_{\tau \tau'}$ contribute equally to the decay width, see \eq{tauH}.
We maximised the product $s_E^2 m_{\tau '}^2$ by taking $s_E= 6\cdot 10^{-2}$ and $m_{\tau '}=100$ GeV.
As the SM total Higgs width is $\Gamma_h\simeq 4.1$ MeV, an enhancement of order 5\% may be possible. 
Note that experimental searches  at the LHC concentrated on  $h\rightarrow \tau \tau$  \cite{Aad:2015vsa,Chatrchyan:2014nva}, 
$\tau \mu$  \cite{Aad:2015gha,Khachatryan:2015kon}   and $\mu\mu$  \cite{Aad:2014xva,Khachatryan:2014aep}.   
These channels are suppressed due to the small masses of the SM leptons, in contrast with the  Higgs decays into a SM lepton plus a heavy lepton.
It would be interesting to perform a dedicated search for this channel.

For the photon-photon channel we find
\begin{equation}
R_{\gamma\gamma}=\frac{\left|{\cal{A}}_{SM}^{\gamma\gamma}+s_{\psi}^2 \left[ A_{1/2}(\tau_{\tau'})-A_{1/2}(\tau_{\tau}) \right] \right|^2}{\left|{\cal{A}}_{SM}^{\gamma\gamma}\right|^2} 
\simeq 1- 2 s_\psi^2 \frac{ A_{1/2}(\tau_{\tau '})}{\left|{\cal{A}}_{SM}^{\gamma\gamma} \right|}~,
\label{1VL-diphoton}
\end{equation}
where the form factors are defined in appendix \ref{Hpp}.
The addition of a VLL amounts to an additional $\tau'$-loop and a modified $\tau$-loop, that interfere destructively with the $W$-loops. 
Maximising the mixing and choosing $m_{\tau'}=100$ GeV (the form factor decreases for larger masses),
we find $\delta R_{\gamma\gamma}\simeq -1.9 \cdot 10^{-3}$, a permil reduction of the signal strength.  
For the $\gamma Z$ channel the relevant $Z$ couplings, $g^{V,A}_{\tau\tau,\tau\tau',\tau'\tau'}$, receive corrections of order $s_\psi^2$ relatively to their unmixed values, 
as follows from \eq{Zmass}. At leading order in the small mixing and neglecting $m_\tau$,
we find
\begin{equation}
R_{\gamma Z} \simeq 1+ 2 s_\psi^2  \frac{\left(T^3_{E,\psi} + s_w^2 \right)  A_{1/2}(\tau_{\tau'}, \lambda_{\tau'}) \pm \dfrac{1}{4 }\sqrt{\dfrac{m_{\tau '}}{m_{\tau}}} A_{1/2}\left(\tau_{\tau}, \lambda_{\tau},\tau_{\tau'},\lambda_{\tau'} \right)}{ c_w^2 \left|{\cal{A}}_{SM}^{\gamma Z}\right|} ~,
\label{gZ1VLL}
\end{equation}
where $T^3_{E,\psi}$ is the isospin of the $Q=-1$ component of the multiplet $\psi$,
and the plus (minus) sign in front of the off-diagonal term corresponds to the case $\psi=L$ ($\psi=E,\Lambda,\Delta$).
As a consequence, the diagonal and off-diagonal terms always interfere destructively. 
The relative magnitude of the form factors is given below \eq{RpZ-seesaw}.
The size of the correction change depending on the VLL under consideration, but it is always very small.
The maximal deviation is obtained for $\Lambda$, with $\delta R_{\gamma Z} \simeq 1.3 \cdot 10^{-3}$.

 \subsection{Two vector-like leptons (including $\tau$ compositeness)}
 \label{2VLL}

Let us consider a SM extension by two VLLs.  They may couple to each other by a Yukawa interaction or not.\\

\noindent $\bullet$ \textbf{\textit{Two VLLs not coupled to each other.}}
In this case, each VLL must be a consistent extension of the SM by itself, therefore it should have the quantum numbers of $E$, $L$, $\Lambda$ or $\Delta$, that are displayed
in Table~\ref{twotable}.
The six possible pairs of VLLs decoupled from each other are $(\psi,\psi')=(E,E'),(L,L'),(\Lambda,\Lambda'),(\Delta,\Delta'),(E,\Delta)$ and$(L,\Lambda) $. In the first four cases the additional mass term $\psi\psi'$ can be rotated away without loss of generality.
The phenomenological effects are a trivial sum of those discussed in section \ref{VLL} for a single VLL, with one noticeable exception.

When $L$ and $\Lambda$ have the same Yukawa coupling to the SM and the same mass, the Lagrangian
\beq 
-{\cal{L}}_{L,\Lambda}= \frac{\lambda_\psi}{\sqrt{2}} \begin{pmatrix} \overline{L_L} & \overline{\Lambda_L} \end{pmatrix}
\begin{pmatrix} H \\ \tilde{H} \end{pmatrix} \tau_R
+ M_\psi \begin{pmatrix} \overline{L_L} & \overline{\Lambda_L} \end{pmatrix}
\begin{pmatrix}
L_R \\ \Lambda_R
\end{pmatrix} + h.c.  ~
\label{Lagrangian-2VLL-custodial}
\eeq
preserves a global $SU(2)_L\times SU(2)_R$ symmetry. In this custodial  limit the corrections to the $T$ parameter vanish, and those to the coupling $Z\tau_R\tau_R$ vanish as well 
\cite{Agashe:2006at}. This is the smallest set of tau custodians \cite{delAguila:2010es,Carmona:2013cq}.  
A linear combination of the charge-one components of $L$ and $\Lambda$, $\tau'' \equiv (E^{(L)}-E^{(\Lambda)})/\sqrt{2}$, does not mix.
The orthogonal combination, $E\equiv (E^{(L)}+E^{(\Lambda)})/\sqrt{2}$, mixes with the SM exactly as shown in Eqs.~(\ref{VLL-mass-matrices})-(\ref{suppressed-mixing}), to form the mass eigenstates $\tau'$ and $\tau$.
The spectrum reads $m_{\tau''}=m_N=m_F= M_\psi \le m_{\tau'} \simeq M_\psi/c_R$.
As discussed in section \ref{VLL}, direct searches already require all these states to be heavier than the Higgs boson. 
Thanks to the custodial symmetry, the $Z$ couplings to leptons do not constrain the right-handed mixing $s_R$ between $\tau$ and $\tau'$: one has $\delta g_{\tau\tau}^R= 0$ and 
$\delta g_{\tau\tau}^L = s_L^2/2=(m_\tau/M_\psi)^2 s_R^2/2$, that is negligibly small. 
The $Z\nu\bar{\nu}$ coupling is SM-like as well.
The $T$ parameter is  almost SM-like as no additional sources of custodial breaking are introduced, and the correction to the $S$ parameter is within the experimental range.

The most stringent constraint on $s_R$ comes from  $R_{\tau\tau}\simeq (1-s_R^2)^2$. 
Using the $3 \sigma$ lower bound $R_{\tau\tau}\gtrsim 0.2$ (see Table~\ref{threetable}),
one finds $s_R\lesssim 0.7$.
Indeed, the mixing can be large and reduce significantly the $h\tau\tau$ coupling. 
As a consequence, the total Higgs width may be slightly reduced and, correspondingly, the signal strength for the other Higgs decay channels, defined in  \eq{SS},  may augment by a factor 
$\Gamma_h^{SM}/\Gamma_h \lesssim 1.04$.
The deviation in the $\gamma\gamma$ channel has the same form as in \eq{1VL-diphoton}: 
imposing the constraint from $h\rightarrow \tau\tau$, one finds a lower bound $\mu_{\gamma\gamma}\gtrsim 0.86$, that is close to the present experimental sensitivity.
Coming to the $\gamma Z$ channel, the loop involving both $\tau$ and $\tau'$ vanishes when one neglects the tiny left-handed mixing $s_L$, 
because the custodial symmetry imposes $g^R_{\tau \tau'}=0$. Then, \eq{gZ1VLL} reduces to 
\begin{equation}
R_{\gamma Z} \simeq 1+ 2 s_R^2  \tan^2\theta_w \frac{  A_{1/2}(\tau_{\tau'}, \lambda_{\tau'}) }{|{\cal{A}}_{SM}^{\gamma Z}|} ~,
\end{equation}
with a maximal correction $\delta \mu_{\gamma Z} \simeq 0.12$.
In the near future the increasing experimental precision on $\mu_{\tau\tau}$ can further constrain or eventually determine the mixing parameter $s_R$ in this custodial limit.
\\

\noindent $\bullet$ \textbf{\textit{Two VLLs coupled to each other,  not mixing with the SM fermions.}}
Next, we have to discuss the case of two VLLs coupled through a Yukawa interaction. 
The most general assignment for their four chiral components is
\begin{equation}
\psi_{1L},\psi_{1R} \sim (1, R_w, Y)~,\qquad 
\psi_{2L},\psi_{2R}\sim(1, R_w+1,Y+\frac{1}{2})~, 
\label{set2VL}
\end{equation}
with the Lagrangian
\begin{equation}
-{\cal {L}}_{\psi_1\psi_2}= \lambda_{12} \overline{\psi_{1L}}\tilde{H}\psi_{2R} + \lambda_{21} \overline{\psi_{2L}} H \psi_{1R} +M_1 \overline{\psi_{1L}}\psi_{1R} + M_{2} \overline{\psi_{2L}}\psi_{2R} + h.c. ~.
\label{2VLLs-Lagrangian}
\end{equation}
The four phases of $\lambda_{12}$, $\lambda_{21}$, $M_1$ and $M_{2}$ cannot be all rotated away: one phase is physical and allows for CP violation.
In the special case $Y=0$ ($Y+1/2=0$) and $R_w$ odd (even), one should add Majorana mass terms for $\psi_{1L,R}$ ($\psi_{2L,R}$): we postpone to section \ref{MVL} the discussion of 
sets formed by one VLL plus Majorana leptons. 
For a few other values of $R_w$ and $Y$, displayed in \eq{mixingL}, interaction terms between $\psi_{1,2}$ and the SM leptons are allowed and should be added to the Lagrangian.
We discuss first the no-mixing case and postpone to the end of this section the discussion of the mixing with the SM.
In the absence of mixing, the lightest new lepton $\psi_{light}$ is stable, at least at the renormalizable level. If $Y$ is an integer multiple of $1/2$, $\psi_{light}$ 
may decay into a SM lepton
through some higher dimensional operator. For all other values of $Y$, this state is absolutely stable and it has non-zero electric charge.  
Collider searches put a lower bound on the mass of stable heavy leptons as a function of their charge $Q$,
see the discussion in section \ref{chiral}
and the limits in Table~\ref{6table}.
The effect of the set of fermions in \eq{set2VL}  on $h\rightarrow \gamma\gamma$ was studied e.g. in Ref.~\cite{Almeida:2012bq}.

Let us begin by analysing the case $R_w=1$. There is one state with $Q=Y+1$ and mass $M_2$, and two states with $Q=Y$ that mix, with mass matrix
\begin{equation}
{\cal{M}}_Y=\begin{pmatrix}
M_{1} & m_{12} \\ m_{21} & M_{2}\end{pmatrix}~, 
\qquad m_{12}=\frac{\lambda_{12} v}{\sqrt{2}}~, ~~ m_{21}=\frac{\lambda_{21} v}{\sqrt{2}}~.
\label{2VLLS-mass-matrix}
\end{equation}
As ${\cal M}_Y$ is the most general $2\times 2$ matrix, it is useful to parametrize it in terms of the five physical parameters,
\begin{equation}
{\cal{M}}_Y=U_L \begin{pmatrix} m_{1} & 0 \\ 0 & m_{2} \end{pmatrix} U_R^\dagger, ~~~~U_L=\begin{pmatrix}
c_L & s_L \\ -s_L & c_L
\end{pmatrix} \begin{pmatrix}
e^{i \varphi} & 0 \\ 0 & 1
\end{pmatrix}, ~~~~~U_R=\begin{pmatrix}
c_R & s_R \\ -s_R & c_R
\end{pmatrix} ,
\label{2VLLS-mixing}
\end{equation}
where $m_{1,2}$ are the real and positive masses of the eigenstates $f_{1,2}$, the mixing angles $\theta_L$ and $\theta_R$ vary between $0$ and $\pi/2$, and the CP violating phase
$\varphi$ varies between $0$ and $2\pi$.
The only restriction comes from the perturbativity of the Yukawa couplings, 
\begin{equation}
|\lambda_{12}|= \frac{\sqrt{2}|m_{2} s_L c_R - m_{1} e^{i \varphi}  c_L s_R |}{v} \ll 4\pi~, ~~~ |\lambda_{21}|=\frac{\sqrt{2}| m_{2} c_L s_R -m_{1} e^{i \varphi} s_L c_R|}{v}\ll 4\pi ~.
\label{2VLLs-Yukawa}
\end{equation}
These relations imply e.g. an upper bound on the masses for fixed values of the mixing angles. Vice versa, as the masses become larger and larger, the mixing angles vanish
and the new fermions decouple from the EW scale.

\begin{figure}[bt]
\begin{center}
\includegraphics[scale=0.35,trim= 0 0 0 0]{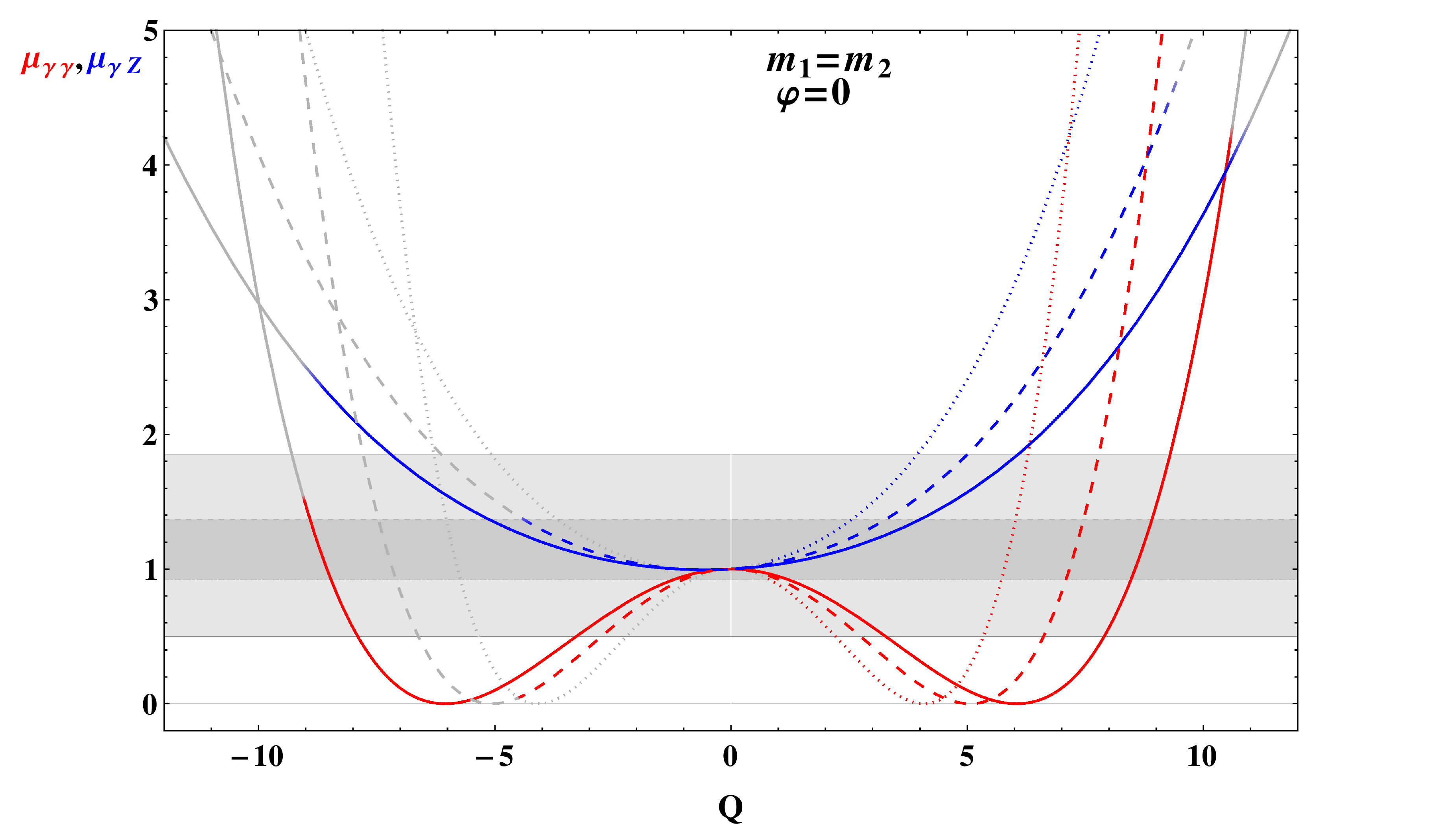}
\caption{Signal strengths $\mu_{\gamma\gamma}$ (red) and $\mu_{\gamma Z}$ (blue) in the case of two 
VLLs  $\psi_1\sim(1,1,Q)$ and $\psi_2\sim(1,2,Q+1/2)$, as a function of $Q$.
We chose the following mass matrix parameters, defined by \eq{2VLLS-mixing}: $m_{1} = m_{2}=800$ GeV, $\varphi=0$, and three values for the relevant mixing angle, 
$\theta_L -\theta_R = \pi/8$ (dotted), $\pi/10$ (dashed), $\pi/12$ (solid). The grey parts of the curves are excluded by $S$ and $T$, see \eq{S-T-2VLLs}.
The shaded horizontal band is the presently allowed range for $\mu_{\gamma\gamma}$ at $1\sigma$ (dark) and $3\sigma$ (light).} 
\label{fig-2VLLs}
\end{center}
\end{figure}

The Higgs boson couplings to $f_{1,2}$ are directly obtained from \eq{hff1}. Taking for illustration
the limit where $f_1$ and $f_2$ are mass-degenerate, one has $|M_1|=|M_2|\equiv M$, $|m_{12}|=|m_{21}|\equiv \mu$ and $m_{1}=m_{2}=\sqrt{M^2+\mu^2}\equiv m_\psi$. 
In this case the contribution to the amplitude for $h\rightarrow \gamma\gamma$ is 
\begin{equation}
{\cal{A}}^{\gamma\gamma}_{f_1,f_2}= 2Q^2  F\left(\theta_L,\theta_R,\varphi \right) A_{1/2} \left(\tau_{\psi} \right)~, ~~~~ F\left(\theta_L,\theta_R,\varphi \right)
= s_L^2 c_R^2+ c_L^2 s_R^2 -2 c_L s_L c_R s_R \cos \varphi ~.
\label{2VLL-pp}
\end{equation}
The CP-odd contribution vanishes because $\tilde{y}_{11}=-\tilde{y}_{22}$ in the degenerate limit.
The perturbativity conditions in \eq{2VLLs-Yukawa} reduce to $F(\theta_L,\theta_R,\pi)\ll 8\pi^2 v^2/m_\psi^2$. 
The interference with the SM is destructive as ${\cal{A}}^{\gamma\gamma}_{f_1,f_2}\geq 0$. There are two allowed regions of parameters:
\begin{itemize}
\item[(i)] A SM-like region for small $Q$: the smallness of the charge ensures a small, negative departure from the SM. 
\item[(ii)] A cancellation region at large $Q$: for $Q^2\simeq 4.8/F(\theta_L,\theta_R,\varphi )$, the rate is accidentally close to the SM 
as ${\cal{A}}^{\gamma\gamma}_{f_1,f_2}\simeq -2 {\cal{A}}^{\gamma\gamma}_{SM}$.   
\end{itemize}
This behaviour is illustrated in Fig.~\ref{fig-2VLLs} for the CP conserving case $\varphi=0$, where
$F(\theta_L,\theta_R,0)=\sin^2(\theta_L - \theta_R)$. 
Indeed the amplitude grows from $\varphi=0$ to $\varphi=\pi$, as $F(\theta_L,\theta_R,\pi)=\sin^2(\theta_L + \theta_R)$.
Further constraints on the parameters come from the EW precision tests.  
In the limit $m_\psi \gg m_Z$ we find 
\begin{equation}
S\simeq\frac{1}{6 \pi} \left[ F\left(\theta_L,\theta_R,\varphi \right)+4 \left(Q+\frac{1}{2}\right) \ln \frac{m_\psi^2}{M^2} \right]~, ~~ T\simeq \frac{1}{8 \pi s_w^2 c_w^2 m_Z^2} \left[m_\psi^2-3 M^2 + \frac{2 M^4}{m_\psi^2-M^2} \ln \frac{m_\psi^2}{M^2} \right]  ~,
\label{S-T-2VLLs}
\end{equation}
where $M$ is the mass of the unmixed state with $Q=Y+1$.
Therefore,  for fixed values of the mixing parameters, $S$ and $T$ constrain the charge $Q$, as illustrated in Fig.~\ref{fig-2VLLs}.

It is interesting to analyse the value of $\mu_{\gamma Z}$ in the allowed space of parameters.
Using Eqs.~(\ref{Zmass}) and (\ref{CP-even}), the CP-even amplitude for $h\rightarrow \gamma Z$ is given by 
\begin{equation}
{\cal{A}}^{\gamma Z}_{f1,f2} = \frac{Q}{c_w^2} F\left(\theta_L,\theta_R,\varphi \right) \left(g^V_{11}+ g^V_{22}\right) 
A_{1/2} \left(\tau_{\psi},\lambda_{\psi} \right)~, ~~~~
g^V_{11} + g^V_{22} = -\frac 12 - 2 Q s_w^2~.
\label{2VLL-Zcouplings}
\end{equation}
Note that the loops involving both $f_1$ and $f_2$ vanish as they are proportional to the Higgs coupling $y_{12}$,  or to the form factor $B_{1/2}$ defined in \eq{B12tot},
and they both vanish for $m_1=m_2$. 
As the SM amplitude is negative, the new contribution interferes constructively as long as  $Q (1+4s_w^2 Q)>0$.  
The CP-odd amplitude $\tilde{\cal{A}}^{\gamma Z}_{f_1,f_2}$ is also non zero,  because $g^V_{11}\neq g^V_{22}$, 
and it can be sizeable for large values of $\sin \varphi$. 
Let us distinguish the two regions of parameters allowed by $\mu_{\gamma\gamma}$:
\begin{itemize}
\item[(i)] In the SM-like region at small $Q$, varying $(\theta_L-\theta_R)$ we find $-0.01\lesssim \delta \mu_{\gamma Z}\lesssim +0.08$. 
\item[(ii)] In the fine-tuned region at large $Q$ we find
$2.5 \lesssim \mu_{\gamma Z}\lesssim 3.2$ for $\varphi=0$. 
This range slightly depends on the sign of $Q$, see Fig.~\ref{fig-2VLLs}.
It is actually possible to obtain an even larger  $\mu_{\gamma Z}$, while keeping $\mu_{\gamma\gamma}$ close to one. 
For example,  taking for simplicity $s_R=0$,
the CP-odd amplitude reads
\begin{equation}
\tilde{\cal{A}}^{\gamma Z}_{f_1,f_2}=-4 Q^2 c_L s_L \tan^2 \theta_w \sin \varphi \,\tilde{A}_{1/2} \left(\tau_\psi ,\lambda_\psi \right) ~.
\label{2VLLs-pZ-odd}
\end{equation}
For $\sin \varphi$ of order one, this contribution becomes important and one can reach $\mu_{\gamma Z}\simeq 7$.
\end{itemize}

When one allows for the two mass eigenstates $f_{1,2}$ to be non-degenerate, $m_1 < m_2$, the amplitudes for the diphoton channel become
\beq\begin{array}{rcl}
{\cal{A}}^{\gamma \gamma}_{f_1,f_2} &\simeq& 2 Q^2 \left[s_L^2 c_R^2+ c_L^2 s_R^2 -\left(\dfrac{m_{1}}{m_{2}}+\dfrac{m_{2}}{m_{1}} \right) c_L s_L c_R s_R \cos \varphi \right] 
A_{1/2}\left(\tau_1 \right), \\
\tilde{\cal{A}}^{\gamma \gamma}_{f_1,f_2} &\simeq & 2 Q^2 \left(\dfrac{m_{2}}{m_{1}}-\dfrac{m_{1}}{m_{2}} \right) c_L s_L c_R s_R \sin \varphi ~\tilde{A}_{1/2}\left(\tau_1 \right).
\end{array}
\label{gg2L}\eeq
where we made the approximation $A_{1/2}\left(\tau_{1} \right)\simeq A_{1/2}\left(\tau_{2} \right)$, that is accurate for $4m_{1,2}^2\gg m_h^2$. 
For sufficiently large mass splitting the interference of ${\cal{A}}^{\gamma \gamma}_{f_1,f_2} $ with the SM can be constructive. 
In the $\gamma Z$ channel the analytic form of the amplitude becomes more involved, in particular the loops involving both $f_1$ and $f_2$ are non-zero, and the interference
with the SM strongly depends on the ratio $m_{1}/m_{2}$. 
One can tune the parameters to cancel the corrections to $\mu_{\gamma\gamma}$ in \eq{gg2L}, e.g. taking $\varphi=0$ and $m_1/m_2=(t_L/t_R)^{\pm 1}$.
For the same set of parameters large contributions to the $\gamma Z$ channel are possible.
For example one can reach $\mu_{\gamma Z}\simeq 2$ for $\theta_L \simeq \pi/6$, $\theta_R \simeq \pi/10$, $m_1/m_2\simeq 1.8$, $m_2\simeq 800$ GeV and $Q \simeq 9$.
This region is compatible with $S$, $T$ and all other constraints.

A similar analysis can be performed when $R_w=2$ or larger in \eq{set2VL}. In this case there are $N_w$ pairs of mixing states, with  $Q=-(N_w-1)/2+Y,\dots,(N_w-1)/2+Y$. 
For each such sector, the mass matrix is
\begin{equation}
{\cal{M}}_{Q}=\begin{pmatrix}
M_{1} & \kappa_{Q} m_{12} \\ \kappa_{Q} m_{21} & M_{2}\end{pmatrix}~,
\label{2VLLS-mass-matrix-Rw}
\end{equation}
where $\kappa_{Q_\alpha}$ is the Clebsch-Gordan coefficient coming from the $SU(2)_w$ contraction, determined by \eq{CGy}, therefore each sector is controlled by the same physical parameters. In other words, the two mass eigenvalues, the two mixing angles and the CP-violating phase of a given sector determine univocally the other sectors too.
The corrections to $\mu_{\gamma\gamma}$ and $\mu_{\gamma Z}$ are obtained summing over the contributions of $N_w$ sectors, each being qualitatively analog to the case $R_w=1$ analysed above.
Note that, however, one cannot take $Q\rightarrow 0$ to recover the SM limit, 
because there are at least two sectors with different values of $Q$. Of course, the SM is still recovered for small values of the mixing angles.
The two fine-tuned regions with $\mu_{\gamma\gamma}\simeq 1$ and large $\mu_{\gamma Z}$ are still possible. On the one hand, the different sectors can add up to realise ${\cal{A}}^{\gamma\gamma}_{new}\simeq -2 {\cal{A}}^{\gamma\gamma}_{SM}$. 
On the other hand, for $R_w=2$ we found a choice of mixing parameters such that 
${\cal{A}}^{\gamma\gamma}_{f_1,f_2}$ vanishes in both the sectors with $Q=Y\pm1/2$. 
\\

\noindent $\bullet$ \textbf{\textit{Two VLLs coupled to each other,  mixing with the SM fermions.}}
Finally, let us discuss the possible interactions between $\psi_{1,2}$ in \eq{set2VL} and the SM leptons. 
A non-zero mixing occurs if and only if $\psi_1$ and/or $\psi_2$ are identified with the states $E$, $L$, $\Lambda$ or $\Delta$ listed in Table~\ref{twotable}. 
There are six such cases,
\beq\ba{ll}
R_w=1: & E+L ~(Y=-1),\quad  E+\Lambda ~(Y=1), \\
R_w=2: & L+ \Delta ~(Y=1/2),\quad  \Lambda + \Delta  ~(Y=-3/2), \quad \Lambda + \Delta_G  ~(Y=3/2), \\
R_w=3: & \Delta+ \Omega ~(Y=-1),\quad  \Delta + \Omega_G ~(Y=1).
\label{mixingL}
\ea\eeq
There are three cases with an additional weak multiplet: a triplet  $\Delta_G =(E,F,G)\sim (1,3,-2)$, and two quartets $\Omega = (E^c,N,E,F)\sim (1,4,-1/2)$ and $\Omega_G = (N,E,F,G)\sim (1,4,-3/2)$, 
with $Q(G)=-3$. 
They do not couple directly to the SM leptons. In these three cases the $Q=2$ sector couples to the Higgs and therefore may contribute significantly to $h\gamma\gamma$ and $h \gamma Z$.
We have shown in section \ref{VLL} that the mixing angles between the SM leptons and $E$, $L$, $\Lambda$ or $\Delta$ must be very small,  
due to the strong constraints coming from the $Z\tau\tau$ couplings. 
With two VLLs the mass matrices become larger, but we expect the phenomenology to be qualitatively the same, up to possible fine-tuned cancellations in some observable. 
A crucial effect of the mixing is to make the new leptons decay into SM leptons.  
The components with $Q=2,3$ decay more slowly, since their decay chains require a virtual exchange of other components of the multiplet.
We already reviewed in section \ref{VLL} the direct bounds on heavy leptons with charge $Q=1,2$, decaying promptly into SM leptons.
We are not aware of any dedicated search for a $Q=3$ heavy lepton.

\begin{figure}[bt]
\begin{center}
\includegraphics[scale=0.9,trim= 70 720 0 75]{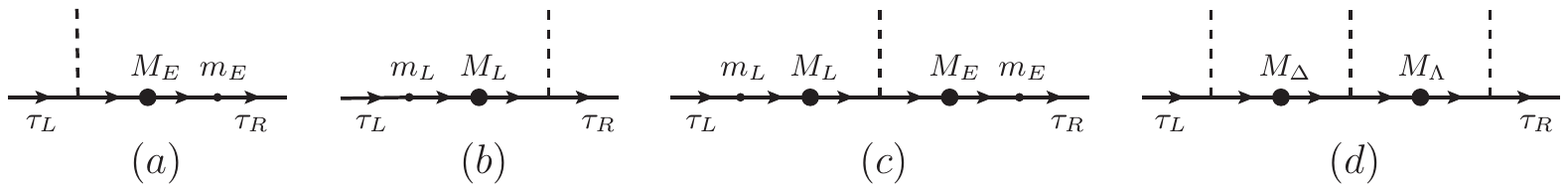}
\caption{Different ways to generate the $\tau$ mass through mixing between the SM leptons and VLLs. 
The dashed lines stand for Higgs vev insertions, the small dots represent a mass mixing between a SM lepton and a VLL, and the big dots correspond to the mass of a VLL. 
The case $(c)$ corresponds to the scenario of $\tau$ partial compositeness.}
\label{2VLLs-Yukawa-generation}
\end{center}
\end{figure}

A detailed analysis of the parameter space is worth only in the context of a specific, well-motivated model, and it goes beyond the scope of this paper. 
The case $E+L$ is analyzed in Ref.~\cite{Kearney:2012zi}.
The phenomenology of a fourth vector-like family of leptons, $L+E+N$, is studied in detail in Ref.~\cite{Joglekar:2012vc}. 
Here we comment only on the  interesting possibility to generate the $\tau$ mass entirely from the mixing with the VLLs, 
in the limit where the SM Yukawa coupling $\overline{l_{\tau L}}H\tau_R$ vanishes.
There are various ways to induce such coupling through mixing, that are illustrated in Fig.~\ref{2VLLs-Yukawa-generation}:
\bit 
\item[$(a)$] In the case of $E$ only, one can proceed through a Yukawa coupling
connecting $l_{\tau L}$ to $E_R$, followed by two singlet vector-like mass terms. 
\item[$(b)$] Analogously, with $L$ only, one employs two doublet vector-like mass terms and a Yukawa connecting $L_L$ to $\tau_R$. 
\item[$(c)$] In the case of $E+L$, one can employ a vector-like mass term both for the singlets and the doublets, with a Yukawa coupling involving only
the new fermions.
This case is particularly interesting, since it corresponds to the scenario of partial compositeness \cite{Kaplan:1991dc} 
in the $\tau$ sector: the SM leptons are elementary fields that mix linearly with composite VLLs, which couple in turn to a composite Higgs doublet. 
The SM leptons feel EWSB only through the mixing
with heavy composite leptons. The $Q=1$ mass matrix and its smallest eigenvalue take the form
\begin{equation}
{\cal{M}}_{e}= \begin{pmatrix}
0 & m_L & 0 \\
0 & M_L & \frac{\lambda_{LE} v}{\sqrt{2}} \\
m_E & \frac{\lambda_{EL}v}{\sqrt{2}} & M_E
\end{pmatrix}~, \quad\quad m_\tau\simeq  \frac{m_L}{M_L} \frac{m_E}{M_E} \frac{\lambda_{LE} v}{ \sqrt{2}}~.
\label{taucomp}
\end{equation}
The mixing with the heavy leptons also controls the deviations in the $Z$ couplings, 
\begin{equation}
 \delta g_{\tau\tau}^R\simeq \frac{1}{2} \left( \frac{m_E}{M_E}\right)^2~, \quad\quad \delta g_{\tau\tau}^L \simeq \frac{1}{2} \left( \frac{m_L}{M_L}\right)^2~.
 \end{equation} 
As these corrections are bounded by $R(Z\rightarrow \tau\tau)$, as shown in \eq{R-definition}, 
 we find that the physical value of $m_\tau$ can be generated for $\lambda_{EL}\gtrsim 2.5$, 
pointing indeed to a strong-coupling regime.
The phenomenology of $\tau$ partial compositeness is studied e.g. in Ref.~\cite{Redi:2013pga}.
\item[$(d)$] Finally, in the case $(\psi_1,\psi_2)=(\Lambda,E)$, $(L,\Delta)$ or $(\Lambda,\Delta)$, the $\tau$ mass can be induced by three Yukawa couplings, connecting $l_{\tau L}$ to $\psi_2$,
$\tau_R$ to $\psi_1$, and $\psi_1$ to $\psi_2$, respectively.
Focusing on $(\Lambda,\Delta)$ for definiteness, one finds
\begin{equation}
m_\tau\simeq    \frac{\lambda_\Delta v}{\sqrt{3} M_\Delta} \frac{\lambda_\Lambda v}{\sqrt{2} M_\Lambda} \frac{\lambda_{\Delta\Lambda } v}{\sqrt{6}}\simeq 
2\sqrt{\delta g_{\tau\tau}^R \delta g_{\tau\tau}^L}    \frac{\lambda_{\Delta\Lambda } v}{\sqrt 6} ~.
\label{mtau-d-case}
\end{equation} 
For $\lambda_{\Delta\Lambda}\gtrsim 4.5$ the physical value of $m_\tau$ can be generated.
\eit

 \subsection{Vector-like plus Majorana leptons (including {\mbox higgsinos} plus gauginos)} 
 \label{MVL}

In this section we consider the interplay between Majorana leptons and VLLs, related by one or more Yukawa couplings.  
If there were no such couplings, the phenomenology would reduce to a trivial addition of the effects of Majorana leptons,
see section \ref{seesaw}, and of VLLs, see sections \ref{VLL} and \ref{2VLL}. 
For reference, the smallest sets of this kind are formed by three (four) chiral multiplets: (two copies of) $N_R$ or $\Sigma_R$, plus a vector-like
pair $(E_L,E_R)$, $(\Lambda_L,\Lambda_R)$ or 
$(\Delta_L,\Delta_R)$.
\\

\noindent $\bullet$ \textbf{\textit{One VLL plus one Majorana lepton.}}
The most general set formed by one Majorana lepton interacting with one VLL pair is
\begin{equation}
\chi_R \sim (1,R_w,0) ~,\qquad \psi_{L},\psi_{R} \sim(1,R_w\pm 1,-1/2)~,
\label{1VL-plus-MAjorana-cases}
\end{equation}
with $N_w\ne 2+4n$ to avoid the global $SU(2)_w$ anomaly.
 The corresponding Lagrangian is
\begin{eqnarray}
-{\cal{L}}_{\chi\psi}&= \frac{1}{2} M_\chi \overline{\chi_R^c}\chi_R + M_{\psi}\overline{\psi_{L}}\psi_{R}  +
\tilde\lambda \overline{\psi_{L}}\tilde{H} \chi_R + \lambda \overline{\chi_R^c} H \psi_{R} + h.c. ~.
\label{1VL-1MAJ-Lagrangian}
\end{eqnarray}
The Majorana mass $M_\chi$ is absent in the case of even $SU(2)_w$ representations, $N_w=4n$: 
in this case there is a conserved
`new lepton' number, and all multiplet components combine into Dirac fermions. 
There is a unique physical complex phase, that can be associated to $M_\chi$, choosing $M_\psi$, $\tilde\lambda$ and $\lambda$ real.
Each component of the new leptons has a (demi-)integer charge $Q$, therefore it can decay to SM leptons, either through renormalizable interactions
or higher dimensional operators.
For $R_w=1$ or $3$, couplings to the SM leptons can be added to the Lagrangian, as we will discuss later.

In the case $\psi \sim (R_w+1)$, after EWSB one identifies: one pair of states with charges $Q=\pm(N_w+1)/2$, that combine in a VLL of mass $M_\psi$; 
three pairs of states with $Q=\pm(N_w-1)/2,\ \pm(N_w-3)/2,\ \dots$, down to $Q=\pm1$ ($\pm1/2$) for $N_w$ odd (even);
when $N_w$ is odd, three additional states with $Q=0$. The $3\times 3$ mass matrix in each sector takes the form
\beq
{\cal M}_{Q} = \left(\ba{ccc}
M_\chi & \kappa_{Q} \tilde\lambda v & \kappa_{-Q} \lambda v \\
\kappa_{-Q} \tilde\lambda v & 0 & M_\psi \\
\kappa_{Q} \lambda v & M_\psi & 0
\ea\right)~,
\label{m33}\eeq
where we chose $Q\ge 0$ and a basis with the components of charge $Q$ ($-Q$) on the left(right)-hand side of ${\cal M}_Q$.
The $SU(2)_w$ Clebsch-Gordan coefficients $\kappa_{\pm Q}$ are defined by \eq{CGy}, and they also include 
for convenience the factor $1/\sqrt{2}$ from the Higgs doublet vev, 
\beq
\kappa_{\pm Q} = \frac{1}{\sqrt 2}\sqrt{\dfrac{N_w+1\pm 2Q}{N_w(N_w+1)}}~.
\label{CG+}\eeq
In the case $\psi \sim (R_w-1)$, there are three pairs of states with $Q=\pm(N_w-3)/2,\pm(N_w-5)/2,\dots$, and three states with $Q=0$ when $N_w$ is odd. 
The $3\times 3$ mass matrix in each of these sectors has the  form of \eq{m33}, but with Clebsch-Gordan coefficients given by
\beq
\kappa_{\pm Q} = \frac{1}{\sqrt 2}\sqrt{\dfrac{N_w-1\mp2Q}{N_w(N_w-1)}}~.
\label{CG-}\eeq
There are also two pairs of states with $Q=\pm(N_w-1)/2$ and mass matrix
\beq
{\cal M}_{Q} = \left(\ba{cc}
M_\chi & \kappa_{-Q} \lambda v  \\
\kappa_{-Q} \tilde\lambda v &  M_\psi 
\ea\right)~.
\label{m22}\eeq
This is the same structure of \eq{2VLLS-mass-matrix}, that was extensively studied in section \ref{2VLL}. 

Note that each $3\times 3$ or $2\times 2$ sector depends on the same five parameters: two masses, two Yukawa couplings, and one physical phase. 
They determine all the mass eigenvalues and the mixing matrices.
We do not attempt a scan of the parameter space here.
The corrections to $S$, $T$, $\mu_{\gamma\gamma}$ and $\mu_{\gamma Z}$  from a sector with the mass matrix 
of \eq{m22} were analysed in section \ref{2VLL}. We expect corrections of the same order from the other sectors.
Coming to collider searches, for $N_w\ne 1,3$ there is no mixing with the SM and the lightest mass eigenstate is stable, at least at the renormalizable level.
When it is charged, one can apply the limits on stable leptons reported in Table~\ref{6table}.
For $N_w$ odd, the lightest state may be neutral, with the typical collider phenomenology of a dark matter candidate.

The phenomenology is radically different when $N_w$ is even. In this case the Majorana mass is absent, $M_\chi=0$, 
and the $\chi$ components are massless before EWSB, therefore one mass eigenvalue for each sector is of the order $\lambda\tilde\lambda v^2/M_\psi$. 
This implies that all masses are bound to the EW scale, whence the situation resembles the one of purely chiral sets of new fermions. 
In particular, one finds a correction to $h\rightarrow \gamma\gamma$ that depends only on $Q$: taking $M_\chi=0$ 
in the mass matrix (\ref{m33}) or (\ref{m22}), and using the LET approximation of \eq{chpp-LET}, we find a CP-even amplitude
${\cal A}^{\gamma\gamma}_{\chi\psi,Q} \simeq 8Q^2/3$ for a $3\times 3$ sector, and the same for a $2\times 2$ sector.
For comparison, when $N_w$ is odd the Majorana mass is allowed and one finds 
\begin{equation}
{\cal{A}}^{\gamma\gamma,3\times 3}_{\chi \psi,Q} \simeq -\frac{8}{3}\frac{\lambda v}{M_\chi} \frac{\widetilde{\lambda} v}{M_\psi}  Q^2 \left(\kappa_{-Q}^2 + \kappa_{Q}^2 \right) ~,\qquad
{\cal{A}}^{\gamma\gamma,2\times2}_{\chi \psi,Q} \simeq -\frac{8}{3}\frac{\lambda v}{M_\chi} \frac{\widetilde{\lambda} v}{M_\psi}  Q^2 \kappa_{-Q}^2 ~,
\label{LET-VLM1pp}
\end{equation}
where we made the approximation $\lambda v, \tilde\lambda v \ll M_\chi, M_\psi$.
The correction grows as the $\chi-\psi$ mixing parameters of the type $\lambda v/M$ increase, 
on the other hand $S$ and $T$ generally put a significant upper bound on these parameters.
Obviously, in each model the total ${\cal A}_{new}^{\gamma\gamma}$ is the  sum over all the sectors with different $Q$.

Let us say a few more words on the cases where $\chi$ and/or $\psi$ mix with the SM leptons. 
\bit
\item $N+L$ : for $R_w=1$, the Majorana fermion is a sterile neutrino  $N$, the vector-like fermion is a lepton doublet $L$, and the Lagrangian in \eq{1VL-1MAJ-Lagrangian}
is extended to include $\lambda_N\overline{l_L}\tilde{H}N_R + \lambda_L\overline{L_L}H\tau_R +h.c.$. The full parameter space includes two real masses, four reals Yukawa couplings and two physical phases, that can be associated e.g. to 
$\lambda$ and $\tilde\lambda$. The $2\times 2$ mass matrix in the $Q=\pm 1$ sector is given by \eq{VLL-mass-matrices}: the mixing with the SM is small due to the strong constraint from the $Z\tau\tau$ couplings, implying  
small deviations in $\mu_{\gamma\gamma}$ and $\mu_{\gamma Z}$. 
The $4\times 4$ mass matrix in the $Q=0$ sector is obtained by adding to \eq{m33} a first row and a first column
of the form $(0,\lambda_N v/\sqrt{2} , 0 , 0)$. 
There are some simple limiting cases. If $\lambda,\tilde\lambda\rightarrow 0$, the matrix reduces to two diagonal blocks, as $N$ and $L$ decouple from each other,
and the phenomenology reduces to the one of the previous sections.
If $M_{L,N}/v$ are much larger than the Yukawa couplings, all the mixing angles are small and the smallness of the light neutrino mass follows from 
the usual seesaw mechanism, $m_{\nu}\simeq\lambda_N^2 v^2 /(2M_N)$. 
Still, $m_\nu$ can be small even in the presence of large Yukawa couplings. In particular, for $\lambda v,\tilde\lambda v \sim M_{N,L}$, large mixing angles are possible,
and, correspondingly, such intricate neutral sector may induce significant corrections to the $S$ and $T$ parameters. 
\item For $R_w=3$, the Majorana fermion is the triplet $\Sigma$ and the vector-like fermion is the doublet $L$ or the quartet $\Omega$. 

$\Sigma+L$ :  the Lagrangian in \eq{1VL-1MAJ-Lagrangian}
is extended to include $\lambda_\Sigma \overline{l_L}\tilde{H}\Sigma_R + \lambda_L\overline{L_L}H\tau_R +h.c.$.
Up to different Clebsch-Gordan coefficients, the $4\times 4$ neutral sector is the same as in the $N+L$ case. 
The mass matrix in the charged sector is
\begin{equation}
{\cal M}_{1}=\begin{pmatrix}
\sqrt{\frac 12}\lambda_\tau v & \sqrt{\frac 13}\lambda_\Sigma v & 0  \\
0  & M_\Sigma & \sqrt{\frac 13}\lambda v  \\
\sqrt{\frac 12}\lambda_L v & \sqrt{\frac 13}\tilde\lambda v& M_L
\end{pmatrix}.
\label{VL+MAj-matrix2}
\end{equation} 
The $Z \tau\tau$ couplings constrain both mixing parameters $\lambda_\Sigma v/M_\Sigma$ and $\lambda_Lv/M_L$ to be small, as explained 
in sections \ref{seesaw} and \ref{VLL}, respectively. 
When the mixing with the SM is neglected, one is left with a special case of \eq{m22}, which corresponds to the chargino  mass matrix in supersymmetry.
Note that, in the limit where $\lambda_\tau$ vanishes, there is still a contribution to the $\tau$ mass, 
$m_\tau \simeq (\lambda_\Sigma \lambda_L \lambda  v^3)/(3 \sqrt 2 M_\Sigma M_L)$, as illustrated in Fig.~\ref{2VLLs-Yukawa-generation}(d). 
Despite the constraint on the mixing from the $Z\tau\tau$ couplings, one can accommodate the correct size of $m_\tau$ for $\lambda \gtrsim 3$, see the discussion below Eqs.~(\ref{taucomp}) and (\ref{mtau-d-case}).

$\Sigma+\Omega$ :  the Lagrangian in \eq{1VL-1MAJ-Lagrangian} is extended to include $\lambda_\Sigma \overline{l_L}\tilde{H}\Sigma_R +h.c.$, as the quartet does not mix with the SM.
The $4\times 4$ neutral sector has the same structure as in the $N+L$ and $\Sigma+L$ cases.
The $Q=1$ sector also has a $4\times 4$ mass matrix, that is obtained from \eq{m33} by adding a first row $(\lambda_\tau v/\sqrt 2,\lambda_\Sigma/\sqrt 3,0,0)$,
and a first column $(\lambda_\tau v/\sqrt 2,0,0,0)$.
In addition, there is a $Q=2$ state with mass $M_\Omega$.
There are no significant phenomenological novelties, as the effects of the SM mixing with $\Sigma$ and of the $\Sigma$ mixing with $\Omega$ mixing do not interfere significantly.
\eit

\noindent $\bullet$ \textbf{\textit{One VLL plus two Majorana leptons.}}
Let us come to sets of two Majorana leptons both interacting with one vector-like pair. One obvious possibility is to take two copies of the same Majorana lepton, that is, 
to replace $\chi_R$ in \eq{1VL-plus-MAjorana-cases} with $\chi_{iR}$,  $i=1,2$, with the obvious doubling of each coupling involving $\chi$ in the Lagrangian.
Note that $N_w$ can be arbitrary and, for even $N_w$, the Majorana mass terms are forbidden but a Dirac mass term $M_\chi \overline{\chi_{1R}^c}\chi_{2R}$ is allowed. 
In all other respects,
the mass matrix structures and the inherent phenomenology are a straightforward generalisation of those discussed above.

The second and last possibility to couple two Majorana leptons to one VLL is provided by the set
\begin{equation}
\chi_{1R} \sim (1,R_w,0)~,\qquad \psi_L,\psi_R \sim (1,R_w + 1,-1/2)~,\qquad  \chi_{2R} \sim (1,R_w+2,0)~,
\end{equation}
with $N_w$ necessarily odd, and Lagrangian
\begin{eqnarray}
-{\cal{L}}_{\chi\psi}&= M_{\psi}\overline{\psi_{L}}\psi_{R} +\sum_{i=1}^2 \left[\frac{1}{2} M_{\chi_i} \overline{\chi_{iR}^c}\chi_{iR}   +
\tilde\lambda_i \overline{\psi_{L}}\tilde{H} \chi_{Ri} + \lambda_i \overline{\chi_{Ri}^c} H \psi_{R}\right] + h.c. ~.
\label{1VL-2MAJ-Lagrangian}
\end{eqnarray}
There are two pairs of states with $Q=\pm(N_w+1)/2$, with a mass matrix given by \eq{m22} with $M_\chi,\lambda,\tilde\lambda \rightarrow M_{\chi_2},\lambda_2,\tilde\lambda_2$.
In addition, there are four pairs of states with $Q=\pm(N_w-1)/2,\pm(N_w-3)/2,\dots,\pm1$, and four states with $Q=0$. The $4\times 4$ mass matrix in each such sector takes the form
\beq
{\cal M}_{Q} = \left(\ba{cccc}
M_{\chi_1} & 0 & \kappa_{1,Q} \tilde\lambda_1 v & \kappa_{1,-Q} \lambda_1 v \\
0 & M_{\chi_2} & \kappa_{2,Q} \tilde\lambda_2 v & \kappa_{2,-Q} \lambda_2 v \\
\kappa_{1,-Q} \tilde\lambda_1 v & \kappa_{2,-Q} \tilde\lambda_2 v & 0 & M_\psi \\
\kappa_{1,Q} \lambda_1 v & \kappa_{2,Q} \lambda_2 v & M_\psi & 0
\ea\right)~,
\label{m44}\eeq
with $\kappa_{1,\pm Q}$ given by \eq{CG+}, and $\kappa_{2,\pm Q}$ given by \eq{CG-} with $N_w\rightarrow N_w+2$. 
As in \eq{LET-VLM1pp}, one can estimate the contribution of this mass matrix to the $h\gamma\gamma$ coupling, by taking the  LET approximation,
\begin{equation}
{\cal{A}}^{\gamma\gamma,4\times4}_{\chi_1\chi_2\psi,Q} \simeq -\frac{8}{3} \sum_{i=1,2} \frac{\lambda_i v}{M_{\chi_i}} \frac{\widetilde{\lambda_i} v}{M_\psi} Q^2\left(\kappa_{i,-Q}^2 + \kappa_{i,Q}^2 \right) ~.
\end{equation}

Let us discuss the minimal cases $R_w=1$ and $R_w=3$,
that admit a mixing with the SM leptons.
\begin{itemize}
\item $N+L+\Sigma$ : for $R_w=1$, the new fermions 
have the gauge quantum numbers of the bino, the higgsinos and the wino in supersymmetry.
Thus, the mass matrices (\ref{m22}) and (\ref{m44}) are a generalisation of the chargino ($Q=\pm1$) and neutralino ($Q=0$) mass matrices, respectively 
(for a review see Ref.~\cite{Martin:1997ns}).  
Supersymmetry restricts the Yukawa couplings to $\tilde\lambda_{1,2}/\lambda_{1,2}=-\tan\beta$, $\lambda_1/\lambda_2=\tilde\lambda_1/\tilde\lambda_2=-\tan\theta_w/\sqrt{3}$, and $\lambda_1=-\cos\beta g'$. 
The effect of charginos and neutralinos on the Higgs boson couplings is analysed e.g. in 
Refs.~\cite{Dreiner:2012ex,Ananthanarayan:2013fga,Casas:2013pta,Batell:2013bka}.
In particular, the chargino loop contributing to $h\gamma\gamma$ and $h \gamma Z$ is controlled by the weak coupling $g$, and it is typically subleading compared to the SM top quark loop. 
Without the supersymmetry constraints, the most general chargino mass matrix has the structure of \eq{2VLLS-mass-matrix}, 
therefore one can apply the results of section \ref{2VLL} for the Higgs decay amplitudes into $\gamma\gamma$ and $\gamma Z$.

In the absence of ($R$-parity conserving) supersymmetry, not only the four Yukawa couplings $\lambda_{1,2}$ and $\tilde\lambda_{1,2}$ are unconstrained, but in addition a mixing with the SM leptons is allowed: one should add to the Lagrangian in \eq{1VL-2MAJ-Lagrangian}
the terms $\lambda_N\overline{l_L}\tilde{H}N_R + \lambda_\Sigma \overline{l_L}\tilde{H}\Sigma_R + \lambda_L\overline{L_L}H\tau_R+h.c.$. The $Q=\pm1$ mass matrix becomes the one in \eq{VL+MAj-matrix2}, and the $Q=0$ mass matrix becomes $5\times 5$, and it
is obtained from \eq{m44} by adding a first row and column of the form $(0,\lambda_Nv/\sqrt{2},\lambda_\Sigma v/\sqrt{6},0,0)$.
Therefore, one can observe the phenomenological effects of $N$, $\Sigma$ and $L$ individually, as analysed in sections \ref{seesaw} and \ref{VLL}, as well as their interplay,
already described above for ($N+L$) and  ($\Sigma+L$).
As usual, the mixing with the SM leptons is typically constrained to be small by the smallness of $m_\nu$ and by the $Z\tau\tau$ couplings,
thus the modifications to the $h\nu\nu$ and $h\tau\tau$ couplings are suppressed. 
However, even a very small mixing with the SM offers decay modes to the heavy fermions,
such that none is stable.
A dedicated analysis of the full parameter space 
would be interesting, to characterise quantitatively the correlations among the different observables, and especially the deviations from the supersymmetric limit.
\item $\Sigma+\Omega+\Xi$: for $R_w=3$, the new fermions are a Majorana triplet, a vector-like quartet, and a Majorana quintuplet $\Xi$. 
There is a $2\times 2$ sector with $Q=\pm2$
given by \eq{m22}. As the triplet mixes with the SM through
$\lambda_\Sigma \overline{l_L}\tilde{H}\Sigma_R + h.c.$, there is a $5\times 5$ sector with $Q=\pm 1$, that is obtained by adding to the matrix in \eq{m44} a first row
$(\lambda_\tau v/\sqrt{2},\lambda_\Sigma v/\sqrt{3},0,0,0)$ and a first column $(\lambda_\tau v/\sqrt{2},0,0,0,0)$. 
This large number of charged states with potentially large mixing can give a significant correction to $\mu_{\gamma\gamma}$ and $\mu_{\gamma Z}$. 
For concreteness, neglecting the mixing with the SM and using the LET approximation, we find 
\begin{equation}
{\cal{A}}^{\gamma\gamma}_{\Sigma \psi \Xi} \simeq -\frac{8}{3} \left(\frac{1}{3}\frac{\lambda_1 v}{M_{\Sigma}} \frac{\widetilde{\lambda}_1 v}{M_\Omega}+ \frac{\lambda_2 v}{M_{\Xi}} \frac{\widetilde{\lambda}_2 v}{M_\Omega} \right)~,
\label{LET-2VLLs-Rw4}
\end{equation}
to be compared with the SM top contribution, ${\cal{A}}^{\gamma\gamma}_t\simeq 16/9$.
One should take into account  the constraints (in particular $S$ and $T$) on the mixing parameters $\sim \lambda v/M$.
The neutral sector has also a $5 \times 5$ mass matrix, obtained by adding a first column and row
$(0,\lambda_\Sigma v/\sqrt{6},0,0,0)$ to to the matrix in \eq{m44}. As usual, the vanishing neutrino mass requires $\lambda_\Sigma v / M_\Sigma$ to be very small.
\end{itemize}

\section{Phenomenology of non-chiral quarks}\label{phenoQ}

In this section we discuss new coloured fermions that either form vector-like pairs, or admit a Majorana mass term.
We will dub them `quarks' even when they are not in the fundamental representation of $SU(3)_c$.

 \subsection{One vector-like quark}\label{VLQ}

There are seven possible VLQs that mix with the SM quarks, as listed in Table \ref{twotable}. 
They have been extensively studied under various respects in the literature  (see e.g. 
Refs.~\cite{delAguila:2000rc,AguilarSaavedra:2009es,Cacciapaglia:2010vn,Gopalakrishna:2011ef,Cacciapaglia:2011fx,Okada:2012gy,Aguilar-Saavedra:2013qpa}). Here we describe in a compact, systematic way
the leading order constraints coming from EW precisions tests, direct searches at colliders, and Higgs couplings. 
As usual we restrict ourselves to mixing with the third family. 
In the top (bottom) sector, a mixing appears whenever the VLQ contains a component $T$ ($B$) with the same charge as $t$ ($b$).
The components of each multiplet are displayed in \eq{newleptons1}.
In the case of weak singlets or triplets, the SM Lagrangian is extended by
\begin{equation}
-{\cal{L}}_\psi=\lambda_\psi  \overline{q_L} \tilde{H}(H) \psi_R +M_\psi \overline{\psi_L} \psi_R +h.c.~, \quad\quad
 {\rm for~}\psi = T, X_Q (B, Y_Q)~,
\label{L1}
\end{equation}
and, in the case of weak doublets, by
\begin{equation}
-{\cal{L}}_\psi =\lambda_\psi^t  \overline{\psi_L} \tilde{H} (H)  t_R + \lambda_\psi^b \overline{\psi_L} H (\tilde{H}) b_R +M_\psi \overline{\psi_L} \psi_R + h.c.~, \quad\quad
{\rm for~} \psi = Q (X_T,Y_B)~,
\label{L2}
\end{equation}
with the further restriction  $\lambda_{X_T}^b=\lambda_{Y_B}^t=0$.
The structure of the top (bottom) sector mass matrix is very close to the charged lepton one
in the case of one VLL, therefore we will frequently refer to section \ref{VLL}.  
In the top sector one has 
 \begin{equation}
 {\cal{M}}_{t}=\begin{pmatrix}
 \lambda_{t}\frac{v}{\sqrt{2}} &  \kappa_\psi^t \lambda_\psi  \frac{v}{\sqrt{2}} \\ 0 & M_\psi
 \end{pmatrix}~,~\psi = T, X_Q, Y_Q ~,
 \quad
  {\cal{M}}_{t}=\begin{pmatrix}
 \lambda_{t} \frac{v}{\sqrt{2}} & 0 \\   \kappa_\psi^t \lambda_\psi^t  \frac{v}{\sqrt{2}}   & M_\psi
 \end{pmatrix}~,~ \psi=Q,X_T ~, 
 \label{VLQ-mass-matrices-T}
 \end{equation}
with Clebsch-Gordan coefficients $\kappa_{T,Q,X_T}^t=1$, $\kappa_{X_Q}^t=\sqrt{1/3}$ and $\kappa_{Y_Q}^t=\sqrt{2/3}$.
The rotation to the mass basis is parametrized as ${\cal M}_t = U_L diag(m_t,m_{t'})U_R^\dag$, in analogy with \eq{VLL-rotation}. 
In the bottom sector one has 
 \begin{equation}
 {\cal{M}}_{b}=\begin{pmatrix}
 \lambda_{b}\frac{v}{\sqrt{2}} &  \kappa_\psi^b \lambda_\psi  \frac{v}{\sqrt{2}} \\ 0 & M_\psi
 \end{pmatrix}~,~\psi = B, X_Q, Y_Q ~,
 \quad
  {\cal{M}}_{b}=\begin{pmatrix}
 \lambda_{b} \frac{v}{\sqrt{2}} & 0 \\   \kappa_\psi^b \lambda_\psi^b  \frac{v}{\sqrt{2}}   & M_\psi
 \end{pmatrix}~,~ \psi=Q,Y_B ~,
 \label{VLQ-mass-matrices-B}
 \end{equation}
with $\kappa_{B,Q,Y_B}^b=1$, $\kappa_{X_Q}^b=\sqrt{2/3}$ and $\kappa_{Y_Q}^b=\sqrt{1/3}$, and one can write 
${\cal M}_b=\widetilde{U}_Ldiag(m_b,m_{b'})\widetilde{U}_R^\dag$. 

In all cases except $Q$, the vector-like mass $M_\psi$ and the three independent Yukawa couplings can be taken to be real. 
In the case of $Q$, there are four Yukawa couplings and one complex phase $\phi$ is physical. In full generality, one can choose $\lambda_{t,b}$ and $\lambda_Q^{t,b}$ real, and add 
a matrix $P_\phi = diag(e^{i\phi},1)$ on the left of ${\cal M}_b$. Then, \eq{Vmass} and \eq{hff1} show that $\phi$ appears in the $W$ couplings to the fermion mass eigenstates, while
the $Z$ and $h$ couplings are independent from $\phi$. It is very difficult to observe such CP-violating effect in the charged current, since it vanishes for $\lambda_b\rightarrow 0$,
that is, it is always suppressed by the small ratio $m_b/v$.
Coming to the mixing angles, the left- and right-hand ones are related 
as in the case of VLLs, see \eq{suppressed-mixing}: 
in the case of weak doublet VLQs one finds
\begin{equation}
\tan \theta_L=\frac{m_t}{m_{t'}} \tan \theta_R < \tan \theta_R ~, 
\quad\quad \tan \widetilde{\theta}_L=\frac{m_b}{m_{b'}} \tan \widetilde{\theta}_R \ll \tan \widetilde{\theta}_R~,
\label{VLQ-suppressed-mixing}
\end{equation}
while for singlet or triplet VLQs, the same relations hold with $L \leftrightarrow R$.
In the following we will drop the subscripts $L,R$ and denote $\theta_\psi$ ($\widetilde\theta_\psi$) the largest mixing angle in the top (bottom) sector, for any given VLQ $\psi$.
For the multiplets with both the $T$ and $B$ components, the mass eigenvalues and the mixing angles in the top and bottom sectors are strictly related,
\beq
\left\{\ba{l}
m_{t'}^2-m_{b'}^2=s_\psi^2(m_{t'}^2-m_t^2)-\tilde{s}_\psi^2(m_{b'}^2-m_b^2)\\
s_\psi c_\psi(m_{t'}^2-m_t^2) r_\psi =\tilde{s}_\psi \tilde{c}_\psi (m_{b'}^2-m_b^2)
\ea\right.~,
\label{bt_relations}\eeq
where $r_{X_Q}=\sqrt{2}$, $r_{Y_Q}=1/\sqrt{2}$, and $r_{Q}\equiv \lambda_Q^b/\lambda_Q^t$.
Therefore, one can determine the bottom sector parameters, $m_{b'}$ and $\widetilde\theta_\psi$, as a function of the top sector one, $m_{t'}$ and $\theta_\psi$,
or vice versa. In the case $\psi=Q$, there is the additional freedom of the choice of $r_{Q}$. Note that the custodial symmetry is preserved in the $Q$ sector
if $r_Q=1$, see  \eq{L2}.
The mass splitting among the heavy quarks is controlled (at tree level) by the mixing with the SM. The mass ordering is determined as
\begin{equation}
\ba{ll}
T:~~M_T \leq m_{t'}~, & B:~~ M_B \leq m_{b'} ~,\\
X_T:~~m_X=M_{X_T} \leq m_{t'} ~, & Y_B:~~m_Y=M_{Y_B} \leq m_{b'} ~,\\
X_Q:~~ m_X=M_{X_Q} \leq m_{t'}\leq m_{b'} ~, & Y_Q:~~ m_Y=M_{Y_Q} \leq m_{b'}\leq m_{t'} ~, \\
\ea
\label{VLQ-spectrum}
\end{equation}
where we took implicitly into account the experimental upper bounds on the mixing and on $m_t/m_{t'}$, when needed to establish the ordering.
In the case of $Q$, for $r_Q=1$ one finds $M_Q\le m_{b'}\le m_{t'}$, but the ordering between $b'$ and $t'$ can change for different  values of $r_Q$.
 
\begin{figure}[btp]
   \begin{minipage}[c]{.1\linewidth}
      \includegraphics[scale=0.3,trim= 0 0 0 0]{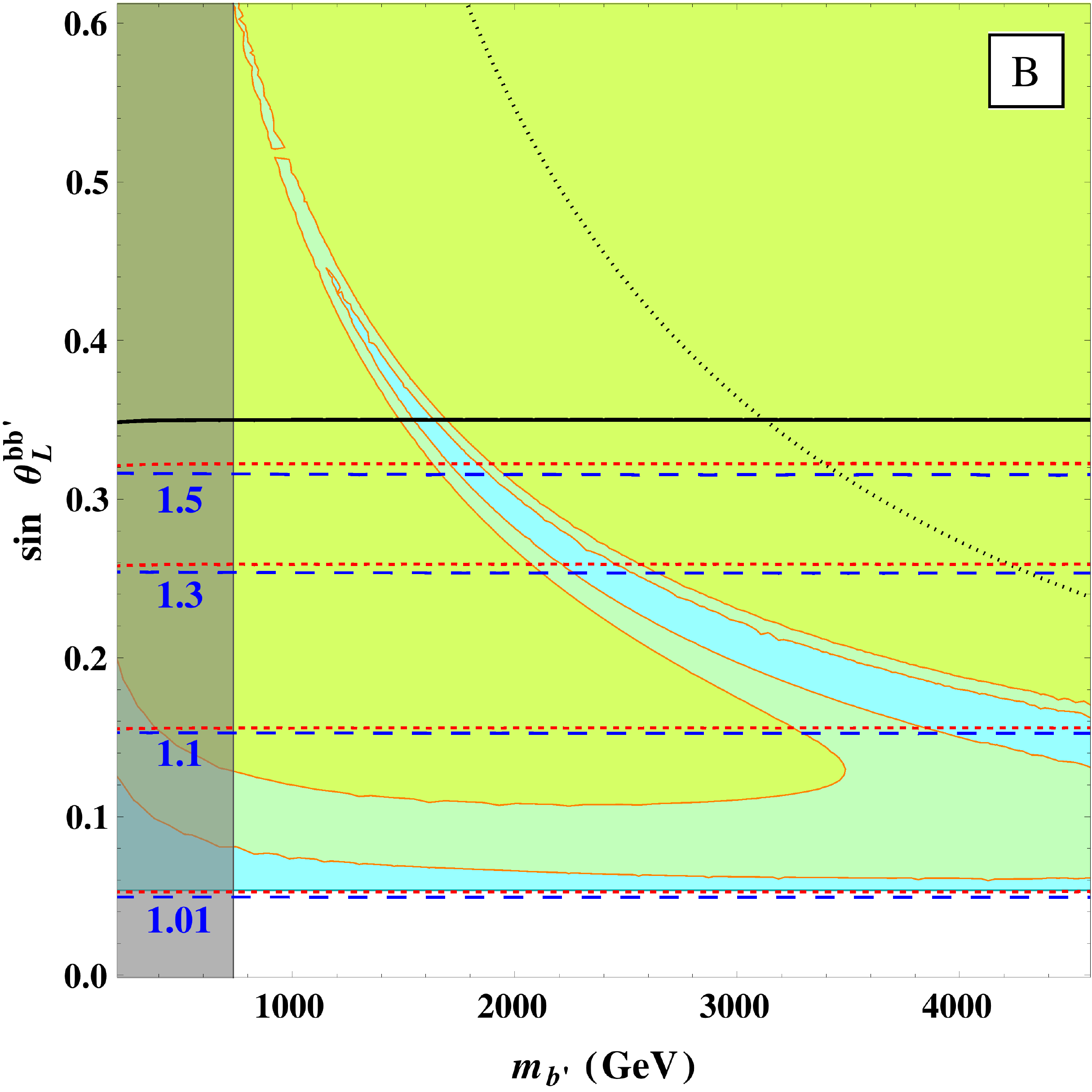}
   \end{minipage} \hfill
   \begin{minipage}[c]{.5\linewidth}
      \includegraphics[scale=0.3,trim= 0 0 0 0]{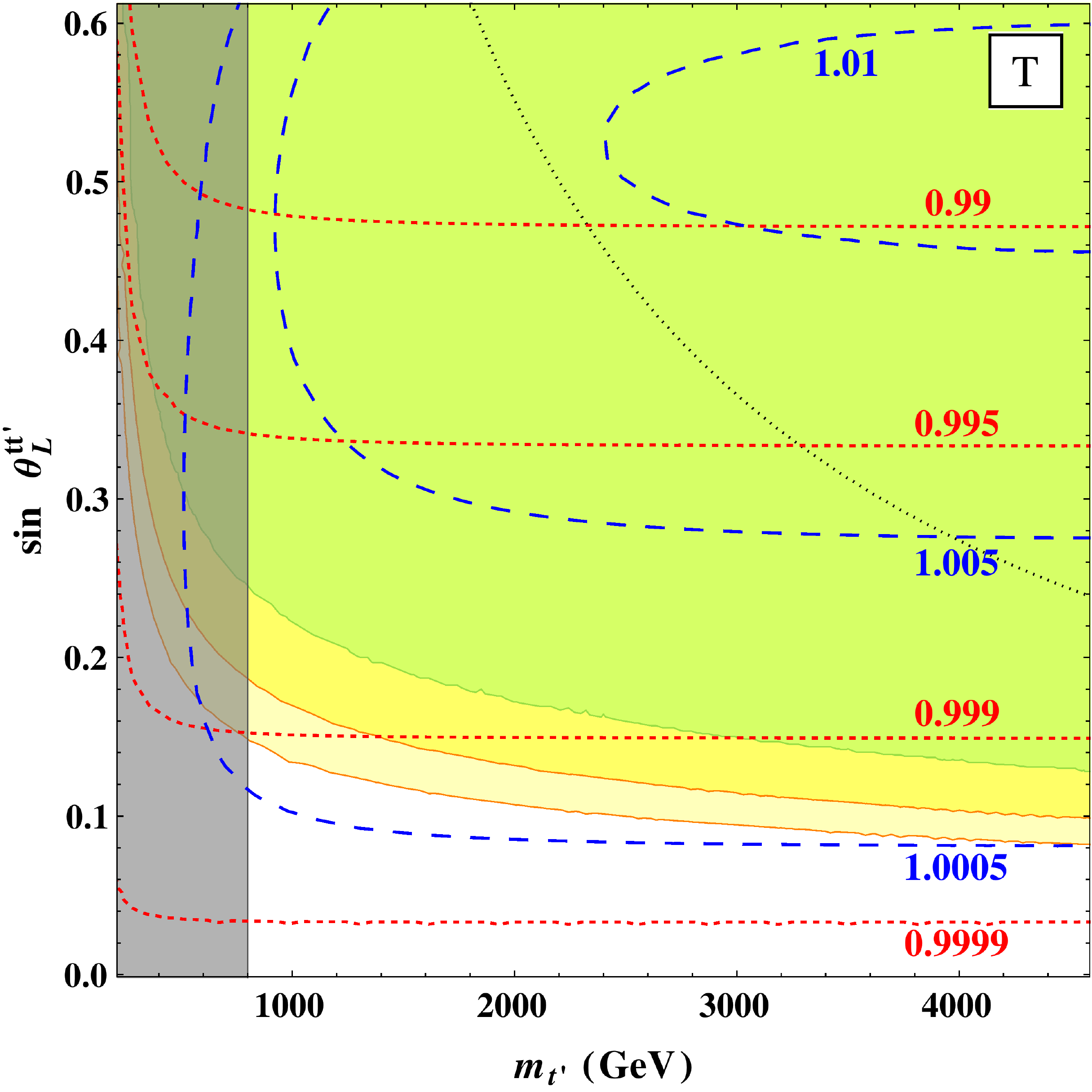}
   \end{minipage}
\caption{Constraints on the weak singlet VLQs $B$ (left panel) and $T$ (right panel), as a function of the mass of $b'$ ($t'$) 
and of the mixing angle between $b_L$ and $b'_L$ ($t_L$ and $t'_L$). 
The region above the dotted black line is excluded by perturbativity. The blue-shaded region is excluded by the $Zb\overline{b}$ couplings.   
The (light) yellow-shaded region is excluded by the $S$ and $T$ parameters at (68\%) 99\% C.L..
The green-shaded region is just the intersection of the previous two.
The grey-shaded region is excluded by the collider searches summarised in Table \ref{5table}. 
The region above the solid black line is excluded by a rough global fit of the Higgs couplings at 99\% C.L..
The dashed blue (dotted red) lines correspond to a few relevant values of the signal strength $\mu_{\gamma Z}$ ($\mu_{\gamma \gamma}$).}
   \label{VLQs-singlets}
\end{figure}

\begin{figure}[btp]
   \begin{minipage}[c]{.1\linewidth}
      \includegraphics[scale=0.3,trim= 0 0 0 0]{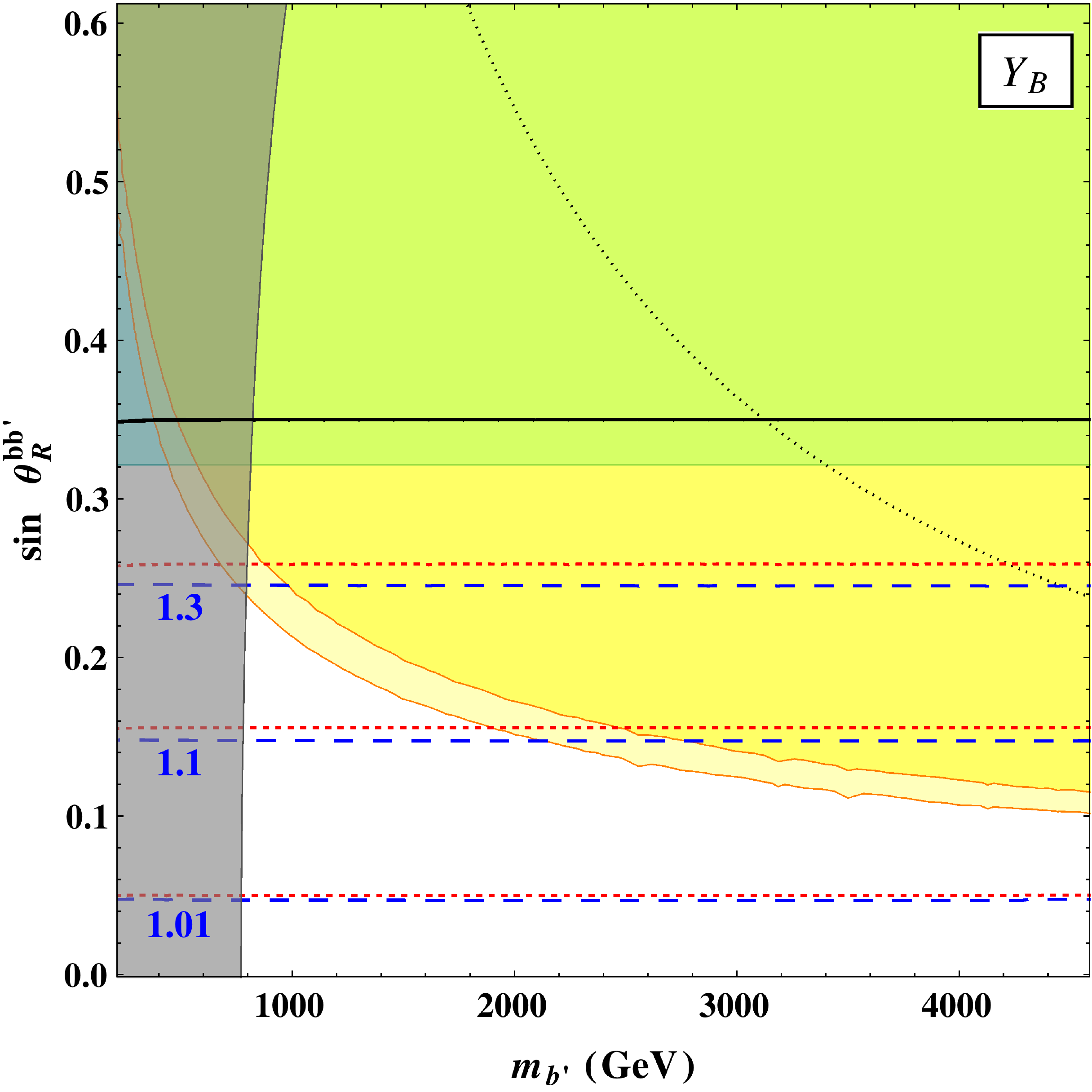}
   \end{minipage} \hfill
   \begin{minipage}[c]{.5\linewidth}
      \includegraphics[scale=0.3,trim= 0 0 0 0]{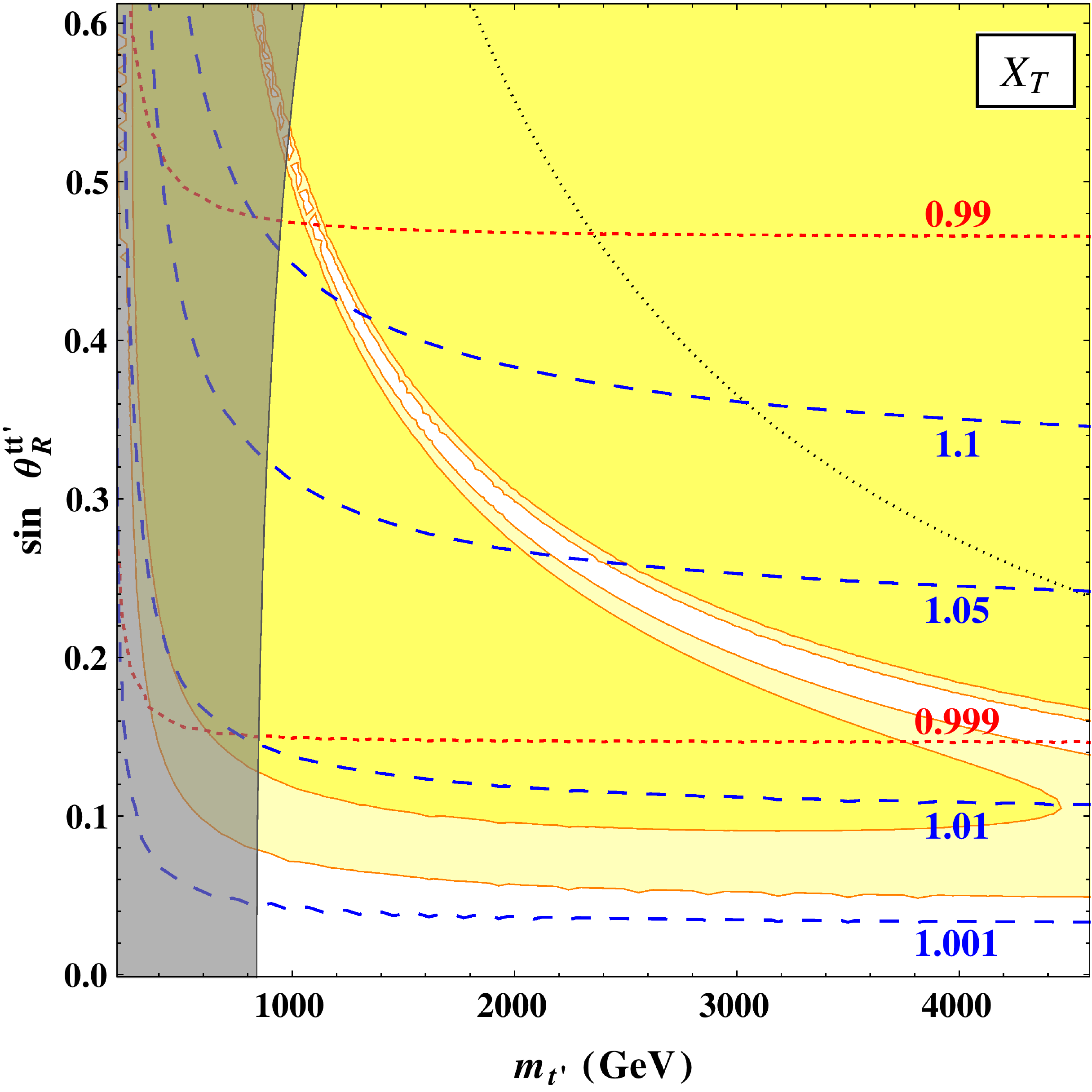}
   \end{minipage}
   \caption{Constraints on the weak doublet VLQs  $Y_B$ ($Y= -5/6$, left panel) and $X_T$ ($Y=+7/6$, right panel) 
as a function of the mass of $b'$ ($t'$) and of the mixing angle between $b_R$ and $b'_R$ ($t_R$ and $t'_R$). 
   The notation is the same as in Fig.~\ref{VLQs-singlets}.}
   \label{VLQs-doublets}
\end{figure}

\begin{figure}[bt]
   \begin{minipage}[c]{.1\linewidth}
      \includegraphics[scale=0.3,trim= 0 0 0 0]{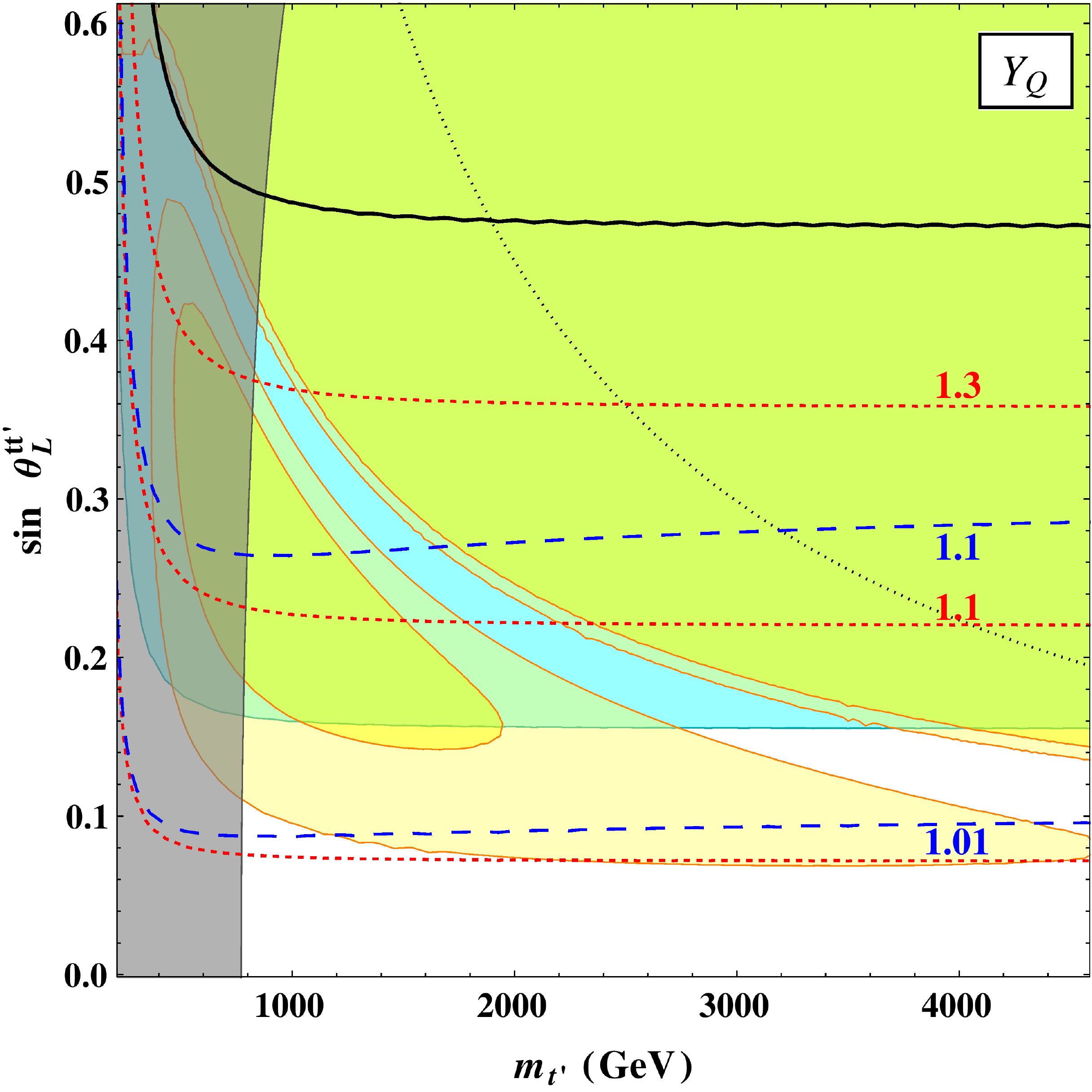}
   \end{minipage} \hfill
   \begin{minipage}[c]{.5\linewidth}
      \includegraphics[scale=0.3,trim= 0 0 0 0]{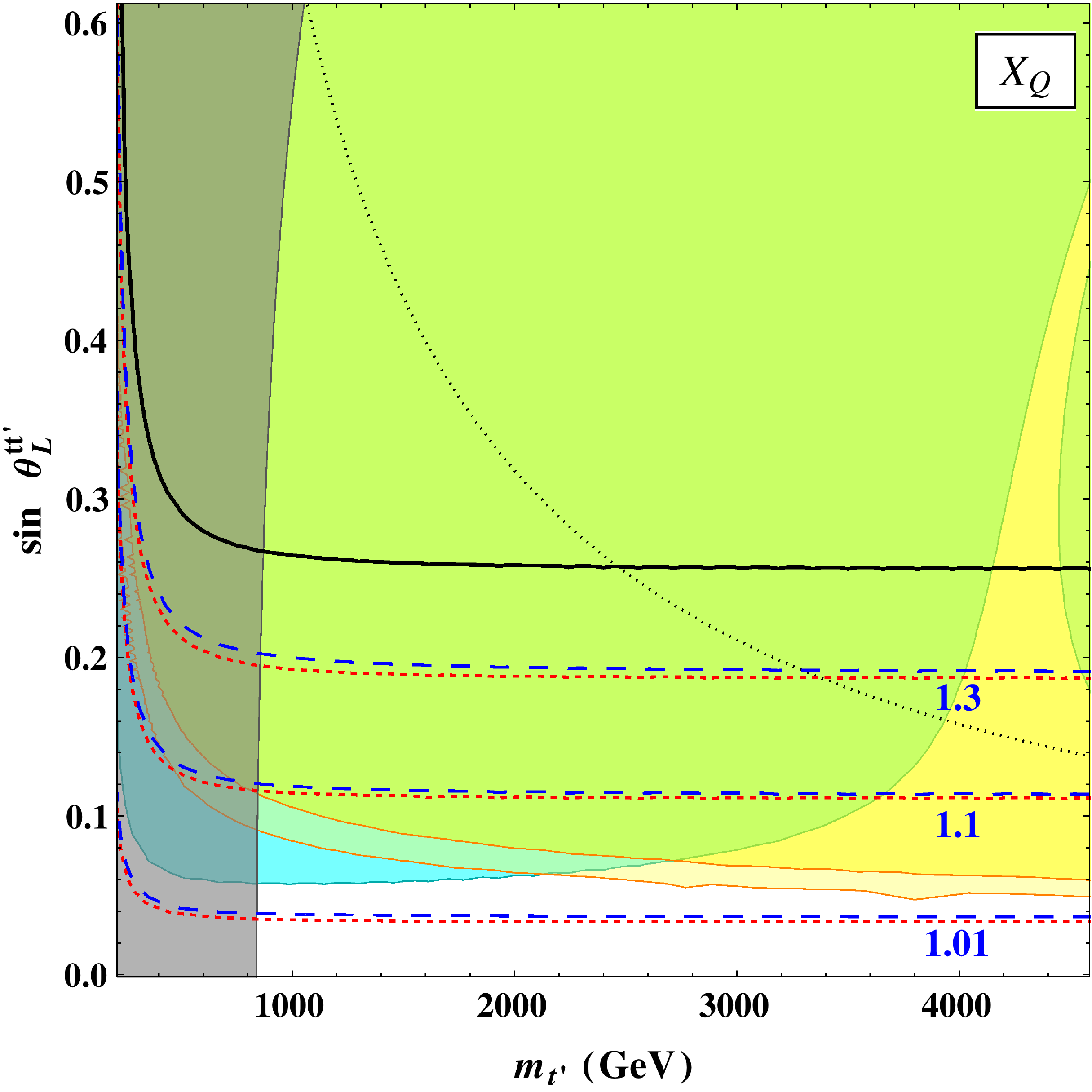}
   \end{minipage}
   \caption{Constraints on the weak triplet VLQs  $Y_Q$ ($Y= -1/3$, left panel) and $X_Q$ ($Y=+2/3$, right panel) as a function of the mass of $t'$ and of the mixing angle between $t_L$ and $t'_L$. 
   The notation is the same as in Fig.~\ref{VLQs-singlets}.}
   \label{VLQs-triplets}
\end{figure}

\begin{figure}[bt]
   \begin{minipage}[c]{.1\linewidth}
      \includegraphics[scale=0.3,trim= 0 0 0 0]{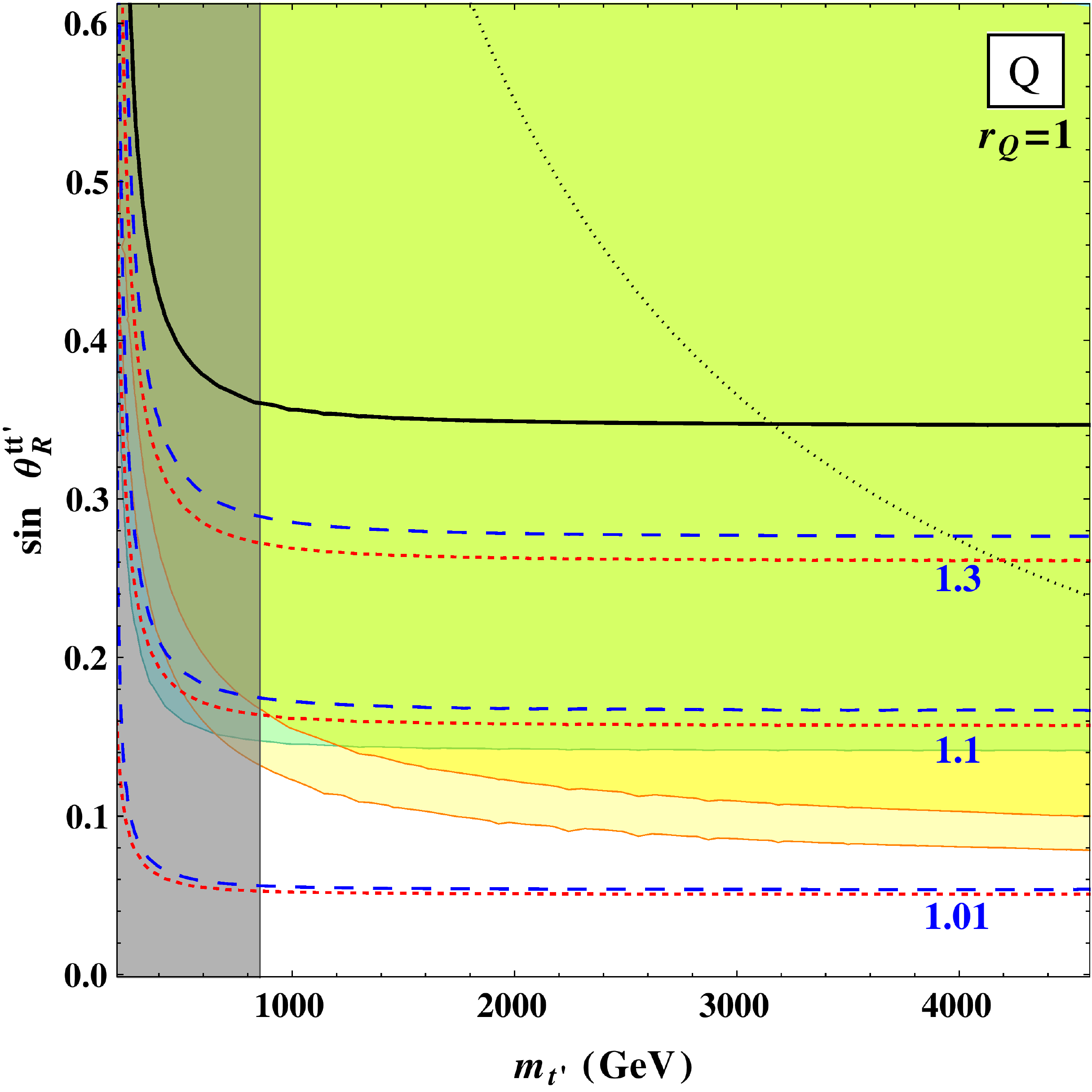}
   \end{minipage} \hfill
   \begin{minipage}[c]{.5\linewidth}
      \includegraphics[scale=0.3,trim= 0 0 0 0]{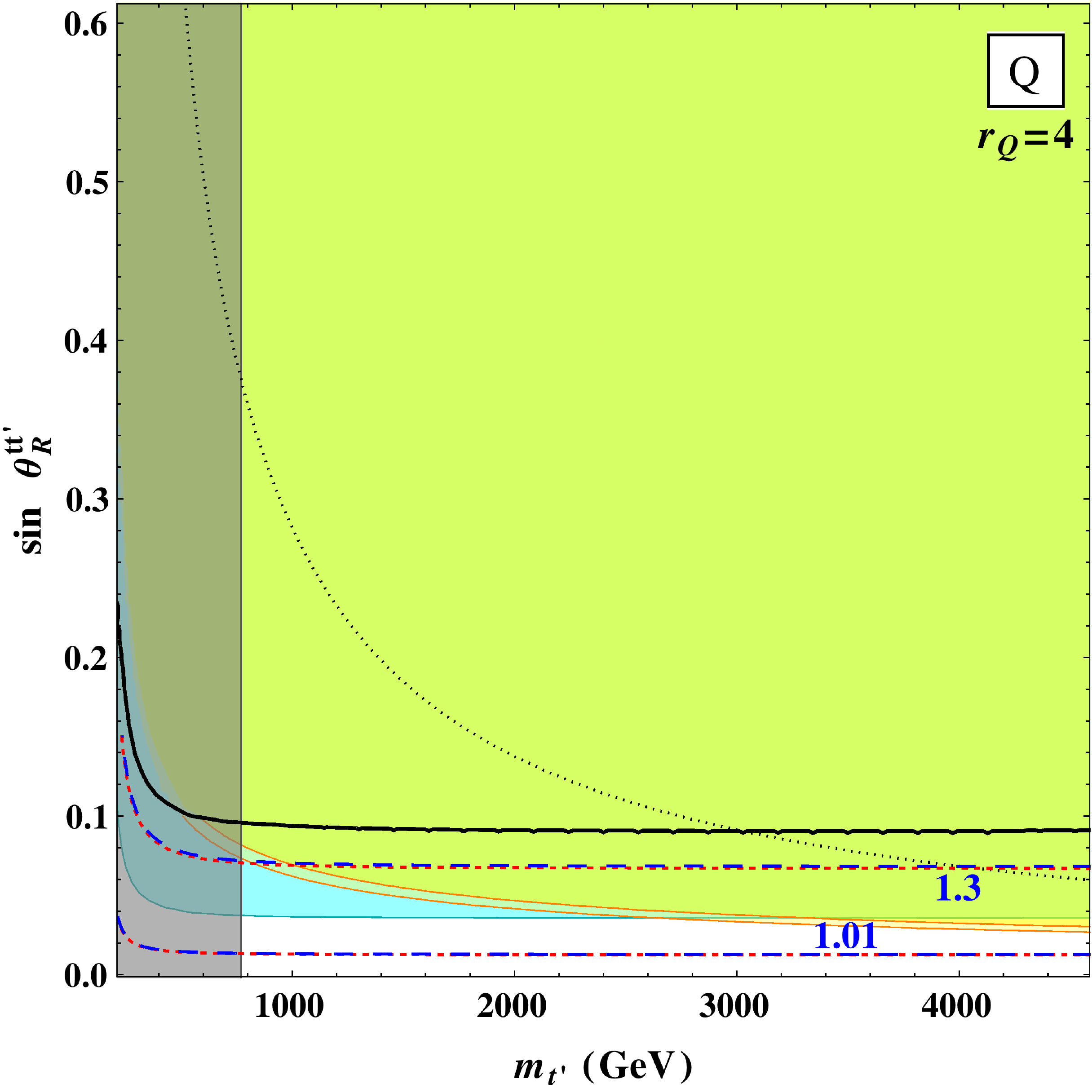}
   \end{minipage}
   \caption{
Constraints on the weak doublet VLQ  $Q$ as a function of the mass of $t'$ and of the mixing angle between $t_R$ and $t'_R$. 
The left panel corresponds to $\lambda_Q^t= \lambda^b_Q$ ($r_Q=1$),  and the right one to $4\lambda^t_Q=\lambda^b_Q$ ($r_Q=4$). 
The notation is the same as in Fig.~\ref{VLQs-singlets}.}
   \label{VLQs-Qdoublet}
\end{figure}

Masses and mixing angles are constrained by the perturbativity of the Yukawa couplings, 
\begin{equation}
\left| \lambda_\psi^t \right| \simeq \left| \frac{\sqrt{2}}{\kappa_\psi^t} s_{ \psi}  \frac{m_{t'}}{v}\right| \ll 4\pi ~,\qquad
\left| \lambda_\psi^b \right| \simeq \left| \frac{\sqrt{2}}{\kappa_\psi^b} \tilde{s}_{ \psi}  \frac{m_{b'}}{v}\right| \ll 4\pi ~.
\label{VLQ-perturbativity}
\end{equation}
{Note that 
we do not impose a stronger upper bound such as $4\pi/\sqrt{N_c}$, 
for reasons discussed in  Appendix \ref{treeH}.}
The perturbativity of the SM  couplings $\lambda_t$ and $\lambda_b$ is guaranteed a fortiori.
For definiteness, in Figures \ref{VLQs-singlets}~-~\ref{VLQs-Qdoublet} we delimit with a black dotted line 
the region of parameters where at least one Yukawa coupling becomes larger than $2\pi$.
In the case of $Q$ both inequalities must be satisfied at the same time, therefore a large departure from $r_Q=1$ leads to a stronger constraint, as illustrated by the comparison
of the left and right panels of Fig.~\ref{VLQs-Qdoublet}.

Important constraints come from the $Z$ couplings to quarks, that are affected by the mixing as shown in \eq{Zmass}.
The tree-level deviations with respect to the SM are given by
\begin{equation}
\delta g_{Z \overline{b} b}^L=\tilde{s}_{L}^2 \left(\frac{1}{2}+T^3_{B} \right)~,
\quad 
\delta g_{Z \overline{b} b}^R=\tilde{s}_{R}^2 T^3_{B} ~, 
\quad\quad
\delta g_{Z \overline{t} t}^L= s_{L}^2 \left(-\frac{1}{2}+T^3_{T} \right)~,
\quad 
\delta g_{Z \overline{t} t}^R= s_{R}^2 T^3_{T} ~, 
\label{VLQ-Zbb-contributions}
\end{equation}
where $T^3_{B}$ ($T^3_{T}$) is the weak isospin of the $B$ ($T$) component of the VLQ under investigation. 
The $Z b\overline{b}$ couplings are  measured less precisely than their leptonic analog, the  $Z \tau^+ \tau^-$ couplings, but nonetheless they are strongly constrained,
especially for $b_L$.
The top couplings to the $Z$ boson are poorly constrained directly, however they also  contribute at one loop to $Z b\overline{b}$.
These constraints are summarised in appendix \ref{Zffapp} and they
exclude the blue-shaded region in Figures \ref{VLQs-singlets}~-~\ref{VLQs-Qdoublet}. 
Deviations in $Z \overline{b_L} b_L$ are present at tree level in the case of $B$, $Y_B$, $Y_Q$ and $X_Q$. 
However, in the case of the doublet $Y_B$, the deviation is suppressed by $(m_b/m_b')^2$ and the most important correction is the one to $Z\overline{b_R} b_R$.
In the case of $T$ and $X_T$, there is no bottom partner and the deviation to $Z \overline{b_L} b_L$ is induced at one loop
 mostly through  $\delta g_{Z \overline{t} t}^L$, leading to a relatively weak constraint.
Finally, in the case of $Q$, $\delta g_{\overline{b} b}^R$ is generated at tree level and $\delta g_{\overline{b} b}^L$ at one loop,
the strongest constraint coming from the  right-handed coupling.

The VLQ couplings to the EW gauge bosons are also constrained by  the EW precision parameters $S$ and $T$,
whose expressions are provided in appendix \ref{ST}. Note that, in contrast with the tau and bottom sectors,  in the top sector contributions proportional to powers of $m_t/m_Z$ are not suppressed.
In Figs.~\ref{VLQs-singlets}~-~\ref{VLQs-Qdoublet} we display in (light) yellow the region corresponding to the ($68\%$) $99\%$ C.L. ellipse
of Fig.~\ref{figST}. 
Since $S$ and $T$ are proportional to the mixing between the SM quarks and the VLQ, 
one typically observes an upper bound $s_\psi,\tilde{s}_\psi\lesssim 0.05 - 0.20$, depending on the VLQ under consideration.
Note that this bound relaxes as the heavy quark mass decreases, because 
$S$ and $T$ eventually vanish in the limit $m_{t'}\rightarrow m_t$ or $m_{b'}\rightarrow m_b$. 
Note also that a cancellation is possible among relatively large contributions to $S$ and $T$, such that large mixing angles may be allowed in a fine-tuned region of parameters.
This is especially relevant in the case of $X_T$, because such region is not excluded by other constraints. 
Indeed, we find
\begin{equation}
T(X_T)\simeq\frac{3 s_{X_T}^2 }{16 \pi c_w^2 s_w^2} \frac{m_{t'}^2}{m_Z^2}\left[\frac 43 s_{X_T}^2 + \frac{m_t^2}{m_{t'}^2} \left(4 \ln \frac{m_t^2}{m_{t'}^2} + 6 \right) \right]~,~~
S(X_T)\simeq \frac{ s_{X_T}^2}{2\pi}  \left( \frac 43 \ln \frac{m_t^2}{m_{t'}^2} +5 \right)~,
\end{equation}
where we dropped terms subleading in $s_{X_T}$ and $m_t/m_{t'}$.
As the logarithm is large and negative, a cancellation is possible in the $T$ parameter even for large mixing:
this explains the allowed strip in Fig.~\ref{VLQs-doublets}, that reaches $s_{X_T} \simeq 0.5$. 
A comment is in order for the case of  $Q$: one would expect a milder constraint from $T$ when the two Yukawa couplings $\lambda_Q^{t,b}$ respect the custodial symmetry, i.e. when $r_Q=1$. 
However, even in this case there is an important deviation from the SM,
 because the residual custodial-breaking parameter, $(\lambda_t- \lambda_b)$, differs from the SM one, $\sqrt{2}(m_t-m_b)/v$, as soon as the mixing is non-zero.

Let us now turn to the direct searches of VLQs at colliders.
As they are coloured, it is easier to produce them at the LHC, relatively to VLLs.
Below $\sim 1$ TeV they are dominantly produced in pairs through strong interactions,
while for higher masses single production by EW interactions can become dominant \cite{AguilarSaavedra:2009es,Gopalakrishna:2013hua}. 
The pair production mechanism, that dominates in the mass range probed at the 8 TeV LHC, is independent
from the VLQ (all are colour triplets) and from the mixing parameters.
The ATLAS and CMS searches focus on the following decay channels for the heavy quark mass eigenstates:
\begin{equation}
X\rightarrow t W^+ ~,\quad\quad
t'\rightarrow t Z, t h, b W^+ ~,\quad\quad
b'\rightarrow b Z, b h,t W^- ~,\quad\quad
Y\rightarrow b W^- ~.
\label{VLQ-decay-modes}
\end{equation}
Note that $t'$ and $b'$ can decay via neutral current at leading order, owing to their vector-like nature.
Decays into another heavy quark coming from the same multiplet, such as $t' \rightarrow X W^-$, are kinematically suppressed; 
decays through loops may also be possible, but they are typically negligible \cite{Cacciapaglia:2010vn}. 
Here we will disregard these sub-leading channels, and assume that the decay channels in \eq{VLQ-decay-modes} have unit branching ratio.
In addition, the decays are assumed to be prompt, that is the case whenever the mixing angles are large enough to have an observable effect on the Higgs couplings.

The relative branching ratios of $t'$ and $b'$ in the three decay channels depend mostly
on the weak isospin of the VLQ and on the mixing angles. Indeed, since the heavy quarks are already constrained to be heavier than a few hundred GeVs,
in good approximation one can neglect the final state masses and find 
\beq\ba{l}
\Gamma(t'\rightarrow h t) \simeq \dfrac{m_{t'}}{16\pi}\left(|y_{tt'}|^2+|\tilde y_{tt'}|^2 \right)~,\\ 
\Gamma(t'\rightarrow Z t) \simeq \dfrac{m_{t'}^3}{32\pi m_Z^2}\left(|(c^Z_L)_{tt'}|^2+|(c^Z_R)_{tt'}|^2 \right)~,\\ 
\Gamma(t'\rightarrow W b) \simeq \dfrac{m_{t'}^3}{32\pi m_W^2}\left(|(c^W_L)_{bt'}|^2+|(c^W_R)_{bt'}|^2 \right)~,
\ea\eeq
and the same for $t'\leftrightarrow b'$ and $t\leftrightarrow b$. Here the Higgs couplings are defined by \eq{hff1}, and the $Z$ and $W$ couplings by \eq{Vmass}. 
By a straightforward computation, one finds $Br \left(t'\rightarrow ht \right)\simeq Br \left(t'\rightarrow Zt \right)$ and
\beq\ba{lll}
T: & Br \left(t'\rightarrow Zt \right)\simeq \dfrac 12 \dfrac{1-s_T^2}{2-s_T^2}~,
& Br \left(t'\rightarrow Wb \right)\simeq \dfrac{1}{2-s_T^2}~;\\
X_T: & Br \left(t'\rightarrow Zt \right)\simeq \dfrac{1}{2}~,
& Br \left(t'\rightarrow Wb \right)\simeq 0~;\\
Q: & Br \left(t'\rightarrow Zt \right)\simeq \dfrac{1}{2} \dfrac{1+s_Q^2(r_Q^2-1)}{r_Q^2+1+s_Q^2(r_Q^2-1)}~, 
& Br \left(t'\rightarrow Wb \right)\simeq \dfrac{r_Q^2}{r_Q^2+1+s_Q^2(r_Q^2-1)}~;\\
X_Q: & Br \left(t'\rightarrow Zt \right)\simeq \dfrac{1}{2} \dfrac{1+ s_{X_Q}^2}{2+ s_{X_Q}^2}~, 
& Br \left(t'\rightarrow Wb \right)\simeq \dfrac{1}{2+s_{X_Q}^2}~; \\
Y_Q: &  Br \left(t'\rightarrow Zt \right)\simeq \dfrac{1}{2} ~, 
& Br \left(t'\rightarrow Wb \right)\simeq 0~.
\ea\label{BRs}\eeq
As before, we neglected the SM masses and, therefore, the subdominant mixing angles in \eq{VLQ-suppressed-mixing}. 
Note however that some branching ratios are proportional to the
SM masses at leading order, for example in the case of $X_T$ one finds $Br \left(t'\rightarrow Wb \right)\simeq m_t^2/(c_{X_T}^4 m_{t'}^2)$.
In the cases where both $t'$ and $b'$ are present, we used the relation  $\tan\tilde\theta_\psi \simeq r_\psi \tan\theta_\psi$,
that follows from  \eq{bt_relations} if one neglects $m_b$ and $m_t$. 
The $b'$ branching ratios are obtained from \eq{BRs} by the replacements $T\rightarrow B$, $X_T\rightarrow Y_B$, $X_Q\leftrightarrow Y_Q$,
$t'\rightarrow b'$, $t\leftrightarrow b$, $r_Q\rightarrow 1/r_Q$ and $s_\psi \rightarrow \tilde s_\psi$
for each $\psi$.

The experimental lower bounds on the mass of $t'$ and $b'$ are presented as a function of two independent branching ratios \cite{ATLAS-CONF-2015-012,Khachatryan:2015axa}. 
We choose, in the plane of branching ratios, their approximate values for the VLQ under consideration. 
To this purpose, we take the limit $s_\psi\rightarrow 0$ in \eq{BRs}  ($\widetilde{s}_\psi\rightarrow 0$ in the case of $b'$), because  the
collider bound is relevant at small mixing angles, see Figs.~\ref{VLQs-singlets}~-~\ref{VLQs-Qdoublet}).
The only exception is $X_T$, where large mixing is possible, but in this case the strongest collider bound is the one on the component $X$.
The lower bounds on each heavy quark mass are collected in Table \ref{5table}, 
and vary between $\sim 600$ and $900$ GeV \cite{ATLAS-CONF-2015-012,Aad:2015mba,Aad:2014efa,CMS:2014dka,CMS:2014bfa,CMS:2012hfa}.
The region excluded at 95 \% C.L. is shaded in grey in Figs.~\ref{VLQs-singlets}~-~\ref{VLQs-Qdoublet}.
A detailed analysis of the lower bound on $m_t'$ is presented in Ref.~\cite{Han:2014qia} for the case of the VLQ $T$, taking also into account indirect constraints from $B$-physics observables.

\begin{table}[bt]
\renewcommand{\arraystretch}{1.3}
\bc\begin{tabular}{|c|c|c|c|}
\hline
heavy quark  & branching ratios & multiplets & mass bound   \\
\hline\hline
$X$  ($Q=5/3$) & $Br_{Wt}=1$  & $X_T$, $X_Q$ & $m_X \geq 840$ GeV ~\cite{Aad:2015mba}   \\
\hline\hline
$t'$ ($Q=2/3$) & $Br_{Wb}=\frac12,~Br_{Zt}= Br_{ht}=\frac14$   & $T$, $X_Q$, $Q(r_Q=1)$ & $m_{t'}\geq 800$ GeV ~\cite{ATLAS-CONF-2015-012}  \\ 
\cline{2-4}
   &  $Br_{Wb}=0,~Br_{Zt}= Br_{ht}=\frac12$ & $X_T$, $Y_Q$, $Q(r_Q\ll 1)$		& $m_{t'}\geq 855$ GeV ~\cite{ATLAS-CONF-2015-012} 	\\ 
\cline{2-4}
       &  $Br_{Wb}=1,~Br_{Zt}= Br_{ht}=0$ &  $Q(r_Q\gg 1)$,  $Q+Y_B$	& $m_{t'}\geq 920$ GeV ~\cite{Khachatryan:2015oba} 	\\ 
\cline{2-4}
           & $Br_{Wb} = Br_{Zt}=0,~Br_{ht}=1$ &  $X_T+Q$	& $m_{t'}\geq 950$ GeV ~\cite{ATLAS-CONF-2015-012} 	\\ 
\cline{2-4}
           & $Br_{Wb} = Br_{ht}=0,~Br_{Zt}=1$ &  $X_T+Q$	& $m_{t'}\geq 800$ GeV ~\cite{ATLAS-CONF-2015-012}	\\ 
\cline{2-4}
       &  $Br_{Wb}+Br_{Zt}+Br_{ht}=1$ &  $T$, $X_T$, $Q$, $Y_Q$, $X_Q$		& $m_{t'}\geq 720$ GeV ~\cite{Khachatryan:2015oba} 	\\ 
\hline\hline
$b'$ ($Q=  -1/3$) & $Br_{Wt}=\frac12,~Br_{Zb}= Br_{hb}=\frac14$  &$B$, $Y_Q$, $Q(r_Q=1)$ & $m_{b'}\geq 735$ GeV  ~\cite{ATLAS-CONF-2015-012}  \\
\cline{2-4}
       &  $Br_{Wt}=0,~Br_{Zb}= Br_{hb}=\frac12$ &$Y_B$, $X_Q$ $Q(r_Q^2\gg 1)$		& $m_{b'}\geq 755$ GeV ~\cite{Aad:2014efa}	\\ 
\cline{2-4}
       &  $Br_{Wt}=1,~Br_{Zb}= Br_{hb}=0$ &  $Q(r_Q^2\ll 1)$,  $X_T+Q$	& $m_{b'}\geq 810$ GeV ~\cite{Aad:2015mba}	\\ 
\cline{2-4}
    &  $Br_{Wt}= Br_{Zb}=0,~Br_{hb}=1$ &  $Q+Y_B$		& $m_{b'}\geq 846$ GeV ~\cite{CMS:2014bfa} 	\\ 
\cline{2-4}
   &  $Br_{Wt}= Br_{hb}=0,~Br_{Zb}=1$ &  $Q+Y_B$		& $m_{b'}\geq 775$ GeV ~\cite{ATLAS-CONF-2015-012}	\\ 
\cline{2-4}
       &  $Br_{Wt}+Br_{Zb}+Br_{hb}=1$ &  $B$, $Y_B$, $Q$, $Y_Q$, $X_Q$		&  $m_{b'}\geq 582$ GeV ~\cite{CMS:2012hfa} 	\\ 
\hline\hline
$Y$  ($Q=-4/3$) & $Br_{Wb}=1$ &$Y_B$, $Y_Q$ & $m_Y \geq 770$ GeV ~\cite{ATLAS-CONF-2015-012} \\
\hline
\end{tabular}\ec
\caption{Lower bounds at 95 \% C.L. on the heavy quark masses $m_X,m_{t'},m_{b'}$ and $m_Y$. 
The experimental searches assume pair production via strong interactions and prompt decays in the indicated channels. 
In the second column we specify the assumption on the heavy quark-decay branching ratios.
Here $Br_{Zt}$  stands for $Br \left(t'\rightarrow Zt \right)$, and so forth. 
In the third column we list the VLQ multiplets that correspond to those branching ratios, in the small mixing approximation.
Here ``$X_T+Q$" and ``$Q+Y_B$" refer to pairs of VLQs with a custodial symmetry, that are discussed in section \ref{2VLQ}.}
\label{5table}
\end{table}

Let us now discuss the corrections induced by the VLQ on the Higgs boson couplings.
The couplings of $t,t',b$ and $b'$ to the Higgs have the same form as those of $\tau$ and $\tau'$ in \eq{tauH}, 
with the obvious  replacement of masses and mixing angles. 
The heavy quarks $X$ and $Y$ do not couple to the Higgs.
The Higgs signal strengths at the LHC $\mu_\alpha$, defined in \eq{SS}, are the product of three factors: Higgs production rate, partial decay rate and lifetime.
While new leptons affect significantly the partial decay rate only, new quarks can modify substantially each factor.
In particular, the Higgs production via gluon fusion  is sensitive to a VLQ,
and the mixing  in the bottom sector can change significantly the total Higgs width $\Gamma_h$.
Let us remind that, as discussed below \eq{VLQ-mass-matrices-B}, the Higgs couplings are CP-conserving for any VLQ.

When the VLQ contains a $B$ component, the Higgs width into $b\overline{b}$ is modified, with respect to the SM, by a factor
$R_{b\overline{b}}=( 1 -\widetilde{s}_\psi^2 )^2$ that, in the light of previously discussed constraints, can be as small as $\sim 0.9$.
Since in the SM $h\rightarrow \overline{b}b$ is the dominant decay channel,
this correction enhances $\mu_\alpha$ for all other decay channels, through the factor  $\Gamma_h^{SM}/\Gamma_h$.
The Higgs production via gluon fusion is modified by a factor 
\begin{equation}
R_{gg} = \frac{\left| {\cal{A}}_{SM}^{gg} +\frac{3}{4} s_\psi^2 \left[A_{1/2}(\tau_{t'})- A_{1/2}(\tau_{t}) \right] + \frac{3}{4} \widetilde{s}_\psi^2 \left[ A_{1/2}(\tau_{b'})- A_{1/2}(\tau_{b}) \right]\right|^2}{\left|{\cal{A}}_{SM}^{gg} \right|^2} ~, 
\label{VLQ-hgg}
\end{equation}
where the form factors are defined in appendix \ref{Hgg}.
The effect of the top sector is qualitatively different from the bottom one:
given the collider lower bound on $m_{t',b'}$, their form factors are very close to the asymptotic value, $A_{1/2}(0)=4/3$.
While the $t$ loop is also close to this value, the $b$ has a small mass and a suppressed form factor:
$A_{1/2}(\tau_{t'}) - A_{1/2}(\tau_{t}) \simeq-0.04$ and $A_{1/2}(\tau_{b'})- A_{1/2}(\tau_{b})\simeq + 1.41$
Therefore, when a $b'$ is present (for $\psi=B,Y_B,Q,X_Q,Y_Q$), its effect dominates and the interference with the SM is constructive.
An exception is possible for $\psi=Q$, where $\tilde{s}_Q/s_Q\ll 1$ for $r_Q\ll 1$, see \eq{bt_relations}.
In the latter case, and when only a $t'$ is present (for $\psi=T,X_T$), there is a slight destructive interference with the SM.
The $t\bar{t}h$ production mode is also modified respect to the SM in the presence of $t-t'$ mixing, with a cross-section
reduced by a factor $c_\psi^4$ .

In the diphoton channel
\begin{equation}
R_{\gamma\gamma}= \frac{\left|{\cal{A}}_{SM}^{\gamma\gamma}+ \frac 43 s_\psi^2 \left[ A_{1/2}(\tau_{t'}) - A_{1/2}(\tau_{t}) \right] 
+ \frac 13 \widetilde{s}_\psi^2 \left[ A_{1/2}(\tau_{b'})- A_{1/2}(\tau_{b}) \right] \right|^2}{\left|{\cal{A}}_{SM}^{\gamma\gamma}\right|^2}~.
\label{VLQ-hpp}
\end{equation}
Here the SM amplitude is negative, therefore the interference pattern is reversed with respect to $R_{gg}$.
A few different values of the signal strength $\mu_{\gamma\gamma}$ are shown in Figs.~\ref{VLQs-singlets}~-~\ref{VLQs-Qdoublet} by dotted red lines.
Once the other constraints are taken into account, one finds at most $\delta\mu_{\gamma\gamma}\sim 0.3$.
Finally, for the Higgs decay into a photon and a $Z$, the new physics amplitude writes
\begin{eqnarray}
{\cal{A}}^{\gamma Z}_{SM+\psi}-{\cal{A}}^{\gamma Z}_{SM} & \simeq  & \sum\limits_{\alpha=t,b} \frac{3 Q_\alpha}{c_w^2} 
\Big\{  \delta g^V_{\alpha\alpha} A_{1/2}(\tau_\alpha,\lambda_\alpha)+s_{\psi,\alpha}^2 \left[g^V_{\alpha'\alpha'} A_{1/2}(\tau_{\alpha'},\lambda_{\alpha'}) -g^V_{\alpha\alpha} A_{1/2}(\tau_\alpha,\lambda_\alpha) \right]  \nonumber \\
& + &  c_{\psi,\alpha} s_{\psi,\alpha} \dfrac{m_\alpha +m_{\alpha'}}{\sqrt{m_\alpha m_{\alpha'}}}
 g^V_{\alpha \alpha'} A_{1/2}(\tau_{\alpha'},\lambda_{\alpha'},\tau_\alpha,\lambda_\alpha)
\Big\}~,
\label{VLQ-hpZ}
\end{eqnarray}
where $s_{\psi,t}\equiv s_\psi$, $s_{\psi,b}\equiv \widetilde{s}_\psi$, and we neglected the terms proportional to the form factor $B_{1/2}$, that are subdominant.
Note also that there is no CP-violating amplitude, as both the $h$ and $Z$ couplings respect CP, see discussion below \eq{VLQ-mass-matrices-B}.
The vector $Z$ couplings are obtained from \eq{Zmass}: one finds $\delta g^V_{\alpha\alpha} \equiv (g^V_{\alpha\alpha}-g^{V,SM}_{\alpha\alpha}) \sim s_{\psi,\alpha}^2$ and 
$g^V_{\alpha \alpha'}\sim s_{\psi,\alpha}$, so that the new physics amplitude of \eq{VLQ-hpZ} is of the order of the mixing squared.
The interference with the SM may be constructive or destructive depending on the sign of the $Z$ couplings. 
A few different values of the signal strength $\mu_{\gamma Z}$ are shown in Figs.~\ref{VLQs-singlets}~-~\ref{VLQs-Qdoublet} by dashed blue lines.
When a $b'$ is present, both $\mu_{\gamma Z}$ and $\mu_{\gamma\gamma}$ receive a similar correction, 
dominated by the increase of $R_{gg}$ and $\Gamma_h^{SM}/\Gamma_h$.
A significant correction is possible for $Y_B$, with both $\delta\mu_{\gamma Z}$ and $\delta\mu_{\gamma\gamma}$ as large as $\sim 0.3$.
On the other hand, in the case of $T$ or $X_T$ the corrections to the two channels are significantly different, because the small factor $A_{1/2}(\tau_{t'}) - A_{1/2}(\tau_{t})$ in \eq{VLQ-hpp} suppresses the correction to $h\gamma\gamma$.
In particular, for $X_T$ one can have $\delta\mu_{\gamma Z}\sim 0.1$ with $\delta\mu_{\gamma\gamma}\sim 0.01$.

In our analysis  we computed the relevant signal strengths $\mu_\alpha$ for $\alpha=b \overline{b}$, 
$\gamma\gamma$, $\gamma Z$, $WW$ and $ZZ$, taking of course into account the corrections to $R_{gg}$ and $\Gamma_h$.
We compared these predictions with the allowed experimental ranges  given in Table \ref{threetable}. By simply computing a $\chi^2$ for these five channels,
we determined the region of parameters disfavoured at $99\%$ C.L., that is delimited by the solid black line in Figs.~\ref{VLQs-singlets}~-~\ref{VLQs-Qdoublet}.

 \subsection{Two vector-like quarks (including $b$ and $t$ compositeness)}\label{2VLQ}

\noindent $\bullet$ \textbf{\textit{Two VLQs not coupled to each other.}}
Let us consider first a pair of VLQs $(\psi,\psi')$ that do not couple to each other via a Yukawa coupling.
In this case $\psi$ and $\psi'$ must be identified with one of the seven VLQs in table \ref{twotable},
and the phenomenological effects are, in most respects, a trivial addition of those of each VLQ separately, already discussed in section \ref{VLQ}.

A noticeable exception occurs when the Yukawa couplings and the vector-like masses of $\psi$ and $\psi'$ respect an additional $SU(2)_R$ global symmetry, that provides custodial protection
for the EW gauge boson couplings: when the parameters approach this custodial limit, the constraints from EW precision tests drastically relax
with respect to the case of a unique VLQ.
There are four pairs that may form a doublet under $SU(2)_R$: the weak singlets $(T,B)$, the weak doublets $(X_T,Q)$ or $(Q,Y_B)$, and the weak triplets $(X_Q,Y_Q)$.
For illustration, we will concentrate on the case of doublets.

The two VLQs transform as bi-doublets under a custodial $SU(2)_L\times SU(2)_R$ symmetry, as long as their Yukawa couplings to the SM fermions 
and their vector-like masses are equal,
\beq\ba{l} 
-{\cal{L}}_{(X_T,Q)}= \frac{\lambda_\psi}{\sqrt{2}} \overline{( X_{T} ~ Q )_L}
\begin{pmatrix} H \\ \tilde{H} \end{pmatrix} t_R
+ M_\psi \overline{( X_{T} ~ Q )_L}
\begin{pmatrix}
X_{T} \\ Q
\end{pmatrix}_R + h.c.  ~,\\
-{\cal{L}}_{(Q,Y_B)}= \frac{\lambda_\psi}{\sqrt{2}} \overline{(Q ~ Y_B)_L}
\begin{pmatrix} H \\ \tilde{H} \end{pmatrix} b_R
+ M_\psi \overline{(Q ~ Y_B)_L}
\begin{pmatrix}
Q \\ Y_{B}
\end{pmatrix}_R + h.c.  ~.
\label{Lagrangian-2VLQ-custodial1}
\ea\eeq
These are the smallest sets of top and bottom quark custodians, respectively \cite{Agashe:2006at,Contino:2006qr,Pomarol:2008bh}.
Note that, in this custodial limit, the additional coupling $\overline{Q}_L H b_R$ ($\overline{Q}_L \widetilde{H} t_R$) must vanish in the top (bottom) case.
Therefore, a mixing occurs only in the top (bottom) sector, and there are no deviations in the bottom (top) couplings, despite the presence of a $b'$ ($t'$) in the spectrum.
The analysis is analogous to the case of $\tau$ custodians, discussed in section \ref{2VLL}.
For example, in the top case the linear combination $t''\equiv (T^{(X_T)}-T^{(Q)})/\sqrt{2}$ does not couple to the Higgs and therefore it does not mix, while the orthogonal combination mixes 
with the SM top quark as in \eq{VLQ-mass-matrices-T}, to form the mass eigenstates $t'$ and $t$. 
The mass spectrum is
\beq\ba{l}
m_{t''}=m_{b'}=m_X=M_\psi \leq m_{t'}\simeq \dfrac{M_\psi}{c_R}~ \qquad (X_T, Q)~,\\
m_{b''}=m_{t'}=m_Y=M_\psi \leq m_{b'}\simeq \dfrac{M_\psi}{\widetilde{c}_R}~\qquad (Q,Y_B)~.
 \label{2VLQs-custodial-spectrum}
\ea\eeq
Due to the custodial symmetry, the values of the heavy quark branching ratios into SM particles differ from the case of a single VLQ, discussed in section \ref{VLQ}.
We assume that the decays to another heavy quark are kinematically suppressed, because of the small mass splitting in \eq{2VLQs-custodial-spectrum},
and once again we neglect the SM masses in the final state, as well as the $t-t'$ ($b-b'$) left-handed mixing angle, that is suppressed by $m_t/m_{t'}$ ($m_b/m_{b'}$). 
Consider for example the top case. The decays $t'/t''\rightarrow Wb$ are suppressed as in \eq{BRs} (here $r_Q=0$). In addition $t'\rightarrow Zt$ vanishes 
because the $Q=2/3$ components of $X_T$ and $Q$ have opposite weak isospin, and $t''\rightarrow ht$ vanishes because $t''$ does not couple to the Higgs.
Similar arguments hold in the bottom case. In summary one finds
\beq\ba{l}
 BR(t''\rightarrow Zt)\simeq  Br(b'\rightarrow W^-t)\simeq Br(X\rightarrow W^+t) \simeq Br(t'\rightarrow ht)\simeq 1\qquad (X_T,Q)~,\\
 BR(b''\rightarrow Zb)\simeq Br(t'\rightarrow W^+b) \simeq Br(Y\rightarrow W^-b) \simeq Br(b'\rightarrow hb)\simeq 1 \qquad (Q,Y_B)~.
\label{2VLQs-custodial-branching}
\ea\eeq
It is amusing that, in these two models, there is one heavy quark decaying exclusively in each of the possible decay channels listed in \eq{VLQ-decay-modes}.
The experimental lower bounds on these heavy quark masses can be read off Table~\ref{5table}.

\begin{figure}[bt]
   \begin{minipage}[c]{.1\linewidth}
      \includegraphics[scale=0.3,trim= 0 0 0 0]{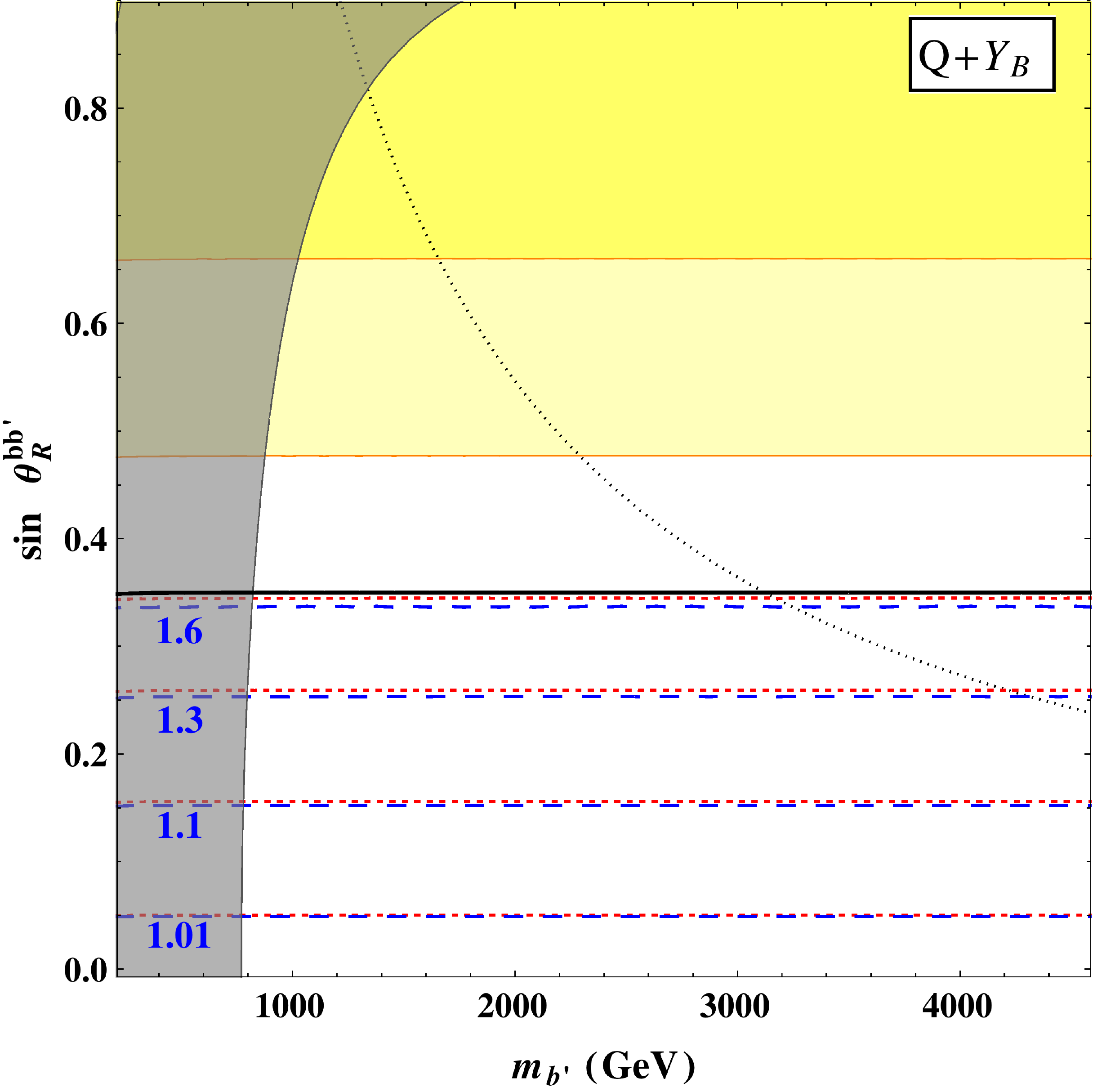}
   \end{minipage} \hfill
   \begin{minipage}[c]{.5\linewidth}
      \includegraphics[scale=0.3,trim= 0 0 0 0]{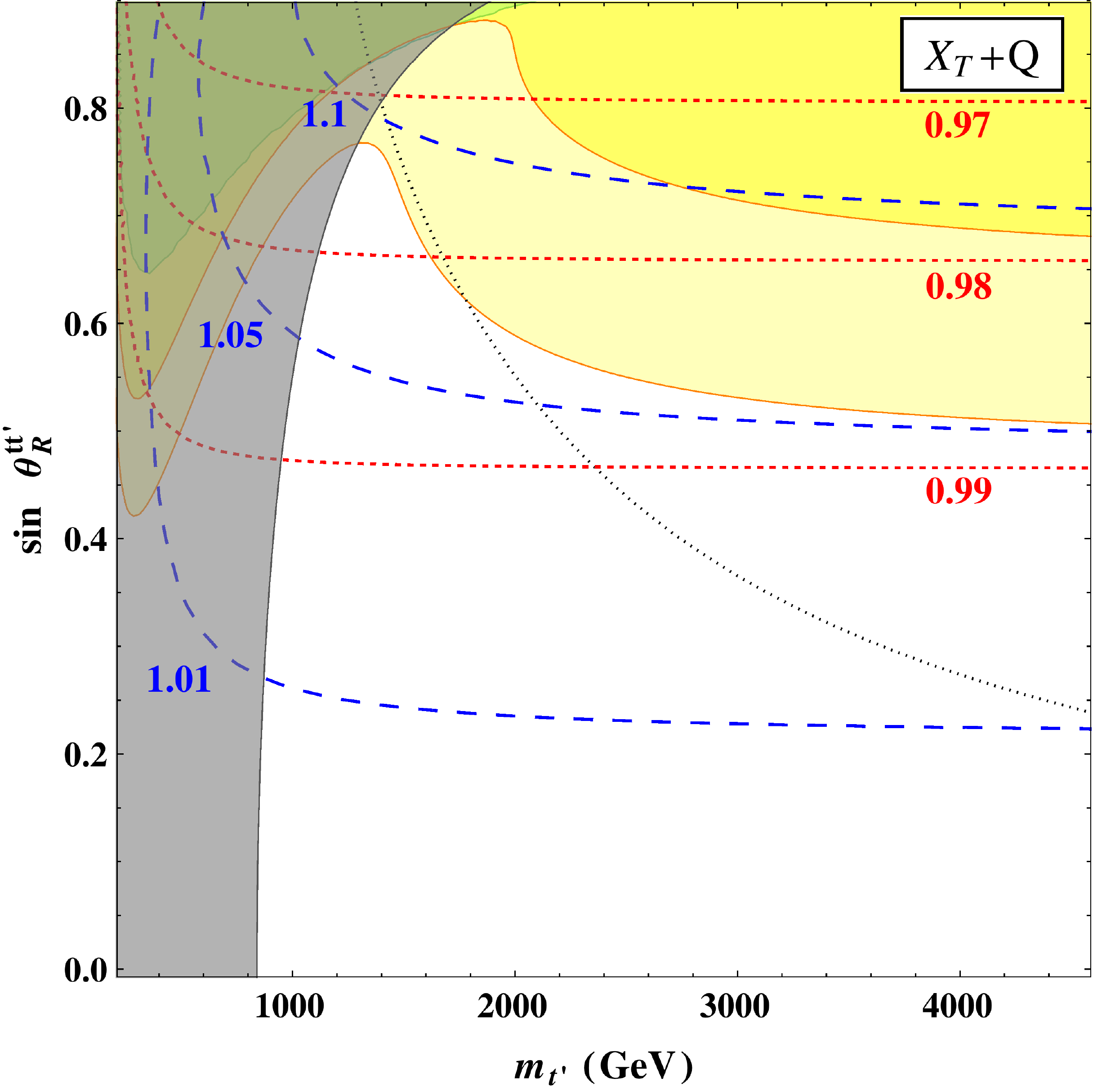}
   \end{minipage}
   \caption{
Constraints on the pairs of VLQs  $(Q, Y_B)$ (left panel) and $(X_T,Q)$ (right panel),
in the custodial limit (equal vector-like masses and Yukawa couplings), 
as a function of the mass of $b'$ ($t'$) and of the mixing angle between $b_R$ and $b'_R$ ($t_R$ and $t'_R$). 
The notation is the same as in Fig.~\ref{VLQs-singlets}.}
   \label{VLQs-custodial}
\end{figure}

The custodial symmetry protects the $Zb\bar b$ couplings: in the top case $\delta g^R_{t\bar t}=0$ and the small $\delta g_{t\overline{t}}^L = (m_t/M_\psi)^2 s_R^2/2$
contributes to $Zb\bar b$ only at one loop; in the bottom case $\delta g^R_{b\bar b}=0$ and
$\delta g_{b\overline{b}}^L=(m_b/M_\psi)^2 \widetilde{s}_R^2/2$ is very suppressed by the bottom mass. 
Thus, in this custodial limit large mixing angles are not excluded, as shown in Fig.~\ref{VLQs-custodial}.
Indeed, one can see that the constraint from the other EW precision parameters is significantly relaxed too,
as $T$ receives a small correction only, from the difference $(\lambda_t-\lambda_b)$, that is not SM-like because of the mixing,
while $S$ acquires a positive correction that remains in the ellipse unless the mixing is very large. 
Note that in the bottom sector the $T$ parameter is almost independent from $m_{b'}$, since the smallness of $m_b$ ensures  $\lambda_b\simeq\lambda_b^{SM}$.
On the other hand, when the mixing occurs in the top sector, the coupling $\lambda_t$ and consequently  the $T$ parameter strongly depend on $m_{t'}$.
The dominant constraints at small and large heavy quark masses come from the direct collider searches and from perturbativity, respectively.
In the top case (right panel of Fig.~\ref{VLQs-custodial}), the mixing is not constrained by the fit of the Higgs coupling, as the bottom sector is SM-like.
As a consequence, for $1.5$ TeV $\lesssim m_{t'} \lesssim 2$ TeV the mixing can be as large as $s_R \simeq 0.8$.
The $\gamma\gamma$ Higgs decay channel can be suppressed at most by $\delta \mu_{\gamma\gamma}\simeq -0.03$,
while the corrections to the $\gamma Z$ channel may be larger, up to $\delta \mu_{\gamma Z}\simeq +0.13$.
In the bottom case (left panel of Fig.~\ref{VLQs-custodial}),
the mixing in the bottom sector  enhances all the other Higgs channels, as $R_{b \overline{b}}= (1-\widetilde{s}_R^2)^2 < 1$.
This leads to an upper bound $\widetilde{s}_R\lesssim 0.35$.
Significant corrections as large as $\delta \mu_{\gamma\gamma}\simeq \delta \mu_{\gamma Z}\simeq 0.6$ are possible.
Note that the Higgs signal strengths in Fig.~\ref{VLQs-custodial} are similar to those with $Y_B$ or $X_T$ only, shown in Fig.~\ref{VLQs-doublets}.
The difference is that the region allowed by EW precision tests largely inflated here, thanks to the custodial symmetry.
\\

\noindent $\bullet$ \textbf{\textit{Two VLQs coupled to each other, not mixing with the SM fermions.}}
Let us move to the case of two VLQs coupled to each other via Yukawa interactions. 
Their chiral components transform as
\begin{equation}
\psi_{1L},\psi_{1R} \sim (R_c, R_w, Y)~,\qquad 
\psi_{2L},\psi_{2R}\sim(R_c, R_w+1,Y+\frac{1}{2})~, 
\label{set2VLQs}
\end{equation}
with $R_c\ne 1$. The corresponding Lagrangian is the same as in the case $R_c=1$ (two VLLs), and it is given in \eq{2VLLs-Lagrangian}.
We bar the special case $R_c=\overline{R_c}$, $Y=0$ and $R_w$ odd ($Y+1/2=0$ and $R_w$ even), 
that allows for a Majorana mass terms for $\psi_{1}$ ($\psi_{2}$) and will be discussed in section \ref{VLMQ}.
We also bar mixing with the SM quarks, that will be discussed 
at the end of the section. 
The effect of two VLQs on $\mu_{\gamma\gamma}$ was discussed in detail in Ref.~\cite{Bertuzzo:2012bt}.

The number of mass eigenstates with a given electric charge $Q$ and the structure of their mass matrices are the same as in the  case of two VLLs, 
see Eqs.~(\ref{2VLLS-mass-matrix}) and (\ref{2VLLS-mass-matrix-Rw}). 
Therefore, there are five physical parameters: two masses $m_{1,2}$, two mixing angles $\theta_{L,R}$ and one phase $\varphi$, defined by \eq{2VLLS-mixing}.
The analysis of the parameter space proceeds exactly as in section \ref{2VLL} and will not be repeated here,
however the phenomenology is strongly modified as the colour representation $R_c$ is non-trivial. 
The main differences are the following:
\begin{itemize}

\item The VLQs are pair-produced via strong interactions and, 
in the absence of mixing with the SM, the lightest state is stable and hadronises. The direct collider bounds on these particle masses are 
above one TeV, as we already described in some more detail in section \ref{chiral}.

\item The contributions of the VLQs to the $S$ and $T$ parameters, as well as to the Higgs decay amplitudes into $\gamma\gamma$ and $\gamma Z$,
have the same form as in Eqs.~(\ref{2VLL-pp}) to (\ref{gg2L}), with an additional factor $N_c$.
As in the case of VLLs, for $R_w=1$ and large values of $Q$ we find two regions of the mixing parameters 
where $\mu_{\gamma\gamma}$ remains SM-like, while $\mu_{\gamma Z}$ can strongly depart from one.
(i) For two degenerate masses $m_1=m_2$, the interference with the SM amplitude is destructive for $\gamma\gamma$
and constructive for $\gamma Z$, as illustrated in Fig.~\ref{fig-2VLLs}. 
Therefore, there are values of the mixing parameters where accidentally $\mu_{\gamma\gamma}$ goes back to the allowed range, while at the same time one can even saturate the present upper bound $\mu_{\gamma Z}\lesssim 10$. 
Note that the gluon-gluon channel remains nearly SM-like, because its amplitude is not enhanced by the large  factor $Q^2$.
(ii) For $m_1\ne m_2$, the amplitude ${\cal A}_{f_1,f_2}^{\gamma\gamma}$ in \eq{2VLL-Zcouplings} can be tuned to zero, while at the same time one can have large contributions to $\mu_{\gamma Z}$ together with sufficiently small corrections to $S$ and $T$. E.g. taking $N_c=3$, $Q \simeq 8$, mixing parameters 
$\varphi=0$, $\theta_L\simeq \pi/8$, $\theta_R \simeq \pi/10$, $m_1/m_2\simeq 1.3$ and $m_2\simeq  1$ TeV, one obtains $\mu_{\gamma Z}\simeq 2$.
For the cases $R_w>1$,  we refer to the discussion below \eq{2VLLS-mass-matrix-Rw}.

\item The VLQs also contribute to the Higgs production by gluon fusion, with an amplitude that can be easily obtained from the $\gamma\gamma$ one.
For the pair of mass eigenstates $f_1,f_2$ of charge $Q$, one has
${\cal A}^{gg}_{f_1,f_2} = [3C(R_c)/2]/(N_c Q^2) {\cal A}^{\gamma\gamma}_{f_1,f_2}$, see appendix \ref{appendix B.2}. 
The interference with the SM is constructive in the gluon case, thus enhancing the Higgs production.
Note that the gluon-gluon channel also receives a non-zero contribution from the $Q=0$ sector.
\end{itemize}

\begin{figure}[bt]
   \begin{minipage}[c]{.1\linewidth}
      \includegraphics[scale=0.3,trim= 0 0 0 0]{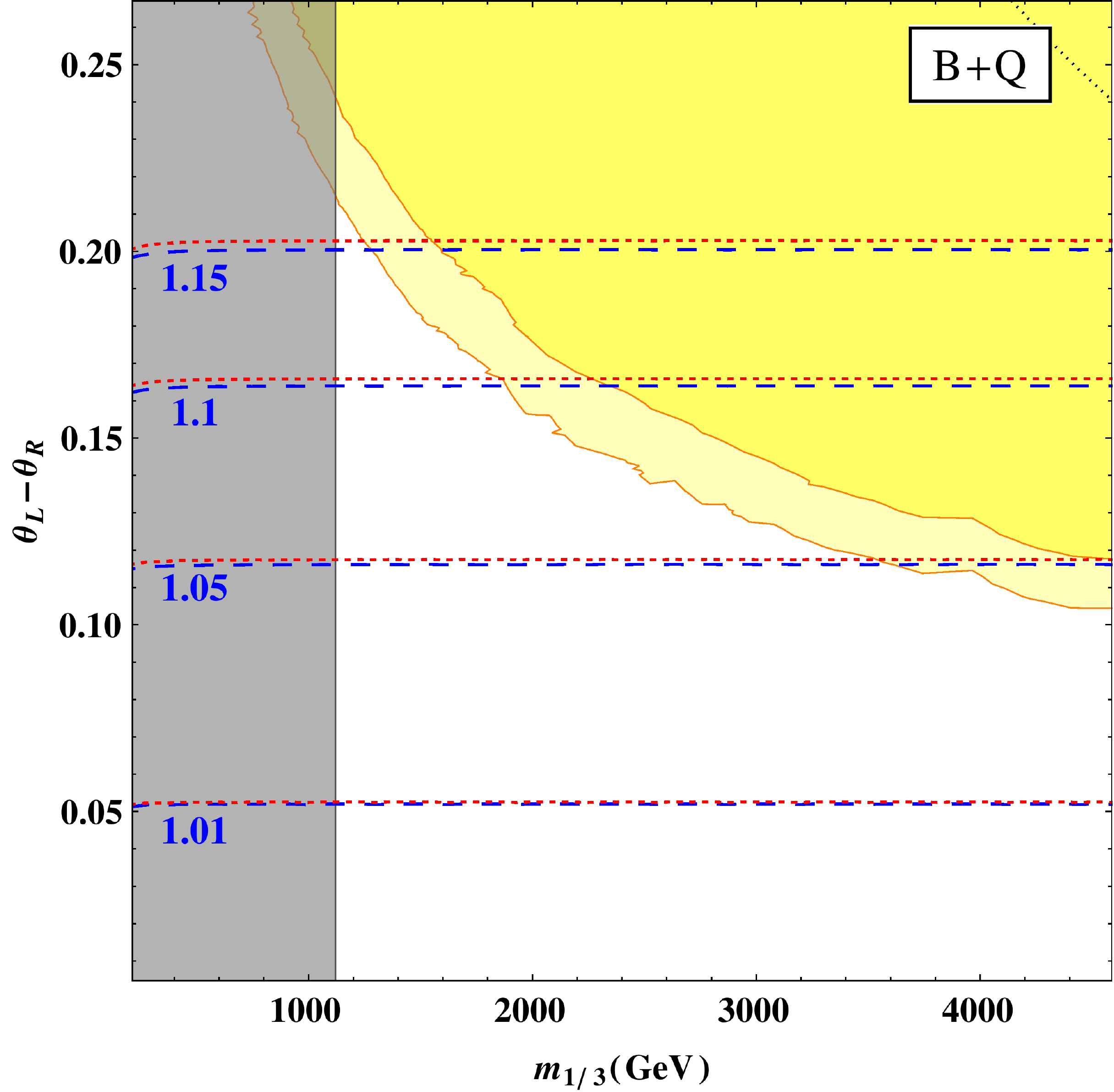}
   \end{minipage} \hfill
   \begin{minipage}[c]{.5\linewidth}
      \includegraphics[scale=0.3,trim= 0 0 0 0]{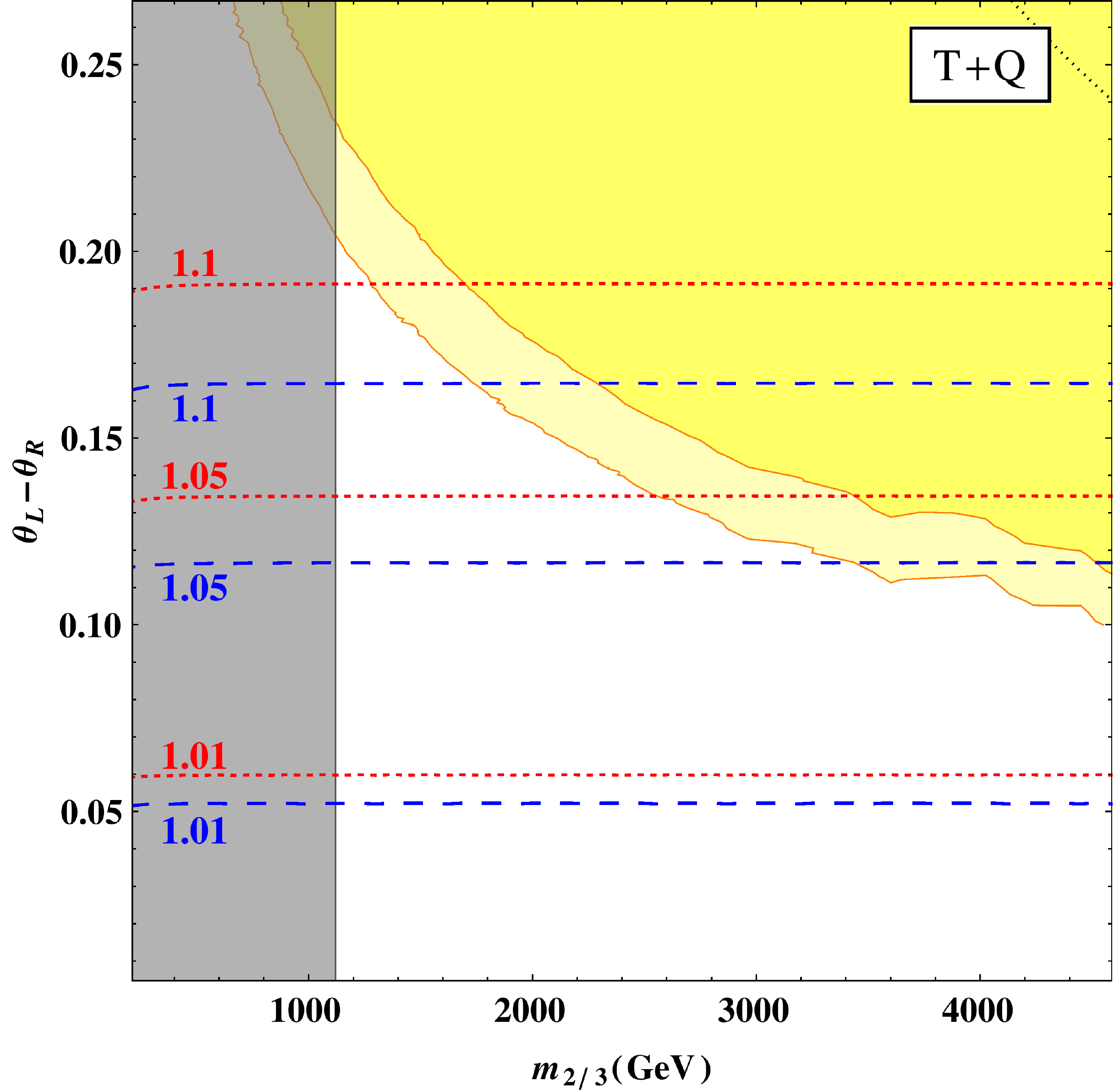}
   \end{minipage}
\caption{In the left (right) panel, we show the constraints on the pair of VLQs $B$ and $Q$  ($T$ and $Q$),
in the limit of vanishing mixing with the SM quarks, degenerate mass eigenvalues $m_{b'}=m_{b''}\equiv m_{1/3}$ ($m_{t'}=m_{t''}\equiv m_{2/3}$) 
and no CP violation, $\varphi=0$. In this case the relevant mixing angle is $\theta_L-\theta_R$, see \eq{2VLL-pp}. 
The notation for the various constraints is the same as in Fig.~\ref{VLQs-singlets}.}
   \label{VLQs-general-bottom}
\end{figure}

\begin{figure}[bt]
   \begin{minipage}[c]{.1\linewidth}
      \includegraphics[scale=0.3,trim= 0 0 0 0]{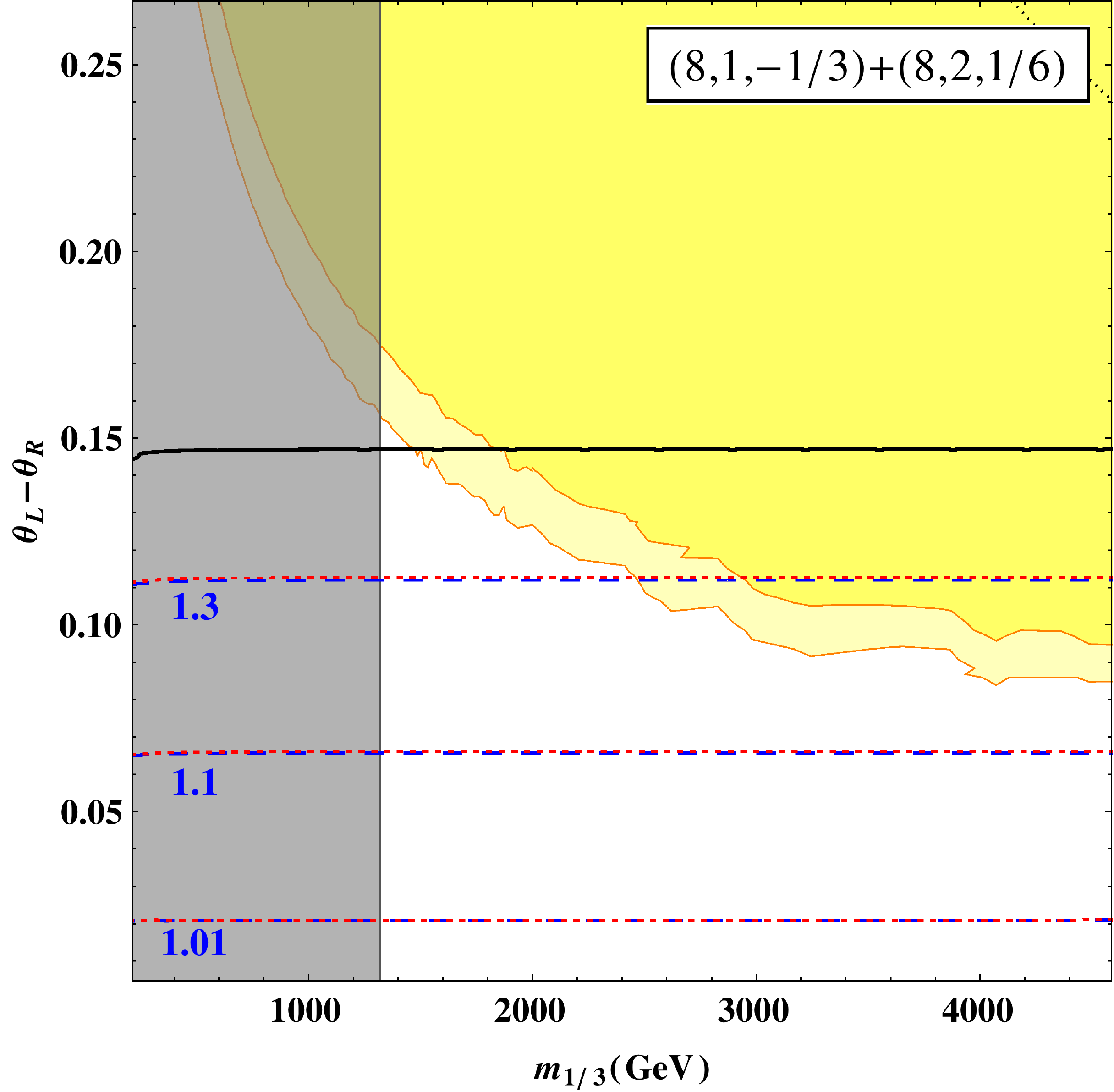}
   \end{minipage} \hfill
   \begin{minipage}[c]{.5\linewidth}
      \includegraphics[scale=0.3,trim= 0 0 0 0]{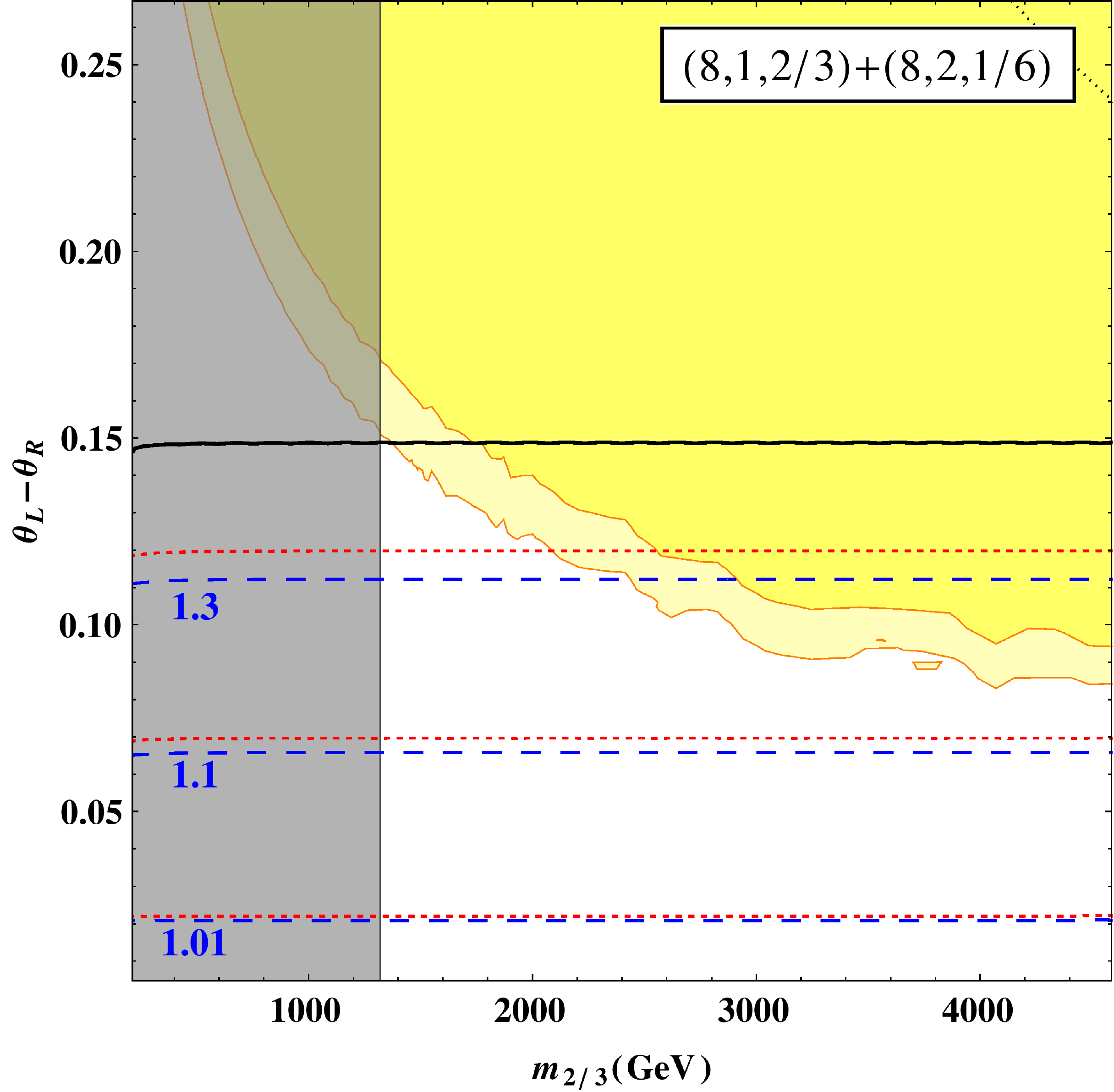}
   \end{minipage}
\caption{The same as in Fig.~\ref{VLQs-general-bottom}, but replacing colour triplets with colour octets.}
   \label{VLQs-general-top}
\end{figure}

For illustration, we display in Fig.~\ref{VLQs-general-bottom}  the parameter space for the case $R_c=3$, $R_w=1$ and $Y=-1/3$ ($Y=-2/3$),
that corresponds to the pair of VLQs $B$ and $Q$ ($T$ and $Q$),
in the limit of no-mixing with the SM quarks. 
In Fig.~\ref{VLQs-general-top}  we illustrate how the constraints change for a larger colour representation, as we replaced $R_c=3$ by $R_c=8$.
For definiteness, we assumed that there is no CP violation, $\varphi=0$, and that the two mixing mass eigenstates $f_{1,2}$ of charge $Q=Y$ 
are degenerate in mass, $m_1=m_2$. 
In the case of colour octets, larger deviations in the Higgs signal strengths are possible, but the various constraints are correspondingly stronger. 
One can reach $ \mu_{\gamma Z}\simeq 1.4$ for the octets and $\mu_{\gamma Z}\simeq 1.2$  for the triplets. 
In the case $Y=-1/3$, $\mu_{\gamma\gamma}$ and $\mu_{\gamma Z}$ are very close to each other, because 
${\cal A}^{\gamma Z}_{f_1,f_2}/{\cal A}^{\gamma \gamma}_{f_1,f_2}= (g^V_{11}+ g^V_{22})/(Y c_w^2)$ is numerically close to one, see Eqs.~(\ref{2VLL-pp}) and (\ref{2VLL-Zcouplings}),
and ${\cal A}^{\gamma Z}_{SM}/{\cal A}^{\gamma \gamma}_{SM}$ is close to one as well.
The strongest constraint on the mixing among the heavy states
comes from the $S$ and $T$ parameters.
In the octet case the fit of the main Higgs decay channels (see the end of section \ref{VLQ} for details) is also a relevant constraint. 
The mass scale $m_1=m_2$  is constrained by the searches of stable coloured particles, discussed in  section \ref{chiral}.
\\

\noindent $\bullet$ \textbf{\textit{Two VLQs coupled to each other, mixing with the SM fermions.}}
Let us briefly discuss the possible interactions between the two VLQs $\psi_{1,2}$ in \eq{set2VLQs} and the SM quarks.
This requires of course $R_c=3$.
A non-zero mixing with the bottom and/or top quark can occur if and only if at least one VLQ belongs to the set of seven VLQs  in Table~\ref{twotable}. 
The complete list is
\beq
{\rm singlet~+~doublet~:~} \left\{
\ba{ll} 
T+X_T ~(2t')~, & T+Q ~(2t'+b')~,\\  
B+Q ~(t'+2b')~, & B+Y_B ~(2b')~,
\ea\right.
\eeq
\beq
{\rm doublet~+~triplet~:~} \left\{
\ba{ll} 
X_T+ Z_{X_T}  ~(2t')~, & X_T+ X_{Q} ~(2t'+b')~, \\
Q+ X_Q  ~(2t'+2b')~, & Q + Y_Q ~(2t'+2b')~, \\
Y_B+ Y_Q ~(t'+2b')~, &  Y_B+ W_{Y_B} ~(2b')~,
\ea\right.
\eeq
\beq
{\rm triplet~+~quartet~:~} \left\{
\ba{ll} 
X_Q+ \Omega_{X_T} ~(2t'+2b')~, & X_Q + \Omega_{Q} ~(2t'+2b')~,\\
Y_Q+ \Omega_{Q} ~(2t'+2b')~, & Y_Q + \Omega_{Y_B} ~(2t'+2b')~,
\ea\right.
\label{mixingquarksL}\eeq
where we indicated in brackets the number of new states mixing with the top and with the bottom quark.
We also introduced a few new multiplets, with no Yukawa couplings to the SM fermions: $Z_{X_T}\sim(3,3,5/3)$, $W_{Y_B}\sim(3,3,-4/3)$,
$\Omega_{X_T}\sim (3,4,7/6)$,  $\Omega_Q \sim (3,4,1/6)$, and  $\Omega_{Y_B}\sim (3,4,-5/6)$.
They can be written in components as
\begin{equation}
Z_{X_T}=\begin{pmatrix}
Z\\ X\\T
\end{pmatrix}~, \quad 
W_{Y_B}=\begin{pmatrix}
B\\ Y \\W
\end{pmatrix}~, \quad \Omega_{X_T}= \begin{pmatrix}
Z\\X\\T\\B
\end{pmatrix}~, \quad
\Omega_Q = \begin{pmatrix}
X\\T\\B\\Y
\end{pmatrix}~, \quad 
\Omega_{Y_B} = \begin{pmatrix}
T\\B\\Y\\W
\end{pmatrix}~,
\label{2VLQs-new-multiplets}
\end{equation}
where the two new exotic states $Z$ and $W$ have charges $Q(Z)=8/3$ and $Q(W)=-7/3$.
Recasting LHC searches, Ref.~\cite{Matsedonskyi:2014lla} puts a lower bound of $940$ GeV on the mass of the $Q=8/3$ state.

When there is only one $t'$ ($b'$) state, the mixing in the top (bottom) sector has the same pattern as in section \ref{VLQ}.
On the other hand, when there are two $t'$ states, the top sector mass matrix takes the form 
\beq
{\cal M}_t = 
\begin{pmatrix}
\lambda_t \frac{v}{\sqrt 2} & \lambda_1 \frac{v}{\sqrt 2} & m_2 \\
m_1 & M_1 & \lambda_{12} \frac{v}{\sqrt 2} \\
\lambda_2 \frac{v}{\sqrt 2} & \lambda_{21} \frac{v}{\sqrt 2} & M_2
\end{pmatrix}~,
\label{2VLQS-mixing-SM}
\eeq
where we dropped possible Clebsch-Gordan coefficients. Here $\psi_1$ ($\psi_2$) is a weak singlet or triplet (doublet or quartet),
$m_1$ ($m_2$) vanishes unless $\psi_1=T$ ($\psi_2=Q$), and $\lambda_{1,2}$ vanishes if $\psi_{1,2}$ is one of the multiplets in \eq{2VLQs-new-multiplets}.
The bottom sector mass matrix in presence of two $b'$ states has an analog structure.
The mixing with the SM is controlled by ratios of the type $\lambda v /M$ or $m/M$.
They are mostly constrained by the $Zb\bar b$ couplings and by $S$ and $T$. 
One typically expects $\left| \lambda v /M \right|$,$\left|m/M \right| \lesssim 0.1 - 0.2$, in analogy with 
Figs.~\ref{VLQs-singlets}~-~\ref{VLQs-Qdoublet}.
Possible cancellations among the various contributions could relax these constraints.
A detailed analysis of the whole parameter space of these models is beyond the scope of this paper.
Some recent study can be found in  Ref.~\cite{Cacciapaglia:2015ixa}, that discusses the phenomenology of the $2t'$ and $2t'+b'$ cases.

Let us focus on the possibility to generate the top mass (and analogously the bottom one) through the mixing with the VLQs,
in the limit where the SM Yukawa coupling $\lambda_t$ ($\lambda_b$) vanishes. 
This is possible whenever the determinant of ${\cal M}_t$ in \eq{2VLQS-mixing-SM} is non-zero for $\lambda_t=0$, that is, 
if and only if the two VLQs both couple directly to the SM. 
The resulting top mass is of order $m_t\sim \lambda v (\lambda v/M)^2$, $\lambda v (m/M)$ or $\lambda v (m/M)^2$.
The latter possibility is motivated by partial compositeness. In this scenario, the SM fermions do not couple directly to the composite Higgs, therefore 
$\lambda_t=\lambda_1=\lambda_2=0$ in \eq{2VLQS-mixing-SM}. Rather, they couple linearly to a composite vector-like fermion with the same quantum numbers.
This corresponds to the VLQs $T$ and $Q$ for the case of the top quark, leading to $m_t\simeq (m_T/M_T)(m_Q/M_Q)\lambda_{QT}v/\sqrt 2$,
and analogously $B$ and $Q$ for the case of the bottom.
The phenomenology of top and bottom partners in composite models, and the associated constraints, are analysed e.g. in Refs.~\cite{Pomarol:2008bh,Gillioz:2008hs,Azatov:2011qy} (see also Ref.~\cite{Gopalakrishna:2013hua} for warped extra dimensional models).

 \subsection{Vector-like plus Majorana quarks}\label{VLMQ}

We define a Majorana quark to be a $Y=0$ fermion multiplet in a non-trivial, real colour representation, $R_c=\overline{R_c}\neq 1$. Such object may
couple to the Higgs only in the presence of a VLQ in the same colour representation.
As a consequence, these new fermions do not mix with the SM ones and the lightest mass eigenstate is stable. 
For the smallest possible representation, $R_c=8$, the searches for a stable gluino lead to a lower bound $\simeq 1.3$ TeV \cite{ATLAS:2014fka}.
For bigger representations, $N_c\ge 27$, one expects an even more stringent limit,  given the larger production cross-section and a similar hadronisation behaviour.
\\

\noindent $\bullet$ \textbf{\textit{One VLQ plus one Majorana quark.}}
The most general set formed by a Majorana quark coupled to a VLQ can be written as 
\begin{equation}
\chi_R \sim (R_c,R_w,0) ~,\quad \psi_{L},\psi_{R} \sim(R_c,R_w\pm 1,-1/2)~, \quad\quad R_c= \overline{R_c} \neq 1~.
\label{1VLQ-plus-MAjorana-cases}
\end{equation}
If $R_c$ is odd, one needs $N_w\ne 2+4n$ to avoid the global $SU(2)_w$ anomaly.
The Lagrangian and the structure of the mass matrices are identical to the analogue leptonic case $R_c=1$, see Eqs.~(\ref{1VL-1MAJ-Lagrangian}) to 
(\ref{VL+MAj-matrix2}).
Here we discuss only the phenomenological differences due to the effect of colour.  
The new states contribute with an additional factor $N_c$ to the one-loop diagrams for the $S$ and $T$ parameters, as well as for the Higgs signal strengths 
$\mu_{\gamma\gamma}$ and $\mu_{\gamma Z}$.
In addition, they also contribute to the Higgs production via gluon fusion, with an  amplitude related to the photon-photon one. For each sector of charge $Q$ that couples to the Higgs,
one has ${\cal A}^{\gamma\gamma}_{\chi\psi,Q} = (N_c Q^2)/[3C(R_c)/2] {\cal A}^{gg}_{\chi\psi,Q}$.
The gluon-gluon channel receives a non-zero contribution even from the $Q=0$ sector, that is present when $N_w$ is odd.

Since the new quarks are necessarily heavy, their loop contributions can be estimated with good accuracy using the LET approximation,
as described in Appendix \ref{appendix B.2} and in section \ref{MVL}.
In particular, ${\cal A}^{\gamma\gamma}_{\chi\psi,Q}$ is 
obtained from the amplitude in \eq{LET-VLM1pp}, times a factor $N_c$. For the gluon-gluon channel, summing over the different sectors one finds
\begin{equation}
{\cal{A}}^{gg}_{\chi\psi}\simeq K_{N_w} C(R_c) \frac{\lambda v}{M_\chi} \frac{\widetilde{\lambda} v}{M_\psi}~, 
\label{VLMQ-Agg}
\end{equation}
where $K_{1}= -4$, $K_{2}= -2$, $K_{3}= -8/3$, $K_{4}= -2$, and so on.
As the top quark amplitude is approximately equal to one, ${\cal{A}}_{gg}^{\chi \psi}$ gives roughly the ratio between the contribution of new fermions and the SM one.
Note that either constructive or destructive interference with the SM amplitude is possible.
The mixing parameters, of the generic form $\lambda v/M$, 
must satisfy the constraints from $R_{gg}$ and $R_{\gamma\gamma}$, whose allowed ranges are given in Appendix \ref{experimental}.
In the minimal case with $R_c=8$ and $R_w=1$, only $R_{gg}$ receives a correction, leading to the upper bound $\left| \lambda v/M \right|\lesssim 0.17$.
One expects similar or even stronger bounds from 
$S$ and $T$, in analogy with the cases of 
Figs.~\ref{VLQs-general-bottom} and \ref{VLQs-general-top}. 
Up to possible cancellations, a larger $R_c$ leads to stronger constraints on the model, and to larger deviations in the Higgs couplings.
\\

\noindent $\bullet$ \textbf{\textit{One VLQ plus two Majorana quarks.}}
Coming to sets formed by two Majorana quarks plus one VLQ, the first obvious possibility is to add a second copy of $\chi_R$ to the previous case. 
The additional Majorana multiplet automatically cancels the global $SU(2)$ anomaly, therefore $N_w$ is arbitrary. The phenomenology
is the generalisation of the one discussed above.

The second and last possibility is provided by the set
\begin{equation}
\chi_{1R} \sim (R_c,R_w,0),~~ \psi_L,\psi_R \sim (R_c,R_w + 1,-1/2),~~  \chi_{2R} \sim (R_c,R_w+2,0),~~R_c= \overline{R_c}\neq 1~,
\label{1VLQ-plus-MAjorana-cases2}
\end{equation}
with $N_w$ necessarily odd if $R_c$ is odd. 
The Lagrangian is the same as in \eq{1VL-2MAJ-Lagrangian}, and the structure of the mass matrices is also the same as in section \ref{MVL}.
Let us just present the amplitude for Higgs production into gluon-gluon fusion, that is obtained by generalising \eq{VLMQ-Agg},
\begin{equation}
{\cal{A}}^{gg}_{\chi_1\chi_2\psi}(R_w) \simeq C(R_c) \left(K_{N_w} \dfrac{\lambda_1 v}{M_{\chi_1}} \dfrac{\widetilde{\lambda}_1 v}{M_\psi}
+ K_{N_w+2} \dfrac{\lambda_2 v}{M_{\chi_2}} \dfrac{\widetilde{\lambda}_2 v}{M_\psi} \right)~.
\label{Agg-2Majorana-VL}
\end{equation}
In the minimal case where $R_c=8$ and $R_w=1$, the allowed range for $R_{gg}$  leads to a bound on the mixing parameters  $|\lambda v/M| \lesssim 0.13$,
where we assumed there is no hierarchy nor cancellations among the various mixing parameters.
The $R_{\gamma\gamma}$ constraint is less restrictive.

\section{Conclusions}\label{conclu}

We undertook a systematic analysis of new fermions interacting with the Higgs boson. Their properties (gauge charges, masses, Yukawa couplings) 
are significantly more constrained after the measurement of the Higgs mass and couplings at the first run of the LHC. It is intriguing
to identify the few extensions of the SM that outlived this test. We especially aimed at those scenarios that may depart from the decoupling limit, 
in which the new fermions become very heavy and/or their mixing with the SM becomes very small.

In section \ref{criteria} we presented the complete classification of sets of $n$ chiral fermions interacting with the Higgs, for $n\le 4$. 
While the minimal possibilities are well-known, already for $n=3$ and $4$ we singled out several exotic sets of fermions with a peculiar phenomenology.
They emerge from a non-trivial interplay of several self-consistency conditions: cancellation of gauge anomalies, absence of
charged massless components, non-zero Yukawa coupling to the SM Higgs doublet.
In our classification we recovered as a special case the fermion content of well-motivated theories beyond the SM, such as the seesaw, supersymmetry, or partial compositeness.
These cases are situated in a more general context, by considering the most general Lagrangian for the new fermions,
not restricted by additional theoretical considerations. Would the evidence of a new particle emerge from data, one should indeed explore
the full parameter space, before endorsing a specific model.
We also argue that larger sets of new fermions, with $n\ge 5$, do not allow for qualitatively different phenomena, as
all the possible building blocks of a fermion mass matrix already appeared in our classification.

In order to examine the phenomenology of the new fermions, in the appendices \ref{appendixA} and \ref{appendix B} 
we derived the general expression of the fermion couplings to the EW gauge bosons and
to the Higgs boson, for fermions in arbitrary SM representations $(R_c,R_w,Y)$. 
We also provided the formalism to define the gauge and Higgs boson couplings to the fermion mass eigenstates, after EWSB.
Besides these tree-level results, we presented the general one-loop amplitudes for the gauge boson vacuum polarisation, $\Pi_{VV'}$, that allows to define the
EW oblique parameters $S$ and $T$, and for the Higgs coupling to gauge bosons, $hVV'$, that allows to compute the rate for $h\rightarrow gg,\gamma\gamma,\gamma Z$. 

Let us summarise the main results of our phenomenological survey of sections \ref{chiral} to \ref{phenoQ}:
\bit
\item Several exotic families of chiral fermions, that receive a mass from EWSB only, are still marginally compatible with EW precision tests and direct collider bounds. 
However, the coloured ones are neatly excluded, as they would greatly enhance the $hgg$ coupling. On the other hand, a colourless family formed by two weak doublets 
and four singlets is still compatible with the measured $h\gamma\gamma$ coupling. 
\item The mixing of two or more sterile neutrinos 
with the SM leptons can have observable effects, despite the smallness of the neutrino masses. 
If the sterile neutrinos are lighter than the EW scale, they may modify
significantly the Higgs invisible width; if heavier, they can appreciably contribute to the $S$ and $T$ parameters. 
\item In general, a heavy charged lepton $\tau'$ cannot mix significantly with the $\tau$ because of the $Z\tau\tau$-coupling constraint. Nonetheless, in a few special regions
of parameters interesting phenomena are possible: (i) If $m_{\tau'}<m_h$, the decay rate for $h\rightarrow \tau'\tau$ can be significant despite the small mixing.
(ii) When both $\tau_L$ and $\tau_R$ mix with heavy leptons, 
it is possible to generate $m_\tau$ entirely through the small mixing permitted by the $Z$ couplings, 
as long as the two heavy leptons are connected by a Yukawa coupling $\lambda\gtrsim 3$. 
(iii) If the new lepton sector is arranged to have an approximate custodial symmetry,
the $Z$ couplings are protected. In this case a large $\tau-\tau'$ mixing is allowed, and it 
may strongly suppress $h\rightarrow \tau\tau$. 
\item There are two extended classes of new fermions that can couple to the Higgs doublet without involving the SM fermions:
either a pair of vector-like fermions, whose components can have an arbitrary charge $Q$, or
a pair formed by a vector-like and a Majorana fermion, whose components have (demi-)integer $Q$. These fermion pairs were not studied in full generality in the previous literature,
and they can produce large observable effects, even when the mixing with the SM fermions is zero. 
By varying their mass matrix  parameters, one can typically scan over the full allowed range for the signal strength $\mu_{\gamma\gamma}$, while remaining in 
agreement with direct collider searches and EW precision tests.
In most cases  $\mu_{\gamma Z}$ receives a correction comparable  to $\mu_{\gamma\gamma}$, but when the latter is accidentally close to one, 
it is possible to have $\delta\mu_{\gamma Z}\gg\delta\mu_{\gamma\gamma}$.
We will discuss this point in detail below.
\item The mixing $\theta$ of a heavy quark $t'$ or $b'$ with its SM partner is constrained by the EW precision tests. 
Nonetheless, the $b-b'$ mixing may significantly suppress the $hb\bar{b}$ coupling, leading to corrections as large as
$\delta\mu_{\gamma\gamma}\simeq \delta \mu_{\gamma Z} \simeq 0.6$.
We also notice two remarkable circumstances that allow for a large mixing:
(i) The corrections to $Zb\bar b$ from $t-t'$ mixing are loop-suppressed and may be also suppressed by $(m_t/m_{t'})^2$.
The $T$ parameter receives opposite sign corrections that cancel each other for a specific value of $\sin\theta \times m_{t'}$.
Both conditions can be realised in the case of the VLQ doublet $(X,T)$, allowing for a large $\sin\theta\lesssim0.5$.
(ii) The Lagrangian of the new quarks
can preserve a custodial symmetry, that suppresses the corrections to $T$ as well as to the $Z b \bar b$ couplings.
In this custodial subspace of parameters, the upper bound on the mixing relaxes, the exact value depending on the model:
for the VLQ doublet $Q$ coupled to $t_R$ and $b_R$ we find $\sin\theta\lesssim 0.15$, while for the two doublets 
$Q+Y_B$ coupled to $b_R$ ($X_T+Q$ coupled to $t_R$) one can reach $\sin\theta\lesssim 0.45$ ($\sin\theta\lesssim 0.8$).
\eit

In the course of our analysis, we paid 
special attention to the relative contribution of the new fermions to $h\rightarrow \gamma\gamma$ and $h\rightarrow \gamma Z$, as the former rate is already 
constrained to be close to the SM prediction, while the latter could still depart strongly from its SM value.
It is commonly believed that new physics cannot provide a large correction to the $\gamma Z$ channel without affecting $\gamma\gamma$
as well. Indeed, let us consider the effective Lagrangian before EWSB, 
that corresponds to the limit where the new fermions are heavier than the EW scale.
There are several dimension-six operators involving the Higgs doublet $H$ and the field-strengths $B_{\mu\nu}$, $W^a_{\mu\nu}$, listed e.g.  in Ref.~\cite{Elias-Miro:2013gya}. The operators contributing to $h\gamma Z$
can be generated, at one loop, only by two fermion multiplets coupled to $H$.
At least one of these fermions has non-zero hypercharge, thus it necessarily induces the operator $H^\dag H B_{\mu\nu}B^{\mu\nu}$ as well, that 
contributes to $h\gamma\gamma$.
One can rephrase the same argument 
in terms of the effective Lagrangian  for the $hVV'$ couplings after EWSB,
that is displayed in \eq{hVVeff}. The coefficients of the dimension-five operators, generated at one loop by the fermion mass eigenstate $f_i$,
are given in \eq{App} for $h\gamma\gamma$ and in \eq{ApZ} for $h\gamma Z$. 
The fermion $f_i$ cannot contribute to the $\gamma Z$ channel only, simply because one needs a charge $Q_i\ne 0$ and a non-zero coupling $y_i$ (or $\tilde{y}_i$) to the Higgs, therefore the $\gamma\gamma$ channel receives a contribution too. 
This argument, however, has some loopholes: first, the sum over all fermion mass eigenstates can lead to a cancellation in the signal strength
$\mu_{\gamma\gamma}$ and not in $\mu_{\gamma Z}$, as the summands in the two channels differ by a factor $\sim g^V_i/Q_i$; second, $h\gamma Z$ receives
an additional contribution from loops involving two fermion mass eigenstates, with off-diagonal couplings to both $h$ and $Z$,
see Eqs.~(\ref{CP-even}) and (\ref{CP-odd}).

As a matter of fact, in our survey of fermionic extensions of the SM,
we encountered a few scenarios where $\delta\mu_{\gamma Z}\gg \delta\mu_{\gamma\gamma}$:
\bit
\item[(i)] One can exploit the order one differences between the $Z$ and $\gamma$ couplings and loop functions. 
For example, in the case of $t-t'$ mixing, $\delta\mu_{\gamma\gamma}$ is
proportional to $A_{1/2}(\tau_{t'})-A_{1/2}(\tau_t)$, that is very small as both form factors are close to the asymptotic value $A_{1/2}(0)$. 
On the contrary,
the correction to $\mu_{\gamma Z}$ is controlled
by $g^V_{t't'} A_{1/2}(\tau_{t'},\lambda_{t'})- g^V_{tt} A_{1/2}(\tau_t,\lambda_\tau)$, that is in general of order one.
Also, off-diagonal loops provide an additional contribution of the same order.
Unfortunately, the absolute size of the correction is too small to be observed,
as the mixing between the SM and new fermions is subject
to the EW precision constraints.
We find
at best $\delta\mu_{\gamma\gamma}\ll\delta \mu_{\gamma Z} \simeq 0.2$,
hardly visible even with 3000~fb$^{-1}$ at 14 TeV, see Table~\ref{fourtable}.
\item[(ii)] A much larger $\delta\mu_{\gamma Z}$ is possible when new fermions couple to each other through the Higgs.
Each sector of heavy states with given charges $N_c$ and $Q$ 
gives a contribution to the $h\gamma\gamma$ amplitude proportional to $\sum_i y_iv/m_i$.
We found that the structure of the fermion mass matrix allows this sum to vanish,
see the discussion below \eq{gg2L}.
At the same time, the $\gamma Z$ amplitude is proportional to $\sum_i g^V_i y_i v/ m_i$. 
One can obtain e.g. $\mu_{\gamma Z}\simeq 2$, by means of a pair of states with $N_c Q s_\psi^2 \simeq 3$, 
where $s_\psi$ is the relevant mixing parameter.
Alternatively, the same effect is produced by several states with smaller charges.
The required set of parameters can be in agreement with $S$ and $T$ as well.
This opens a discovery opportunity for the second run of the LHC.
\item[(iii)] There is a second possibility to achieve a large $\mu_{\gamma Z}$.
The signal strength $\mu_{\gamma\gamma}$ may be accidentally close to the SM, because the amplitude generated by two
new fermion multiplets coupled to the Higgs has sign opposite to the SM one, and for
${\cal A}_{f}^{\gamma\gamma}\simeq - 2 {\cal A}_{SM}^{\gamma\gamma}$ one recovers $\mu_{\gamma\gamma}\simeq 1$.
One needs either small weak multiplets with large charges, $N_c Q^2s_\psi^2 \simeq 5$, or larger multiplets with smaller charges.
In this region of parameters the $S$ and $T$ constraint can be satisfied and, moreover,
one generically expects $\mu_{\gamma Z}$ much larger than one, 
because for large values of $Q$ the amplitude ${\cal A}_{f}^{\gamma Z}$ interferes constructively with the SM.
We find that one can almost saturate the present experimental bound $\mu_{\gamma Z}\lesssim 10$,
therefore coming LHC data will be able to quickly probe this scenario.
\eit
In the cases listed above, a mild tuning of the parameters is sufficient to comply with 
the presently allowed range for $\mu_{\gamma\gamma}$, shown in Table~\ref{threetable}. 
In the future, the room for a large
$\mu_{\gamma Z}$ will progressively shrink.

The second run of the LHC, that recently started data taking at 13 TeV, will close in on most of the scenarios we have been considering.
The allowed regions of parameters at low masses will be covered by direct searches for new fermionic resonances.
The islands that survive at  large mixing between the SM and new fermions will be probed by the increasing precision in the Higgs coupling measurements,
even though there are models where one needs to wait for a high accuracy.
In the absence of a signal, 
we shall be virtually cornered to the region of very heavy masses and/or very small mixing. 
Even when the new fermions are too heavy to be 
directly produced and mix negligibly with the SM, their Yukawa couplings to the Higgs can be effectively constrained by the radiative Higgs couplings.


\acknowledgments{NB and MF acknowledge the partial support of the OCEVU
Labex (ANR-11-LABX-0060) and the A*MIDEX project (ANR-11-IDEX-0001-02),
funded by the ``Investissements d'Avenir" French government program managed by the ANR.
MF acknowledges the partial support of the  European Union FP7  ITN INVISIBLES (Marie Curie Actions, PITN-GA-2011-289442).
MF thanks SISSA, Trieste for hospitality during the accomplishment of this project.
We thank C.~Delaunay, L.~Di~Luzio, M.~Knecht, G.~Moultaka, M.~Nardecchia, E.~Salvioni for relevant discussions.}

\appendix
\numberwithin{equation}{section}

\section{Electroweak precision tests in presence of new fermions}
\label{appendixA}

In this appendix we provide general formulas for the EW gauge boson couplings to fermions, as well as for their vacuum polarisation amplitudes. 
This allows to define and compute
the oblique parameters $S$ and $T$ \cite{Peskin:1990zt,Altarelli:1990zd,Peskin:1991sw,Barbieri:2004qk}.
We discuss the experimental constraints on these parameters, 
as well as on  the $Z$ couplings to light SM fermions, such as $Zb\overline{b}$.

\subsection{Electroweak gauge boson couplings}
\label{bo-fe}

The couplings of the EW gauge bosons to a chiral fermion multiplet in a given representation $(R_w,Y)$ of the EW gauge group are determined
by the covariant derivative 
\beq
D_\mu =\partial_\mu -i g T^a W_\mu^a-i g' Y B_\mu = \partial_\mu -i \frac{g}{\sqrt{2}}(T^+ W_\mu^++T^- W_\mu^-) -i g T^3 W_\mu^3-ig' Y B_\mu ~, 
\eeq
where $T^a$ are the $SU(2)_w$ generators for $R_w$, $T^\pm=T^1\pm iT^2$ and $W^\pm_\mu=(W^1_\mu\mp iW^2_\mu)/\sqrt{2}$. 
In full generality, the resulting non-vanishing couplings are 
\begin{equation}\ba{l}
c^{W^\pm}(\overline{f_{m'}},f_m)=\frac{g}{\sqrt{2}}(T^\pm)_{m'm}= \frac{g}{\sqrt{2}}\sqrt{j(j+1)-T^3_m(T^3_m \pm 1)}\delta_{m', m\pm 1} ~,
\\
c^{W^3}(\overline{f_{m'}},f_m) = g T^3_m \delta_{m'm}~,\qquad\qquad
c^{B}(\overline{f_{m'}},f_m)= g' Y \delta_{m'm}~,
\label{W3B}
\ea\end{equation}
where $j=(N_w-1)/2$ is the weak isospin, and the $N_w$ components of $R_w$ are labelled by $m,m'=-j,-j+1,\dots,j-1,j$,
and have electric charge $Q_m = T^3_m+Y$.
It is straightforward to derive from \eq{W3B} the couplings of $Z_\mu = c_wW^3_\mu-s_wB_\mu$ and $A_\mu = s_w W^3_\mu+c_w B_\mu$,  
\beq
c^Z(\overline{f_{m'}},f_m)=\frac{g}{c_w}(T^3_m-s_w^2 Q_m)\delta_{m'm}~,\quad\quad
c^A(\overline{f_{m'}},f_m)=e Q_m \delta_{m'm}~.
\eeq

After EWSB, for each value of the charge $Q$,  the fermion mass term can be written as 
$\overline{f_{L\alpha}} ({\cal M}_Q)_{\alpha\beta} f_{R\beta}$, where $\alpha,\beta=1,\dots,n_Q$ run over the $n_Q$ fermions of charge $Q$
(in a given colour representation). 
In general the mass matrix is not diagonal and the mixing can be described by
$f_{L\alpha} = (U_Q^L)_{\alpha i} f_{Li}$ and $f_{R\alpha} = (U_Q^R)_{\alpha i} f_{Ri}$, where $f_i$ are the mass eigenstates. 
Therefore, the couplings of the gauge bosons to the mass eigenstates are 
\beq
(c^V_{L,R})_{ij} = (c^V_{L,R})_{\alpha\beta} (U^{L,R}_Q)^*_{\alpha i} (U^{L,R}_{Q'})_{\beta j}~,\quad V = W^\pm, W^3,B~,
\label{Vmass}\eeq
where the $(c^V_{L,R})_{\alpha\beta}$ are given in \eq{W3B}, and $Q=Q'\pm1$ for $V=W^\pm$, $Q=Q'$ for
$V=W^3,B$.

The mixing cancels out in the photon couplings, because $U(1)_{em}$ is unbroken, and one finds immediately $(c^A_{L,R})_{ij}=eQ\delta_{ij}$.
The $Z$-boson couplings to the mass eigenstates, instead, do depend on the mixing. Using the parametrisation
\begin{equation}
\mathcal{L}_{\bar{f}fZ}= \frac{g}{c_w} Z_\mu \sum \limits_{i,j} \overline{f_i} \gamma^\mu \left(g^L_{ij}P_L + g^R_{ij}P_R \right)f_j 
\equiv\frac{g}{c_w} Z_\mu \sum \limits_{i,j} \overline{f_i} \gamma^\mu \left(g^V_{ij}-g^A_{ij} \gamma_5 \right)f_j 
~,
\label{Zff}
\end{equation}
one finds
\beq
g^{L,R}_{ij} =  T^3_\alpha \delta_{\alpha\beta} (U_Q^{L,R})^*_{\alpha i} (U^{L,R}_Q)_{\beta j} - s_w^2 Q \delta_{ij}~,\qquad
g^{V,A}\equiv\frac{g^L\pm g^R}{2} ~.
\label{Zmass}\eeq
The matrices $g^{L,R}$ ($g^{V,A}$) are hermitian, with possibly non-vanishing off-diagonal entries. 
Note that the mixing of fermions with equal EW charges does not affect the couplings to the neutral gauge bosons: if $T^3_\alpha$ (or, equivalently, $Y_\alpha$) is the same for all $\alpha$,
then one can use $(U^{L,R}_Q)^*_{\alpha i} (U^{L,R}_{Q})_{\alpha j} = \delta_{ij}$, and the couplings to  $Z$ (as well as to $W^3$ and $B$) reduce to their unmixed values. 
Thus, the neutral-current couplings of SM fermions receive a correction, only when they mix with new fermions with a different value of $T^3$ (of $Y$). 
For example, the mixing of two left-handed fermions $f_{La}$ and $f_{Lb}$ of charge $Q$ amounts to
$$
U_Q^L=\begin{pmatrix} c & s \\ -s & c \end{pmatrix}~,\quad\quad g^L= \begin{pmatrix}
T^3_a -s_w^2 Q & 0 \\
0 & T^3_b-s_w^2 Q \end{pmatrix} + (T^3_a-T^3_b) \begin{pmatrix}
-s^2 & s c \\
s c  &  s^2 
\end{pmatrix}~.
$$
In this paper we assume that new fermions mix with the third SM family only.  Indeed,  
flavour-changing neutral currents among the different SM families are strongly constrained experimentally.

\subsection{Constraints from $S$ and $T$}\label{ST}

\begin{figure}[bt]
\begin{center}
\includegraphics[scale=0.3,trim= 0 0 0 0]{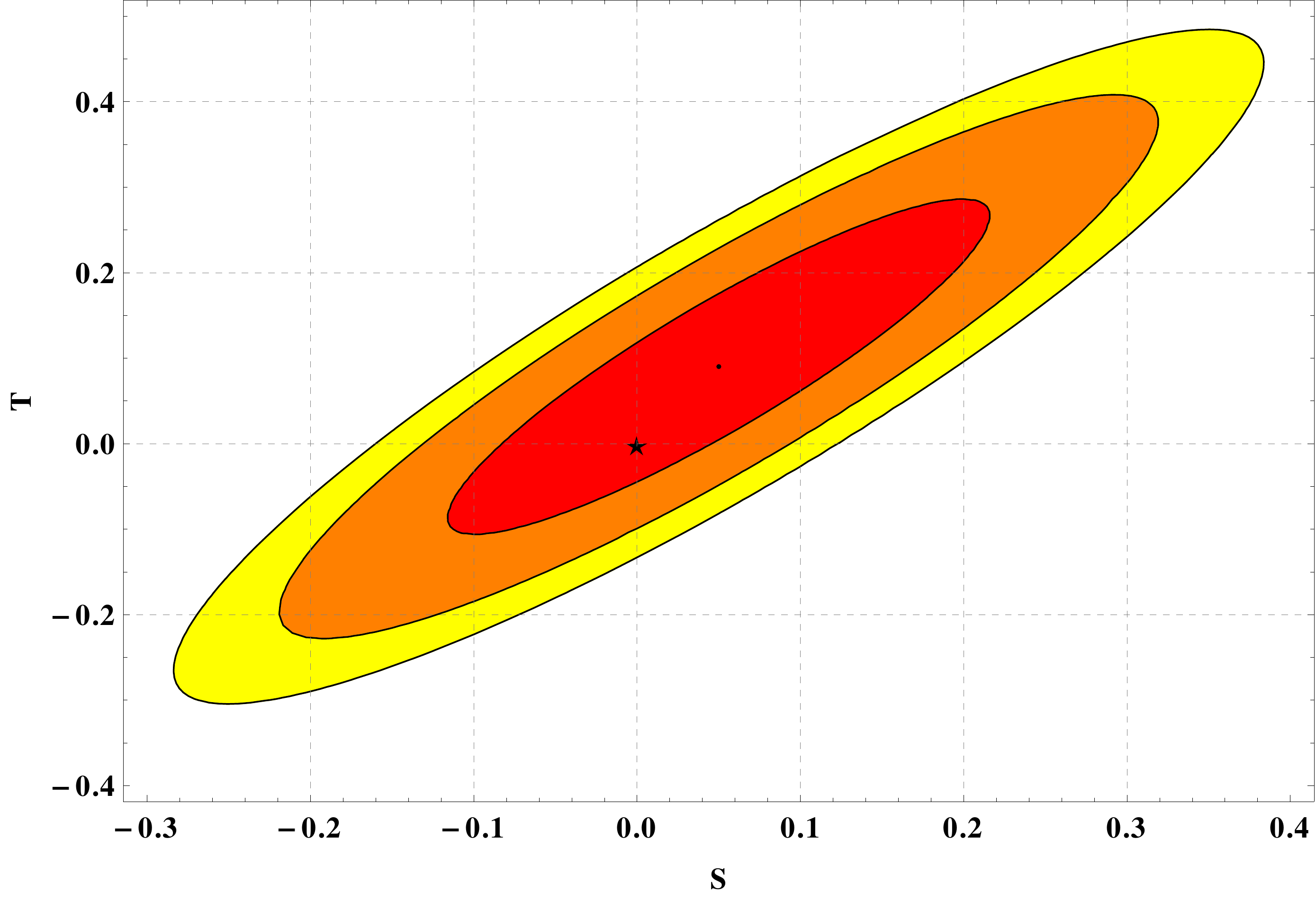}
\caption{The 68 \% (red), 95 \% (orange) and 99 \% (yellow)  C.L. ellipses in the $S-T$ plane, from the fit of Ref.~\cite{Baak:2014ora}, with the other EW parameter $U$ left free. 
The black dot indicates the best fit, while the star at $S=T=0$ is  the SM point, with  $m_{t,ref}=173$ GeV and $m_{h,ref}=125$ GeV.}
\label{figST}
\end{center}
\end{figure}

The vacuum polarisation amplitudes for the EW gauge bosons, defined by the effective momentum space Lagrangian 
\begin{equation}
{\cal L}_{\Pi} = -W_\mu^+ \Pi_{WW}^{\mu\nu}(p) W_\nu^- -\frac{1}{2}B_\mu \Pi_{00}^{\mu\nu}(p)B_\nu - W_\mu^3\Pi_{30}^{\mu\nu}(p) B_\nu
-\frac12W_\mu^3\Pi^{\mu\nu}_{33}(p)W_\nu^3~,
\end{equation}
can be decomposed into transverse and longitudinal parts,
\begin{equation}
\Pi_{VV'}^{\mu\nu}(p)=\Pi_{VV'}^T(p) (p^2 g^{\mu\nu}-p^\mu p^\nu)+\Pi_{VV'}^L(p)p^\mu p^\nu
= \Pi_{V V'}(p^2) g^{\mu\nu} + (p^\mu p^\nu {\rm -~terms})~.
\label{pVV}\end{equation}
As in the experiments the mass of the external fermions is much smaller than the EW scale, $m_f^2\ll m_Z^2$, one can drop  the $p^\mu p^\nu$-terms and expand in $p^2$,

\begin{equation}
\Pi_{V V'}(p^2)= \Pi_{V V'}(0)+p^2 \Pi'_{V V'}(0)+ {\cal O}(p^4)~.
\end{equation}
The lowest terms in this expansion are sufficient to describe accurately the effect of heavy new physics:
when new particles at scale $m_F$ contribute to the vacuum polarisation amplitudes,
the higher order corrections are suppressed by powers of $m_Z^2/m_F^2$.
Taking into account that three coefficients can be traded for the experimental values of 
$\alpha$, $s_w$ and $m_Z$,
and two others are determined by 
the Ward identities for the photon, one finds that two parameters are sufficient to characterise the effect of new physics at leading order in $m_Z^2/m_F^2$ \cite{Peskin:1991sw,Barbieri:2004qk}.
The combination that describes the custodial symmetry breaking at leading order is given by
\begin{equation}
T\equiv \frac{1}{\alpha c_w^2 m_Z^2}\left[(\Pi_{33}(0)-\Pi_{33}^{SM}(0))-(\Pi_{WW}(0)-\Pi_{WW}^{SM}(0))\right]~.
\label{T}\end{equation}
The combination that breaks the weak isospin at leading order, but respects the custodial symmetry, is given by
\begin{equation}
S\equiv \frac{4 s_w c_w}{\alpha m_Z^2} \left[(\Pi_{30}(m_Z^2)-(\Pi_{30}(0))-(\Pi_{30}^{SM}(m_Z^2)-{\Pi}_{30}^{SM}(0)) \right]
\simeq \frac{4 s_w c_w}{\alpha} [\Pi'_{30}(0)-{\Pi'}_{30}^{SM}(0)]~.
\label{S}\end{equation}
The approximation in terms of amplitude derivatives evaluated at $p^2=0$ is appropriate only for new physics much heavier than $m_Z$ that does not mix with light SM particles; 
in the general case one should keep the definition of $S$ in terms of amplitude differences, to avoid unphysical singularities that may appear in the derivative.
The subtracted SM contribution is evaluated at a reference point for the SM parameters. 
Following Ref.~\cite{Baak:2014ora}, if one takes $m_{t,ref}=173$ GeV and $m_{h,ref}=125$ GeV, the present 
experimental allowed ranges are given by
\begin{equation}
S=0.05\pm 0.11 ~, \quad\quad\quad T=0.09\pm 0.13~,
\end{equation}
with a correlation coefficient $\simeq 0.9$. 
In Fig.~\ref{figST} we display the allowed region in the $S-T$ plane, that we adopt in the rest of the paper to constrain the parameter space of each model.

\begin{figure}[bt]
\begin{center}
\includegraphics[scale=0.6,trim= 0 600 0 80]{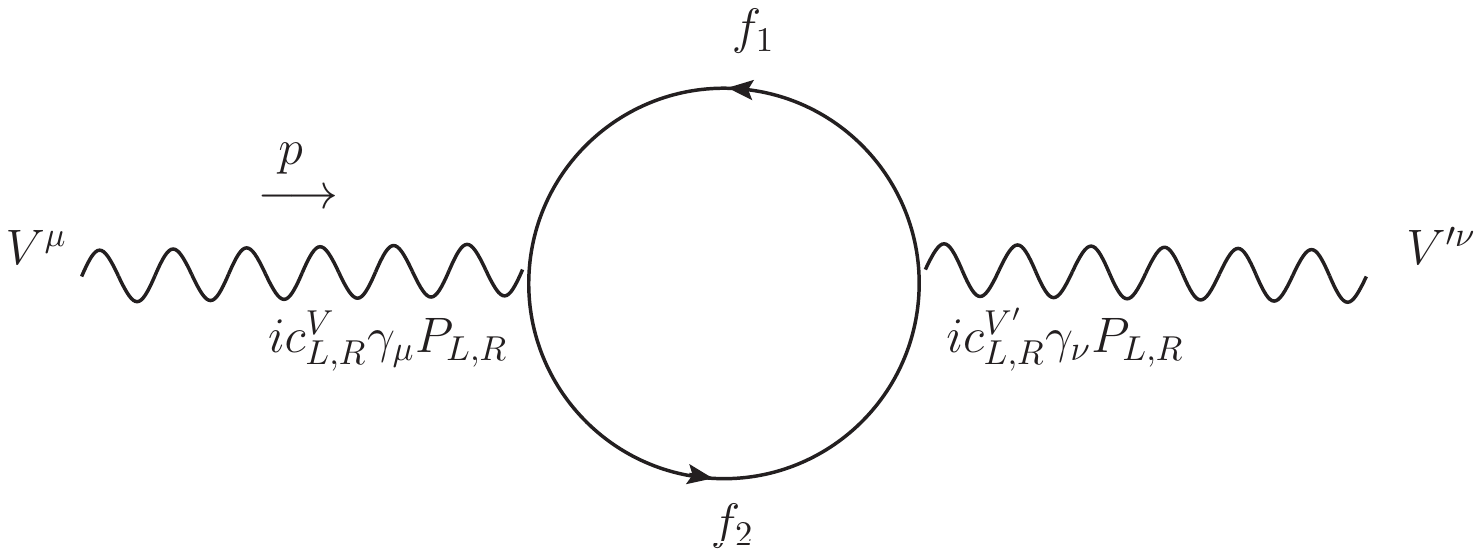}
\caption{Contribution of two fermion mass eigenstates $f_1$ and $f_2$ to the vacuum polarisation amplitude for the gauge bosons $V^\mu$ and $V'^\nu$. 
The relevant couplings are defined by ${\cal L}_{\bar{f} fV} = V_\mu \overline{f_i}\gamma^\mu[(c^V_L)_{ij} P_L + (c^V_R)_{ij} P_R] f_j$.}
\label{figPVV}
\end{center}
\end{figure}

In order to estimate the contributions to $S$ and $T$, in any theory where the new physics is weakly coupled, one should just compute the one-loop diagram 
contributing to the EW gauge boson vacuum polarisation amplitudes, shown in Fig.~\ref{figPVV}. 
The most general couplings of the EW gauge bosons to the fermion mass eigenstates are defined in \eq{Vmass}. 
The functions $\Pi_{VV'}(p^2)$ defined in \eq{pVV} will receive different contributions from the left- and right-handed
couplings,
 \begin{equation}
 \Pi_{VV'}=c_L^V c_L^{V'}\Pi_{LL}+c_R^V c_R^{V'}\Pi_{RR}+c_L^V c_R^{V'}\Pi_{LR}+c_R^V c_L^{V'}\Pi_{RL}~.
 \label{VVprime}\end{equation}
Performing the computation with dimensional regularisation, one finds the following general result:
\begin{eqnarray}
\Pi_{LL,RR}(p^2)&=&\frac{N_c} 
{(4\pi)^2}\left[ \frac{(m_1^2-m_2^2)^3}{6 p^4}\ln\frac{m_1^2}{m_2^2} - \frac{(m_1^2-m_2^2)^2}{3 p^2}
-(m_1^2+m_2^2)\left(\frac{1}{\bar{\epsilon}}+ \ln \frac{4\pi^2 \mu^2}{m_1 m_2} +\frac{1}{6}  \right) 
  \right. 
\nonumber\\
& + & \left. \frac{2}{3}p^2 \left(\frac{1}{\bar{\epsilon}}+\ln\frac{4\pi^2 \mu^2}{m_1 m_2} +\frac{7}{6} \right) 
+\left( \frac{(m_1^2-m_2^2)^2}{3 p^2}+ \frac{m_1^2+m_2^2}{3}-\frac{2}{3}p^2\right)R(p^2)   \right] ,
\label{piLL}\end{eqnarray}
\begin{equation}
\Pi_{LR,RL}(p^2)= \frac{2N_c m_1 m_2} 
{(4\pi)^2}  \left[
-\frac{m_1^2-m_2^2}{2 p^2}\ln\frac{m_1^2}{m_2^2} + \frac{1}{\bar{\epsilon}}+\ln\frac{4\pi^2 \mu^2}{m_1 m_2}+\frac{3}{2}-R(p^2)
\right] ,
\label{piLR}\end{equation}
where $N_c$ is the dimension of the $SU(3)_c$ representation of the fermions $f_{1,2}$, $m_{1,2}$ are the masses of $f_{1,2}$,
$\bar\epsilon$ and $\mu$ are defined by ${\rm d}^4k\equiv \mu^{4-n}{\rm d}^n k$ and 
$1/\bar\epsilon \equiv 2/(4-n)-\gamma-\ln\pi$ with $\gamma \simeq 0.5772$, and finally
\begin{equation}
R(p^2) \equiv \frac{\sqrt{\lambda}}{p^2}\log\frac{m_1^2+m_2^2-\sqrt{\lambda}-(p^2+i\epsilon)}{2 m_1 m_2} ~,~~~~~
\lambda\equiv(m_1^2 -m_2^2)^2 -2 p^2 (m_1^2 + m_2^2) +p^4~.
\end{equation}
Note that $R(p^2)$ is invariant for $\sqrt{\lambda} \rightarrow -\sqrt{\lambda}$ and it has a non-vanishing imaginary part for $p^2>(m_1+m_2)^2$. 
When evaluating $S$ in \eq{S}, one must include only the real part of $R(m_Z^2)$, while the imaginary part contributes to the decay width of the EW gauge bosons,
that may be also modified with respect to its SM value.
Note also that, when $f_{1,2}$ are Majorana fermions, one must include in the amplitude an additional symmetry factor $1/2$. 

Let us provide for convenience the form of $\Pi_{VV'}(p^2)$ in some relevant limits. For $m_1=m_2=0$ one has
\beq
\Pi_{LL,RR}(p^2) = \frac{2N_c}{3(4\pi)^2} p^2 \left(\frac{1}{\bar{\epsilon}}+\log\frac{4\pi^2\mu^2}{-(p^2+i\epsilon)} +\frac 76\right)~,\qquad \Pi_{LR,RL}(p^2) =0~.
\eeq
This limit is relevant for loops involving the light SM fermions (all but the top quark), whose mass can be neglected.
In order to compute $T$ and $S$ (in the derivative approximation), it is sufficient to compute the
first and second term in the $p^2$-expansion of $\Pi_{VV'}(p^2)$, respectively:
\begin{eqnarray}
\Pi_{LL,RR}(0) &=& -\frac{N_c}{(4\pi)^2}  \left[(m_1^2+m_2^2)\left(\frac{1}{\bar{\epsilon}}+\ln\frac{4\pi^2 \mu^2}{m_1 m_2}\right)
+\frac{m_1^4+m_2^4}{m_1^2-m_2^2}\ln\frac{m_2}{m_1} \right], \nonumber\\
\Pi_{LR,RL}(0)&=&\frac{2N_c  m_1 m_2}{(4\pi)^2}  
\left[\frac{1}{\bar{\epsilon}}+\ln\frac{4\pi^2 \mu^2}{m_1 m_2}+\frac{1}{2} 
+\frac{m_1^2+m_2^2}{m_1^2-m_2^2}\ln\frac{m_2}{m_1}\right],
\label{Pzero}\\
\Pi'_{LL,RR}(0) &=& \frac{2N_c}{3(4\pi)^2}  \left[\frac{1}{\bar{\epsilon}}+\ln\frac{4\pi^2 \mu^2}{m_1 m_2}
-\frac{1}{6} -\frac{2m_1^2 m_2^2}{(m_1^2-m_2^2)^2}
+\frac{m_1^6+m_2^6-3m_1^4m_2^2-3m_1^2m_2^4}{(m_1^2-m_2^2)^3}\ln\frac{m_2}{m_1}
\right] , \nonumber\\ 
\Pi'_{LR,RL}(0) &=& \frac{N_c}{(4\pi)^2}    \frac{m_1 m_2}{(m_1^2-m_2^2)^3}\left(m_1^4-m_2^4 +4 m_1^2 m_2^2 \ln \frac{m_2}{m_1}\right).
\label{PPzero}
\end{eqnarray}
For $m_1=0$ and $m_2=m$ (relevant e.g. for the bottom-top quark loop) one obtains
\begin{equation}
\Pi_{LL,RR}(0)= -\frac{N_c m^2}{(4\pi)^2}   \left(\frac{1}{\bar{\epsilon}}+\ln\frac{4\pi^2 \mu^2}{m^2}\right)~,~~~~~
\Pi'_{LL,RR}(0)=\frac{2N_c}{3(4\pi)^2}   \left(\frac{1}{\bar{\epsilon}}+\ln\frac{4\pi^2 \mu^2}{m^2}-\frac{1}{6}\right)~,
\end{equation}
and $\Pi_{LR,RL}(0)=\Pi'_{LR,RL}(0)=0$.
For  $m_1=m_2=m\ne0$ (loops involving a unique fermion mass eigenstate) one reduces to
\begin{equation}
\Pi_{LL,RR}(0) = -\Pi_{LR,RL}(0) = -\frac{2N_c m^2}{(4\pi)^2}  \left(\frac{1}{\bar{\epsilon}}+\ln\frac{4\pi^2 \mu^2}{m^2}-\frac{1}{2}\right)~,
\label{mmzero}\end{equation}
\begin{equation}
\Pi'_{LL,RR}(0)=\frac{2N_c }{3(4\pi)^2} \left(\frac{1}{\bar{\epsilon}}+\ln\frac{4\pi^2 \mu^2}{m^2}-1\right)~,~~~~~
\Pi'_{LR,RL}(0)=\frac{N_c}{3(4\pi)^2}~.   
\label{mmprime}\end{equation}

As an illustrative example, consider the case of a fermion ``family" with no mixing with the SM fermions, formed by one weak doublet $Q_L=(T_L,B_L)\sim(R_c,2,Y)$ and two 
singlets $T_R\sim(R_c,1,Y+1/2)$ and $B_R\sim(R_c,1,Y-1/2)$, with $\dim(R_c)=N_c$. After EWSB they combine into two mass eigenstates
$T$ and $B$ with masses $m_T$ and $m_B$; their non-zero couplings to EW gauge bosons are obtained from  \eq{W3B},
\begin{equation}\ba{c}
c^{W^\pm}_L(T,B)=\dfrac{g}{\sqrt 2}~,~~c^{W^3}_L(T)=\dfrac g2~,~~c_L^{W^3}(B)=- \dfrac g2 ~,\\
c^B_L(T)=c^B_L(B)=g'Y~,~~c^B_R(T)=g' \left(Y+\dfrac12\right)~,~~c^B_R(B)=g'\left(Y-\dfrac12\right)~.
\ea\end{equation}
To compute the correction to $S$, one should evaluate Eqs.~(\ref{piLL}-\ref{piLR})  for $p^2=m_Z^2$ and $m_1=m_2=m_{T,B}$, while for $p^2=0$ 
one can use directly \eq{mmzero} with $m=m_{T,B}$.  
Adding the various contributions to $\Pi_{30}$ as shown in Eq.~(\ref{VVprime}), and replacing into Eq.~(\ref{S}), the result is 
\begin{equation}
S_{T,B}=\frac{N_c}{6 \pi}\left[\left( 1- 2 Y \ln \frac{m_T^2}{m_B^2}\right) +\frac{m_Z^2}{m_T^2} \left(\frac 12 + \frac{4 Y}{3}\right) 
+ \frac{m_Z^2}{m_B^2} \left(\frac 12 -\frac{4 Y}{3}\right) +\mathcal{O}\left( \frac{m_Z^4}{m_{T,B}^4}\right)  \right]~.
\label{S-TB}
\end{equation}
If one adopted the approximate expression for $S$ in terms of derivatives, given by the right-hand side of \eq{S}, then using  \eq{mmprime} one finds only
the first term in the squared bracket of \eq{S-TB},
which is accurate for $m_{T,B}\gg m_Z$.
To compute the correction to $T$, one should use Eq.~(\ref{mmzero}) for the $T$ and $B$ loops that contribute to $\Pi_{33}$,
and Eq.~(\ref{Pzero}) for the $T/B$ loop  that contributes to $\Pi_{WW}$. Replacing into Eq.~(\ref{T}) one obtains
\begin{equation}
T_{T,B}= \frac{N_c}{16\pi c_w^2s_w^2m_Z^2}
\left(m_T^2 +m_B^2-2\frac{m_T^2 m_B^2}{m_T^2 -m_B^2}\ln\frac{m_T^2}{m_B^2}\right)~.
\label{T-TB}
\end{equation}
Particularising these results to the case of the SM top and bottom quarks ($N_c=3$, $Y=1/6$), and neglecting the uncertainty on $m_b$ as well
as $(m_b^2/m_t^2)$-corrections, we can immediately extract 
the well-known dependence of $S$ and $T$ on the value of the top quark mass, 
\begin{equation}
S_{top} = -\frac{1}{6 \pi} \ln\frac{m_t^2}{m_{t,ref}^2}~, \quad\quad\quad 
T_{top}= \frac{3}{16 \pi c_w^2 s_w^2}\frac{m_t^2-m^2_{t,ref}}{m_Z^2}~.
\end{equation}

\subsection{Constraints from  $Zf\overline{f}$}
\label{Zffapp}

The $Z$-boson couplings to the SM fermions are precisely measured. We discuss only those with the third family, since in this paper we assume that the mixing of the new fermions with the light
families is negligible. The deviations with respect to the SM can be expressed in terms of the $Z$ partial decay width into any given final state $f\overline{f}$,
\beq
R(Z\rightarrow f\overline{f})\equiv \frac{\Gamma(Z\rightarrow f\overline{f})}{\Gamma_{SM}(Z\rightarrow f\overline{f})} \equiv 1+ \delta R(Z\rightarrow f\overline{f})~,\quad
f=\nu_\tau,\tau,b~.
\label{R-definition}
\end{equation}
For each fermion $f$ there are two independent couplings $g^{L,R}_{f\overline{f}}$ as shown in \eq{Zmass}, that can be separately constrained
if the angular distribution of the fermions is measured.

Beginning from leptons, 
the $Z$ invisible width and its width into taus  are determined at the per mil level \cite{Agashe:2014kda},
$\Gamma(Z\rightarrow inv)=499.0\pm 1.5$ MeV and $\Gamma(Z\rightarrow \tau^+ \tau^-)= 84.08\pm 0.22$ MeV.
Given this precision and the relatively good agreement between the central values and the SM predictions, we constrain the mixing with the new leptons by imposing a rough $3 \sigma$ upper bound, 
\beq
\left| \delta R(Z\rightarrow inv) \right| \leq 9 \cdot 10^{-3}~,\qquad \left| \delta R(Z\rightarrow \tau^+ \tau^-) \right|\leq 8 \cdot 10^{-3}~.
\label{Rlep}
\eeq

\begin{figure}[bt]
\begin{center}
\includegraphics[scale=0.3,trim= 0 0 0 0]{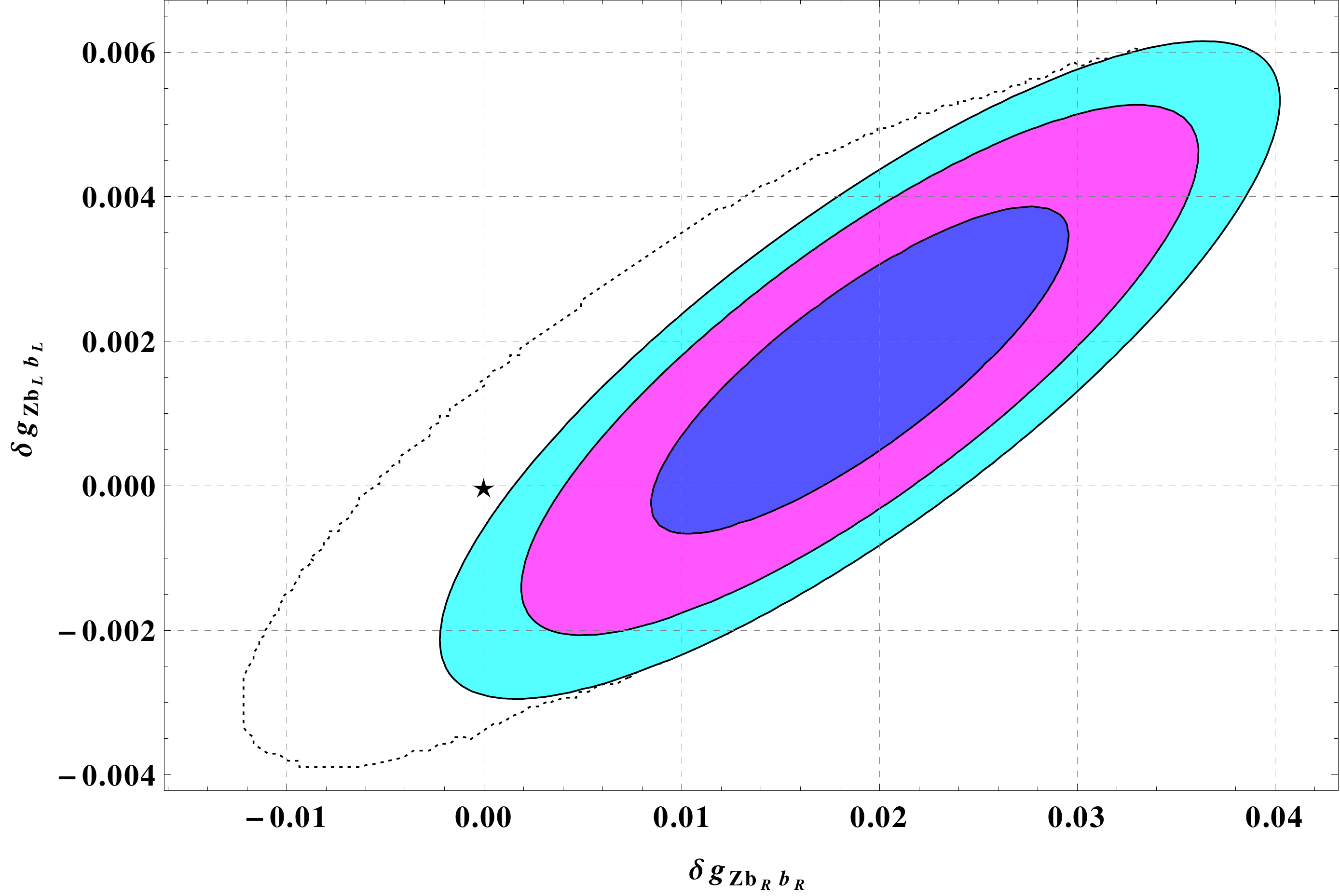}
\caption{The 68 \% (blue), 95 \% (magenta) and 99 \% (cyan)  C.L. ellipses in the $\delta g^{R}_{b\bar{b}}-\delta g^{L}_{b\bar{b}}$ plane,  extracted from Ref.~\cite{Ciuchini:2013pca}. 
The black dot indicates the best fit, while the star at the origin represents the SM.
In our analysis, we allow for a larger parameter space, delimited by the dotted line, that is obtained by shifting the $99 \%$ ellipse towards the origin, till the SM point enters the $68 \%$ region.}
\label{fig-ellipse-Z}
\end{center}
\end{figure}

Coming to the $Z$ coupling to bottom quarks, a more detailed discussion is worth, to fairly gauge the resulting constraint on the new fermions.
The $Z b \bar{b}$ Lagrangian can be written as
\begin{equation}
{\cal{L}}_{Zb\bar{b}}=\frac{g}{c_w}Z_\mu \overline{b}\ \gamma^\mu \left[ \left(g^L_{b\bar{b},SM} + \delta g^L_{b\bar{b}}\right) P_L + \left(g^R_{b\bar{b},SM} + \delta g^R_{b\bar{b}} \right) P_R \right] b ~,
\label{Zbb}\end{equation} 
where the SM couplings at tree-level are given by $g^L_{b\bar{b},SM}= -1/2 + s_w^2/3$ and $g^R_{b\bar{b},SM}= s_w^2/3$.
Deviations at tree-level occur when the bottom quark mixes with a new fermion with 
a different value of $T^3$ (and $Y$).
The present experimentally allowed range 
is given by \cite{Ciuchini:2013pca} 
\begin{equation}
\delta g^{L}_{b\bar{b}}=0.0016 \pm 0.0015 ~, \quad\quad\quad \delta g^{R}_{b\bar{b}}= 0.019 \pm 0.007 ~,
\end{equation}
with a correlation coefficient $\simeq 0.8$.
In Fig.~\ref{fig-ellipse-Z} we display the allowed region in the $\delta g^R_{b\bar{b}}-\delta g^L_{b\bar{b}}$ plane.
Note that the left- and right-handed couplings are determined with per mil and per cent precision, respectively, and the best fit region is incompatible with the SM at about $99 \%$ C.L.. 
In some analyses, a slightly better agreement is obtained, at about $95 \%$, due to different details in the global electroweak fit, see e.g. Ref.~\cite{Gori:2015nqa}. 
The discrepancy with the SM comes mostly from the measurement of the forward-backward asymmetry $A_{FB}^b$, that may be just an upward statistical fluctuation, 
an unidentified systematic error, or alternatively an indication for 
a significant new-physics contribution. 
In this paper, in order to constrain the new fermions that modify the $Zb\bar{b}$-couplings,
we will conservatively enlarge the $99 \%$ C.L. region, by allowing it to shift towards the SM point, till the latter touches the $68 \%$ ellipse, as illustrated in Fig.~\ref{fig-ellipse-Z}.
Such a shift roughly corresponds to introduce a systematic error in the measurement of $A_{FB}^b$.

The mixing with new fermions can be such that no tree-level deviations occur in the $Zb\overline{b}$-couplings, but they do occur in the $Wt\overline{b}$ coupling 
and/or  in the $Zt\overline{t}$ couplings. 
These deviations may affect significantly $g^{L,R}_{b\bar{b}}$,
because the contribution of one-loop diagrams involving the top quark and the $W$-boson is  larger than the experimental uncertainty.
Also new fermions may correct significantly $Zb\overline{b}$ at the one-loop level, if they are not much heavier than the top. 
The detailed structure of the one-loop corrections to $Zb\overline{b}$ can be found e.g. in Ref.~\cite{Bamert:1996px}, that we employ for our analysis of models with modified top couplings.
For example, the correction to the left-handed coupling, from the top loops only, can be written as 
\beq
\delta g^{L}_{b\bar{b}} = \delta g^{L}_{t\bar{t}} f_L(m_W/m_t) + \delta g^{R}_{t\bar{t}} f_R(m_W/m_t) + \delta g^L_{t\overline{b}} f_W(m_W/m_t) ~, 
\eeq
where $\delta g^{L,R}_{t\bar{t}}$ are defined in analogy to \eq{Zbb} with $t\leftrightarrow b$, the $Wt_L\overline{b_L}$ coupling is given by $(g/\sqrt{2})(1+\delta g^L_{t\overline{b}})$,
and the functions $f_{L,R,W}$ can be extracted from Ref.~\cite{Bamert:1996px}.

\section{Higgs boson couplings in presence of new fermions}
\label{appendix B}

In this appendix we present a general parametrisation for the Yukawa couplings among the SM Higgs doublet and two arbitrary fermion multiplets.
We then analyse the resulting modifications in the Higgs couplings to the SM particles, at leading order: corrections at tree-level to the Higgs-fermions couplings, 
and one-loop corrections to the Higgs-gauge bosons couplings. Finally, we briefly review the present experimental constraints on these couplings.

\subsection{Tree-level Higgs couplings \label{treeH}}

The SM Yukawa couplings are given by
\beq
-{\cal L}_Y^{SM} = y_u \overline{q_{L}} u_{R} \tilde{H} + y_d \overline{q_L}d_{R} H 
+ y_e \overline{l_L}e_{R} H + h.c. ~,
\label{SMy}\eeq
where $q_L=(u_L\,d_L)^T$, $H=(H^+\,H^0)^T$ and $\tilde H \equiv i\sigma_2H^*$ are $SU(2)_w$ doublets, while $u_R$, $d_R$ and $e_R$ are singlets, and 
flavour indexes are understood.

In full generality, the Higgs doublet may have a non-zero Yukawa interaction with any pair of chiral fermions that transform  under $SU(3)_c\times SU(2)_w\times U(1)_Y$
as $\psi_L \sim (R_c,R_w,Y)$ 
and $\psi^{u,d}_{R}\sim (R_c, R_w - 1,Y \pm \frac{1}{2})$:
\beq
-{\cal L}_Y= \sum_{\psi_L}\left[
\sum_{\psi^u_R}y_{\psi_L\psi^u_R} \left(\overline{\psi_{L}} \psi_{R}^u \tilde{H}\right) + \sum_{\psi^d_R} y_{\psi_L\psi^d_R} \left(\overline{\psi_L} \psi_R^d H \right)
\right] + h.c. ~.
\label{totaly}\eeq
Here the parentheses stand for the appropriate contraction of the $SU(2)_w$ indexes.
Let us denote the components of $\psi_L$ by the index $m=j,j-1,\dots,-j+1,-j$, where $j=(N_w-1)/2$. Then, the multiplet $\overline{\psi_L}$
with components $(\overline{\psi_L})_m \equiv \overline{(\psi_L)_{m}}$ transforms in the conjugate representation $R_w^*$.
It is possible to define a multiplet $\overline{\psi_L}'$ that properly transforms in the representation $R_w$, by using the $SU(2)_w$ conjugation
matrix $R$,
$(\overline{\psi_L}')_m = R_{mn} (\overline{\psi_L})_n \equiv (-1)^{j-m} \overline{(\psi_L)_{-m}}$.
The $2j$ components of $\psi_R^{u}$ pair with the upper (lower) $2j$ components of $\overline{\psi_L}'$ to form the upper (lower) component of a weak doublet $D$.
The corresponding Clebsch-Gordan coefficients are given by
\beq
\left\langle j,j-\frac12;m,-m\pm\frac12\right|\left.\frac12,\pm\frac12\right\rangle = \pm (-1)^{j-m} \sqrt{\frac{j\pm m}{j(2j+1)}}~.
\eeq
Contracting $D$ with $\tilde{H}$ into a weak singlet, $D_a (i\sigma_2)_{ab} \tilde{H}_b$, one finds
\beq
\left(\overline{ \psi_L} \psi_R^u \tilde H \right) \equiv 
\sum_{m=-j+1}^j 
\sqrt{\frac{j+m}{j(2j+1)}} \left[
\overline{(\psi_L)_m}(\psi_R^u)_{m-\frac12} \tilde{H}_{\frac12}
+ \overline{(\psi_L)_{-m}} (\psi_R^u)_{-m+\frac12} \tilde{H}_{-\frac12}
\right].
\label{CGy}\eeq
The same expression holds for $\psi^u_R \leftrightarrow \psi_R^d$ and $\tilde{H}\leftrightarrow H$ as well. 
Of course, all the results above also apply
when one makes everywhere the replacement $L\leftrightarrow R$.

The relative size of the Clebsch-Gordan coefficients has important phenomenological consequences, e.g. the different components of the fermion multiplets acquire a different 
mass after EWSB. The overall normalisation of the $SU(2)_w$ contraction is also important, to establish the perturbative range for a Yukawa coupling $y$: 
for instance, the contribution of $y$ to the Higgs
wavefunction renormalisation at one-loop goes as $y^2/(16\pi^2)$ times the sum of the Clebsch-Gordan coefficients squared, taken over all possible isospin components in the loop. 
Adopting the above conventions,  such a sum is normalised to one, and we can easily define the region where perturbation theory can be trusted, by requiring $y/(4\pi)\ll1$. 
However, we keep the conventional normalisation for the doublet-doublet contraction into a singlet, with no overall factor $1/\sqrt{2}$, that strictly-speaking should be included:
the issue of $SU(2)_w$ normalisation is more relevant for large weak multiplets.

{
We note that the perturbative upper bound on a Yukawa coupling $y$ depends on the process under consideration. Schematically, the next-to-leading order amplitude is given by the
leading order one times a factor $y^n g^m F_c/(16\pi^2)$, where $g$ stands for other couplings such as gauge couplings, and $F_c$ is the colour factor,
with typical values $F_c=1,N_c,C(R_c)$.
In the example of the Higgs wavefunction normalisation adopted above, one has $n=2$, $m=0$ and $F_c=N_c$, therefore we could have taken into account the 
$SU(3)_c$ contraction by adding a factor $1/\sqrt{N_c}$ on the right-hand side of \eq{CGy}, or alternatively
requiring $y\sqrt{N_c}/(4\pi)\ll1$. 
However, the one-loop  amplitudes relevant in our analysis (EW precision tests, Higgs couplings to fermions and gauge bosons, etc.) behave differently from each other,
and the perturbativity criterion varies correspondingly. In some cases the colour enhancement is absent, or compensated by small gauge couplings, or by a small
mixing between new and SM fermions. Therefore, we find more conservative to stick to the bound $y\ll 4\pi$.}

At EWSB, $H^0$ can be replaced by $(v+h)/\sqrt{2}$, to obtain the couplings of the physical Higgs boson $h$ to the fermions in the interaction basis. 
All fermions with equal charge and in the same colour representation may mix, 
and their mass matrix ${\cal M}=U_L diag(m_1,\dots,m_n) U_R^\dagger$ may include both $v$-independent vector-like mass terms, and the EWSB contributions $\sim yv$. Thus, one can derive the $h$-couplings 
to the fermion mass eigenstates as follows:
\bea
-{\cal L}&\supset& \overline{f_{L\alpha}} [{\cal M}(v)]_{\alpha\beta} f_{R\beta}
+\overline{f_{L\alpha}} \frac{\partial}{\partial v}\left[{\cal M}(v)\right]_{\alpha\beta} f_{R\beta} \,h + h.c.\nonumber\\
&=& \sum_i m_i \overline{f_{Li}}f_{Ri}+\overline{f_{Lj}} \left[U_L^\dagger \frac{\partial {\cal M}(v)}{\partial v} U_R\right]_{jk} f_{Rk} \,h + h.c.\nonumber\\
&=&\sum_i m_i \overline{f_i}f_i+
\overline{f_j} \left(y_{jk}+i \gamma_5 \tilde{y}_{jk}
\right)   f_k \, h~.
\label{hff1}
\eea
where $y$ and $\tilde{y}$ are hermitian matrices defined by
\beq
y=\frac{\lambda+\lambda^\dagger}{2}~,\qquad 
\tilde{y}=\frac{\lambda-\lambda^\dagger}{2i}~,\qquad 
\lambda\equiv U_L^\dagger \frac{\partial {\cal M}(v)}{\partial v} U_R~.
\label{y-ytilde}
\eeq 
In the CP-conserving case $\lambda$ is real, therefore $y=y^T$ is real and $\tilde y=-\tilde y^T$ is imaginary.
In the case of purely chiral masses (e.g.~in the SM), one has ${\cal M}(v) \propto v$, therefore $\partial {\cal M}(v)/\partial v = {\cal M}/v$, 
$y=\lambda = diag(m_1,\dots,m_n)/v$ and $\tilde y=0$.
On the other hand, in presence of both chiral and vector-like masses, the Higgs boson can have both scalar and pseudo-scalar, CP-even and CP-odd, diagonal and off-diagonal couplings to the fermions mass eigenstates, 
and its couplings are not proportional to the fermion masses. 
A simplification occurs in those SM extensions such that 
 $(\partial {\cal M}/\partial v)_{\alpha\beta} = {\cal M}_{\alpha\beta} c_\beta$, 
 that is, each row of the mass matrix has the same dependence on $v$.
In this case $U_L^\dagger$ and $U_L$ cancel out in $\lambda$ and one finds $\lambda_{jk} =  m_j \sum_\beta c_\beta (U_R^*)_{\beta j}  (U_R)_{\beta k}$. Similarly,
when $(\partial {\cal M}/\partial v)_{\alpha\beta} = c_{\alpha} {\cal M}_{\alpha\beta}$, 
one finds $\lambda_{jk} =  m_k \sum_\alpha c_\alpha (U_L^*)_{\alpha j}  (U_L)_{\alpha k}$.

The mixing with new fermions modifies the Higgs boson decay width into SM fermions at the tree-level.
In addition, there may be new Higgs decay channels, with one or more new fermions in the final state, as long as they are lighter than $h$.
In full generality, the Higgs decay width into two fermions at leading order is given by
\begin{equation}
\Gamma_{tree} (h\rightarrow \overline{f_j}f_k) =\frac{N_c  \Delta_{jk}}{8\pi}  m_h \left[
|y_{jk}|^2  (\beta^+_{jk})^3 \beta^-_{jk} 
+ |\tilde y_{jk}|^2 \beta^+_{jk} (\beta^-_{jk})^3  \right] \theta(m_h-m_j-m_k)~,
\label{Gammaff}
\end{equation}
where $\beta^\pm_{jk}\equiv [1-(m_j\pm m_k)^2/m_h^2]^{1/2}$, $\Delta_{jk}=2$ if the final state particles are identical Majorana fermions ($j=k$ and $f_j=f^c_j$), and  $\Delta_{jk}=1$ otherwise.

Besides the Yukawa couplings to fermions, the Higgs boson has tree-level couplings to $WW$, $ZZ$ and to itself. 
In the presence of new fermions, all the tree-level couplings may receive
corrections at the one-loop level. Even though these corrections may become relevant in view of precision 
measurements of the Higgs couplings, they represent in general
a sub-leading effect and we will not discuss them further. In the following subsection we will focus instead on a 
more sensitive probe of new physics: those Higgs couplings that are absent at tree-level in the SM.

\subsection{Loop-induced Higgs couplings}
\label{appendix B.2}

At one-loop new couplings are induced between the Higgs boson and the SM particles, that are absent at tree-level. 
In particular, $U(1)_{em}$ and $SU(3)_c$ gauge invariance prevents renormalizable couplings to photons and gluons, therefore
the tree-level amplitudes for $h\rightarrow gg $, $h\rightarrow \gamma \gamma$ and $h\rightarrow Z\gamma$ are zero, 
and in addition the one-loop amplitudes for these processes are free from divergences.
The effective Higgs boson couplings generated at one-loop can be described in full generality by two dimension-five operators,
\beq
{\cal L}_{hVV'} = \left(c_{hVV'}  V_{\mu\nu}V'^{\mu\nu} + \frac12 \tilde c_{hVV'}  V_{\mu\nu} V'_{\rho\sigma} \epsilon^{\mu\nu\rho\sigma}\right)h ~,
\label{hVVeff}\eeq
where $VV'=gg,\gamma\gamma,\gamma Z$ and $V_{\mu\nu}=G^a_{\mu\nu},F_{\mu\nu},Z_{\mu\nu}$ are the
field strength tensors for the gluon, the photon and the $Z$, respectively.
The CP-even (odd) coefficients $c_{hVV'}$ ($\tilde c_{hVV'}$) have mass dimension minus one,
and may receive contributions from loops of both SM and new particles. One should be careful to avoid double-counting:
if one considers  ${\cal L}_{hVV'}$ as part of the effective Lagrangian valid below the EWSB scale $v$, 
$c_{hVV'}$ and $\tilde c_{hVV'}$ should include only the contributions of particles heavier than $v$.

In momentum space, the $hVV'$ couplings are given by
\begin{equation}
\mathcal{L}_{hVV'}(p,p')=
2 \left[ c_{h VV'} (p'^\mu p^\nu - p \cdot p' g^{\mu\nu} ) + \tilde{c}_{h VV'} \epsilon^{\mu\nu\rho\sigma}p_\rho p'_\sigma \right] 
h(p+p') V_\mu (p) {V'}_\nu (p') ~.
\end{equation}
The decay width of the Higgs boson into two vector bosons is then given by
\begin{equation}
\Gamma (h\rightarrow V V') = \frac{N_c \Delta_{V V'}}{8\pi} m_h^3 \,\beta^+_{VV'}\beta^-_{VV'}  \left[ 
c_{h V V'}^2  \left(\beta_{VV'}^{+2} \beta^{-2}_{VV'} +\frac{6 m_V^2m_{V'}^2}{m_h^4}\right) 
+ \tilde{c}_{h V V'}^2  \beta^{+2}_{VV'}\beta^{-2}_{VV'}       
\right] ,
\label{GammaVV}
\end{equation}
where $N_c=8\,(1)$ for gluons (for $\gamma$ and $Z$),
$\Delta_{V V'}= 2$ for $V=V'$ and $\Delta_{VV'}=1$ for $V\ne V'$, and finally 
$\beta^{\pm}_{VV'}\equiv [1-(m_V\pm m_{V'})^2/m_h^2]^{1/2}$. 
In the following, we will match the explicit loop computation with the effective coefficients $c_{hVV'}$ and $\tilde c_{hVV'}$.

\begin{figure}[bt]
\begin{center}
\includegraphics[scale=0.6,trim= 70 540 0 100]{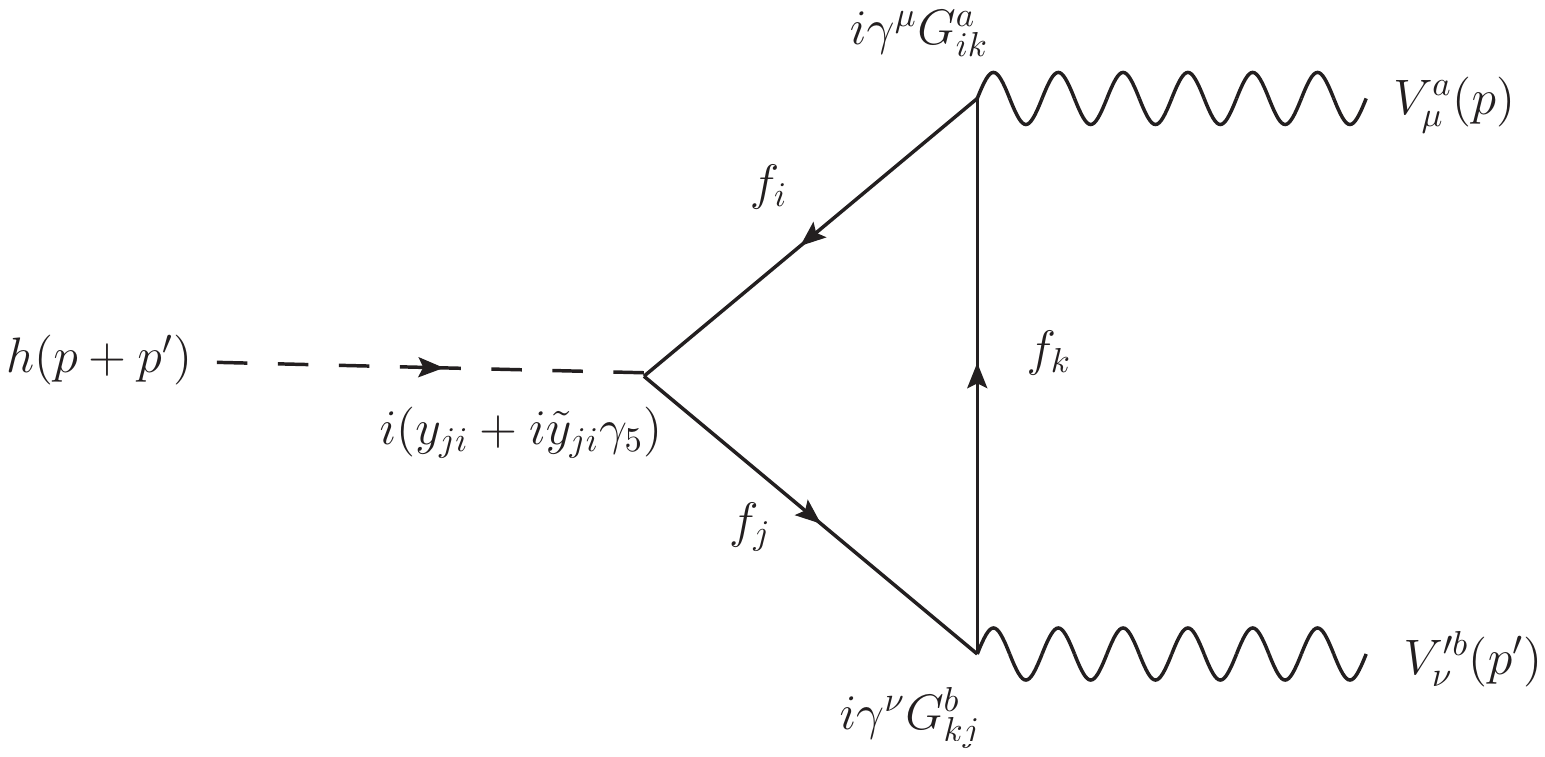}
\caption{Fermionic triangle loop contributing to the coupling of the Higgs boson to two gauge bosons. 
A crossed diagram with the two gauge boson insertions interchanged has to be added too, 
and one must sum over all possible sets $(f_i,f_j,f_k)$ of fermion mass eigenstates.
 For the gluons one has $G^a_{ik}=g_s\delta_{ik}T^a_i$,
for the photon $G^\gamma_{ik} = e\delta_{ik}Q_i$, and for the $Z$ boson $G^Z_{ik}=g(g^V_{ik}-\gamma_5 g^A_{ik})/c_w$.} 
\label{triangle}
\end{center}
\end{figure}

\subsubsection{Higgs coupling to two gluons}
\label{Hgg}

The Higgs-gluon-gluon coupling $c_{h gg}$ ($\tilde c_{h gg}$) receives a one-loop contribution from each coloured fermion with 
a non-zero CP-conserving (CP-violating) Yukawa coupling to the Higgs boson.
The gluon coupling to fermions is determined by $SU(3)_c$ gauge invariance,
\beq
{\cal L}_{g\bar{f}f} =  
g_s \sum_i \overline{f_{ib}}  \gamma^\mu A^a_\mu (T^a_i)_{bc} f_{ic} ~,
\eeq 
where $T^a_i $ are the $SU(3)_c$ generators in the representation $R_{ci}$ of the fermion $f_i$. 
As the $SU(3)_c$ symmetry is unbroken, there are no `off-diagonal' gluon couplings to two different mass eigenstates. 
Recall that the Higgs is mostly produced via gluon fusion, with a partonic cross-section that can be expressed as a function of the partial decay width,
\beq
\sigma(gg\rightarrow h;\hat s)=\frac{\pi^2}{8m_h}\delta(\hat{s}-m_h^2)\Gamma(h\rightarrow gg)~.
\eeq

\begin{figure}[bt]
\begin{center}
\includegraphics[scale=0.3,trim= 0 0 0 0]{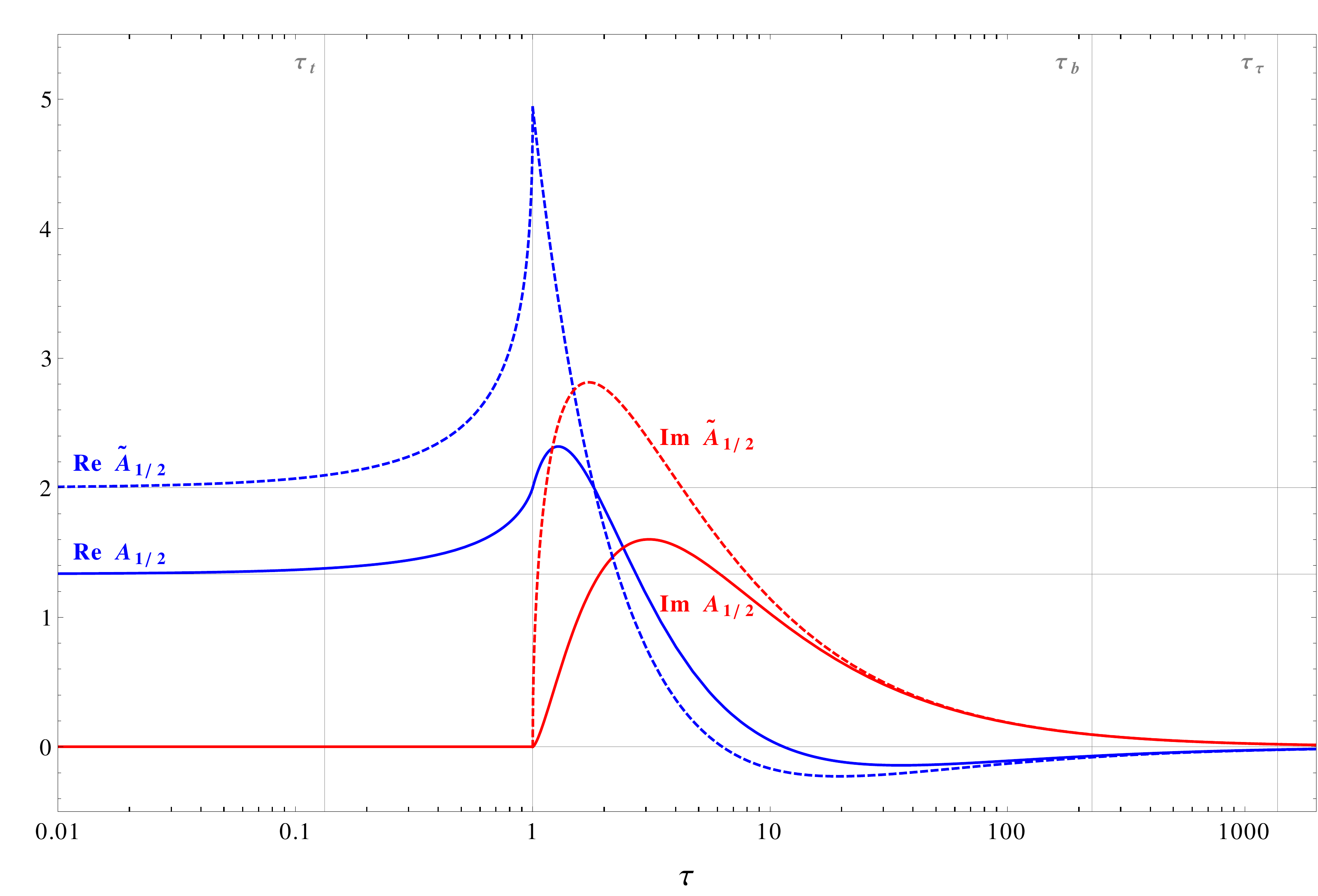}
\caption{Real (blue lines) and imaginary (red lines) parts of the form factors $A_{1/2}(\tau)$ (solid lines) and 
$\tilde{A}_{1/2}(\tau)$ (dashed lines). The horizontal lines correspond to the
asymptotic values $2$, $4/3$ and $0$. The vertical lines correspond to the reference values $\tau_t \simeq 0.13$, $\tau=1$, $\tau_b\simeq 230$ and $\tau_\tau \simeq 1300$.
}
\label{fig2}
\end{center}
\end{figure}

In the SM, the contribution of the quarks triangle loop 
to $h\rightarrow gg $ reads 
\begin{equation}
\Gamma_{SM} (h\rightarrow g g) =\frac{\alpha_s^2 m_h^3}{72\pi^3v^2}  
\left| \frac 34 \sum_q A_{1/2}(\tau_q)  \right|^2 ~,
\end{equation}
where $\tau_q\equiv m_h^2/(4m_q^2)$ and the form factor is given by
\begin{equation}
A_{1/2}(\tau) = \frac{2[\tau +(\tau-1) f(\tau)]}{ \tau^2}~,\quad 
f(\tau)  = \left\lbrace
\begin{array}{cc}
\arcsin^2 \sqrt{\tau}  & \mbox{for}~\tau\leqslant 1~,\\
-\dfrac{1}{4}\left( \log\dfrac{1+\sqrt{1-\tau^{-1}}}{1- \sqrt{1-\tau^{-1}}} -i\pi\right)^2 & \mbox{for}~\tau >1~.
\end{array}\right.
\end{equation}
As illustrated in Fig.~\ref{fig2}, the top quark gives the dominant contribution, because $\tau_t \ll 1$ and $\tau_{b,c,\dots} \gg 1$.
In a generic extension of the SM, the fermions will couple to the Higgs as in \eq{hff1}, but only the diagonal, real couplings
$y_i\equiv y_{ii}$ and $\tilde{y}_i\equiv \tilde y_{ii}$ are relevant for $h\rightarrow gg$. One obtains
\begin{equation}
\Gamma (h\rightarrow g g)  =\frac{\alpha_s^2 m_h^3}{72\pi^3v^2} \left( \mid \mathcal{A}_f^{g g}  \mid^2  + \mid \mathcal{\tilde{A}}_f ^{g g}\mid^2 \right)~,
\end{equation}
where the 
CP-even and CP-odd amplitudes 
are 
\begin{equation}
\mathcal{A}_f^{g g}= \frac 32 \sum_i C(R_{ci}) \frac{y_i v}{m_i} A_{1/2}(\tau_i)~,~~~~~
\mathcal{\tilde{A}}_f^{g g}= \frac 32 \sum_i C(R_{ci})\frac{\tilde y_i  v}{m_i}   \tilde{A}_{1/2}(\tau_i)~,~~~~~ 
\tilde{A}_{1/2}(\tau)= 2\frac{f(\tau)}{\tau}~,
\end{equation}
with  the Dynkin index $C(R_{ci})$ defined below \eq{anom}.

By matching with \eq{GammaVV}, one finds that the contribution of a fermion loop 
to the effective $hgg$-couplings is 
\beq
c_{hgg}^i= \frac{\alpha_s }{8\pi v}  C(R_{ci}) \frac{y_iv}{m_i} A_{1/2}(\tau_i)~,~~~~~
\tilde c_{hgg}^i= \frac{\alpha_s}{8\pi v}  C(R_{ci}) \frac{\tilde y_iv}{m_i} \tilde{A}_{1/2}(\tau_i) ~,
\label{chgg}
\eeq 
where $c_{hgg}\equiv |\sum_i c_{hgg}^i|$ and $\tilde c_{hgg}\equiv |\sum_i \tilde c_{hgg}^i|$.
In the heavy fermion limit, $2 m_i \gg m_h$, one can use  $A_{1/2}(0)=4/3$ and $\tilde A_{1/2}(0)=2$.
We note that, in the literature, a factor $1/2$ is sometimes missing in the expression for $\tilde c^t_{hgg}$.

One can use the Low Energy Theorem (LET) \cite{Ellis:1975ap,Shifman:1979eb} (see also \cite{Kniehl:1995tn,Carena:2012xa,Gillioz:2012se})
to evaluate the effective $h V V'$ couplings induced by states much heavier than the EW scale. 
For a given sector of mixing states in the representation $(R_c,Q)$ of $SU(3)_c\times U(1)_{em}$, the low
energy result is a function of their mass matrix ${\cal M}$ only. 
For the $CP$-conserving and $CP$-violating \cite{Voloshin:2012tv} gluon-gluon case one finds, respectively,
\begin{equation}
c_{hgg}^{LET} = \frac{\alpha_s}{12 \pi}   C(R_{c}) \frac{\partial}{\partial v} \ln \left[ \det \left(  {\cal M} {\cal M}^\dagger \right) \right]~,
\quad\quad 
\widetilde{c}_{hgg}^{LET} = \frac{\alpha_s}{4 \pi}   C(R_{c}) \frac{\partial}{\partial v} \arg \left[ \det \left(  {\cal{ M}}  \right) \right]~.
\label{chgg-LET}
\end{equation}
This is very useful in the case of a large, complicated mass matrix ${\cal M}$, because this expression is much easier to evaluate,
with respect to an explicit computation of the mass eigenvalues $m_i$ and of the mass eigenstate couplings $y_i$ and $\tilde y_i$. 
Note, however, that this approximation requires all the mass eigenstates in a given sector to be heavy, $2 m_{i} \gg m_h$.
It is easy to check the consistency of \eq{chgg} and \eq{chgg-LET} for one heavy chiral fermion (e.g. the SM top quark),
as ${\cal M}_i=m_i=y_iv$ and $A_{1/2}(\tau_i)\simeq 4/3$.

\subsubsection{Higgs coupling to two photons}
\label{Hpp}

The fermions charged under $U(1)_{em}$ contribute to the Higgs-photon-photon couplings $c_{h \gamma\gamma}$ and $\tilde{c}_{h \gamma\gamma}$ at one loop. 
In the SM, there is also the contribution from $W$-boson loops, that we include using the SM tree-level couplings of the $W$ to the Higgs and to the photon. 
The SM decay width is given by \cite{Shifman:1979eb,Ellis:1975ap,Gavela:1981ri}

\begin{equation}
\Gamma_{SM} (h\rightarrow \gamma\gamma) =\frac{\alpha^2 m_h^3}{256 \pi^3 v^2}  \left | A_1(\tau_W) + 
\sum_{f_i \in SM} N_{ci} Q^2_i  A_{1/2}(\tau_i)
\right |^2 ~,
\end{equation}
where  $\tau_W\equiv m_h^2/(4m_W^2)$, $\tau_i\equiv m_h^2/(4m_i^2)$,
and the form factor for the $W$-loops reads
\begin{equation}
A_1(\tau) = -\frac{2\tau^2 +3 \tau + 3(2\tau-1)f(\tau)}{\tau^2}~.
\end{equation}
The $W$ contribution is dominant, $A_1(\tau_W)\simeq-8.36$, and it interferes destructively with the top-quark loop,
$N_{ct} Q^2_t  A_{1/2}(\tau_t)\simeq1.83$.

In presence of extra fermions, there are new contributions to $h\rightarrow\gamma\gamma$ that depend, as in the case of $h\rightarrow gg$,
on the Yukawa couplings $y_i$ and $\tilde{y}_i$, 
\begin{equation}
\Gamma (h\rightarrow \gamma\gamma)  = \frac{\alpha^2 m_h^3}{256 \pi^3 v^2} \left[ \mid A_1(\tau_W) +\mathcal{A}_f^{\gamma \gamma}\mid^2  +\mid \mathcal{\tilde{A}}_f^{\gamma \gamma}\mid^2 \right]~,
\label{hpp}
\end{equation}
where
\begin{equation}
\mathcal{A}_f^{\gamma \gamma}= \sum_i \frac{y_iv}{m_i} N_{ci}Q^2_i A_{1/2}(\tau_i)~,\qquad
\mathcal{\tilde{A}}_f^{\gamma \gamma}= \sum_i \frac{\tilde y_iv}{m_i} N_{ci}Q^2_i \tilde{A}_{1/2}(\tau_i)~.
\label{App}
\end{equation}
In terms of the coefficients of the effective Lagrangian \eq{hVVeff}, the contribution of each fermion is given by 
\beq
c_{h\gamma\gamma}^i= \frac{\alpha}{8\pi v}\frac{y_iv}{m_i}N_{ci}Q^2_i A_{1/2}(\tau_i) ~,\qquad
\tilde c_{h\gamma\gamma}^i= \frac{\alpha}{8\pi v}\frac{\tilde y_iv}{m_i}N_{ci}Q^2_i \tilde{A}_{1/2}(\tau_i)~.
\eeq 
 The LET approximation, for a set of heavy fermions in the representation $(R_c,Q)$ with a mass matrix 
${\cal{ M}}$, is given by
\begin{equation}
c_{h\gamma\gamma}^{LET}=  \frac{\alpha}{12 \pi}   Q^2 N_{c} \frac{\partial}{\partial v} \ln \left[ \det \left(  {\cal{ M}} {\cal{ M}}^\dagger \right) \right]~, 
\quad\quad
\widetilde{c}_{h\gamma\gamma}^{LET}=  \frac{\alpha}{4 \pi}   Q^2 N_{c} \frac{\partial}{\partial v} \arg \left[ \det \left(  {\cal{ M}}  \right) \right] ~,
\label{chpp-LET}
\end{equation}
in analogy with \eq{chgg-LET}.

\subsubsection{Higgs coupling to a $Z$ boson and a photon}
\label{HpZ}

The last loop-induced coupling to be considered is $hZ\gamma$. It is generated by $W$-boson loops, that we take to be SM-like, as well as by fermionic triangle loops.
The $Z$-boson couplings to fermion mass eigenstates are defined in \eq{Zmass}.

\begin{figure}[tb]
\begin{center}
\includegraphics[scale=0.3,trim= 0 0 0 0]{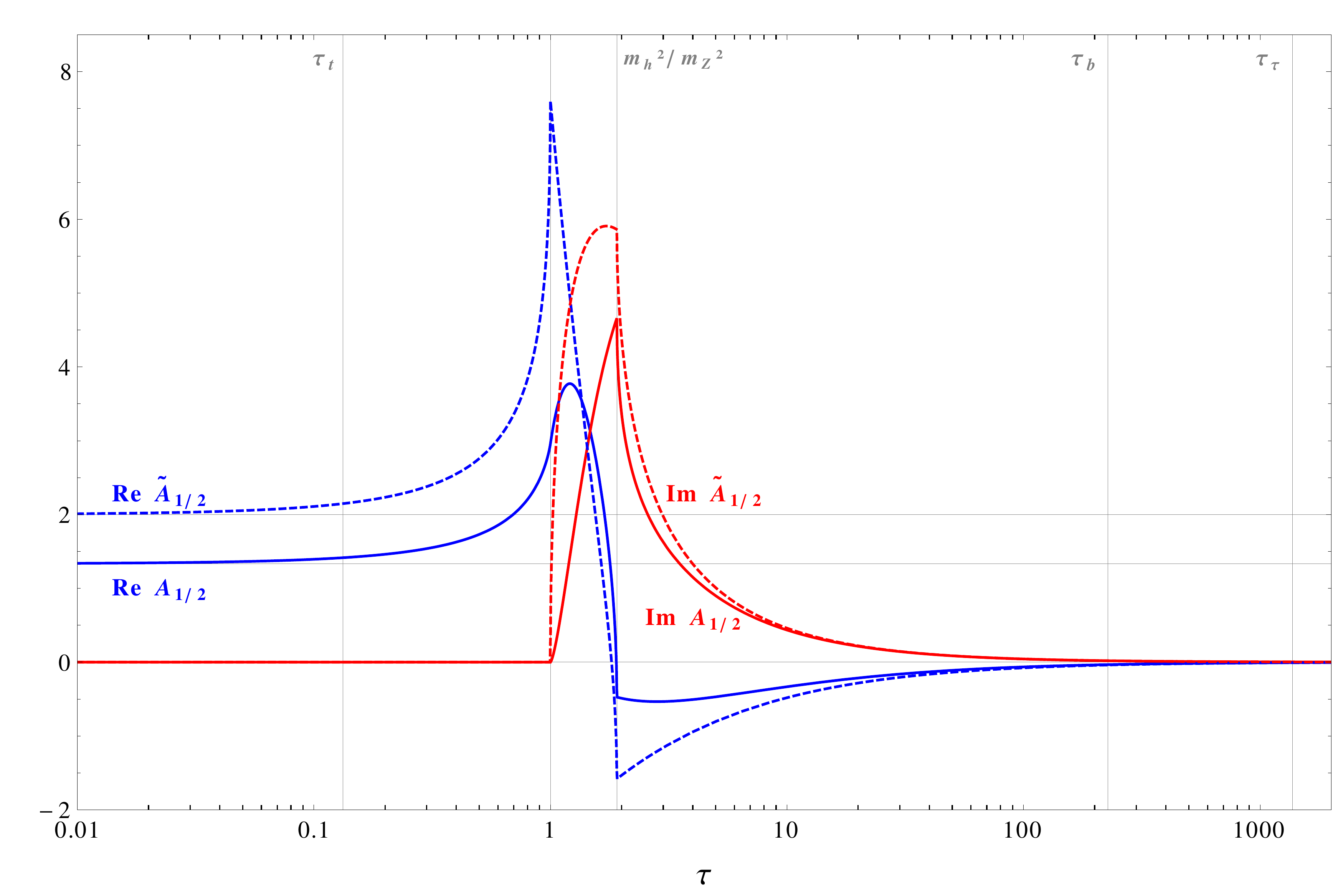}
\caption{
Real (blue lines) and imaginary (red lines) parts of the form factors $A_{1/2}(\tau,\lambda)$ (solid lines) and 
$\tilde{A}_{1/2}(\tau,\lambda)$ (dashed lines), for $\lambda = (m_Z/m_h)^2\tau \simeq 0.52 \tau$. The horizontal lines correspond to the
asymptotic values $2$, $4/3$ and $0$. The vertical lines correspond to $\tau_t \simeq 0.13$, $\tau=1$, $\tau \simeq 1.9$ ($\lambda=1$), $\tau_b\simeq 230$ and $\tau_\tau \simeq 1300$.}
\label{fig3}
\end{center}
\end{figure}

The Higgs decay width into a photon and a $Z$ in the SM is given by \cite{Cahn:1978nz,Bergstrom:1985hp} 
\begin{equation}
\Gamma_{SM} (h\rightarrow \gamma Z) =
\frac{\alpha g^2c_w^2 m_h^3 }{512  \pi^4 v^2} \left( 1-\frac{m_Z^2}{m_h^2} \right)^3  \left| 
A_1(\tau_W,\lambda_W) + \sum_{f_i\in SM} \frac{N_{ci} Q_i g^V_i}{c_w^2}A_{1/2}(\tau_i,\lambda_i)
\right| ^2 ~,
\label{gammaZ-SM}\end{equation}
where $\lambda_W \equiv m_Z^2/(4m_W^2)=(m_Z/m_h)^2\tau_W\simeq 0.52 \tau_W$, and analogously $\lambda_i \equiv m_Z^2/(4m_i^2)\simeq 0.52 \tau_i$.
Note that only the $Zf_i\overline{f_i}$ vector coupling contributes, $g^V_i =T_3(f_{Li})/2-Q_is_w^2$.
The form factors are given by 
\begin{equation}
A_1(\tau,\lambda)=2 \left[3+2\tau-2\lambda(1+2\tau)\right] I_1(\tau,\lambda)-16(1-\lambda)I_2(\tau,\lambda)~,
\end{equation}
\begin{equation}
A_{1/2}(\tau,\lambda)= 4 \left[ I_2(\tau,\lambda) -I_1(\tau,\lambda)\right] ~, 
\end{equation}
where
\begin{equation}
I_1(\tau,\lambda) =-\frac{1}{2(\tau-\lambda)}+\frac{f(\tau)-f(\lambda)}{2(\tau-\lambda)^2}+\frac{\lambda[g(\tau)-g(\lambda)]}{(\tau-\lambda)^2}~,~~~~~
I_2(\tau,\lambda) =\frac{f(\tau)-f(\lambda)}{2(\tau-\lambda)}~,
\end{equation}
\begin{equation}
g(\tau)  = \left\lbrace
\begin{array}{ccc}
\sqrt{\tau^{-1}-1} \, \arcsin \sqrt{\tau}  & \mbox{for} & \tau\leqslant 1 ~,\\
\dfrac{\sqrt{1-\tau^{-1}}}{2}\left( \log\dfrac{1+\sqrt{1-\tau^{-1}}}{1- \sqrt{1-\tau^{-1}}} -i\pi\right) & \mbox{for} & \tau > 1~.
\end{array}\right.
\end{equation}
The normalisation is chosen to match with the $\gamma\gamma$ form factors:
$A_{1}(\tau,0)=A_{1}(\tau)$ and $A_{1/2}(\tau,0)=A_{1/2}(\tau)$.  The behaviour of $A_{1/2}(\tau,\lambda)$ is displayed in Fig.~\ref{fig3},
for the relevant case $\lambda = (m_Z/m_h)^2 \tau$.
The $W$-boson and $t$-quark summands in \eq{gammaZ-SM} take the value $A_1(\tau_W,\lambda_W)\simeq -6.64$ and 
$N_{ct}Q_t g^V_t A_{1/2}(\tau_t,\lambda_t)/c_w^2 \simeq 0.37$,  with the lighter SM fermions adding a very small contribution.

In a generic fermionic extension of the SM, we find a decay width
\begin{equation}
\Gamma(h\rightarrow \gamma Z) =\frac{\alpha g^2 c_w^2 m_h^3}{512 \pi^4 v^2}\left(1-\frac{m_Z^2}{m_h^2} \right)^3 
\left[ \mid A_1(\tau_W,\lambda_W) +\mathcal{A}_f^{Z \gamma} \mid^2  +\mid \mathcal{\tilde{A}}_f^{Z \gamma}\mid^2 \right]~,
\end{equation}
with the $CP$-even and odd fermionic amplitudes given by 
\beq
\mathcal{A}_f^{Z \gamma} = \sum_{j,k} \frac{N_{ck} Q_k v}{c_w^2 \sqrt{m_jm_k}} 
\left[{\rm Re}( g^V_{kj} y_{jk}) a_{1/2}(m_j,m_k,m_k) 
+i {\rm Im}( g^A_{kj} \tilde y_{jk}) b_{1/2}(m_j,m_k,m_k) 
\right]~,
\label{CP-even}
\eeq
\beq
\mathcal{\tilde{A}}_f^{Z \gamma} = \sum_{j,k} \frac{N_{ck} Q_k v }{c_w^2\sqrt{m_jm_k}}  
\left[
{\rm Re}  (g^V_{kj} \tilde{y}_{jk}) \tilde a_{1/2}(m_j,m_k,m_k) 
+i {\rm Im} (g^A_{kj} y_{jk}) \tilde b_{1/2}(m_j,m_k,m_k) 
\right]~.
\label{CP-odd}
\eeq
The explicit expression of the four independent form factors will be given below. As far as we know, this expression for $\Gamma(h\rightarrow \gamma Z)$,
corresponding to a generic set of fermions, was not available in the literature.
Here the sum runs 
over all pairs of fermion mass eigenstates: the triangular fermion loop is formed by one $f_j$ propagator from the $h$ vertex to the $Z$ vertex, 
and two $f_k$ propagators from $Z$ to $\gamma$, and from $\gamma$ to $h$. As both $h$ and $Z$ can have off-diagonal couplings, $j$ and $k$ can be different.
Note that only those combination of couplings that are even under the charge conjugation C contribute,
because the transition $h\rightarrow Z\gamma$ is even: under C, one has $Q\rightarrow -Q$,  
$g^V_{ij}\rightarrow -g^{V*}_{ij}$ and $x_{ij}\rightarrow x_{ji}=x_{ij}^*$, for $x=g^A, y, \tilde{y}$.
The P and CP-even (odd) amplitude corresponds to an even (odd) number
of axial-vector and pseudo-scalar couplings $g^A$ and $\tilde{y}$.

Let us discuss first the loops involving one fermion mass eigenstate only ($j=k$). The diagonal couplings $g^{V,A}_{i}$, $y_{i}$ and $\tilde y_{i}$ are all real, therefore 
the form factors $b_{1/2}$ and $\tilde b_{1/2}$ are irrelevant,  while the others reduce to 
\beq
a_{1/2}(m,m,m) 
= A_{1/2}(\tau,\lambda) ~,~~~~~~
\tilde a_{1/2}(m,m,m) 
= \tilde A_{1/2}(\tau,\lambda) \equiv 4I_2(\tau,\lambda) ~.
\eeq
These two form factors are  displayed in Fig.~\ref{fig3} as a function of $\tau$.
As usual the normalisation matches with the $\gamma\gamma$ form factors, in particular $\tilde A_{1/2}(\tau,0)=\tilde A_{1/2}(\tau)$.
Comparing with \eq{GammaVV}, one finds that the contributions of such fermion `diagonal' loop to the effective $h\gamma Z$ couplings are
\beq
c_{h\gamma Z}^i= \frac{\alpha}{4\pi s_w c_w v} \frac{y_i v}{m_i}  N_{c_i} Q_i  g^V_{i} A_{1/2}(\tau_i,\lambda_i)~,~~~~~~
\tilde c_{h\gamma Z}^i= \frac{\alpha}{4\pi s_w c_w v} \frac{\tilde y_i v}{m_i}  N_{cf} Q_i  g^V_{i} \tilde A_{1/2}(\tau_i,\lambda_i) ~. 
\label{ApZ}
\eeq 

\begin{figure}[tbp]
   \begin{minipage}[c]{.1\linewidth}
      \includegraphics[scale=0.21,trim= 0 0 0 0]{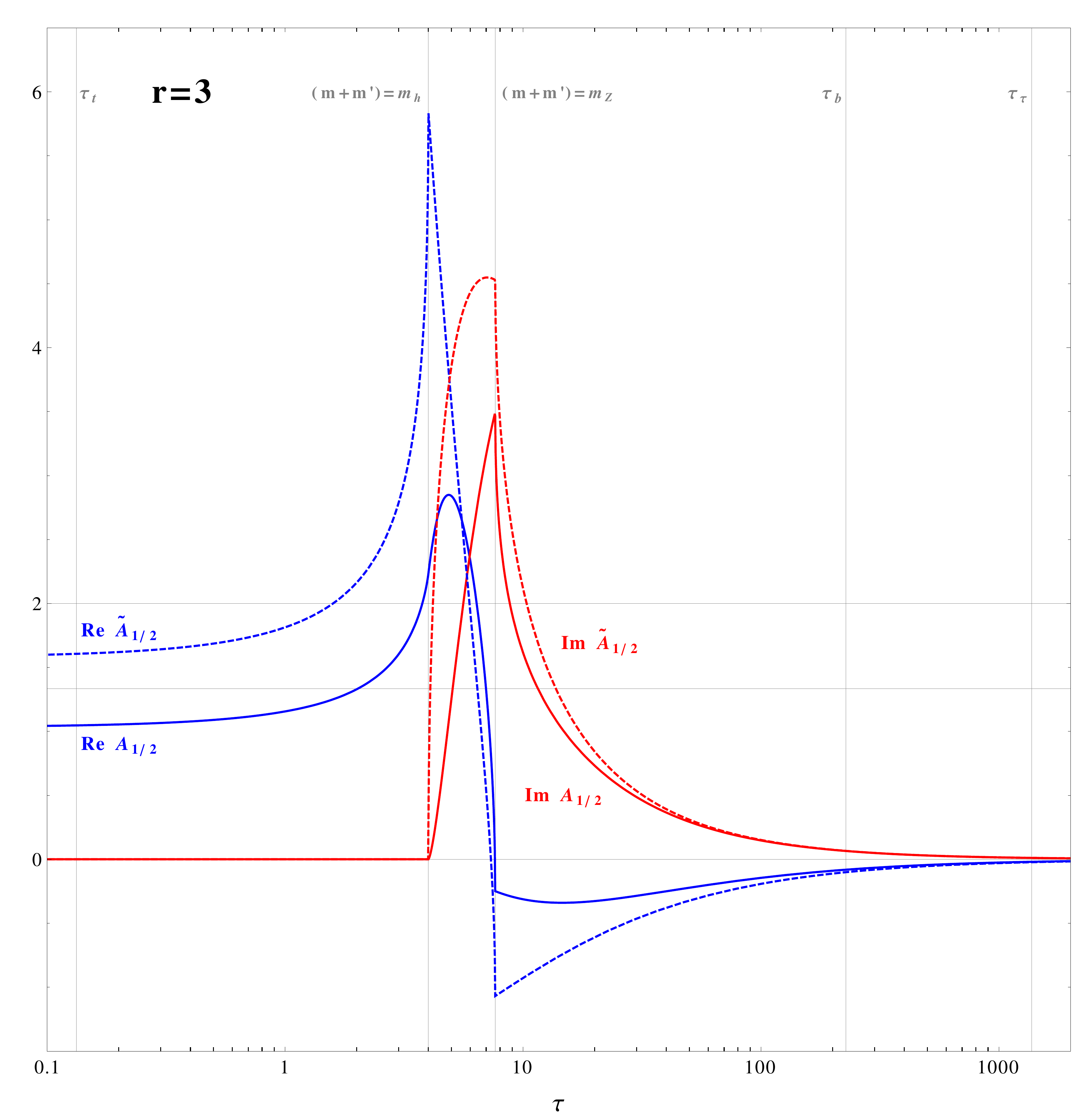} 
   \end{minipage} \hfill
   \begin{minipage}[c]{.5\linewidth}
      \includegraphics[scale=0.21,trim= 0 0 0 0]{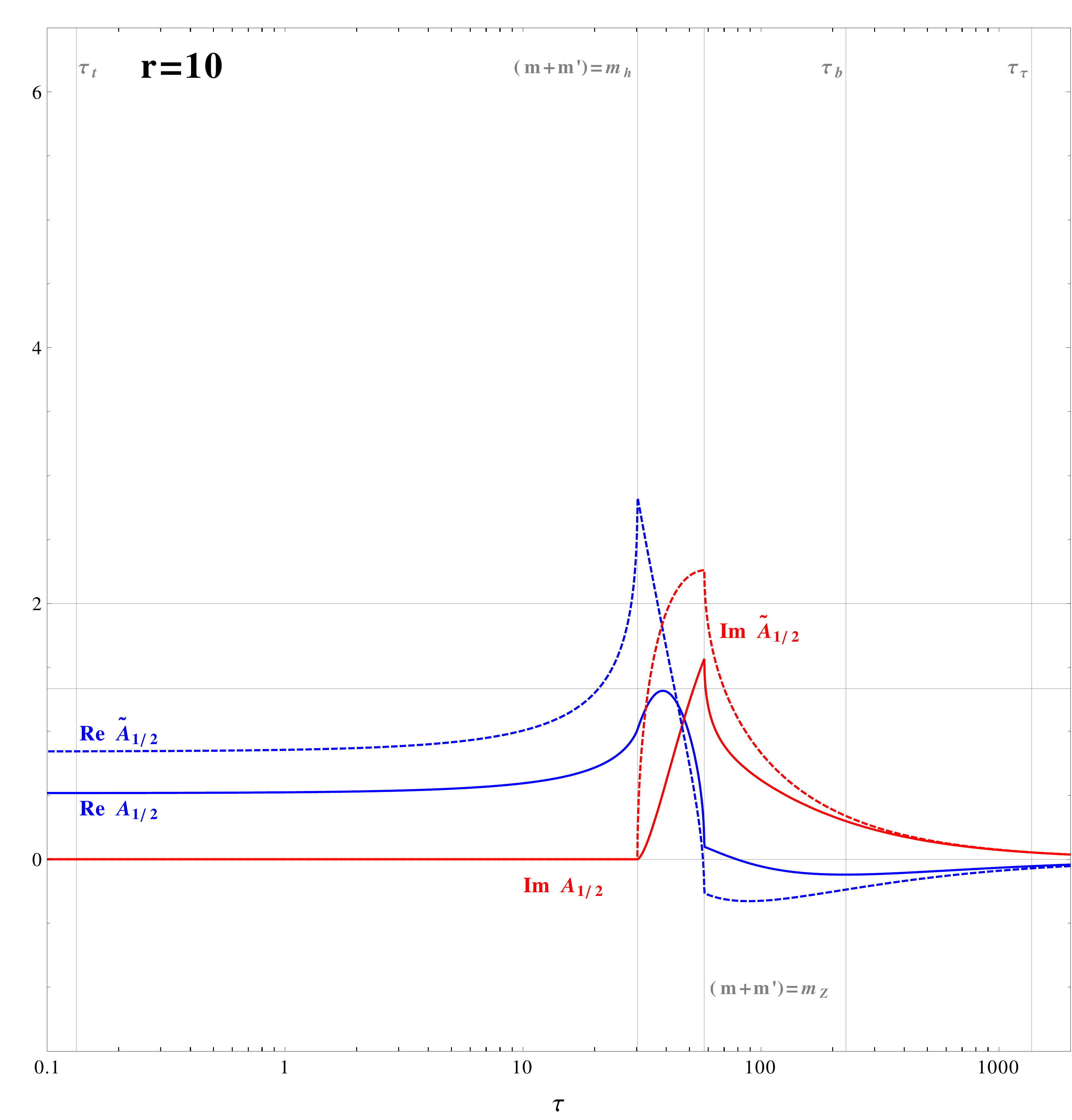}
   \end{minipage}\\
      \begin{minipage}[c]{.1\linewidth}
      \includegraphics[scale=0.215,trim= 25 0 0 0]{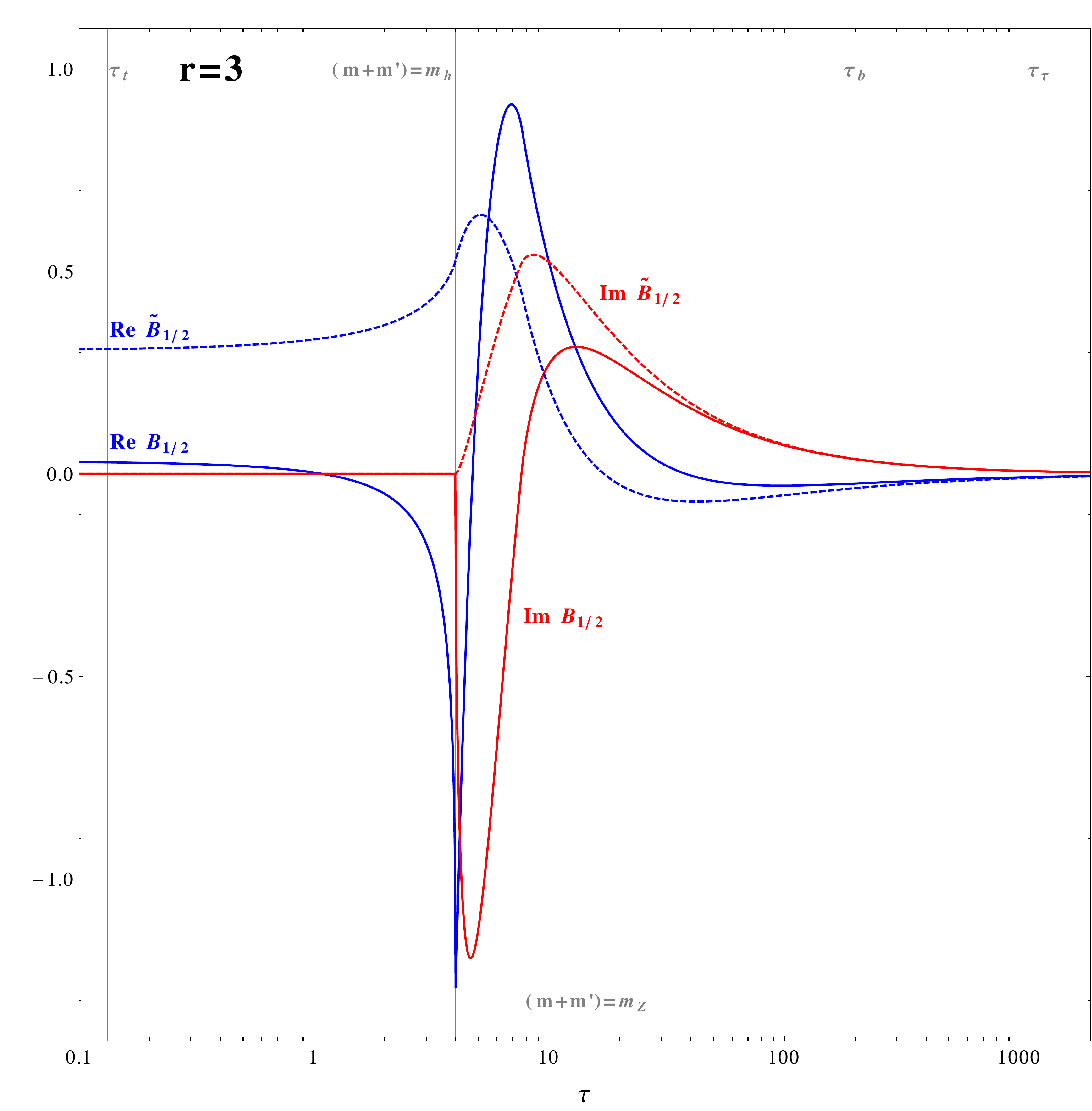}
   \end{minipage} \hfill
   \begin{minipage}[c]{.5\linewidth}
      \includegraphics[scale=0.215,trim= 25 0 0 0]{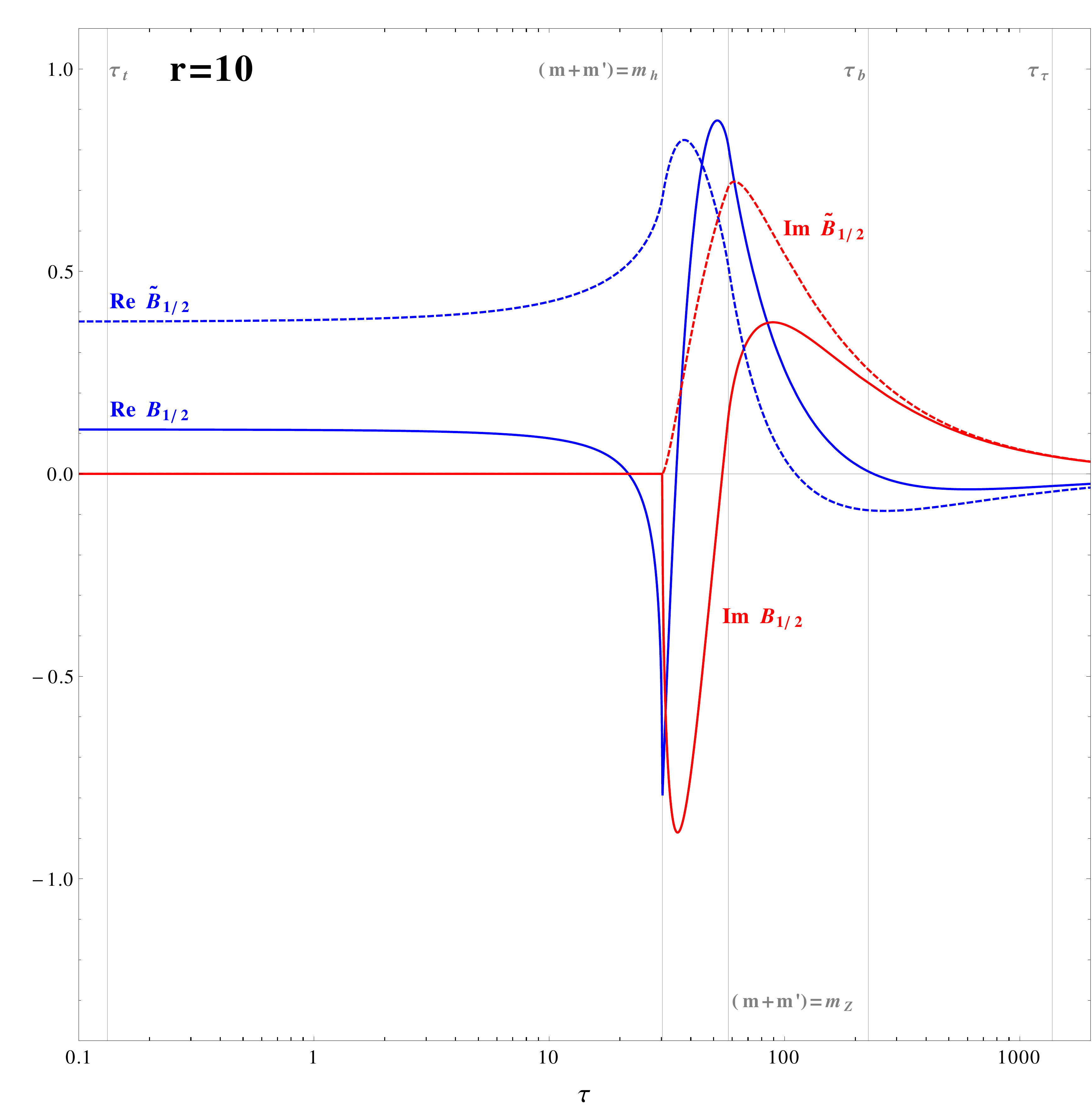}
   \end{minipage}
   \caption{
Form factors for the triangle loops involving two fermions of mass $m$ and $m'=r\,m$,
for $r=3$ and $r=10$, in the left and right-hand panels, respectively.
We show the real (in blue) and imaginary (in red) parts of 
$A_{1/2}(\tau,\lambda,\tau/r^2,\lambda/r^2)$ (solid lines, upper panels),
$\tilde A_{1/2}(\tau,\lambda,\tau/r^2,\lambda/r^2)$ (dashed, upper),
$B_{1/2}(\tau,\lambda,\tau/r^2,\lambda/r^2)$ (solid, lower)
and $\tilde B_{1/2}(\tau,\lambda,\tau/r^2,\lambda/r^2)$ (dashed, lower),
as a function of $\tau\equiv 4m_h^2/m^2$, for a 
fixed value of  $\lambda \equiv (m_Z/m_h)^2 \tau \simeq 0.52\ \tau$.
The horizontal lines correspond to the asymptotic values for the case $r=1$ (see Fig.~\ref{fig3}). 
The vertical lines correspond to the third family masses, $\tau_t \simeq 0.13$, $\tau_b\simeq 230$ and $\tau_\tau \simeq 1300$, and to the 
threshold values,
$m+m'=m_h$ [$\tau=(1+r)^2/4$], and $m+m'=m_Z$ [$\tau= (m_h/m_Z)^2(1+r)^2/4$]. }
\label{offdiag}
\end{figure}

Let us now discuss the loops involving two fermion mass eigenstates ($j\ne k$).
The form factors are given by 
\begin{eqnarray}  
&& \frac{a_{1/2}(m_j,m_k,m_k)}{\sqrt{m_j m_k}} 
=  - \frac{4}{m_Z^2-m_h^2} \bigg\{ \left[ \frac{A_0(m_j)-A_0(m_k)}{m_h^2} +1 \right](m_j+m_k)   \nonumber \\&& 
+ \frac{B_0(m_h^2;m_j,m_k)}{m_Z^2-m_h^2} \left[ (m_j+m_k) \left(2 m_j^2-2m_k^2-m_h^2 +\frac{m_Z^2}{m_h^2}(m_k^2-m_j^2)\right) + 2 m_k(m_h^2-m_Z^2) \right] \nonumber  \\&& 
+ \frac{B_0(m_Z^2;m_j,m_k)}{m_Z^2-m_h^2} \left[ (m_j+m_k)(m_k^2-m_j^2) +m_k (2 m_Z^2- m_h^2)+m_j m_h^2 \right] \nonumber  \\ && 
+ C_0(m_Z^2,0,m_h^2;m_j,m_k,m_k) \left[ 2 m_k^2(m_j +m_k) +m_k(m_Z^2-m_h^2) \right] \bigg\}~,
\label{A12}
\end{eqnarray}
\begin{eqnarray} 
\frac{\tilde a_{1/2}(m_j,m_k,m_k) }{\sqrt{m_j m_k} }
 & = & -4  \bigg[   \frac{B_0(m_h^2;m_j,m_k)-B_0(m_Z^2;m_j,m_k)}{m_Z^2-m_h^2}  (m_j-m_k) \nonumber  \\
&& +   C_0(m_Z^2,0,m_h^2;m_j,m_k,m_k) m_k \bigg]~,
\label{A12tilde}
\end{eqnarray}
\beq
\frac{b_{1/2}(m_j,m_k,m_k)}{\sqrt{m_j m_k} } =   \frac{a_{1/2}(-m_j,m_k,m_k)}{\sqrt{m_j m_k}} ~,\qquad
\frac{\tilde b_{1/2}(m_j,m_k,m_k)}{\sqrt{m_jm_k}} =  \frac{\tilde a_{1/2}(-m_j,m_k,m_k)}{\sqrt{m_j m_k}}~,
\label{B12}
\eeq
where $A_0, B_0$ and $C_{0}$ are the standard Passarino-Veltman scalar functions ~\cite{Passarino:1978jh,Bardin:1999ak}, in the convention of \Ref{Denner:1991kt}. 
In the literature, these `off-diagonal' loops are often neglected but, in models with significant $h$ and $Z$ off-diagonal couplings, they may provide a contribution to the decay width
of the same order as the `diagonal' loops.
For example, the  form factors in Eqs.~(\ref{A12})-(\ref{B12}) 
have been already employed in supersymmetric models to compute the charginos loops ~\cite{Djouadi:1996ws,Djouadi:1996yq,Cao:2013ur}.

It is useful to combine all the loops involving two given fermions $f_j$ and $f_k$. 
As Re$( g^V_{kj} y_{jk} ) =$ Re$(g^V_{jk} y_{kj})$ and Im$( g^A_{kj} \tilde{y}_{jk} ) = -$Im$(g^A_{jk} \tilde{y}_{kj})$, the combinations that appear in the total amplitude are
\begin{equation}
A_{1/2}(\tau_j,\lambda_j,\tau_k,\lambda_k) \equiv \frac{1}{2} \left[ a_{1/2}(m_j,m_k,m_k) + a_{1/2}(m_k,m_j,m_j)
\right]~,
\end{equation}
\begin{equation}
B_{1/2}(\tau_j,\lambda_j,\tau_k,\lambda_k)\equiv \frac{1}{2} \left[b_{1/2}(m_j,m_k,m_k)-b_{1/2}(m_k,m_j,m_j)
\right]~,
\label{B12tot}
\end{equation} 
and analog definitions for the CP-odd counterparts $\tilde{A}_{1/2}$ and $\tilde{B}_{1/2}$.
In Fig.~\ref{offdiag} we illustrate the behaviour of these four form factors 
as a function of $\tau = m_h^2/(4m^2)$, where $m$ is the mass of the lightest fermion, 
for a fixed value of the ratio $r = m'/m$, where $m'$ is the mass of the heaviest fermion.
In the limit $r=1$ one recovers the diagonal form factors, shown in Fig.~\ref{fig3}. 
Since $m'$ is the mass of a new charged fermion, it should be sufficiently large to comply with experimental lower bounds; requiring for example $m'>m_h$, one
finds  that only the region $\tau < r^2/4$ is relevant for phenomenology.
Note that  the behaviour of the form factors as $\tau\rightarrow 0$ is sensitive to the mass ratio: 
as $r$ increases from $1$ to infinity, the asymptotic regime settles at larger values of $\tau$, and the asymptotic value of the form factors $A_{1/2}$ and $\tilde A_{1/2}$ 
tends to zero as $1/\sqrt{r}$.
The form factors $B_{1/2}$ and $\tilde B_{1/2}$ are zero for $r=1$, then become of order one as $r$ grows, then tend to zero in the large-$r$ limit.

\subsection{Experimental constraints on the Higgs couplings}\label{experimental}

Here we collect the constraints on the Higgs couplings that we use in our analysis. For a given Higgs-decay final state $\alpha$, the LHC measures the signal strength $\mu_\alpha$ defined as
\begin{equation}
\mu_\alpha\equiv \frac{\sigma(pp\rightarrow h)}{\sigma^{SM}(pp\rightarrow h)} \frac{Br(h\rightarrow \alpha)}{Br^{SM}(h\rightarrow \alpha)}=\frac{\sigma(pp\rightarrow h)}{\sigma^{SM}(pp\rightarrow h)} \frac{\Gamma(h\rightarrow \alpha)}{\Gamma^{SM}(h\rightarrow \alpha)} \frac{\Gamma_{h}^{SM}}{\Gamma_{h}} ~,
\label{SS}
\end{equation}
where $\Gamma_h$ is the total Higgs width. 
In Table~\ref{threetable} we report the present determination of $\mu_\alpha$ for $\alpha=\gamma\gamma,Z Z^*, WW^*,b\overline{b},\tau\tau,\gamma Z,\mu\mu$ as well as on the invisible width, by the ATLAS and CMS collaborations. 
The expected precision for a luminosity of $300-3000$ fb$^{-1}$  at 14 TeV \cite{ATL-PHYS-PUB-2014-016,CMS:2013xfa} is reported for these same channels in Table~\ref{fourtable}.

From a global fit of the Higgs data, one can also extract information on the Higgs coupling to gluons.
Taking a rough extrapolation from the fit in Fig.~16 of Ref.~\cite{ATLAS-CONF-2015-044}, we find $0.5 \lesssim R_{gg}\lesssim 1.8$  at $99 \%$ C.L., where $R_{gg}$ is defined in \eq{Rgg}.
In the same way we extract the analogue quantity for the diphoton channel,  $0.5 \lesssim R_{\gamma\gamma}\lesssim 1.9$  at $99 \%$ C.L., that we employ throughout this paper.

\begin{table}[btp]
\renewcommand{\arraystretch}{1.3}
\bc\begin{tabular}{|c|c|c|c|}
\hline
 &  ATLAS & CMS & ATLAS+CMS\\
\hline\hline 
$\sqrt{s}$&  7+8 TeV &  7+8 TeV &  7+8 TeV \\
\hline
 $\cal{L}$ & $4.7$ + $20.3$ $fb^{-1}$ & $5.1$ + $19.7$  $fb^{-1}$ &  \\
\hline\hline
$m_h$ & $125.36\pm 0.41$ GeV ~\cite{Aad:2015gba} & $125.02^{+0.29}_{-0.31}$ GeV ~\cite{Khachatryan:2014jba} & $125.09\pm 0.24$ GeV \cite{Aad:2015zhl} \\
\hline\hline 
$\mu_{\gamma\gamma}$ & $1.17^{+0.28}_{-0.26}$  ~\cite{Aad:2015gba} & $1.12\pm 0.24$  ~\cite{Khachatryan:2014jba}  & $1.13^{+0.24}_{-0.21}$ ~\cite{ATLAS-CONF-2015-044}\\ 
\hline 
$\mu_{ZZ^*\rightarrow 4l}$ & $1.46^{+0.40}_{-0.34}$   \cite{Aad:2015gba} & $1.00\pm0.29 $  ~\cite{Khachatryan:2014jba}  & $1.29^{+0.29}_{-0.25}$ ~\cite{ATLAS-CONF-2015-044} \\
\hline 
$\mu_{WW^*}$ & $1.18^{+0.24}_{-0.21}$ ~\cite{Aad:2015gba} & $0.83\pm 0.21 $ ~\cite{Khachatryan:2014jba} &  $1.08^{+0.22}_{-0.19}$ ~\cite{ATLAS-CONF-2015-044} \\
\hline\hline 
$\mu_{\tau \tau}$ & $ 1.44^{+0.42}_{-0.37}$ ~\cite{Aad:2015gba}  &$0.91\pm 0.28$  ~\cite{Khachatryan:2014jba} & $1.07^{+0.35}_{-0.28}$ ~\cite{ATLAS-CONF-2015-044} \\ 
\hline
$\mu_{b\bar{b}}$ & $ 0.63^{+0.39}_{-0.37}$ ~\cite{Aad:2015gba}  & $0.84\pm 0.44$  ~\cite{Khachatryan:2014jba} & $0.65^{+0.37}_{-0.28}$ ~\cite{ATLAS-CONF-2015-044} \\
\hline \hline 
$\mu_{global}$ 
& $1.18^{+0.15}_{-0.14}$ ~\cite{Aad:2015gba} & $1.00\pm 0.14$ ~\cite{Khachatryan:2014jba} & $1.09^{+0.11}_{-0.10}$ ~\cite{ATLAS-CONF-2015-044}\\
\hline
\hline
$\mu_{\gamma Z}$ & $<11.0$   ~\cite{Aad:2014fia} & $<9.5$ ~\cite{Chatrchyan:2013vaa} &\\
\hline 
$\Gamma_{invisible}/\Gamma_h$ &  $ <0.29$ ~\cite{ATLAS-CONF-2015-004}    &  $<0.58$ ~\cite{Chatrchyan:2014tja} & \\
\hline
\end{tabular}\ec
\caption{ Higgs signal strengths measured by the ATLAS  and CMS  collaborations at $\sqrt{s}=7$ and $8$ TeV and their combination. The error bars and upper limits correspond respectively to $\pm 1\sigma$ and $95 \%$ C.L.. 
In this paper we adopt the value of the Higgs mass and of the signal strengths from the combined fit of the ATLAS and CMS data, reported in the last column.}
\label{threetable}
\end{table}

The new fermions may or may not affect each of the three factors on the right-hand side of \eq{SS}. Let us discuss first the Higgs production cross section. 
The dominant gluon fusion channel can be modified at leading order only by coloured fermions. 
The  weak-vector-boson fusion and associated production can be modified, at tree level, only by fermions mixing with the initial state quarks: as we limit our analysis to mixing with the third family, we neglect these modifications.
Finally, the $t\bar t$ associated production can be corrected by those fermions mixing with the top quark.
Concerning the total Higgs decay width $\Gamma_h$, the dominant branching ratio into $b\bar b$ is affected by fermions mixing with the bottom quark, and  the second
dominant decay channel into $W W^*$ is not modified by new fermions at leading order. Finally, $\Gamma_h$ may be modified significantly by new invisible decays, that are possible in the presence of sterile neutrinos.
When both $\sigma(pp\rightarrow h)$ and $\Gamma_h$ are close to their SM value, the signal strength in \eq{SS} reduces to $\mu_\alpha\simeq R_\alpha$, where 
\begin{equation}
R_\alpha \equiv \frac{\Gamma(h\rightarrow \alpha)}{\Gamma^{SM}(h\rightarrow \alpha)}=\frac{ \left| {\cal A}^\alpha_{SM} + {\cal A}^\alpha_{new} \right|^2+ |  \widetilde{{\cal A}}^\alpha_{new} |^2 }{\left| {\cal A}^\alpha_{SM} \right|^2}~.
\label{R}
\end{equation} 
Here ${\cal A}^\alpha_{new}$ and $\widetilde{{\cal A}}^\alpha_{new}$ are the parity-even and odd new physics amplitudes, respectively.
The approximation $\mu_\alpha\simeq R_\alpha$ holds for all colourless new fermions, with the possible exception of light sterile neutrinos.

Several groups performed global fits of the Higgs couplings to the SM particles, allowing for deviations in both the fermionic and bosonic decay channels, see e.g. Refs.~\cite{Espinosa:2012im,Carmi:2012in,Cacciapaglia:2012wb,Azatov:2012qz,Falkowski:2013dza,
Giardino:2013bma,Djouadi:2013qya,Belanger:2013xza}.
The fit of Ref.~\cite{Moreau:2012da} analyzed deviations in the Higgs couplings in the presence of new fermions only.

\begin{table}[btp]
\renewcommand{\arraystretch}{1.3}
\bc\begin{tabular}{|c|c c|c c|}
\hline
 &  ATLAS & \cite{ATL-PHYS-PUB-2014-016}    & CMS  & \cite{CMS:2013xfa}   \\
\hline
\hline $\Delta \mu_\alpha /\mu_\alpha$  & 300 $fb^{-1}$ & 3000 $fb^{-1}$ & 300 $fb^{-1}$ &3000 $fb^{-1}$\\
\hline 
\hline $\gamma\gamma$ & 0.13 & 0.09 & [0.06,0.12] & [0.04,0.08] \\
\hline $ZZ$ & 0.11 & 0.09 & [0.07,0.11] & [0.04,0.07] \\
\hline $WW$ & 0.13 & 0.11 & [0.06,0.11] & [0.04,0.07]\\
\hline
\hline $\tau\tau$ & 0.21 & 0.19 & [0.08,0.14] & [0.05,0.08]\\
\hline $b \bar{b}$ & 0.26 & 0.14 & [0.11,0.14] & [0.05,0.07] \\
\hline
\hline $\gamma Z$ & 0.46 & 0.30 & [0.62,0.62] & [0.20,0.24] \\
\hline $\mu\mu$ & 0.39 & 0.16 & [0.40,0.42] & [0.20,0.24] \\
\hline
\hline $\Gamma_{invisible}/\Gamma_h$ & $< 0.22$ & $< 0.14$ & $< [0.17,0.28]$ & $< [0.06,0.17]$ \\
\hline
\end{tabular}\ec
\caption{Expected relative uncertainty at $1\sigma$ on the signal strengths $\mu_\alpha$ for ATLAS and CMS. The expected precisions correspond to $\sqrt{s}= 14$ TeV and ${\cal{L}}= 300$ and $3000$ $fb^{-1}$. We also display the expected limit at $95 \%$ C.L. on $\Gamma_{invisible}/\Gamma_h$ for the same luminosities. 
For CMS the two numbers correspond to two different estimations of the future uncertainties \cite{CMS:2013xfa}.}
\label{fourtable}
\end{table}

\bibliographystyle{JHEP}
\bibliography{biblio}

\providecommand{\href}[2]{#2}\begingroup\raggedright\begin{thebibliography}{100}

\bibitem{Flacher:2008zq}
H.~Flacher, M.~Goebel, J.~Haller, A.~Hocker, K.~Monig, et~al., {\it {Revisiting
  the Global Electroweak Fit of the Standard Model and Beyond with Gfitter}},
  {\em Eur.Phys.J.} {\bf C60} (2009) 543--583,
  [\href{http://arxiv.org/abs/0811.0009}{{\tt arXiv:0811.0009}}].

\bibitem{Schael:2013ita}
{\bf ALEPH, DELPHI, L3, OPAL, LEP Electroweak} Collaboration, S.~Schael et~al.,
  {\it {Electroweak Measurements in Electron-Positron Collisions at
  W-Boson-Pair Energies at LEP}},  {\em Phys.Rept.} {\bf 532} (2013) 119--244,
  [\href{http://arxiv.org/abs/1302.3415}{{\tt arXiv:1302.3415}}].

\bibitem{Agashe:2014kda}
{\bf Particle Data Group} Collaboration, K.~Olive et~al., {\it {Review of
  Particle Physics}},  {\em Chin.Phys.} {\bf C38} (2014) 090001.

\bibitem{Raidal:2008jk}
M.~Raidal et~al., {\it {Flavour physics of leptons and dipole moments}},  {\em
  Eur. Phys. J.} {\bf C57} (2008) 13--182,
  [\href{http://arxiv.org/abs/0801.1826}{{\tt arXiv:0801.1826}}].

\bibitem{Isidori:2010kg}
G.~Isidori, Y.~Nir, and G.~Perez, {\it {Flavor Physics Constraints for Physics
  Beyond the Standard Model}},  {\em Ann.Rev.Nucl.Part.Sci.} {\bf 60} (2010)
  355, [\href{http://arxiv.org/abs/1002.0900}{{\tt arXiv:1002.0900}}].

\bibitem{Toback:2014tea}
D.~Toback and L.~ŽIvković, {\it {Review of Physics Results from the Tevatron:
  Searches for New Particles and Interactions}},  {\em Int.J.Mod.Phys.} {\bf
  A30} (2015), no.~06 1541007, [\href{http://arxiv.org/abs/1409.4910}{{\tt
  arXiv:1409.4910}}].

\bibitem{Nakamura:2010zzi}
{\bf Particle Data Group} Collaboration, K.~Nakamura et~al., {\it {Review of
  particle physics}},  {\em J. Phys.} {\bf G37} (2010) 075021.

\bibitem{twiki-ATLAS}
{\bf ATLAS} Collaboration,
  \url{https://twiki.cern.ch/twiki/bin/view/AtlasPublic}.

\bibitem{twiki-CMS}
{\bf CMS} Collaboration,
  \url{http://cms-results.web.cern.ch/cms-results/public-results/publications/}.

\bibitem{Aad:2012tfa}
{\bf ATLAS} Collaboration, G.~Aad et~al., {\it {Observation of a new particle
  in the search for the Standard Model Higgs boson with the ATLAS detector at
  the LHC}},  {\em Phys.Lett.} {\bf B716} (2012) 1--29,
  [\href{http://arxiv.org/abs/1207.7214}{{\tt arXiv:1207.7214}}].

\bibitem{Chatrchyan:2012ufa}
{\bf CMS} Collaboration, S.~Chatrchyan et~al., {\it {Observation of a new boson
  at a mass of 125 GeV with the CMS experiment at the LHC}},  {\em Phys.Lett.}
  {\bf B716} (2012) 30--61, [\href{http://arxiv.org/abs/1207.7235}{{\tt
  arXiv:1207.7235}}].

\bibitem{Khachatryan:2014jba}
{\bf CMS} Collaboration, V.~Khachatryan et~al., {\it {Precise determination of
  the mass of the Higgs boson and tests of compatibility of its couplings with
  the standard model predictions using proton collisions at 7 and 8 TeV}},
  \href{http://arxiv.org/abs/1412.8662}{{\tt arXiv:1412.8662}}.

\bibitem{Aad:2015zhl}
{\bf ATLAS, CMS} Collaboration, G.~Aad et~al., {\it {Combined Measurement of
  the Higgs Boson Mass in $pp$ Collisions at $\sqrt{s}=7$ and 8 TeV with the
  ATLAS and CMS Experiments}},  \href{http://arxiv.org/abs/1503.07589}{{\tt
  arXiv:1503.07589}}.

\bibitem{Aad:2015gba}
{\bf ATLAS} Collaboration, G.~Aad et~al., {\it {Measurements of the Higgs boson
  production and decay rates and coupling strengths using $pp$ collision data
  at $\sqrt{s}=7$ and $8$ TeV in the ATLAS experiment}},
  \href{http://arxiv.org/abs/1507.04548}{{\tt arXiv:1507.04548}}.

\bibitem{ALEPH:2005ab}
{\bf ALEPH, DELPHI, L3, OPAL, SLD, LEP Electroweak Working Group, SLD
  Electroweak Group, SLD Heavy Flavour Group} Collaboration, S.~Schael et~al.,
  {\it {Precision electroweak measurements on the $Z$ resonance}},  {\em
  Phys.Rept.} {\bf 427} (2006) 257--454,
  [\href{http://arxiv.org/abs/hep-ex/0509008}{{\tt hep-ex/0509008}}].

\bibitem{Kaplan:1991dc}
D.~B. Kaplan, {\it {Flavor at SSC energies: A New mechanism for dynamically
  generated fermion masses}},  {\em Nucl.Phys.} {\bf B365} (1991) 259--278.

\bibitem{Contino:2004vy}
R.~Contino and A.~Pomarol, {\it {Holography for fermions}},  {\em JHEP} {\bf
  11} (2004) 058, [\href{http://arxiv.org/abs/hep-th/0406257}{{\tt
  hep-th/0406257}}].

\bibitem{Cirelli:2005uq}
M.~Cirelli, N.~Fornengo, and A.~Strumia, {\it {Minimal dark matter}},  {\em
  Nucl.Phys.} {\bf B753} (2006) 178--194,
  [\href{http://arxiv.org/abs/hep-ph/0512090}{{\tt hep-ph/0512090}}].

\bibitem{DelNobile:2009st}
E.~Del~Nobile, R.~Franceschini, D.~Pappadopulo, and A.~Strumia, {\it {Minimal
  Matter at the Large Hadron Collider}},  {\em Nucl. Phys.} {\bf B826} (2010)
  217--234, [\href{http://arxiv.org/abs/0908.1567}{{\tt arXiv:0908.1567}}].

\bibitem{DiLuzio:2015oha}
L.~Di~Luzio, R.~Grober, J.~F. Kamenik, and M.~Nardecchia, {\it {Accidental
  matter at the LHC}},  \href{http://arxiv.org/abs/1504.00359}{{\tt
  arXiv:1504.00359}}.

\bibitem{Picek:2008dd}
I.~Picek and B.~Radovcic, {\it {Nondecoupling of terascale isosinglet quark and
  rare K- and B-decays}},  {\em Phys. Rev.} {\bf D78} (2008) 015014,
  [\href{http://arxiv.org/abs/0804.2216}{{\tt arXiv:0804.2216}}].

\bibitem{Cacciapaglia:2011fx}
G.~Cacciapaglia, A.~Deandrea, L.~Panizzi, N.~Gaur, D.~Harada, et~al., {\it
  {Heavy Vector-like Top Partners at the LHC and flavour constraints}},  {\em
  JHEP} {\bf 1203} (2012) 070, [\href{http://arxiv.org/abs/1108.6329}{{\tt
  arXiv:1108.6329}}].

\bibitem{Han:2014qia}
C.~Han, A.~Kobakhidze, N.~Liu, L.~Wu, and B.~Yang, {\it {Constraining Top
  partner and Naturalness at the LHC and TLEP}},  {\em Nucl. Phys.} {\bf B890}
  (2014) 388--399, [\href{http://arxiv.org/abs/1405.1498}{{\tt
  arXiv:1405.1498}}].

\bibitem{Witten:1982fp}
E.~Witten, {\it {An SU(2) Anomaly}},  {\em Phys.Lett.} {\bf B117} (1982)
  324--328.

\bibitem{Eberhardt:2012gv}
O.~Eberhardt, G.~Herbert, H.~Lacker, A.~Lenz, A.~Menzel, et~al., {\it {Impact
  of a Higgs boson at a mass of 126 GeV on the standard model with three and
  four fermion generations}},  {\em Phys.Rev.Lett.} {\bf 109} (2012) 241802,
  [\href{http://arxiv.org/abs/1209.1101}{{\tt arXiv:1209.1101}}].

\bibitem{Azatov:2012rj}
A.~Azatov, O.~Bondu, A.~Falkowski, M.~Felcini, S.~Gascon-Shotkin, et~al., {\it
  {Higgs boson production via vector-like top-partner decays: Diphoton or
  multilepton plus multijets channels at the LHC}},  {\em Phys.Rev.} {\bf D85}
  (2012) 115022, [\href{http://arxiv.org/abs/1204.0455}{{\tt
  arXiv:1204.0455}}].

\bibitem{Montull:2013mla}
M.~Montull, F.~Riva, E.~Salvioni, and R.~Torre, {\it {Higgs Couplings in
  Composite Models}},  {\em Phys.Rev.} {\bf D88} (2013) 095006,
  [\href{http://arxiv.org/abs/1308.0559}{{\tt arXiv:1308.0559}}].

\bibitem{Flacke:2013fya}
T.~Flacke, J.~H. Kim, S.~J. Lee, and S.~H. Lim, {\it {Constraints on composite
  quark partners from Higgs searches}},  {\em JHEP} {\bf 1405} (2014) 123,
  [\href{http://arxiv.org/abs/1312.5316}{{\tt arXiv:1312.5316}}].

\bibitem{Chen:2014xwa}
C.-Y. Chen, S.~Dawson, and I.~Lewis, {\it {Top Partners and Higgs Boson
  Production}},  {\em Phys.Rev.} {\bf D90} (2014), no.~3 035016,
  [\href{http://arxiv.org/abs/1406.3349}{{\tt arXiv:1406.3349}}].

\bibitem{Matsedonskyi:2014mna}
O.~Matsedonskyi, G.~Panico, and A.~Wulzer, {\it {On the Interpretation of Top
  Partners Searches}},  {\em JHEP} {\bf 1412} (2014) 097,
  [\href{http://arxiv.org/abs/1409.0100}{{\tt arXiv:1409.0100}}].

\bibitem{Aad:2012pra}
{\bf ATLAS Collaboration} Collaboration, G.~Aad et~al., {\it {Searches for
  heavy long-lived sleptons and R-Hadrons with the ATLAS detector in $pp$
  collisions at $\sqrt{s}=7$ TeV}},  {\em Phys.Lett.} {\bf B720} (2013)
  277--308, [\href{http://arxiv.org/abs/1211.1597}{{\tt arXiv:1211.1597}}].

\bibitem{Chatrchyan:2013oca}
{\bf CMS Collaboration} Collaboration, S.~Chatrchyan et~al., {\it {Searches for
  long-lived charged particles in pp collisions at $\sqrt{s}$=7 and 8 TeV}},
  {\em JHEP} {\bf 1307} (2013) 122, [\href{http://arxiv.org/abs/1305.0491}{{\tt
  arXiv:1305.0491}}].

\bibitem{ATLAS:2014fka}
{\bf ATLAS} Collaboration, G.~Aad et~al., {\it {Searches for heavy long-lived
  charged particles with the ATLAS detector in proton-proton collisions at $
  \sqrt{s}=8 $ TeV}},  {\em JHEP} {\bf 1501} (2015) 068,
  [\href{http://arxiv.org/abs/1411.6795}{{\tt arXiv:1411.6795}}].

\bibitem{Fairbairn:2006gg}
M.~Fairbairn, A.~C. Kraan, D.~A. Milstead, T.~Sjostrand, P.~Z. Skands, and
  T.~Sloan, {\it {Stable massive particles at colliders}},  {\em Phys. Rept.}
  {\bf 438} (2007) 1--63, [\href{http://arxiv.org/abs/hep-ph/0611040}{{\tt
  hep-ph/0611040}}].

\bibitem{ATLAS-CONF-2015-044}
{\it {Measurements of the Higgs boson production and decay rates and
  constraints on its couplings from a combined ATLAS and CMS analysis of the
  LHC pp collision data at $\sqrt{s}$ = 7 and 8 TeV}},  Tech. Rep.
  ATLAS-CONF-2015-044, CMS-PAS-HIG-15-002, CERN, Geneva, Sep, 2015.

\bibitem{Banerjee:2013hxa}
S.~Banerjee, M.~Frank, and S.~K. Rai, {\it {Higgs data confronts Sequential
  Fourth Generation Fermions in the Higgs Triplet Model}},  {\em Phys. Rev.}
  {\bf D89} (2014), no.~7 075005, [\href{http://arxiv.org/abs/1312.4249}{{\tt
  arXiv:1312.4249}}].

\bibitem{Alves:2013dga}
A.~Alves, E.~Ramirez~Barreto, D.~A. Camargo, and A.~G. Dias, {\it {A Model with
  Chiral Quarks of Electric Charges -4/3 and 5/3}},  {\em JHEP} {\bf 07} (2013)
  129, [\href{http://arxiv.org/abs/1306.1275}{{\tt arXiv:1306.1275}}].

\bibitem{Aad:2015oga}
{\bf ATLAS} Collaboration, G.~Aad et~al., {\it {Search for heavy long-lived
  multi-charged particles in $pp$ collisions at $\sqrt{s}$ = 8 TeV using the
  ATLAS detector}},  \href{http://arxiv.org/abs/1504.04188}{{\tt
  arXiv:1504.04188}}.

\bibitem{Aad:2011mb}
{\bf ATLAS} Collaboration, G.~Aad et~al., {\it {Search for Massive Long-lived
  Highly Ionising Particles with the ATLAS Detector at the LHC}},  {\em
  Phys.Lett.} {\bf B698} (2011) 353--370,
  [\href{http://arxiv.org/abs/1102.0459}{{\tt arXiv:1102.0459}}].

\bibitem{Wyler:1982dd}
D.~Wyler and L.~Wolfenstein, {\it {Massless Neutrinos in Left-Right Symmetric
  Models}},  {\em Nucl. Phys.} {\bf B218} (1983) 205.

\bibitem{Mohapatra:1986bd}
R.~N. Mohapatra and J.~W.~F. Valle, {\it {Neutrino Mass and Baryon Number
  Nonconservation in Superstring Models}},  {\em Phys. Rev.} {\bf D34} (1986)
  1642.

\bibitem{GonzalezGarcia:1988rw}
M.~C. Gonzalez-Garcia and J.~W.~F. Valle, {\it {Fast Decaying Neutrinos and
  Observable Flavor Violation in a New Class of Majoron Models}},  {\em Phys.
  Lett.} {\bf B216} (1989) 360.

\bibitem{Law:2013gma}
S.~S.~C. Law and K.~L. McDonald, {\it {Generalized inverse seesaw mechanisms}},
   {\em Phys. Rev.} {\bf D87} (2013), no.~11 113003,
  [\href{http://arxiv.org/abs/1303.4887}{{\tt arXiv:1303.4887}}].

\bibitem{deGouvea:2007uz}
A.~de~Gouvea, {\it {GeV seesaw, accidentally small neutrino masses, and Higgs
  decays to neutrinos}},  \href{http://arxiv.org/abs/0706.1732}{{\tt
  arXiv:0706.1732}}.

\bibitem{Chen:2010wn}
J.-H. Chen, X.-G. He, J.~Tandean, and L.-H. Tsai, {\it {Effect on Higgs Boson
  Decays from Large Light-Heavy Neutrino Mixing in Seesaw Models}},  {\em
  Phys.Rev.} {\bf D81} (2010) 113004,
  [\href{http://arxiv.org/abs/1001.5215}{{\tt arXiv:1001.5215}}].

\bibitem{BhupalDev:2012zg}
P.~Bhupal~Dev, R.~Franceschini, and R.~Mohapatra, {\it {Bounds on TeV Seesaw
  Models from LHC Higgs Data}},  {\em Phys.Rev.} {\bf D86} (2012) 093010,
  [\href{http://arxiv.org/abs/1207.2756}{{\tt arXiv:1207.2756}}].

\bibitem{Cely:2012bz}
C.~G. Cely, A.~Ibarra, E.~Molinaro, and S.~Petcov, {\it {Higgs Decays in the
  Low Scale Type I See-Saw Model}},  {\em Phys.Lett.} {\bf B718} (2013)
  957--964, [\href{http://arxiv.org/abs/1208.3654}{{\tt arXiv:1208.3654}}].

\bibitem{Antusch:2015mia}
S.~Antusch and O.~Fischer, {\it {Testing sterile neutrino extensions of the
  Standard Model at future lepton colliders}},  {\em JHEP} {\bf 05} (2015) 053,
  [\href{http://arxiv.org/abs/1502.05915}{{\tt arXiv:1502.05915}}].

\bibitem{ATLAS-CONF-2012-139}
{\bf ATLAS Collaboration} Collaboration, {\it {Search for Majorana neutrino
  production in pp collisions at sqrt(s)=7 TeV in dimuon final states with the
  ATLAS detector}},  Tech. Rep. ATLAS-CONF-2012-139, CERN, Geneva, Sep, 2012.

\bibitem{Aad:2015xaa}
{\bf ATLAS} Collaboration, G.~Aad et~al., {\it {Search for heavy Majorana
  neutrinos with the ATLAS detector in pp collisions at $\sqrt{s} = 8$ TeV}},
  \href{http://arxiv.org/abs/1506.06020}{{\tt arXiv:1506.06020}}.

\bibitem{Chatrchyan:2012fla}
{\bf CMS Collaboration} Collaboration, S.~Chatrchyan et~al., {\it {Search for
  heavy Majorana neutrinos in $\mu^+\mu^+[\mu^-\mu^-]$ and $e^+e^+[e^-e^-]$
  events in $pp$ collisions at $\sqrt{s} = 7$ TeV}},  {\em Phys.Lett.} {\bf
  B717} (2012) 109--128, [\href{http://arxiv.org/abs/1207.6079}{{\tt
  arXiv:1207.6079}}].

\bibitem{Khachatryan:2015gha}
{\bf CMS Collaboration} Collaboration, V.~Khachatryan et~al., {\it {Search for
  heavy Majorana neutrinos in $\mu^\pm \mu^\pm$+jets events in proton-proton
  collisions at $\sqrt{s}$ = 8 TeV}},
  \href{http://arxiv.org/abs/1501.05566}{{\tt arXiv:1501.05566}}.

\bibitem{Abada:2008ea}
A.~Abada, C.~Biggio, F.~Bonnet, M.~Gavela, and T.~Hambye, {\it {mu e gamma and
  tau l gamma decays in the fermion triplet seesaw model}},  {\em Phys.Rev.}
  {\bf D78} (2008) 033007, [\href{http://arxiv.org/abs/0803.0481}{{\tt
  arXiv:0803.0481}}].

\bibitem{Abada:2007ux}
A.~Abada, C.~Biggio, F.~Bonnet, M.~Gavela, and T.~Hambye, {\it {Low energy
  effects of neutrino masses}},  {\em JHEP} {\bf 0712} (2007) 061,
  [\href{http://arxiv.org/abs/0707.4058}{{\tt arXiv:0707.4058}}].

\bibitem{Franceschini:2008pz}
R.~Franceschini, T.~Hambye, and A.~Strumia, {\it {Type-III see-saw at LHC}},
  {\em Phys.Rev.} {\bf D78} (2008) 033002,
  [\href{http://arxiv.org/abs/0805.1613}{{\tt arXiv:0805.1613}}].

\bibitem{Achard:2001qw}
{\bf L3 Collaboration} Collaboration, P.~Achard et~al., {\it {Search for heavy
  neutral and charged leptons in $e^{+} e^{-}$ annihilation at LEP}},  {\em
  Phys.Lett.} {\bf B517} (2001) 75--85,
  [\href{http://arxiv.org/abs/hep-ex/0107015}{{\tt hep-ex/0107015}}].

\bibitem{CMS:2012ra}
{\bf CMS Collaboration} Collaboration, S.~Chatrchyan et~al., {\it {Search for
  heavy lepton partners of neutrinos in proton-proton collisions in the context
  of the type III seesaw mechanism}},  {\em Phys.Lett.} {\bf B718} (2012)
  348--368, [\href{http://arxiv.org/abs/1210.1797}{{\tt arXiv:1210.1797}}].

\bibitem{Aad:2015cxa}
{\bf ATLAS} Collaboration, G.~Aad et~al., {\it {Search for type-III Seesaw
  heavy leptons in $pp$ collisions at $\sqrt{s}= 8$ TeV with the ATLAS
  Detector}},  \href{http://arxiv.org/abs/1506.01839}{{\tt arXiv:1506.01839}}.

\bibitem{Falkowski:2013jya}
A.~Falkowski, D.~M. Straub, and A.~Vicente, {\it {Vector-like leptons: Higgs
  decays and collider phenomenology}},  {\em JHEP} {\bf 1405} (2014) 092,
  [\href{http://arxiv.org/abs/1312.5329}{{\tt arXiv:1312.5329}}].

\bibitem{Altmannshofer:2013zba}
W.~Altmannshofer, M.~Bauer, and M.~Carena, {\it {Exotic Leptons: Higgs, Flavor
  and Collider Phenomenology}},  {\em JHEP} {\bf 1401} (2014) 060,
  [\href{http://arxiv.org/abs/1308.1987}{{\tt arXiv:1308.1987}}].

\bibitem{Ma:2014zda}
T.~Ma, B.~Zhang, and G.~Cacciapaglia, {\it {Doubly Charged Lepton from an
  Exotic Doublet at the LHC}},  {\em Phys. Rev.} {\bf D89} (2014), no.~9
  093022, [\href{http://arxiv.org/abs/1404.2375}{{\tt arXiv:1404.2375}}].

\bibitem{Delgado:2011iz}
A.~Delgado, C.~Garcia~Cely, T.~Han, and Z.~Wang, {\it {Phenomenology of a
  lepton triplet}},  {\em Phys. Rev.} {\bf D84} (2011) 073007,
  [\href{http://arxiv.org/abs/1105.5417}{{\tt arXiv:1105.5417}}].

\bibitem{Ma:2013tda}
T.~Ma, B.~Zhang, and G.~Cacciapaglia, {\it {Triplet with a doubly-charged
  lepton at the LHC}},  {\em Phys. Rev.} {\bf D89} (2014), no.~1 015020,
  [\href{http://arxiv.org/abs/1309.7396}{{\tt arXiv:1309.7396}}].

\bibitem{Aad:2015vsa}
{\bf ATLAS} Collaboration, G.~Aad et~al., {\it {Evidence for the Higgs-boson
  Yukawa coupling to tau leptons with the ATLAS detector}},  {\em JHEP} {\bf
  04} (2015) 117, [\href{http://arxiv.org/abs/1501.04943}{{\tt
  arXiv:1501.04943}}].

\bibitem{Chatrchyan:2014nva}
{\bf CMS Collaboration} Collaboration, S.~Chatrchyan et~al., {\it {Evidence for
  the 125 GeV Higgs boson decaying to a pair of $\tau$ leptons}},
  \href{http://arxiv.org/abs/1401.5041}{{\tt arXiv:1401.5041}}.

\bibitem{Aad:2015gha}
{\bf ATLAS} Collaboration, G.~Aad et~al., {\it {Search for
  lepton-flavour-violating $H\to\mu\tau$ decays of the Higgs boson with the
  ATLAS detector}},  \href{http://arxiv.org/abs/1508.03372}{{\tt
  arXiv:1508.03372}}.

\bibitem{Khachatryan:2015kon}
{\bf CMS} Collaboration, V.~Khachatryan et~al., {\it {Search for
  lepton-flavour-violating decays of the Higgs boson}},
  \href{http://arxiv.org/abs/1502.07400}{{\tt arXiv:1502.07400}}.

\bibitem{Aad:2014xva}
{\bf ATLAS Collaboration} Collaboration, G.~Aad et~al., {\it {Search for the
  Standard Model Higgs boson decay to $\mu^{+}\mu^{-}$ with the ATLAS
  detector}},  \href{http://arxiv.org/abs/1406.7663}{{\tt arXiv:1406.7663}}.

\bibitem{Khachatryan:2014aep}
{\bf CMS} Collaboration, V.~Khachatryan et~al., {\it {Search for a standard
  model-like Higgs boson in the $\mu^+$ $\mu^{−}$ and $e^+$ $e^{−}$ decay
  channels at the LHC}},  {\em Phys. Lett.} {\bf B744} (2015) 184--207,
  [\href{http://arxiv.org/abs/1410.6679}{{\tt arXiv:1410.6679}}].

\bibitem{Agashe:2006at}
K.~Agashe, R.~Contino, L.~Da~Rold, and A.~Pomarol, {\it {A Custodial symmetry
  for Zb anti-b}},  {\em Phys.Lett.} {\bf B641} (2006) 62--66,
  [\href{http://arxiv.org/abs/hep-ph/0605341}{{\tt hep-ph/0605341}}].

\bibitem{delAguila:2010es}
F.~del Aguila, A.~Carmona, and J.~Santiago, {\it {Tau Custodian searches at the
  LHC}},  {\em Phys.Lett.} {\bf B695} (2011) 449--453,
  [\href{http://arxiv.org/abs/1007.4206}{{\tt arXiv:1007.4206}}].

\bibitem{Carmona:2013cq}
A.~Carmona and F.~Goertz, {\it {Custodial Leptons and Higgs Decays}},  {\em
  JHEP} {\bf 1304} (2013) 163, [\href{http://arxiv.org/abs/1301.5856}{{\tt
  arXiv:1301.5856}}].

\bibitem{Almeida:2012bq}
L.~G. Almeida, E.~Bertuzzo, P.~A. Machado, and R.~Z. Funchal, {\it {Does $H \to
  \gamma \gamma$ Taste like vanilla New Physics?}},  {\em JHEP} {\bf 1211}
  (2012) 085, [\href{http://arxiv.org/abs/1207.5254}{{\tt arXiv:1207.5254}}].

\bibitem{Kearney:2012zi}
J.~Kearney, A.~Pierce, and N.~Weiner, {\it {Vectorlike Fermions and Higgs
  Couplings}},  {\em Phys. Rev.} {\bf D86} (2012) 113005,
  [\href{http://arxiv.org/abs/1207.7062}{{\tt arXiv:1207.7062}}].

\bibitem{Joglekar:2012vc}
A.~Joglekar, P.~Schwaller, and C.~E.~M. Wagner, {\it {Dark Matter and Enhanced
  Higgs to Di-photon Rate from Vector-like Leptons}},  {\em JHEP} {\bf 12}
  (2012) 064, [\href{http://arxiv.org/abs/1207.4235}{{\tt arXiv:1207.4235}}].

\bibitem{Redi:2013pga}
M.~Redi, {\it {Leptons in Composite MFV}},  {\em JHEP} {\bf 1309} (2013) 060,
  [\href{http://arxiv.org/abs/1306.1525}{{\tt arXiv:1306.1525}}].

\bibitem{Martin:1997ns}
S.~P. Martin, {\it {A Supersymmetry primer}},  {\em Adv.Ser.Direct.High Energy
  Phys.} {\bf 21} (2010) 1--153,
  [\href{http://arxiv.org/abs/hep-ph/9709356}{{\tt hep-ph/9709356}}].

\bibitem{Dreiner:2012ex}
H.~K. Dreiner, J.~S. Kim, and O.~Lebedev, {\it {First LHC Constraints on
  Neutralinos}},  {\em Phys. Lett.} {\bf B715} (2012) 199--202,
  [\href{http://arxiv.org/abs/1206.3096}{{\tt arXiv:1206.3096}}].

\bibitem{Ananthanarayan:2013fga}
B.~Ananthanarayan, J.~Lahiri, P.~Pandita, and M.~Patra, {\it {Invisible decays
  of the lightest Higgs boson in supersymmetric models}},  {\em Phys.Rev.} {\bf
  D87} (2013), no.~11 115021, [\href{http://arxiv.org/abs/1306.1291}{{\tt
  arXiv:1306.1291}}].

\bibitem{Casas:2013pta}
J.~A. Casas, J.~M. Moreno, K.~Rolbiecki, and B.~Zaldivar, {\it {Implications of
  light charginos for Higgs observables, LHC searches and dark matter}},  {\em
  JHEP} {\bf 1309} (2013) 099, [\href{http://arxiv.org/abs/1305.3274}{{\tt
  arXiv:1305.3274}}].

\bibitem{Batell:2013bka}
B.~Batell, S.~Jung, and C.~E. Wagner, {\it {Very Light Charginos and Higgs
  Decays}},  {\em JHEP} {\bf 1312} (2013) 075,
  [\href{http://arxiv.org/abs/1309.2297}{{\tt arXiv:1309.2297}}].

\bibitem{delAguila:2000rc}
F.~del Aguila, M.~Perez-Victoria, and J.~Santiago, {\it {Observable
  contributions of new exotic quarks to quark mixing}},  {\em JHEP} {\bf 09}
  (2000) 011, [\href{http://arxiv.org/abs/hep-ph/0007316}{{\tt
  hep-ph/0007316}}].

\bibitem{AguilarSaavedra:2009es}
J.~Aguilar-Saavedra, {\it {Identifying top partners at LHC}},  {\em JHEP} {\bf
  0911} (2009) 030, [\href{http://arxiv.org/abs/0907.3155}{{\tt
  arXiv:0907.3155}}].

\bibitem{Cacciapaglia:2010vn}
G.~Cacciapaglia, A.~Deandrea, D.~Harada, and Y.~Okada, {\it {Bounds and Decays
  of New Heavy Vector-like Top Partners}},  {\em JHEP} {\bf 1011} (2010) 159,
  [\href{http://arxiv.org/abs/1007.2933}{{\tt arXiv:1007.2933}}].

\bibitem{Gopalakrishna:2011ef}
S.~Gopalakrishna, T.~Mandal, S.~Mitra, and R.~Tibrewala, {\it {LHC Signatures
  of a Vector-like b'}},  {\em Phys. Rev.} {\bf D84} (2011) 055001,
  [\href{http://arxiv.org/abs/1107.4306}{{\tt arXiv:1107.4306}}].

\bibitem{Okada:2012gy}
Y.~Okada and L.~Panizzi, {\it {LHC signatures of vector-like quarks}},  {\em
  Adv.High Energy Phys.} {\bf 2013} (2013) 364936,
  [\href{http://arxiv.org/abs/1207.5607}{{\tt arXiv:1207.5607}}].

\bibitem{Aguilar-Saavedra:2013qpa}
J.~Aguilar-Saavedra, R.~Benbrik, S.~Heinemeyer, and M.~Pérez-Victoria, {\it
  {Handbook of vectorlike quarks: Mixing and single production}},  {\em
  Phys.Rev.} {\bf D88} (2013), no.~9 094010,
  [\href{http://arxiv.org/abs/1306.0572}{{\tt arXiv:1306.0572}}].

\bibitem{Gopalakrishna:2013hua}
S.~Gopalakrishna, T.~Mandal, S.~Mitra, and G.~Moreau, {\it {LHC Signatures of
  Warped-space Vectorlike Quarks}},  {\em JHEP} {\bf 08} (2014) 079,
  [\href{http://arxiv.org/abs/1306.2656}{{\tt arXiv:1306.2656}}].

\bibitem{ATLAS-CONF-2015-012}
T.~A. collaboration, {\it {Search for production of vector-like quark pairs and
  of four top quarks in the lepton plus jets final state in $pp$ collisions at
  $\sqrt{s}=8$ TeV with the ATLAS detector}}, .

\bibitem{Khachatryan:2015axa}
{\bf CMS} Collaboration, V.~Khachatryan et~al., {\it {Search for vector-like T
  quarks decaying to top quarks and Higgs bosons in the all-hadronic channel
  using jet substructure}},  {\em JHEP} {\bf 06} (2015) 080,
  [\href{http://arxiv.org/abs/1503.01952}{{\tt arXiv:1503.01952}}].

\bibitem{Aad:2015mba}
{\bf ATLAS} Collaboration, G.~Aad et~al., {\it {Search for vector-like $B$
  quarks in events with one isolated lepton, missing transverse momentum and
  jets at $\sqrt{s}=$ 8 TeV with the ATLAS detector}},
  \href{http://arxiv.org/abs/1503.05425}{{\tt arXiv:1503.05425}}.

\bibitem{Aad:2014efa}
{\bf ATLAS Collaboration} Collaboration, G.~Aad et~al., {\it {Search for pair
  and single production of new heavy quarks that decay to a $Z$ boson and a
  third-generation quark in $pp$ collisions at $\sqrt{s}=8$ TeV with the ATLAS
  detector}},  \href{http://arxiv.org/abs/1409.5500}{{\tt arXiv:1409.5500}}.

\bibitem{CMS:2014dka}
{\bf CMS} Collaboration, C.~Collaboration, {\it {Search for vector-like quarks
  in final states with a single lepton and jets in pp collisions at sqrt s = 8
  TeV}}, .

\bibitem{CMS:2014bfa}
{\bf CMS} Collaboration, C.~Collaboration, {\it {Search for pair-produced
  vector-like quarks of charge -1/3 decaying to bH using boosted Higgs
  jet-tagging in pp collisions at sqrt(s) = 8 TeV}}, .

\bibitem{CMS:2012hfa}
{\bf CMS} Collaboration, C.~Collaboration, {\it {Search for pair-produced
  vector-like quarks of charge -1/3 in lepton+jets final state in pp collisions
  at sqrt(s) = 8 TeV}}, .

\bibitem{Khachatryan:2015oba}
{\bf CMS} Collaboration, V.~Khachatryan et~al., {\it {Search for vector-like
  charge 2/3 T quarks in proton-proton collisions at sqrt(s) = 8 TeV}},
  \href{http://arxiv.org/abs/1509.04177}{{\tt arXiv:1509.04177}}.

\bibitem{Contino:2006qr}
R.~Contino, L.~Da~Rold, and A.~Pomarol, {\it {Light custodians in natural
  composite Higgs models}},  {\em Phys.Rev.} {\bf D75} (2007) 055014,
  [\href{http://arxiv.org/abs/hep-ph/0612048}{{\tt hep-ph/0612048}}].

\bibitem{Pomarol:2008bh}
A.~Pomarol and J.~Serra, {\it {Top Quark Compositeness: Feasibility and
  Implications}},  {\em Phys. Rev.} {\bf D78} (2008) 074026,
  [\href{http://arxiv.org/abs/0806.3247}{{\tt arXiv:0806.3247}}].

\bibitem{Bertuzzo:2012bt}
E.~Bertuzzo, P.~A. Machado, and R.~Zukanovich~Funchal, {\it {Can New Colored
  Particles Illuminate the Higgs?}},  {\em JHEP} {\bf 1302} (2013) 086,
  [\href{http://arxiv.org/abs/1209.6359}{{\tt arXiv:1209.6359}}].

\bibitem{Matsedonskyi:2014lla}
O.~Matsedonskyi, F.~Riva, and T.~Vantalon, {\it {Composite Charge 8/3
  Resonances at the LHC}},  {\em JHEP} {\bf 04} (2014) 059,
  [\href{http://arxiv.org/abs/1401.3740}{{\tt arXiv:1401.3740}}].

\bibitem{Cacciapaglia:2015ixa}
G.~Cacciapaglia, A.~Deandrea, N.~Gaur, D.~Harada, Y.~Okada, et~al., {\it
  {Interplay of vector-like top partner multiplets in a realistic mixing
  set-up}},  \href{http://arxiv.org/abs/1502.00370}{{\tt arXiv:1502.00370}}.

\bibitem{Gillioz:2008hs}
M.~Gillioz, {\it {A Light composite Higgs boson facing electroweak precision
  tests}},  {\em Phys. Rev.} {\bf D80} (2009) 055003,
  [\href{http://arxiv.org/abs/0806.3450}{{\tt arXiv:0806.3450}}].

\bibitem{Azatov:2011qy}
A.~Azatov and J.~Galloway, {\it {Light Custodians and Higgs Physics in
  Composite Models}},  {\em Phys. Rev.} {\bf D85} (2012) 055013,
  [\href{http://arxiv.org/abs/1110.5646}{{\tt arXiv:1110.5646}}].

\bibitem{Elias-Miro:2013gya}
J.~Elias-Miró, J.~Espinosa, E.~Masso, and A.~Pomarol, {\it {Renormalization of
  dimension-six operators relevant for the Higgs decays $h\rightarrow
  \gamma\gamma,\gamma Z$}},  {\em JHEP} {\bf 1308} (2013) 033,
  [\href{http://arxiv.org/abs/1302.5661}{{\tt arXiv:1302.5661}}].

\bibitem{Peskin:1990zt}
M.~E. Peskin and T.~Takeuchi, {\it {A New constraint on a strongly interacting
  Higgs sector}},  {\em Phys.Rev.Lett.} {\bf 65} (1990) 964--967.

\bibitem{Altarelli:1990zd}
G.~Altarelli and R.~Barbieri, {\it {Vacuum polarization effects of new physics
  on electroweak processes}},  {\em Phys.Lett.} {\bf B253} (1991) 161--167.

\bibitem{Peskin:1991sw}
M.~E. Peskin and T.~Takeuchi, {\it {Estimation of oblique electroweak
  corrections}},  {\em Phys.Rev.} {\bf D46} (1992) 381--409.

\bibitem{Barbieri:2004qk}
R.~Barbieri, A.~Pomarol, R.~Rattazzi, and A.~Strumia, {\it {Electroweak
  symmetry breaking after LEP-1 and LEP-2}},  {\em Nucl.Phys.} {\bf B703}
  (2004) 127--146, [\href{http://arxiv.org/abs/hep-ph/0405040}{{\tt
  hep-ph/0405040}}].

\bibitem{Baak:2014ora}
{\bf Gfitter Group} Collaboration, M.~Baak et~al., {\it {The global electroweak
  fit at NNLO and prospects for the LHC and ILC}},  {\em Eur.Phys.J.} {\bf C74}
  (2014) 3046, [\href{http://arxiv.org/abs/1407.3792}{{\tt arXiv:1407.3792}}].

\bibitem{Ciuchini:2013pca}
M.~Ciuchini, E.~Franco, S.~Mishima, and L.~Silvestrini, {\it {Electroweak
  Precision Observables, New Physics and the Nature of a 126 GeV Higgs Boson}},
   {\em JHEP} {\bf 1308} (2013) 106,
  [\href{http://arxiv.org/abs/1306.4644}{{\tt arXiv:1306.4644}}].

\bibitem{Gori:2015nqa}
S.~Gori, J.~Gu, and L.-T. Wang, {\it {The Zbb Couplings at Future e+e-
  Colliders}},  \href{http://arxiv.org/abs/1508.07010}{{\tt arXiv:1508.07010}}.

\bibitem{Bamert:1996px}
P.~Bamert, C.~Burgess, J.~M. Cline, D.~London, and E.~Nardi, {\it {R($b$) and
  new physics: A Comprehensive analysis}},  {\em Phys.Rev.} {\bf D54} (1996)
  4275--4300, [\href{http://arxiv.org/abs/hep-ph/9602438}{{\tt
  hep-ph/9602438}}].

\bibitem{Ellis:1975ap}
J.~R. Ellis, M.~K. Gaillard, and D.~V. Nanopoulos, {\it {A Phenomenological
  Profile of the Higgs Boson}},  {\em Nucl.Phys.} {\bf B106} (1976) 292.

\bibitem{Shifman:1979eb}
M.~A. Shifman, A.~Vainshtein, M.~Voloshin, and V.~I. Zakharov, {\it {Low-Energy
  Theorems for Higgs Boson Couplings to Photons}},  {\em Sov.J.Nucl.Phys.} {\bf
  30} (1979) 711--716.

\bibitem{Kniehl:1995tn}
B.~A. Kniehl and M.~Spira, {\it {Low-energy theorems in Higgs physics}},  {\em
  Z.Phys.} {\bf C69} (1995) 77--88,
  [\href{http://arxiv.org/abs/hep-ph/9505225}{{\tt hep-ph/9505225}}].

\bibitem{Carena:2012xa}
M.~Carena, I.~Low, and C.~E.~M. Wagner, {\it {Implications of a Modified Higgs
  to Diphoton Decay Width}},  {\em JHEP} {\bf 08} (2012) 060,
  [\href{http://arxiv.org/abs/1206.1082}{{\tt arXiv:1206.1082}}].

\bibitem{Gillioz:2012se}
M.~Gillioz, R.~Grober, C.~Grojean, M.~Muhlleitner, and E.~Salvioni, {\it {Higgs
  Low-Energy Theorem (and its corrections) in Composite Models}},  {\em JHEP}
  {\bf 1210} (2012) 004, [\href{http://arxiv.org/abs/1206.7120}{{\tt
  arXiv:1206.7120}}].

\bibitem{Voloshin:2012tv}
M.~B. Voloshin, {\it {CP Violation in Higgs Diphoton Decay in Models with
  Vectorlike Heavy Fermions}},  {\em Phys. Rev.} {\bf D86} (2012) 093016,
  [\href{http://arxiv.org/abs/1208.4303}{{\tt arXiv:1208.4303}}].

\bibitem{Gavela:1981ri}
M.~Gavela, G.~Girardi, C.~Malleville, and P.~Sorba, {\it {A Nonlinear R(xi)
  Gauge Condition for the Electroweak SU(2) X U(1) Model}},  {\em Nucl.Phys.}
  {\bf B193} (1981) 257.

\bibitem{Cahn:1978nz}
R.~Cahn, M.~S. Chanowitz, and N.~Fleishon, {\it {Higgs Particle Production by Z
  H Gamma}},  {\em Phys.Lett.} {\bf B82} (1979) 113.

\bibitem{Bergstrom:1985hp}
L.~Bergstrom and G.~Hulth, {\it {Induced Higgs Couplings to Neutral Bosons in
  $e^+ e^-$ Collisions}},  {\em Nucl.Phys.} {\bf B259} (1985) 137.

\bibitem{Passarino:1978jh}
G.~Passarino and M.~Veltman, {\it {One Loop Corrections for e+ e- Annihilation
  Into mu+ mu- in the Weinberg Model}},  {\em Nucl.Phys.} {\bf B160} (1979)
  151.

\bibitem{Bardin:1999ak}
D.~Y. Bardin and G.~Passarino, {\it {The standard model in the making:
  Precision study of the electroweak interactions}}, .

\bibitem{Denner:1991kt}
A.~Denner, {\it {Techniques for calculation of electroweak radiative
  corrections at the one loop level and results for W physics at LEP-200}},
  {\em Fortsch.Phys.} {\bf 41} (1993) 307--420,
  [\href{http://arxiv.org/abs/0709.1075}{{\tt arXiv:0709.1075}}].

\bibitem{Djouadi:1996ws}
A.~Djouadi, V.~Driesen, W.~Hollik, and J.~Rosiek, {\it {Associated production
  of Higgs bosons and a photon in high-energy e+ e- collisions}},  {\em
  Nucl.Phys.} {\bf B491} (1997) 68--102,
  [\href{http://arxiv.org/abs/hep-ph/9609420}{{\tt hep-ph/9609420}}].

\bibitem{Djouadi:1996yq}
A.~Djouadi, V.~Driesen, W.~Hollik, and A.~Kraft, {\it {The Higgs photon - Z
  boson coupling revisited}},  {\em Eur.Phys.J.} {\bf C1} (1998) 163--175,
  [\href{http://arxiv.org/abs/hep-ph/9701342}{{\tt hep-ph/9701342}}].

\bibitem{Cao:2013ur}
J.~Cao, L.~Wu, P.~Wu, and J.~M. Yang, {\it {The Z+photon and diphoton decays of
  the Higgs boson as a joint probe of low energy SUSY models}},  {\em JHEP}
  {\bf 1309} (2013) 043, [\href{http://arxiv.org/abs/1301.4641}{{\tt
  arXiv:1301.4641}}].

\bibitem{ATL-PHYS-PUB-2014-016}
{\it {Projections for measurements of Higgs boson signal strengths and coupling
  parameters with the ATLAS detector at a HL-LHC}},  Tech. Rep.
  ATL-PHYS-PUB-2014-016, CERN, Geneva, Oct, 2014.

\bibitem{CMS:2013xfa}
{\bf CMS} Collaboration, {\it {Projected Performance of an Upgraded CMS
  Detector at the LHC and HL-LHC: Contribution to the Snowmass Process}},
  \href{http://arxiv.org/abs/1307.7135}{{\tt arXiv:1307.7135}}.

\bibitem{Aad:2014fia}
{\bf ATLAS Collaboration} Collaboration, G.~Aad et~al., {\it {Search for Higgs
  boson decays to a photon and a Z boson in pp collisions at $\sqrt{s}$=7 and 8
  TeV with the ATLAS detector}},  {\em Phys.Lett.} {\bf B732} (2014) 8--27,
  [\href{http://arxiv.org/abs/1402.3051}{{\tt arXiv:1402.3051}}].

\bibitem{Chatrchyan:2013vaa}
{\bf CMS Collaboration} Collaboration, S.~Chatrchyan et~al., {\it {Search for a
  Higgs boson decaying into a Z and a photon in pp collisions at $\sqrt{s}$ = 7
  and 8 TeV}},  {\em Phys.Lett.} {\bf B726} (2013) 587--609,
  [\href{http://arxiv.org/abs/1307.5515}{{\tt arXiv:1307.5515}}].

\bibitem{ATLAS-CONF-2015-004}
{\it {Search for an Invisibly Decaying Higgs Boson Produced via Vector Boson
  Fusion in $pp$ Collisions at $\sqrt{s}=8$ TeV using the ATLAS Detector at the
  LHC}},  Tech. Rep. ATLAS-CONF-2015-004, CERN, Geneva, Mar, 2015.

\bibitem{Chatrchyan:2014tja}
{\bf CMS} Collaboration, S.~Chatrchyan et~al., {\it {Search for invisible
  decays of Higgs bosons in the vector boson fusion and associated ZH
  production modes}},  {\em Eur.Phys.J.} {\bf C74} (2014) 2980,
  [\href{http://arxiv.org/abs/1404.1344}{{\tt arXiv:1404.1344}}].

\bibitem{Espinosa:2012im}
J.~R. Espinosa, C.~Grojean, M.~Muhlleitner, and M.~Trott, {\it {First Glimpses
  at Higgs' face}},  {\em JHEP} {\bf 12} (2012) 045,
  [\href{http://arxiv.org/abs/1207.1717}{{\tt arXiv:1207.1717}}].

\bibitem{Carmi:2012in}
D.~Carmi, A.~Falkowski, E.~Kuflik, T.~Volansky, and J.~Zupan, {\it {Higgs After
  the Discovery: A Status Report}},  {\em JHEP} {\bf 10} (2012) 196,
  [\href{http://arxiv.org/abs/1207.1718}{{\tt arXiv:1207.1718}}].

\bibitem{Cacciapaglia:2012wb}
G.~Cacciapaglia, A.~Deandrea, G.~D. La~Rochelle, and J.-B. Flament, {\it {Higgs
  couplings beyond the Standard Model}},  {\em JHEP} {\bf 03} (2013) 029,
  [\href{http://arxiv.org/abs/1210.8120}{{\tt arXiv:1210.8120}}].

\bibitem{Azatov:2012qz}
A.~Azatov and J.~Galloway, {\it {Electroweak Symmetry Breaking and the Higgs
  Boson: Confronting Theories at Colliders}},  {\em Int. J. Mod. Phys.} {\bf
  A28} (2013) 1330004, [\href{http://arxiv.org/abs/1212.1380}{{\tt
  arXiv:1212.1380}}].

\bibitem{Falkowski:2013dza}
A.~Falkowski, F.~Riva, and A.~Urbano, {\it {Higgs at last}},  {\em JHEP} {\bf
  11} (2013) 111, [\href{http://arxiv.org/abs/1303.1812}{{\tt
  arXiv:1303.1812}}].

\bibitem{Giardino:2013bma}
P.~P. Giardino, K.~Kannike, I.~Masina, M.~Raidal, and A.~Strumia, {\it {The
  universal Higgs fit}},  {\em JHEP} {\bf 05} (2014) 046,
  [\href{http://arxiv.org/abs/1303.3570}{{\tt arXiv:1303.3570}}].

\bibitem{Djouadi:2013qya}
A.~Djouadi and G.~Moreau, {\it {The couplings of the Higgs boson and its CP
  properties from fits of the signal strengths and their ratios at the 7+8 TeV
  LHC}},  {\em Eur. Phys. J.} {\bf C73} (2013), no.~9 2512,
  [\href{http://arxiv.org/abs/1303.6591}{{\tt arXiv:1303.6591}}].

\bibitem{Belanger:2013xza}
G.~Belanger, B.~Dumont, U.~Ellwanger, J.~F. Gunion, and S.~Kraml, {\it {Global
  fit to Higgs signal strengths and couplings and implications for extended
  Higgs sectors}},  {\em Phys. Rev.} {\bf D88} (2013) 075008,
  [\href{http://arxiv.org/abs/1306.2941}{{\tt arXiv:1306.2941}}].

\bibitem{Moreau:2012da}
G.~Moreau, {\it {Constraining extra-fermion(s) from the Higgs boson data}},
  {\em Phys. Rev.} {\bf D87} (2013), no.~1 015027,
  [\href{http://arxiv.org/abs/1210.3977}{{\tt arXiv:1210.3977}}].

\end{thebibliography}\endgroup

\end{document}